\documentclass[11pt,a4paper, oneside, openright]{book}

\usepackage [left=3cm,bottom=3cm,right=3cm,top=3cm]{geometry}
\usepackage{amsmath} 
\usepackage[utf8x]{inputenc} 
\usepackage{graphicx} 
\usepackage{float} 
\usepackage[footnotesize, bf]{caption} 
\usepackage{pdfpages} 
\usepackage{booktabs}
\usepackage[plainpages=false, pdfpagelabels]{hyperref} 
\hypersetup{colorlinks=true,linkcolor=black,citecolor=black,urlcolor=black,breaklinks}
\usepackage{mathrsfs}
\usepackage{empheq}
\usepackage{multirow}
\usepackage{lscape} 
\usepackage{eurosym}
\usepackage{bookmark}
\usepackage{amssymb}

\makeindex

\newcommand{\virg}{``}

\begin{document}

\begin{titlepage}
 \begin{center}
 \includegraphics[width=1\textwidth]{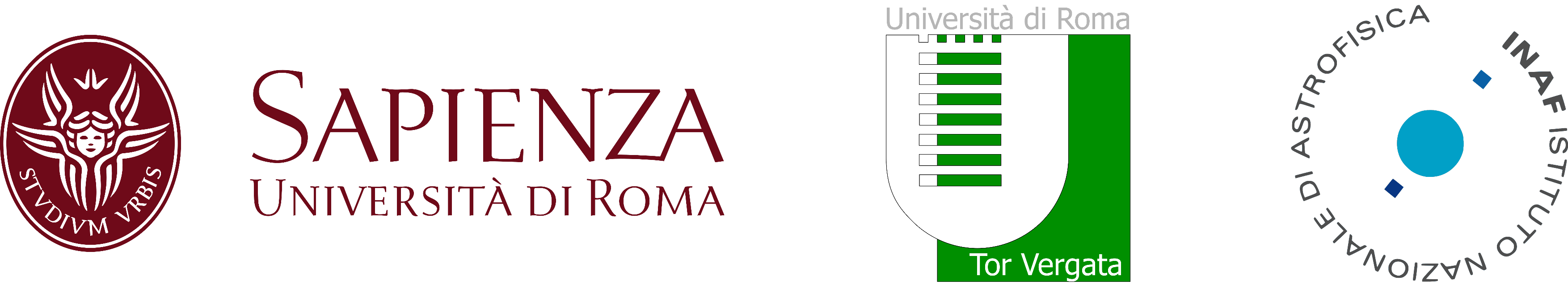}\\
 \vspace{1.5cm}
 \LARGE 
 PH.D IN ASTRONOMY, ASTROPHYSICS\\
 AND SPACE SCIENCE\\  
 \vspace{0.5cm}
 \Large 
 CYCLE XXXI\\
 \vspace{1.5cm}
 \LARGE 
 \textbf{\Large Opening the low-background and high-spectral-resolution}\\
 \textbf{\Large domain with the ATHENA large X-ray observatory:}\\ 
 \textbf{ }\\ 
 \textbf{Development of the Cryogenic AntiCoincidence}\\ 
 \textbf{Detector for the X-ray Integral Field Unit}\\ 
 \textbf{}\\
 \vspace{1.0cm} 
 \Large 
 \textbf{Matteo D'Andrea}\\
 \textbf{A.Y. 2018/2019}\\
 \end{center}
 \vspace{5.0cm}
 \large 
 \textbf{Supervisor: Luigi Piro}\\
 \textbf{Co-supervisor: Claudio Macculi}\\
 \textbf{Coordinator: Paolo De Bernardis}\\
 \textbf{Deputy Coordinator: Pasquale Mazzotta}
 \vfill
\end{titlepage}

\thispagestyle{empty}
\noindent The research activities presented in this thesis have been carried out at the \textit{Cryogenic Laboratory for X-ray Astrophysics} of the \textit{Institute for Space Astrophysics and Planetology} of Roma (IAPS Roma), a department of the \textit{Italian National Institute for Astrophysics} (INAF).

\bigskip

\noindent The work has been partially supported by the University of Roma \virg Tor Vergata'' (XXXI PhD cycle scolarship), by ASI (Italian Space Agency) through the Contract no. 2015-046-R.0, by ESA (European Space Agency) through the Contract no. 4000114932/15/NL/BW, and it has also received funding from the European Union's Horizon 2020 Programme under the AHEAD project (grant agreement n. 654215).

$$ $$

$$ $$

$$ $$

\begin{figure}[H]
\centering
\includegraphics[width=0.6\textwidth]{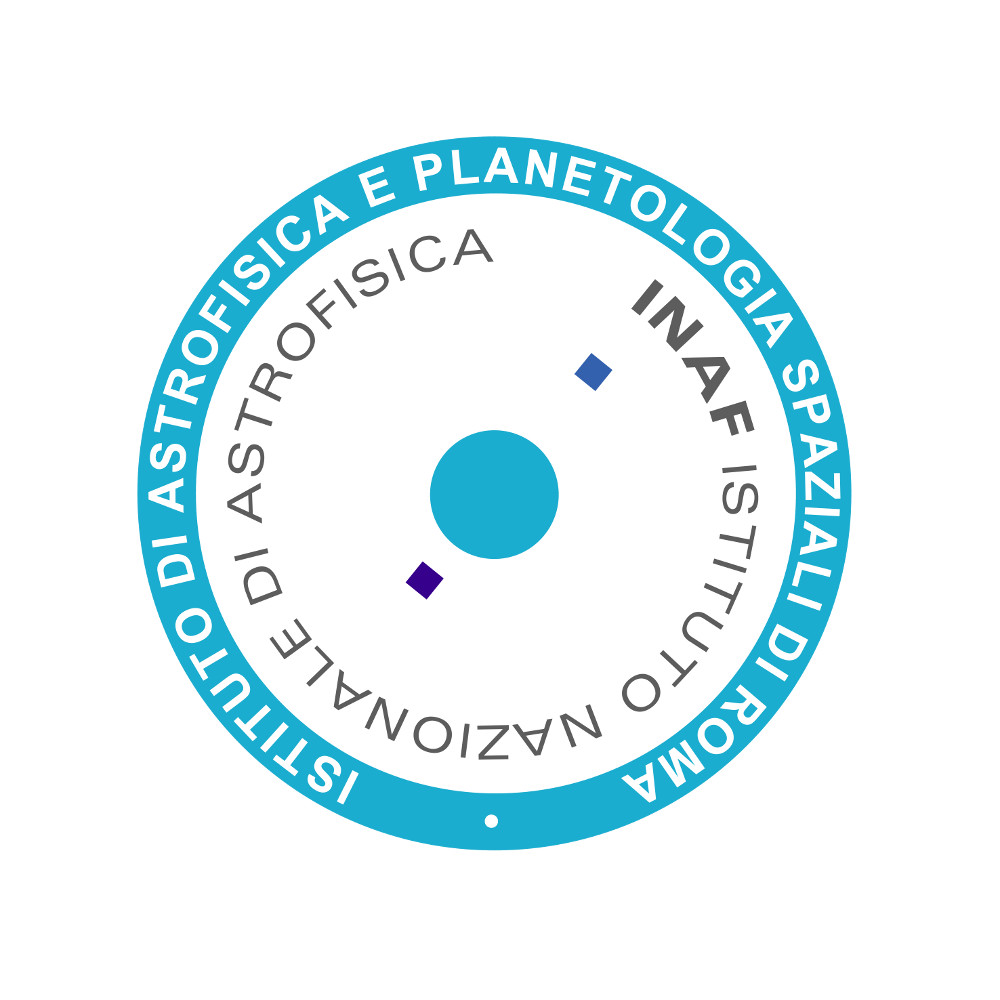}
\end{figure}










\tableofcontents

\chapter*{Introduction}
\addcontentsline{toc}{chapter}{Introduction}
\markboth{INTRODUCTION}{INTRODUCTION}

In the last sixty years, the development of scientific space missions has enabled the observation of the sky in the whole range of the electromagnetic spectrum, overcoming the barrier due to the opacity of the Earth's atmosphere that until then had limited the observations to the visible, near infrared and radio wavelengths. 

The birth of the X-ray Astronomy, traditionally dated 1962 with the discovery of the first cosmic X-ray source (Scorpius X-1), has in particular opened the study of the most hot environments and the most energetic phenomena in the Universe, revolutioning our understanding of the Cosmos. Typical mechanisms responsible for the production of cosmic X-rays are the thermal emission from hot plasma (at the temperature of tens of millions degree), and the accretion of matter onto compact objects, i.e. Neutron Stars and Black Holes. In addition, X-rays are a powerful tool to observe luminous transient phenomena, which are often huge sources of high-energy photons. Among these events, Gamma Ray Bursts (GRB) play a unique role in the study of the distant (old) Universe, being the brightest light sources at all redshift.

Thanks to X-ray observation we know now that galaxies are gathered togheter into larger structures, i.e. groups and clusters, in which most of the baryonic matter is not in form of stars, but in the state of a hot X-ray emitting plasma filling the intracluster medium. We also know that most of the galaxies host in their center a Super-Massive Black Hole (SMBH), whose properties are strongly related to those of the host galaxy. The SMBH growth is probably able to drive the evolution of the whole galaxy, influencing phenomena like the star formation. This is just the tip of the iceberg of the \virg Hot and Energetic Universe'' that has been discovered out there. Despite the outstanding progresses of the last decades, our understanding of the Universe leaves key observational and theoretical issues un-answered. X-ray observations can address some of these pressing question in Astropysics and Cosmology.

Scientific progress can not be separated from a parallel technological advancement in the field of instrumentation. The need to perform more and more accurate observations relies on the possibility to exploit space observatories based on new technologies, with instrumentation able to guarantee an order of magnitude improvement in the performances required to address the specific science questions, including e.g. efficiency, collecting area, spectral and angular resolution, response times and particles background rejection.

\bigskip

The research activities I have carried out during my PhD fit within this framework. My work is indeed placed in the context of the developement of the Advanced Telescope for High Energy Astrophysics (ATHENA). ATHENA is the second large-class mission selected in the scientific programme of the European Space Agency (ESA), with the launch foreseen in 2031 toward the second Lagrangian point of the Earth-Sun system (L2). The mission is designed to address the \virg Hot and Energetic Universe'' science theme, looking for an answer to two key questions in Astrophysics: 1) How does ordinary matter assemble into the large scale structures (galaxy clusters and groups) that we see today?, and 2) How do black holes grow and shape the Universe? Aside from these topics, ATHENA is designed to explore high energy phenomena in all the astrophysical contexts, including the transient and multimessenger Universe, providing a unique contribution to astrophysics in the 2030s.

The mission consists of a large X-ray telescope with unprecedent collecting area (1.4 m$^2$ at 1 keV) and good angular resolution (5 arcsec Half Energy Width), coupled with two complementary focal plane instruments operating in the soft X-ray band: the Wide Field Imager (WFI), and the X-ray Integral Field Unit (X-IFU). The first will offer wide field spectral imaging (40' $\times$ 40' Field of View, 0.2-15 keV energy band) with moderate energy resolution ($\Delta$E$_{FWHM}$ $<$170 eV @ 7 keV), while the second will deliver spatially resolved high-resolution spectroscopy ($\Delta$E$_{FWHM}$ $<$2.5 eV @ 6 keV) over a smaller FoV (5 arcmin equivalent diameter, 0.2-12 keV energy band).

The X-IFU is the most technologically challenging instrument of the payload. It is a cryogenic instrument based on a large array (about 4000 pixels) of Transition Edge Sensor (TES) microcalorimeters, operated at a bath temperature around 50 mK. TES microcalorimeters are cryogenic detectors that converts the individual incident X-ray photons into heat pulses and measure their energy via precise thermometry, using the narrow superconducting-to-normal-state transition of a thin film as an extremely sensitive thermometer. Since these detectors are not able to distinguish among photons and particles that release energy inside them, the X-IFU shall include a particle rejection subsystem. 

The X-IFU performances would be indeed strongly degraded by the particle background expected in the L2 environment, thus advanced reduction techniques are needed to reduce this contribution by a factor $\sim$50 down to the requirement of 0.005 cts cm$^{-2}$ s$^{-1}$ keV$^{-1}$ (between 2 and 10 keV). This is mandatory to enable many core science objectives of the mission, including the study of the hot plasma in galaxy cluster outskirts and the characterization of highly obscured distant Active Galactic Nuclei (AGNs). Most of the background reduction ($\sim 80\%$) is achieved thanks to the Cryogenic AntiCoincidence detector (CryoAC), a 4 pixels TES microcalorimeter which is placed less than 1 mm below the TES array. The CryoAC is a sort of instrument-inside-the-instrument, with independent cold and warm electronics and a dedicated data processing chain. To satisfy the requirements about its rejection efficiency ($> 98 \%$ for primaries) and its intrinsic deadtime ($\leq 1\%$), it shall have a wide energy band (from 20 keV to $\sim$ 750 keV) and achieve strict timing requirements (rise time $< 30$ $\mu$s, effective decay time  $\lesssim 250$ $\mu$s, thermal decay time  $\lesssim 2.5$ ms), while respecting several constraints to ensure mechanical, thermal and electromagnetic compatibility with the TES array. The major technological challenge in the CryoAC development is to obtain a so fast detector despite the large pixel size ($\sim$ 1 cm$^2$ area per pixel). To do this the detector is based on a Silicon absorber, and it is designed to work in a so-called athermal regime, collecting the first out-of-equilibrium family of phonons generated when a particle deposit energy into the crystal.

\bigskip

The INAF/IAPS research group I worked with during my PhD has the Co-PIship of X-IFU, and it is especially in charge of the estimation of the X-IFU background and the development of the Cryogenic Anticoincidence Detector. In this context, my research activity has been mainly focused on the development of the CryoAC Demonstration Model (DM), a single pixel prototype requested by ESA before the mission adoption, and aimed at probing the detector critical technologies (i.e. the operation with a 50 mK thermal bath and the threshold energy at 20 keV, with a representative pixel area of $\sim$ 1 cm$^2$). 

In particular, I have been in charge of the integration and test of the CryoAC prototypes that we have produced to better understand the physics of the detector and to finally fix the DM design. These activities include, beside the proper test procedures, the design of the cryogenic test setup, its installation and optimization, the development of the software for the detector readout, the development of the routines for the data analysis and the final study of the observed detector performances, relevant to the instrument design and the background simulations. I have been then involved in the development of the X-IFU mass model used in the background simulations, and in carrying out studies related to the X-IFU backgroud issue and the CryoAC scientific capabilities.

\bigskip
This thesis is divided in two main parts. In the first one I will mainly present the astrophysical framework of my research, and in the second one I will focus on the experimental activities carried out towards the CryoAC DM development.

\bigskip
The first part of the thesis deals with the key ATHENA science topics and outlines the requirement of a high sensitivity and low-background instrumentation. It is titled \virg Opening the low-background and high-spectral resolution domain with the ATHENA Large X-ray Telescope'', and it is structured as follows.

In the first chapter, I will review the \textit{Hot and Energetic Universe} science theme, then presenting the ATHENA mission concept. Finally, I will show how the background could potentially limit the science achievable by the X-IFU observations without the implementation of dedicated reduction techiniques and instrumentation.

In the second chapter, after a brief introduction about the background issue for X-rays detectors, I will review the latest estimates for the X-IFU background, introducing the reduction techniques that have been adopted in the instrument design. Finally, I will present the CryoAC concept design, showing how the top levels requirements translate into the detector specifications.

In the third chapter, I will present two studies aimed of understanding how the baseline design of the CryoAC could be further improved in order to enhance its performance thus providing additional capabilities. In the first section, it is investigated the possibility of exploiting the CryoAC not only as anticoincidence, but also as an hard X-ray detector, in order to extend the scientific band of the X-IFU up to 20 keV. In the second, it is instead studied the possibility of insert additional vertical CryoAC pixels in the Focal Plane Assembly, in order to form a sort of \virg anticoincidence box'' around the detector and further decrease the residual particle background of the instrument.

\bigskip

The second part of the thesis is titled \virg The X-IFU Cryogenic Anticoincidence Detector: the experimental path towards the Demonstration Model'', and it is structured as follows.

In the fourth chapter, I will provide a basic introduction to the physics underlying the CryoAC detector. First, I will present the standard model describing TES microcalorimeters, and then I will focus on the \virg athermal phonon mediated'' TES detectors (which the CryoAC belongs to), developing their electrothermal model.

In the fifth chapter, I will present the cryogenic setup where the CryoAC prototypes are integrated and tested, introducing the working principles of the employed cryogenic refrigerators, and finally showing the typical measurements that are performed to characterize the CryoAC samples and to verify their performance. 

In the sixth capter, I will report the experimental activity performed with two CryoAC pre-DM prototypes, namely AC-S7 and AC-S8. I will report the main results obtained with them, showing how they helped us to better understand the detector thermal/athermal dynamics and to fix the final DM design.

In the seventh chapter, I will present a non-standard method of pulses processing, based upon Principal Component Analysis (PCA), whose goal is to obtain the optimal energy spectrum from pulses with severe shape variations. I will report the first implementation of this method on CryoAC data, which has been performed to process the pulses acquired by the AC-S7 prototype. Then, I will show the validation of the developed procedures by applying this method on simulated CryoAC signals.

In the eighth chapter, I will report the main results of two activities carried out to improve the cryogenic test setup in preparation to the CryoAC DM integration. The first activity has mainly consisted in the development of a cryogenic magnetic shielding system and in the assessment of its effectiveness by means of FEM simulation and a measurement at
warm. The latter has instead concerned the possibility to add also a signal filtering stage at cold, and the evaluation of its potential influence on the detector readout dynamic. 

In the ninth chapter, I will finally present the CryoAC DM, describing the detector design, its fabrication process and the development of its cryogenic test setup. Then I will present the main results obtained with AC-S9, a pre-DM prototype that we have integrated and tested in preparation to the proper DM development, that will be finally reported showing the preliminary results obtained with the first CryoAC DM prototype (AC-S10).

\bigskip
Finally, in the Conclusions I will summarize the main results obtained.

\part{Opening the low-background and high-spectral-resolution domain with the ATHENA large X-ray observatory}

\chapter{Exploring the Hot and Energetic Universe with ATHENA}
\chaptermark{Exploring the Hot and Energetic Universe}

The astrophysical and cosmological study of the Universe requires the observation of the celestial objects and phenomena over the full range of the electromagnetic spectrum. Since certain wavelenghts do not reach the earth surface due to the atmospheric absorption (i.e. X and Gamma rays and most of the infrared and ultraviolet radiations), it is necessary to develop missions on stratospheric balloons, rockets and satellites in order to observe the sky from the space. In this context, the major space agencies have long-term science programmes dedicated to the development of space missions for the observation of the Universe. These programmes provide the stability needed for activities which could take more than two decades to go from initial concept to the production of the first scientific results.

The current ESA science programme is \virg Cosmic Vision 2015-2025'', which has been designed in the early 2000s to address four main questions:

\begin{itemize}
\item[1)] What are the conditions for planet formation and the emergence of life?
\item[2)] How does the Solar System work?
\item[3)] What are the fundamental physical laws of the Universe?
\item[4)] How did the Universe originate and what is it made of?
\end{itemize}

In March 2013, a Call for White Papers asked to the science community to propose science themes associated to these questions, that could be addressed by the second and the third large class missions of the programme (L2 and L3 missions, about 1 billion \euro $ $ as Cost-at-Completion cap each one).  The theme \textit{The Hot and Energetic Universe} (related to the Cosmic Vision questions 3 and 4) has been so selected for the L2 slot, and a second Call for a large collecting area X-ray observatory addressing this subject finally ended with the selection of the \textit{Advanced Telescope for High-Energy Astrophysics} (ATHENA) in June 2014. Due to this two-stage selection process, ATHENA and its science theme are strongly related each other. 

In this Chapter, I will first review the \textit{Hot and Energetic Universe} science theme, highligting the main ATHENA scientific objectives. Then, I will present the mission concept, focusing on the on-board high resolution spectrometer: the X-ray Integral Field Unit (X-IFU). Finally, addressing the topic of this thesis, I will show how the background issue could potentially limit the science achievable by the X-IFU observations, highligting the fundamental role of the Cryogenic AntiCoincidence Detector and the others background reduction techniques.

\section{The Hot and Energetic Universe}

ATHENA \cite{athenaprop} is the large X-ray observatory selected by ESA to address the science theme \textit{The Hot and Energetic Universe} \cite{heuniverse}, undertaking the three main goals summarized in the Tab. \ref{toplevel}.

\bigskip

\begin{table}[H]
\centering
\includegraphics[width=0.9\textwidth]{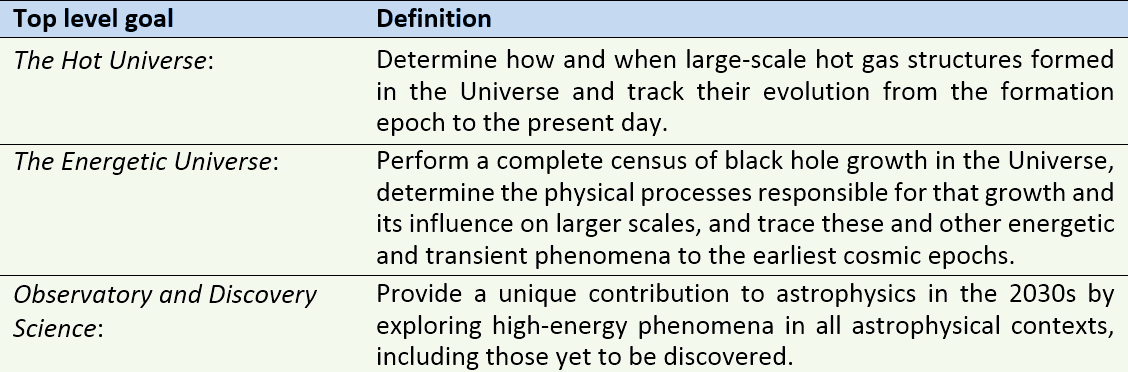}
\caption{ATHENA Level 0 Requirements (top level scientific goals), from: \cite{athenascireq}.}
\label{toplevel}
\end{table}

\textit{The Hot Universe} subject regards the most massive bound structures in the Universe, i.e. galaxies groups and clusters. They are a fundamental components of the Universe, but the astrophysical processes that drive their formation and evolution are still poorly known. The baryonic content of these structures is totally dominated by hot gas (millions of degrees of temperature), and it is important to understand how and when this gas has been trapped in the dark matter haloes and what drives its chemical and thermodynamic evolution from the formation of these structures (z $\sim$ 2 - 3) to the present day. Moreover, we want to unequivocally detect and characterize the filaments of Warm Hot Intergalactic Medium (WHIM) that, in our current understanding, connect these structures, hosting most of the baryonic matter in the local Universe. These questions can be uniquely tackled by observations in the X-ray band, combining high angular resolution imaging with high resolution spectroscopy, in order to simultaneously track the hot gas and characterize its physical and thermodinamical state. 

\textit{The Energetic Universe} topic is mainly dedicated to understand how the accretion of matter into Supermassive Black Holes (SMBH) drives the evolution of galaxies, influencing processes that happens on much larger scales, as the star formation. To do this it is necessary to detect and characterize distant AGN (even if they are heavily obscured) up to z = 6 - 10, where the first galaxies are forming. This is possible only combining wide field X-ray imaging, to pinpoint the AGNs in the deep sky, and high resolution spectroscopy, to study the inner part of the AGN engine in the local Universe (i.e. studying the outflows of ionised gas). Finally, it is also important to trace other high energy and transient phenomena as Gamma Ray Bursts to the earliest cosmic epochs.

Aside from these topics, it is also of primary importance to exploit the capabilities of ATHENA as a proper \textit{X-ray Observatory}, in order to provide an unique contribution to all the fields of Astrophysics in the 2030s, working also in synergy with the multi-wavelength astronomical facilities then in exploitation (e.g. LOFAR, SKA, ALMA, JWST, E-ELT, LSST, CTA, ...).

In the next sections the science objective included in these topics will be reported in more detail.

\newpage
\subsection{The Hot Universe}

The ATHENA science objectives referring to the \virg Hot Universe'' topic are reported in Tab. \ref{hotuniversereq} (from \cite{athenascireq}). In the following paragraphs I will focus on two of them: the \textit{Cluster bluk motions and turbolence (R-SCIOBJ-112)} and the \textit{Missing Baryons (R-SCIOBJ-141)}.

\begin{table}[H]
\centering
\includegraphics[width=1\textwidth]{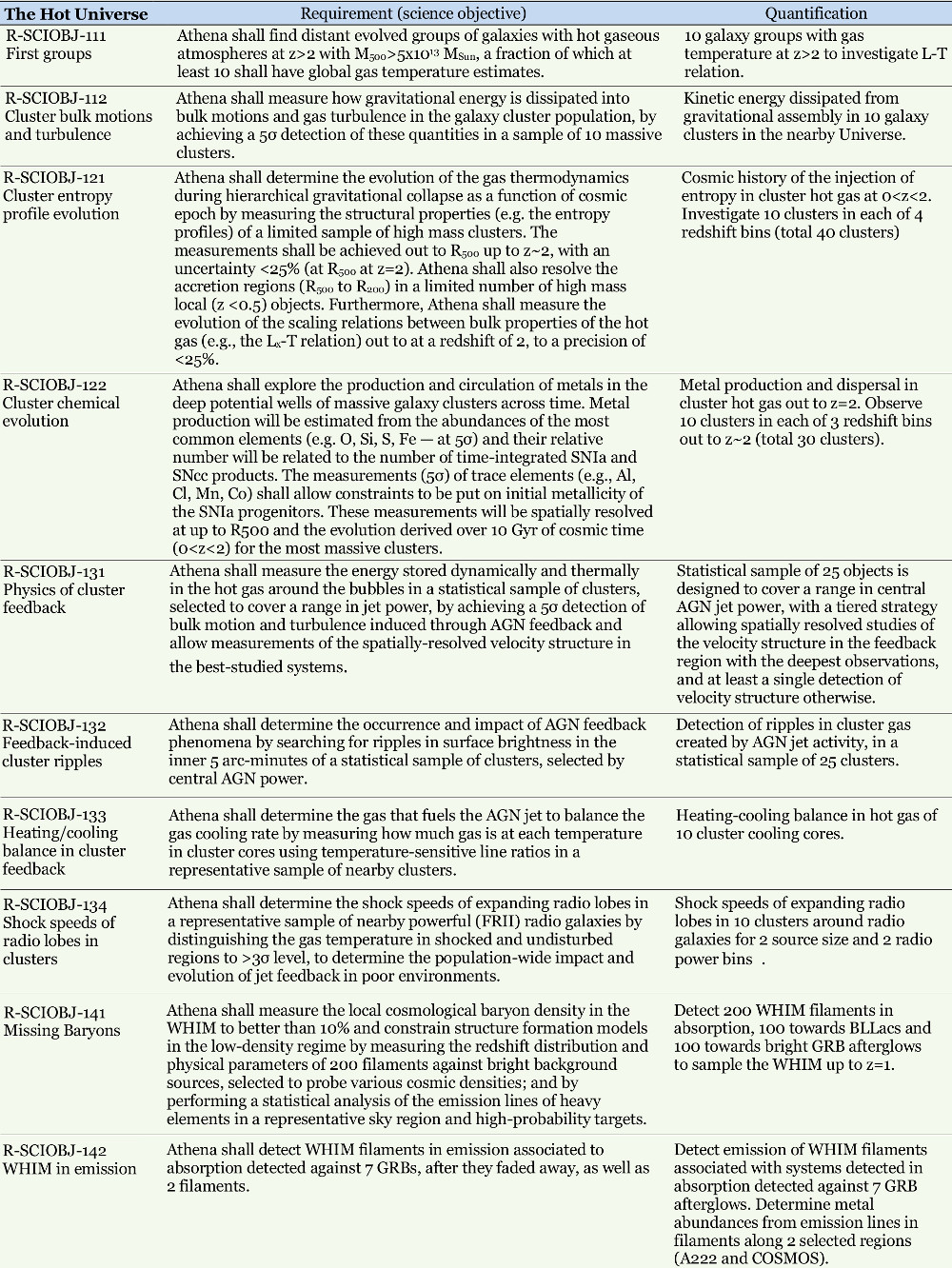}
\caption{ATHENA science objectives addressing the \virg Hot Universe'' theme, from \cite{athenascireq}.}
\label{hotuniversereq}
\end{table}

\subsubsection{The Hitomi spectrum of the Perseus cluster: a tasty aperitif for the ATHENA study of Cluster bulk motions and turbolence}

\begin{figure}[H]
\centering
\includegraphics[width=1\textwidth]{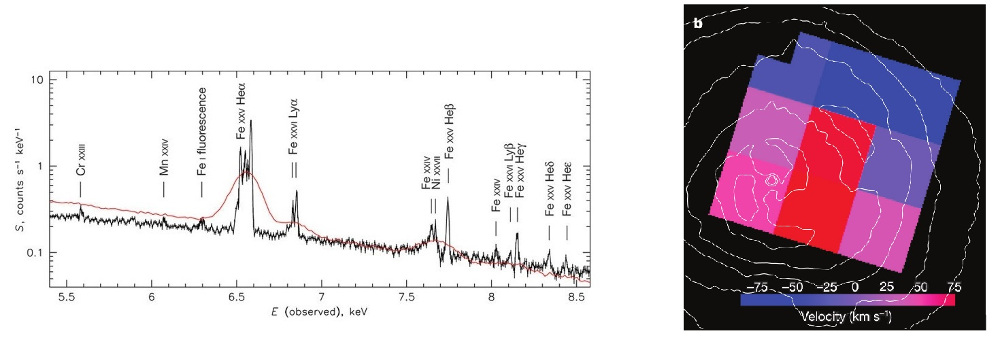}
\caption{Hitomi observation of the Perseus cluster. \textit{Left:} Hitomi spectrum of the full field (black line) compared with the preovius best observation of the same region obtained by the CCD detector on-board Suzaku (red line). From \cite{hitominature}. \textit{Right:} The bulk velocity field across the imaged region. Colours show the difference from the velocity of the central galaxy NGC 1275; positive difference means gas receding faster than the galaxy. The 1-arcmin pixels of the map correspond approximately to the angular resolution. From \cite{hitominature}.}
\label{hitomi}
\end{figure}

Hitomi \cite{hitomispie} (formerly ASTRO-H) has been a X-ray mission developed by the JAXA (Japan Aerospace Exploration Agency) that has exploited for the first time the cryogenic micro-calorimeter technology to perform high-resolution X-ray spectroscopy for Astrophysics. This is an equivalent technology to the one that will be used by the ATHENA spectrometer (the X-IFU), so Hitomi can help us to really understand the future ATHENA capabilties. In the few days of operation before the satellite loss\footnote{The loss has been due to multiple incidents with the attitude control system leading to an uncontrolled spin rate and the breakup of structurally weak elements.}, the Hitomi microcalorimeter has observed the central part of the Perseus cluster, for a total exposure of about 3 days \cite{hitominature}. The main result of the observation has been to determine the level of turbolence in the cluster hot gas, by measuring its velocity from the Doppler broadening of the strong X-ray emission lines due to highly ionized iron (Fig. \ref{hitomi} - Left). The result has revealed a remarkably quiescent atmosphere in which the gas has a line-of-sight velocity dispersion of 164 $\pm$ 10 km/s in the region 30–60 kiloparsecs from the central nucleus (Fig. \ref{hitomi} - Right). This means that the level of turbulence in the cluster core is surprisingly low, representing an energy density and pressure of only $\sim$ 4\% of thermal values. This may imply that turbulence in the intracluster medium is difficult to generate and/or easy to damp, and therefore that the clusters evolution could be not seriously disturbed by the effects of turbulence.

\bigskip

\begin{figure}[H]
\centering
\includegraphics[width=1\textwidth]{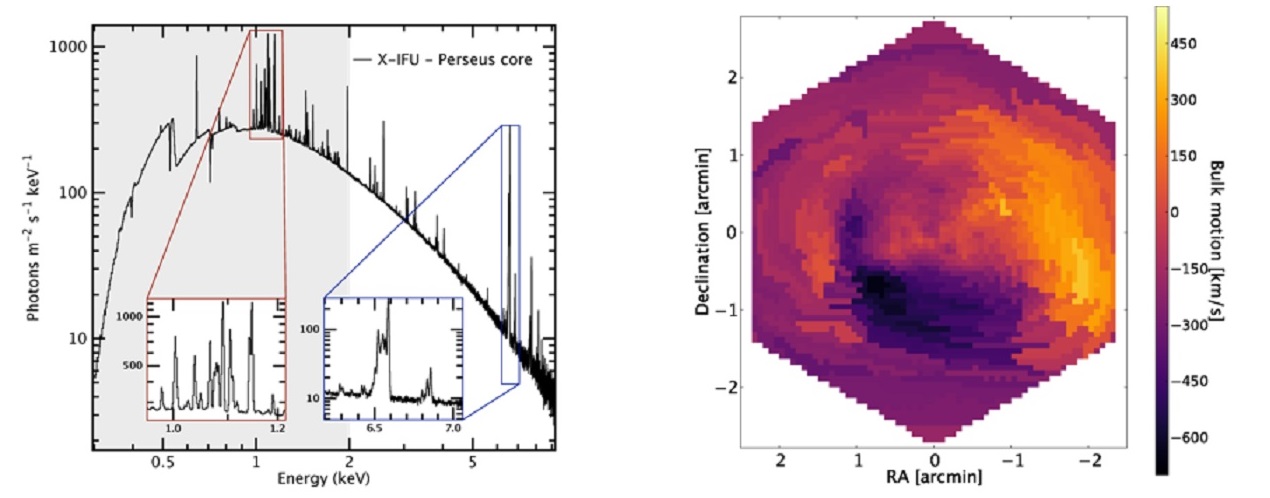}
\caption{\textit{Left:} The simulated ATHENA X-IFU spectrum of the core of the Perseus cluster, based on the Hitomi SXS observations. The exposure time is 100 kilo-seconds. The grey area indicates the region not fully explored by Hitomi, due to the fact that the Hitomi SXS Perseus observation was performed with gate valve closed (blocking most X-rays below 2 keV). From \cite{xifuspie2016}. \textit{Right:} Reconstructed bulk motion induced velocity field (in km/s) of the hot intra-cluster gas for a 50 kilo-seconds ATHENA X-IFU observation of the central parts of a Perseus like cluster. The cluster has the luminosity of Perseus but is considered at a redshift of 0.1. From \cite{xifuspie2016}. }
\label{hitomi2}
\end{figure}

This has been only a preview of the ATHENA X-IFU capabilities, which will exploit a spectral resolution twice that of Hitomi, a two orders of magnitude larger number of imaging pixels, a vastly greater collecting area of the telescope and a spatial resolution one order of magnitude better. The breakthrough capabilities of the ATHENA X-IFU are illustrated in Fig. \ref{hitomi2}, which shows the expected X-IFU spectrum of the core of the Perseus cluster, and a reconstructed bulk motion map of the central parts of a Perseus like cluster at z = 0.1 (both from \cite{xifuspie2016}). According to the science objective R-SCIOBJ-112 (Tab. \ref{hotuniversereq}), ATHENA shall obtain this kind of map for a sample of 10 massive cluster.

\subsubsection{The missing baryons and the WHIM}

\begin{figure}[H]
\centering
\includegraphics[width=0.47\textwidth]{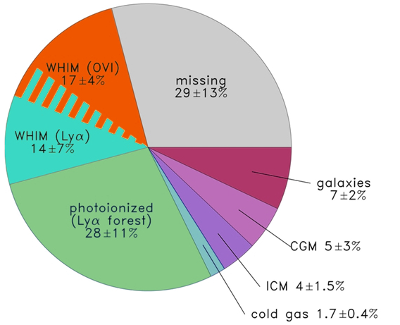}
\includegraphics[width=0.52\textwidth]{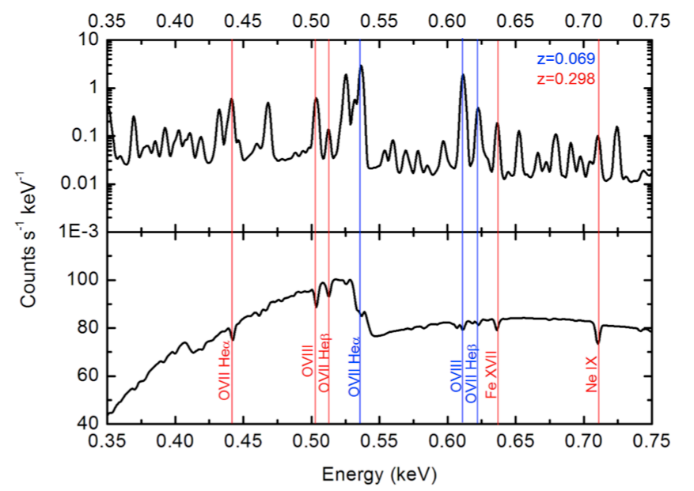}
\caption{
\textbf{Left:} Compilation of current observational measurements of the low-redshift baryon census. Slices of the pie-chart show baryons in collapsed form (galaxies, groups, clusters), in the circumgalactic medium (CGM) and intercluster medium (ICM) and in cold gas (H I and He I). Reservoirs include diffuse photoionized Ly$\alpha$ forest and WHIM traced by O VI and broad Ly$\alpha$ absorbers. Formally, 29 $\pm$ 13\% of the baryons remain unaccounted for. Additional baryons may be found in weaker lines of low-column density O VI and Ly$\alpha$ absorbers. Deeper spectroscopic UV and X-ray surveys are needed to resolve this issue. From: \cite{whimarxiv}. 
\textbf{Rigth:} Simulated emission and absorption line spectra captured in a single ATHENA X-IFU observation for two filaments at different redshifts. Lower panel: absorption spectrum from a line of sight where two different filamentary systems are illuminated by a bright background source. Upper panel: corresponding emission from a 2’x2’ region from the same filaments for 1 Ms exposure time. The high spectral resolution allows us to distinguish both components. ATHENA will be able to study dozens of these sight lines in detail. From: \cite{whimathena}.}
\label{whim}
\end{figure}

The standard cosmological model indicates that the baryons constitute only $\sim$ 5\% of the mass-energy budget of the Universe, while the rest is composed of dark matter and dark energy. Even within this small fraction there is a $\sim$ 30\% of baryons that still eludes detections. Today's observations can indeed only account for $<$ 70\% of the baryons that the cosmological models predict in our local Universe, and that have been detected in the high-z Universe (Fig. \ref{whim} - Left).

One of the key scientific objectives of ATHENA is to detect this missing fraction of baryons that it is supposed to be found in a hot, rarefied state connecting clusters in a cosmic web: the Warm Hot Intergalactic Medium (WHIM). According to hydrodynamical simulations (e.g. \cite{whimsim}), the missing baryons should be indeed into the long diffuse filaments of low density gas at intermediate temperatures (T $\sim$ $10^6$ K) that connect clusters forming a cosmic web \cite{whimathena}. The low density and intermediate temperatures make their continuum emission very faint and dominated by the background. Since at these temperatures H and He are completely ionized, the only way to detect and characterize the WHIM and its physical state comes from characteristic radiation related to highly ionized metals present in it (C, N O, Ne and Fe).

The high resolution ATHENA X-IFU spectrometer, with its capability to perform spectroscopy of extended sources, will allow to resolve spectral lines of metals, determining temperature, density and the chemical composition of the WHIM (for such a measurements an energy resolution of 2.5 eV below 1 keV is required \cite{heuniverse}). This will allow to construct a WHIM 3D map through redshift measurements of the same lines, and to discriminate between different models of formation. ATHENA X-IFU shall achieve these goals by measuring the WHIM lines in absorbtion against bright background sources (R-SCIOBJ-141 in Tab. \ref{hotuniversereq}), but also in emission (R-SCIOBJ-142), as shown in Fig. \ref{whim} - Right. Note that multiple observation of the same filament will provide a model independent estimate of the density $\rho$: observations against a bright background transient source (i.e. GRBs at z $>$ 1) will allow indeed first to measure absorption lines, for which the Equivalent Width is proportional to $\rho$, and after, when the transient will fade, to measure the same filament also in emission, for which line strength is proportional to $\rho^2$.

$$ $$

\newpage
\subsection{The Energetic Universe}
\label{energeticuniverselab}

The ATHENA science objectives addressing the \virg Energetic Universe'' science theme are reported in Tab. \ref{energeticuniversereq}. In the following paragraphs I will focus on two of them: the \textit{Complete AGN census (R-SCIOBJ-221)} and the \textit{High z GRBs (R-SCIOBJ-261)}.

\begin{table}[H]
\centering
\includegraphics[width=1\textwidth]{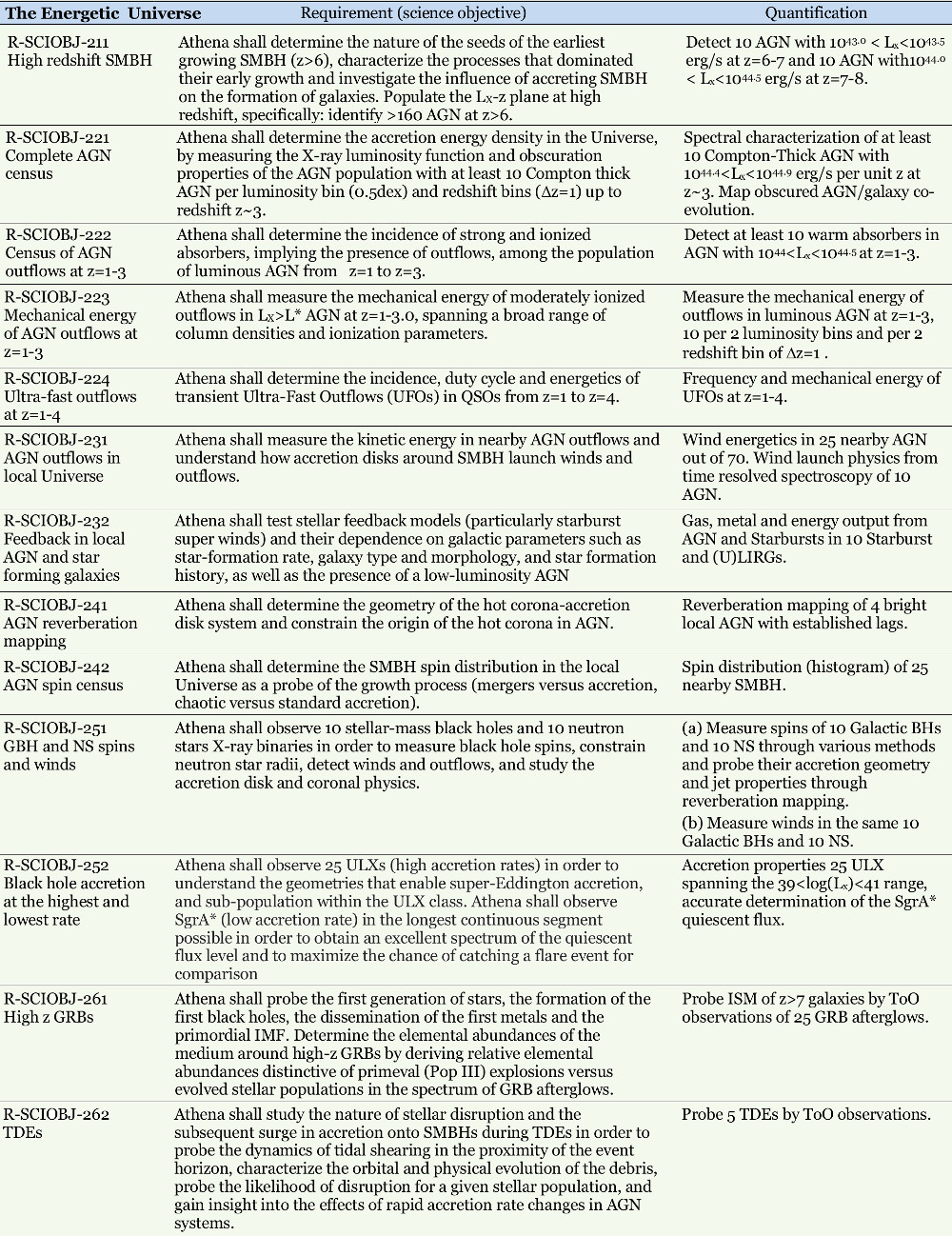}
\caption{ATHENA Science Objectives addressing the \virg Energetic Universe'' theme. From: \cite{athenascireq}.}
\label{energeticuniversereq}
\end{table}

\subsubsection{Understanding the build-up of SMBH and galaxies}

Some galaxies have a compact nucleus that outshines the total light coming from the billions of stars in it. They are referred to as Active Galactic Nuclei (AGN). This emission can't be explained by stellar emission models, but it is thought to be due to mass accretion on a supermassive black hole (SMBH). When the matter falls into a black hole, due to its momentum it tends to form an accretion disk, where it warms up due to the dynamical friction emitting electromagnetic radiation also in the X-Ray part of the spectrum.

One of the main scientific goals of the ATHENA mission is to perform a demographic analysis of the AGN population in the Universe (R-SCIOBJ-221 in Tab. \ref{energeticuniversereq}), in order to determine how the supermassive black holes properties are related to the cosmic time and the evolutionary history of the host galaxies. Observations in the last decades have indeed provided strong evidence that the growth of supermassive black holes at the centres of galaxies is among the most influential process in galaxy evolution. The generic scenario proposed involves an early phase of intense black hole growth that takes place behind large obscuring columns of inflowing dust and gas clouds. It is postulated that this is followed by a blow-out stage during which some form of AGN feedback controls the fate of the interstellar medium and hence, the evolution of the galaxy \cite{smbhathena}.

\begin{figure}[H]
\centering
\includegraphics[width=0.90\textwidth]{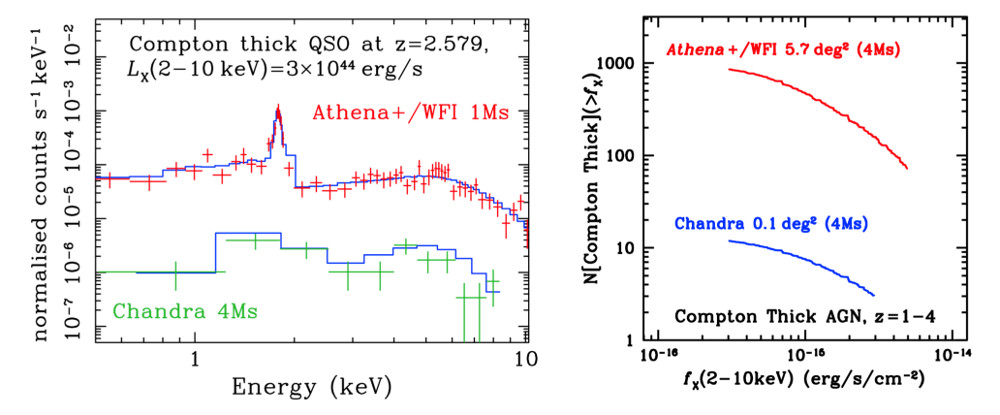}
\caption{\textbf{Left:} Spectral quality that the ATHENA WFI surveys will deliver for heavily obscured AGN. The red data-points are the simulated ATHENA WFI 1Ms spectrum of a CT AGN at z=2.579. The most prominent feature in that spectrum is the Fe Ka emission line. The simulated source has been previously identified in the 4Ms Chandra Deep Field South (green points). From: \cite{smbhathena}.
\textbf{Right:} Predicted cumulative counts of CT AGN at redshifts z=1–4 as a function of hard (2-10keV) flux. The blue curve corresponds to the 4Ms Chandra Deep Field South, the deepest X-ray image currently available. ATHENA WFI (red curve) yield, for the same time investment, it is up to a factor of $\sim$100 more CT AGN compared to the current Chandra X-ray surveys. From: \cite{smbhathena}.
}
\label{agn}
\end{figure}

The most important epoch for investigating the relation between AGN and galaxies is the redshift range z $\sim$ 1 - 4, when most black holes and stars we see in the present-day Universe were put in place. Unfortunately, at these redshifts a significant fraction of AGNs ($\sim$ 40\%) is strongly obscured by the large amount of gas and dust in the galaxies, which are still forming stellar populations. When the column density of the obscuration exceeds the inverse of the cross-section for Thompson scattering ($N_H \sim 10^{24} \; cm^{-2}$), these source are called \textit{Compton Thick AGN} (CT AGN). ATHENA will be able to detect and characterize (up to redshift  z $\sim$ 4) also this heavily obscured sources (Fig. \ref{agn}- Left), performing their census mainly by the Wide Field Imager (WFI) instrument, thanks to its combination of high throughput, large field of view and wide collecting area. Note that exhaustive efforts with current high-energy telescopes only scrape the tip of the iceberg of the most obscured AGN population (Fig. \ref{agn} - Right).

\subsubsection{High-z GRBs and the first stars}

The cosmological standard model foresees the existence of a first generation of stars, known as \textit{Population III}, formed directly by the primordial material from the Big Bang nucleosynthesis (H, He and small traces of Li and Be), and thus characterized by an extremely low metallicity (Z $\sim$ 0). The theoretical models predict that these stars would have been very massive (several hundred times more massive than the Sun) and so have burned out quickly, finally exploding as Supernovae enriching the primordial ISM with the first metals. Finding these stars directly it is very difficult, since they are extremely short-lived and they formed when the Universe was largely opaque to their light. As a consequence, a conclusive proof of their existence has not yet been found. 

One of the science objectives of ATHENA (R-SCIOBJ-261 in Tab. \ref{energeticuniversereq}) is to probe the existence of these primeval stars by detecting the chemical \virg fingerprint'' of their explosion in the Interstellar Medium (ISM) at high redshift (up to z $\sim$ 7). The relative abundance of the elements generated by the population III stars is indeed strongly distinct from that of the later stellar generations, and it is also strongly related to the typical masses of the first stars, opening the possibility to probe the Initial Mass Function (IMF) of the Universe \cite{transientathena}.

\begin{figure}[H]
\centering
\includegraphics[width=0.7\textwidth]{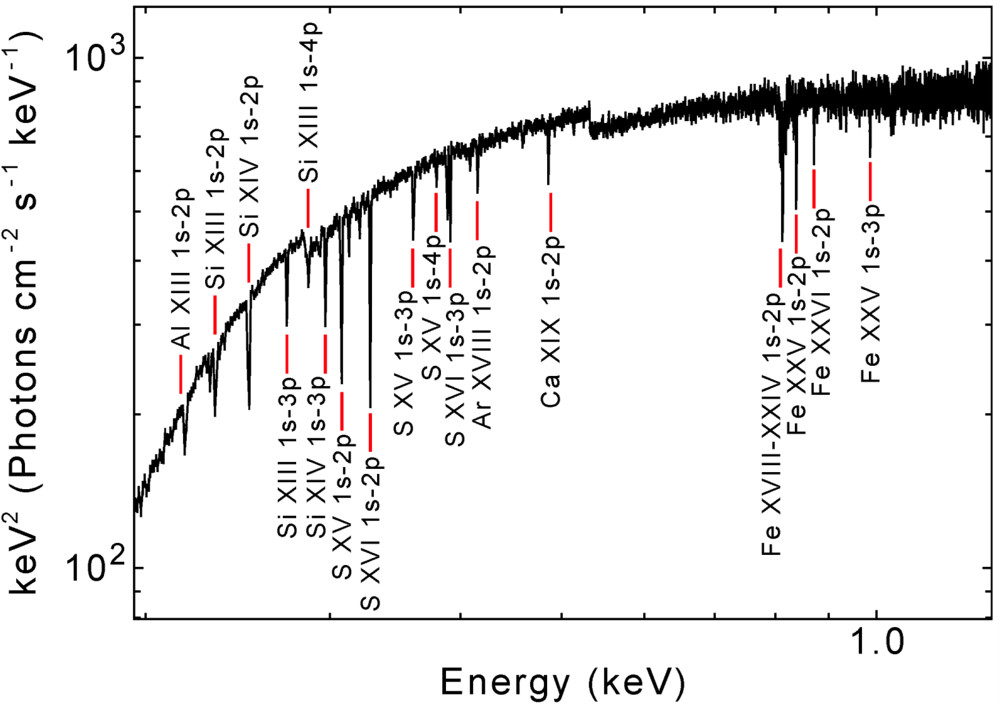}
\caption{A simulated ATHENA X-IFU X-ray spectrum of a GRB afterglow at z=7, showing the capability of ATHENA in tracing the primordial stellar populations. This medium bright afterglow (fluence=0.4 $\times$ 10$^{-6}$ erg cm$^{-2}$) is characterized by deep narrow resonant lines of Fe, Si, S, Ar, Mg, from the ionized gas in the environment of the GRB. An effective column density of 2 $\times$ 10$^{22}$ cm$^{-2}$ has been adopted. The abundance pattern measured by ATHENA can distinguish Population III from Population II star forming regions. From: \cite{transientathena}.}
\label{grbpop3}
\end{figure}

In this context, Gamma Ray Bursts (GRBs) play a unique role in the study of metal enrichment in the primordial Universe, as they are the brightest light sources at all redshifts. High-resolution X-ray spectroscopy of GRB afterglows can simultaneously probe all the elements (C through Ni) in all their ionization stages, thus providing a model-independent survey of the metals in the GRB environment. ATHENA X-IFU would extend the frontier of such studies into the high-redshift Universe, tracing the early Population III stars and their environment (Fig. \ref{grbpop3}). 

\bigskip

\newpage
\subsection{Observatory science} 

The transformational capabilities of ATHENA are finally expected to have a profound impact in essentially all fields of Astrophysics. In this context, the science objectives related to the ATHENA operation as observatory are reported in Tab. \ref{observatoryreq}.

\begin{table}[H]
\centering
\includegraphics[width=1\textwidth]{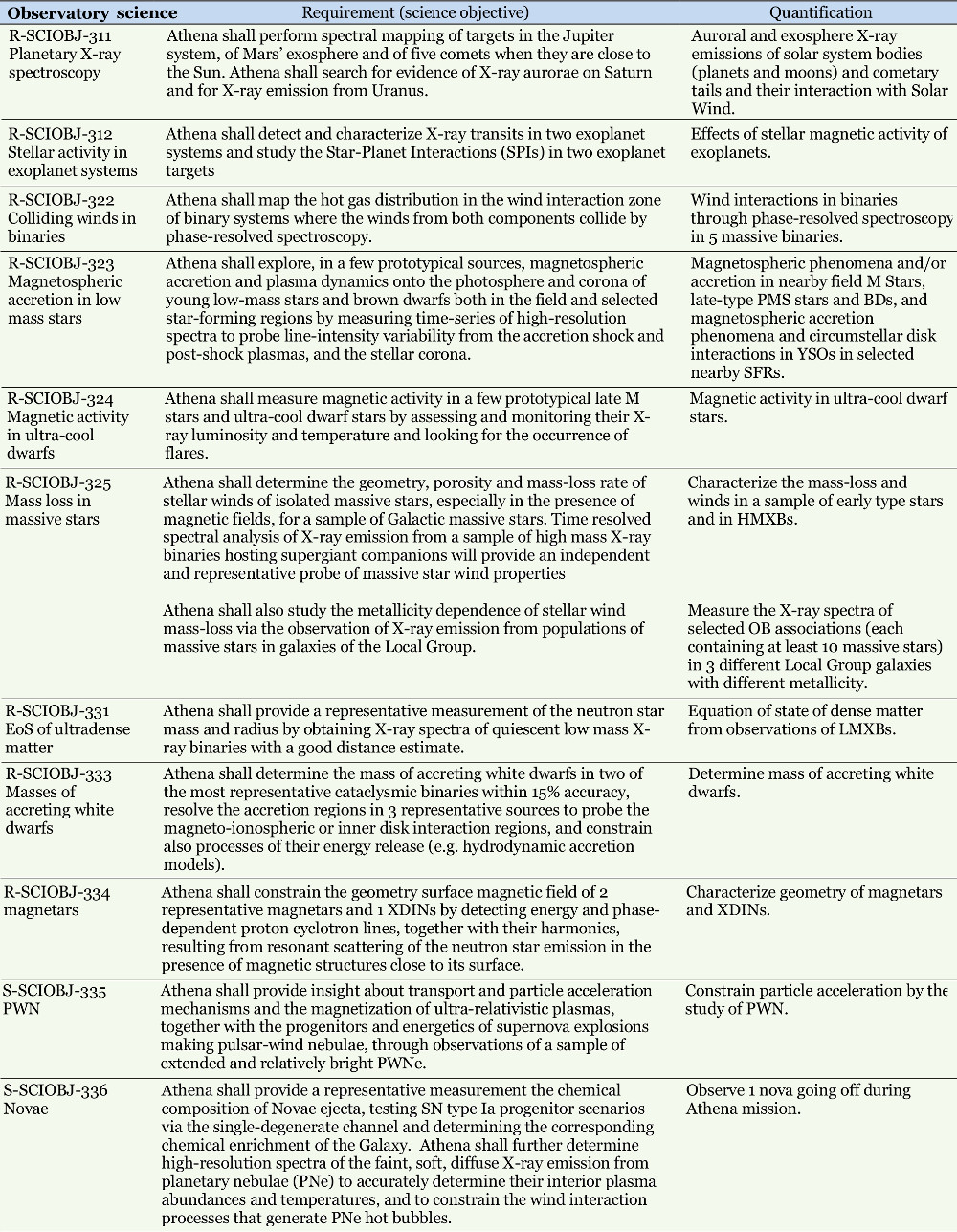}
\end{table}

\begin{table}[H]
\centering
\includegraphics[width=1\textwidth]{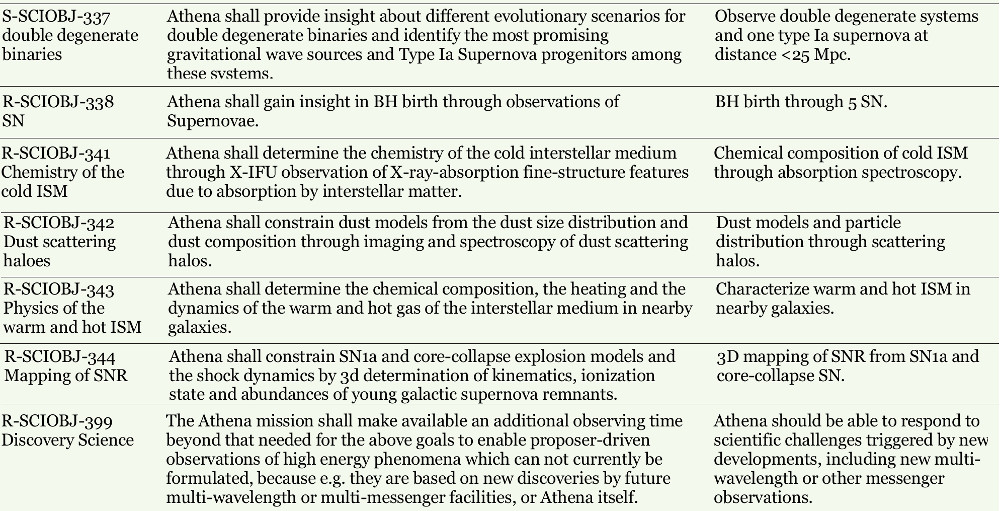}
\caption{ATHENA Science Objectives in the context of its observatory operations. From: \cite{athenascireq}.}
\label{observatoryreq}
\end{table}

\newpage
\section{The Advanced Telescope for High Energy Astrophysics}

\begin{figure}[H]
\centering
\includegraphics[width=1\textwidth]{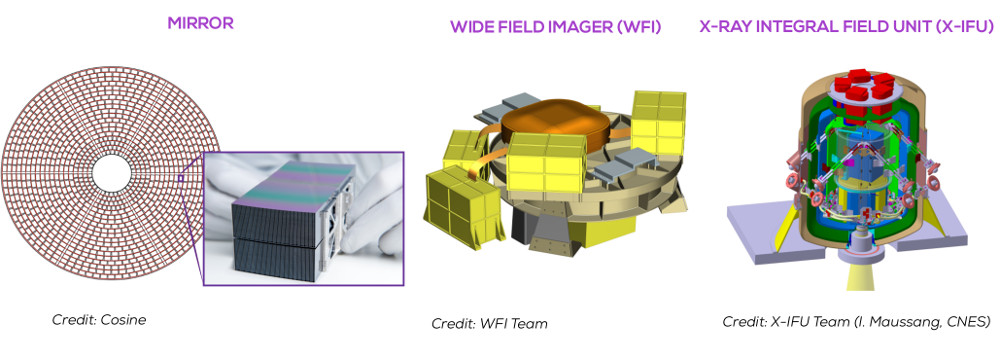}
\caption{The key elements of the ATHENA mission. \textit{Left:} The large-aperture grazing-incidence telescope, utilizing a novel high-performance Silicon Pore Optics technology developed in Europe. \textit{Center}: the WFI, providing sensitive wide field imaging and spectroscopy and high count-rate capability with a 40'$\times$40' field of view. \textit{Right:} The X-IFU, delivering spatially-resolved high-resolution X-ray spectroscopy over a field of view of 5' effective diameter. From: \cite{athenasite}.}
\label{instruments}
\end{figure}

The ATHENA observatory has been designed to address the science themes presented in the previous sections, offering X-ray deep wide-field spectral imaging and spatially-resolved high-resolution spectroscopy with performance greatly exceeding that offered by current X-ray missions. To do this, the ATHENA spacecraft will be constituted by three key elements (Fig. \ref{instruments}):

\begin{itemize}
\item An X-ray Telescope with a focal length of 12 m, with large collecting area (1.4 m$^2$ @ 1 keV) and good angular resolution (5 arcsec Half Energy Width) \cite{athenaoptics};
\item The Wide Field Imager (WFI), an instrument for high count rate offering moderate resolution spectroscopy ($\Delta$E$_{FWHM}$ $<$ 170 eV @ 7 keV ) over a large Field Of View (40 $\times$ 40 arcmin$^2$) \cite{wfisite};
\item The X-ray Integral Field Unit (X-IFU), a cryogenic spectrometer offering spatially resolved high-resolution spectroscopy ($\Delta$E$_{FWHM}$ $<$ 2.5 eV @ 6 keV) over a small Field Of View (5 arcmin equivalent diameter) \cite{xifusite}.
\end{itemize}

The mission development is currently at the end of the Phase A (Feasibility study), which will end in the early 2019 with the Instrument Preliminary Requirement Reviews (IPRRs). The Phase B1 (Preliminary definition) will then follow until late 2019, ending with the Mission Formulation Review (MFR). The mission adoption by the ESA Science Programme Committee (SPC) is finally expected at the end of 2021, leading to launch in 2031.

\bigskip

In the next sections, after an introduction about the foreseen satellite operations, I will quickly introduce the three key elements of the ATHENA payload.

\newpage
\subsection{Operations}

ATHENA is planned to be launched by using an Ariane 6 or another launch vehicle with equivalent lift capability and fairing size to that of the Ariane 5 ECA. The observatory will operate in a large halo orbit (radius $\sim$ 10$^6$ km) around L2, the second Lagrange point of the Sun-Earth system, at a distance of more than 1.5 millions of km from the Earth. The L2 orbit is preferred to alternative scenarios (e.g. a low inclination Low Earth Orbit or a Highly Elliptical Orbit) since it provides a very stable thermal environment and a good instantaneous sky visibility, thus ensuring a high observing efficiency. Note that the possibility of an L1 halo orbit it is also being assessed with respect to the particle environment affecting the intruments sensitivity\footnote{The particle background is induced by two different particle populations: the Galactic Cosmic Rays (GCR) and the Soft Solar Protons. The GCR contribution is expected to be the same in L1 and L2. Concerning the soft protons, L1 is instead a more known environment, since there is a high data coverage of about 1.5 solar cycles (more than ten years, to compare to the few months available for L2), enabling to make more representative predictions about the background level \cite{L1}.
}. 

ATHENA will be operated as an observatory, like previous missions such as XMM-Newton and Herschel, being accesible from the users by open proposal Calls. It will predominantly perform pointed observations of celestial targets. There will be about 300 such observations per year, with durations from 10$^3$ to 10$^6$ seconds (typically 100 ks per pointing). The routine observing plan will be sometimes interrupted by Target of Opportunity (ToO) observations (e.g. to point Gamma Ray Bursts or other transient events), at an expected rate of twice per month. The ToO reaction time of the telescope it is foreseen to be $<$ 4 hours.

The baseline mission duration is four years, with a possible four-year extension. 

\subsection{Telescope and X-ray optics}

The ATHENA observatory will consist of a single X-ray telescope with a focal length of 12 meters and an unprecedented collecting area (1.4 m$^2$ at 1 keV). The X-ray telescope employs Silicon Pore Optics (SPO), a new technology that has been developed in Europe over the last decade. SPO are based on a highly modular concept, being composed by a set of compact individual mirror modules able to provide a combination of a large field of view, while meeting the stringent mass budget. Informations about the SPO thecnology and its development can be found in \cite{athenaoptics}.

\begin{figure}[H]
\centering
\includegraphics[width=0.9\textwidth]{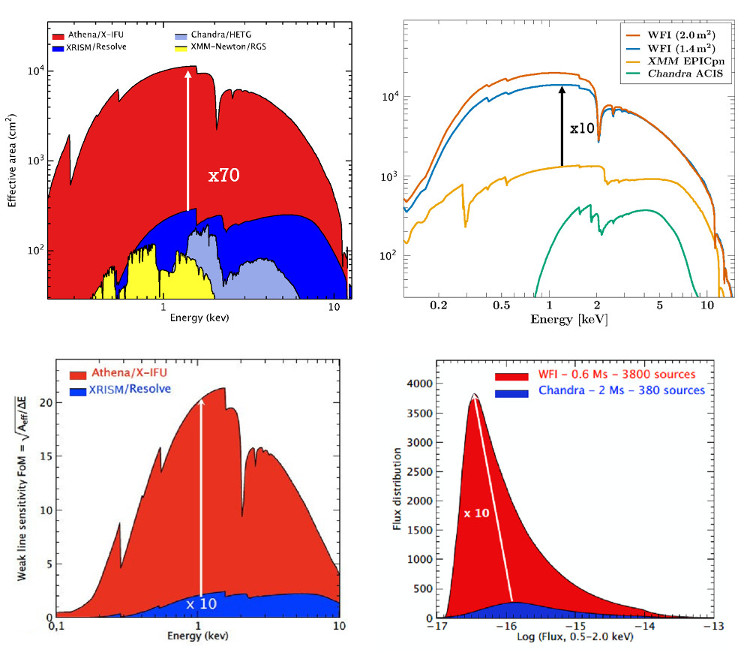}
\caption{\textit{Top left:} ATHENA X-IFU Effective Area (red), compared with that of XRISM/Resolve (blue), Chandra/HETG (light blue) and XMM-Newton RGS (yellow). \textit{Top right:} ATHENA WFI Effective Area (blue line), compared with that of XMM EPICpn (yellow) and Chandra ACIS (green) \textit{Bottom left:} Weak line sensitivity comparison between X-IFU and XRISM (Figure of Merit for weak unresolved line detection) \textit{Bottom right:} Flux distribution comparison between WFI and Chandra deep pointing (Figure of merit for survey speed, i.e number of sources per unit flux detected in single pointing). From: \cite{athenasite}.}
\label{area}
\end{figure}

Exploiting such a telescope configuration in combination with its two focal plane instruments, ATHENA will provide transformational capabilities with respect to that offered by current X-ray missions, as illustrated in Fig. \ref{area}. 

The X-IFU will exploit an effective area a factor 70 higher than the X-Ray Imaging and Spectroscopy Mission (XRISM, the JAXA \virg recovery mission'' resulting from the loss of Hitomi, Fig. \ref{area} - Top Left), while the WFI will provide an improvement of a factor 10 compared to the effective area of the XMM-Newton EPICpn (Fig. \ref{area} - Top Right). 

With regard to the line spectroscopy, the large effective area of the X-IFU (in combination with its energy resolution) will translate in a weak line sensitivity\footnote{The weak line sensitivity is a Figure of Merit (FoM) for weak unresolved line
detection, combining the effective area and the spectral resolution of the instrument: Weak line sensitivity FoM = $\sqrt{A_{eff}/\Delta E}$.} one order of magnitude better with respect to XRISM/Resolve (Fig. \ref{area} - Bottom Left), opening the high-resolution X-ray spectroscopy to the high redshift Universe. Concerning the WFI, the combination of the combination of effective area, field of view, angular resolution (and its flatness across the field of view) will deliver a significant improvement in the survey speed compared to Chandra (Fig. \ref{area} - Bottom Right).

\subsection{The Wide Field Imager (WFI)}

The WFI \cite{wfispie} is very powerful survey instrument, based on a Silicon detector using the Depleted P-channel Field Effect Transistor (DEPFET) active sensor technology. It is designed to provide images in the 0.2-15 keV band, simultaneously performing spectrally and time-resolved photon counting.

The WFI focal plane includes a Large Detector Array (LDA), with over 1 million pixels of 130 $\mu$m $\times$ 130 $\mu$m size, permitting oversampling of the PSF by a factor $>$2 and spanning a 40 $\times$ 40 arcmin$^2$ Field of View. It is complemented by a smaller Fast Detector (FD) optimized for high count rate applications, up to and beyond 1 Crab\footnote{The Crab is a photometrical unit commonly used in X-ray astronomy to measure the intensity of astrophysical X-ray sources. One Crab is defined as the intensity of the Crab Nebula at the corresponding X-ray photon energy. Because of its stable and intense emission and its well-known spectrum, the Crab Nebula is indeed commonly used in X-ray astronomy as a standard candle for in-flight calibration of instruments response. In the range from 2 to 10 keV, 1 Crab equals 2.4 $\cdot$ 10$^{−8}$ erg cm$^{−2}$ s$^{−1}$.
} source intensity.

The instrument is developed by an international proto-consortium led by the Max Planck Institute for extraterrestrial Physics (Germany) and composed by several institutes from ESA member states and external partners.

\begin{figure}[H]
\centering
\includegraphics[width=1\textwidth]{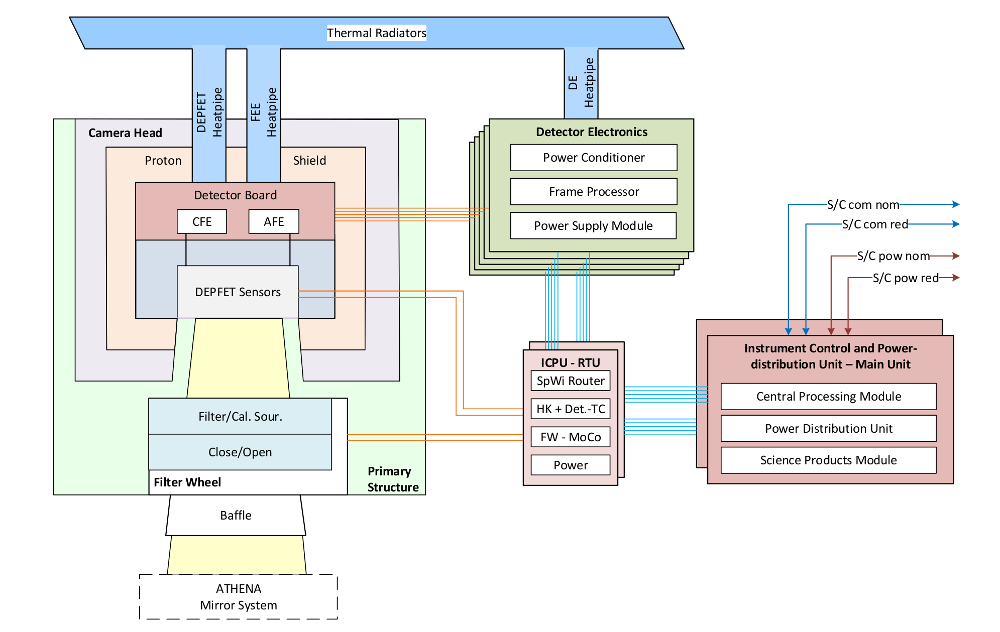}
\caption{The WFI functional diagram, showing the main components of the instrument. The filter wheel controls the photon flux arriving from the Athena mirror on the focal plane. The energy, incidence position, and arrival time of the X-ray photons are measured by the DEPFET active pixel sensors which are controlled and read out by the front-end ASICs. Together they form the camera head, surrounded by a graded Z-shield and a proton shield to minimize instrumental background and radiation damage, respectively. The camera head is cooled passively via heat pipes and radiators. The analog signal from the camera head is transferred to the six detector electronic boxes (one for each large DEPFET and for each half of the fast DEPFET) where it is digitized and the events are pre-processed. The Interface Control and Power Distribution Unit (ICPU) is the communication interface between the Athena spacecraft and the WFI. It also controls the operation of the six detector units, it merges their six data streams, and it is responsible for the filter wheel and thermal control. From: \cite{wfisite}.}
\label{wfi}
\end{figure}

The main characteristics of the WFI are summarized in the Tab. \ref{table:wfitab}, and the functional block diagram of the instrument is reported in Fig. \ref{wfi}.

\begin{table}[H]
\footnotesize
\centering
\begin{tabular}{ll}
\toprule
\textbf{Parameter} & \textbf{Value} \\
\midrule
Energy Range & 0.2 - 15 keV \\
Pixel Size & 130 $\mu$m $\times$ 130 $\mu$m (2.2 arcsec $\times$ 2.2 arcsec) \\
Operating Modes & rolling shutter (min. power consumption), full frame mode,\\
& window mode (optional)\\

High-count rate detector & 64 x 64 pixel, mounted defocussed \\
               & split full frame readout, time resolution: 80 $\mu$s\\
               & 1 Crab: $>$90\% throughput and $<$1\% pileup \\

Large-area DEPFET & 4 quadrants, each with 512x512 pixel (total FOV 40'x40'),\\ 
& time resolution: $<$ 5 ms\\

Quantum efficiency with filters & $>$ 20\% @ 277 eV , $>$ 80\% @ 1 keV, $>$ 90\% @ 10 keV \\ 
& transmissivity for optical photons: 3 $\times$ 10$^{-7}$ \\
& transmissivity for UV photons: $<$ 10$^{-9}$ \\

Energy Resolution & $\Delta$E$_{FWHM}$ @ 7 keV $<$ 170 eV \\

Non-X-ray Background & $<$ 5$\times$10$^{-3}$ counts/cm$^2$/s/keV \\
\bottomrule
\end{tabular}
\caption{WFI instrument characteristics. From: \cite{wfisite}.}
\label{table:wfitab}
\end{table}

\newpage
\subsection{The X-ray Integral Field Unit (X-IFU)}

The X-IFU \cite{xifuspie} is a cryogenic X-ray spectrometer for high-spectral resolution imaging. It will be the first X-ray instrument able to perform the so-called integral field spectroscopy, simultaneously providing detailed images of its Field Of View (5 arcmin equivalent diameter) with an angular resolution of 5 arcsec, and high-resolution energy spectra ($\Delta$ E $<$ 2.5eV at 6 keV). 

The X-IFU is based on a large array of 3840 Transition Edge Sensor (TES) microcalorimeters, working at a temperature of $\sim$ 90 mK and read-out by SQUIDs amplifiers. A Frequency Domain Multiplexing enables to read out 40 pixels in one single channel and a Cryogenic AntiCoincidence detector (CryoAC) located underneath the prime TES array enables to reduce by a factor $\sim$ 50 the Non X-ray Background. A bath temperature of $\sim$ 50 mK is obtained combining a series of mechanical pre-coolers (15K Pulse Tubes, 4K and 2K Joule-Thomson) with a sub Kelvin cooler made of a $^3$He sorption cooler coupled with an Adiabatic Demagnetization Refrigerator. The defocusing capability of the ATHENA movable mirror assembly enables the X-IFU to observe also the brightest X-ray sources of the sky (up to Crab-like intensities) by spreading the telescope point spread function over hundreds of pixels.

The instrument will be provided by an international consortium led by France, the Netherlands and Italy, with further ESA member state contributions from Belgium, Czech Republic, Finland, Germany, Ireland, Poland, Spain, Switzerland and contributions from Japan and the United States.

\begin{figure}[H]
\centering
\includegraphics[width=1\textwidth]{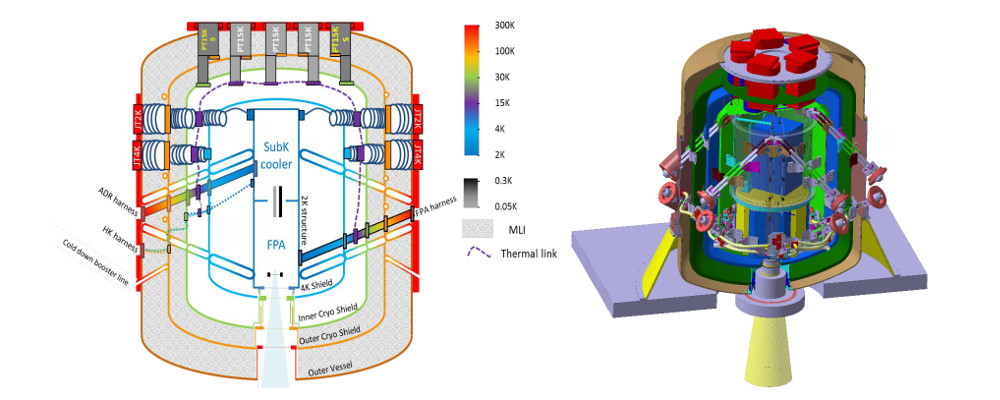}
\caption{\textit{Left:} The X-IFU cryostat, emphasizing the different mechanical coolers used and the different thermal shields considered. The outer vessel has a temperature of 200K, assumed to be acheived through passive cooling. How to best achieved this interface temperature on the SIM is currently being studied. \textit{Right:} An open view of the X-IFU cryostat, emphasizing the 2K core and FPA (in blue) and the mechanical straps (courtesy of CNES). From: \cite{xifuspie}.}
\label{xifucryostat}
\end{figure}

\begin{figure}[H]
\centering
\includegraphics[width=0.9\textwidth]{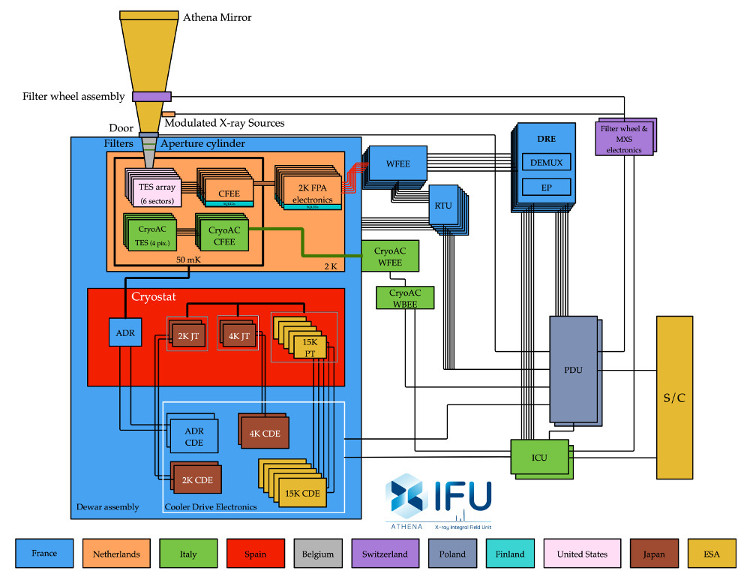}
\caption{The X-IFU functional diagram, showing the main components of the instrument. The color code indicates the corresponding work packages country leads. FPA stands for Focal Plane Assembly, CryoAC for Cryogenic AntiCoincidence detector, CFEE for Cold Front End Electronics, WFEE for Warm Front End Electronics, DRE for Digital Readout Electronics, Demux for Demultiplexing, EP for Event Processor, WBEE for Warm Back-End Electronics, RTU for Remote Terminal Unit, ICU for Instrument Control Unit, PDU for Power Distribution Unit, JT for Joule-Thomson, PT for Pulse-Tube and ADR for Adiabatic Demagnetization Refrigerator. From: \cite{xifusite}.}
\label{xifu}
\end{figure}

The X-IFU key performance requirements are listed in Tab. \ref{table:xifuspec}, a schematic view of the X-IFU cryostat is given in Fig. \ref{xifucryostat} and the functional diagram of the instrument is shown in Fig. \ref{xifu}.

In the next paragraphs I will quickly introduce the X-IFU photon detection principle and its particle background rejection subsystem. Note that both these topics will be then examined in depth in the following Chapters of this thesis.

\bigskip

\begin{table}[H]
\centering
\begin{tabular}{ll}
\toprule
\textbf{Parameter} & \textbf{Value} \\
\midrule
Energy range & 0.2 - 12 keV \\
Energy resolution E $<$ 7 keV & 2.5 eV \\
Energy resolution E $>$ 7 keV & E/$\Delta$E = 2800 \\
Field of view & 5 arcmin (equivalent diameter) \\
Effective area @ 0.3 keV & $>$ 1500 cm$^2$ \\
Effective area @ 1.0 keV & $>$ 15000 cm$^2$ \\
Effective area @ 7.0 keV & $>$ 1600 cm$^2$ \\
Gain calibration error (RMS, 7 keV) & 0.4 eV \\
Count rate capability & 1 mCrab  \\
& ($>$80\% high resolution events)\\
Count rate capability \small{(brightest point sources)} & $>$30\% throughput\\
Time resolution & 10 $\mu$s \\
Non X-ray background (2-10 keV) & $<$5$\times$10$^{-3}$ \small{counts/cm$^2$/s/keV}\\
& (80\% of the time) \\
\bottomrule
\end{tabular}
\caption{Baseline X-IFU top level performance requirements. From: \cite{xifusite}.}
\label{table:xifuspec}
\end{table}

\newpage

\subsubsection{Detection principle}
\begin{figure}[htbp]
\centering
\includegraphics[width=1\textwidth]{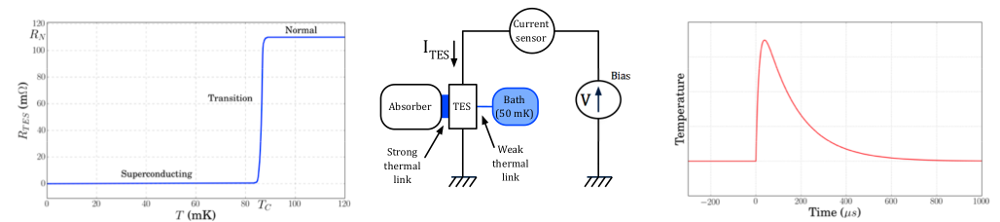}
\caption{Principle of a TES (Transition Edge Sensor) acting as a microcalorimeter. \textit{Left:} The TES is cooled within its trasition. \textit{Middle:} The absorption of an X-ray photon heats both the absorber and the TES through the strong thermal link. \textit{Right:} The change in temperature (or resistance) with time shows a typical pulse shape, with an height proportional to the energy of the absorbed photon. From: \cite{xifuproposal}.}
\label{tess}
\end{figure}

A Transition Edge Sensor (TES) microcalorimeter is a cryogenic detector that senses the heat pulses generated by the absorbtion of X-ray photons in an absorber. The temperature of the absorber increases sharply with the deposited photon energy and it is measured by the change in the electrical resistence of the TES, which is a superconducting film cooled at temperature less than 100 mK and biased within its transition between superconducting and normal states (Fig. \ref{tess}). The TES is usually voltage biased, and a photon event results as a small change in the TES current, which is readout using an extremely sensitive low noise amplifier: the Superconducting Quantum Interference Device (SQUID).

The X-IFU TES array will use molybdenum-gold TES coupled with 249 $\mu$m squared absorbers made of one layer of gold (1.7 $\mu$m thick) and one layer of bismuth (4.2 $\mu$ thick).

\subsubsection{Particle background rejection}
\begin{figure}[htbp]
\centering
\includegraphics[width=0.6\textwidth]{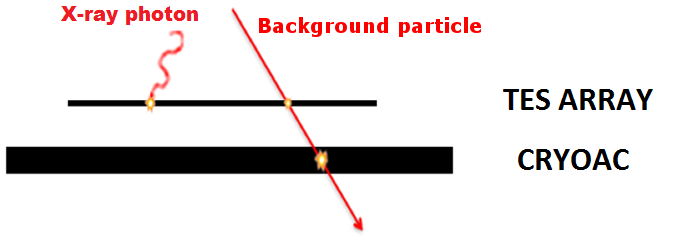}
\caption{Functioning principle of the cryogenic anticoincidence detector: particles cross through the TES array and the CryoAC (unlike the \virg scientific'' photons that are completely absorbed inside the main detector), causing a simultaneous signal that allows to discriminate the event.}
\label{cryoacscheme}
\end{figure}

TES microcalorimeters do not distinguish among photons and particles that release energy inside the absorber, therefore the X-IFU includes an active particle background rejection subsystem. It is composed by a Cryogenic AntiCoincidence detector (CryoAC) placed 1 mm below the TES array. This way particles will deposit energy on both the main TES array and the CryoAC, generating simultaneous signals that will allow to discriminate these events (Fig. \ref{cryoacscheme}).

\bigskip

As will be shown in the Chapt. \ref{beckmark} of this thesis, advanced Monte Carlo simulations estimate the X-IFU Non X-ray background (NXB), driving the Focal Plan Assembly (FPA) and the CryoAC design in order to reach the scientific requirement of a NXB $\leq$ 0.005 cts cm$^{-2}$ s$^{-1}$ keV$^{-1}$ (between 2 and 10 keV).

\section{The role of the background in the X-IFU observations}

In this section, I want to show how the particle background could potentially limit the science achievable by the X-IFU observations without the implementation of dedicated reduction techiniques and instrumentation. Initially, I will show a rough evaluation of the scientific objectives affected by the particle background when no reduction technique is applied to the instrument. Then, I will report a case study simulating the observation of an interesting astrophysical source (a distant Compton Thick AGN) with different background levels.

\subsection{The X-IFU science affected by the background}

To roughly evaluate what science objectives need the reduction of the particle background level in order to be achieved by X-IFU observations, I have analyzed the latest ATHENA Mock Observation Plan \cite{athenaMOP}. First, I have selected the X-IFU observations marked as containing scientific information at energies $>$ 2 keV (where the particle background is dominant with respect to the X-ray background), and not foreseeing the defocusing of the ATHENA optics (i.e. excluding the brightest sources, certainly not affected by the background). For each science objective, I have then evaluated, from the average flux of the indicated sources, the typical expected count rate on the TES array in the band 2-10 keV. Finally, I have compared this value with the expected particle background count rate in the same band, assuming that no background reduction techniques was applied to the X-IFU. In that case we expect a particle background level about 50 times the requirement, i.e. $\sim$ 0.23 cts/cm$^{-2}$ s$^{-1}$ keV$^{-1}$ (the estimation of the background will be presented in the Chapt. \ref{beckmark} of this thesis). This level corresponds to $\sim$ 5$\times$10$^{-3}$ cts/s for a point source (5 arcsec extraction radius) and $\sim$ 5 cts/s for a source diffused over the whole detector surface (2.5 arcmin radius). 
 
The result of this comparison is summarized in Tab. \ref{xibkg}, where the color code shows when the background rate is dominant with respect to the source one (red: background domination, green: source domination, yellow: sources partially or possibly background dominated). Note that for the science objectives needing the observation of galaxy cluster outskirts, I have estimated that less than a tenth of the total source counts resides in the outskirts.

\bigskip
From the table, we can see that in the absence of background reduction techniques only the $\sim$ 50 $\%$ of the X-IFU observing time would not be dominated by the particle background, with this affecting especially the \virg Hot Universe'' related observations. This is just a rough evaluation, but it is able to highlight the importance of the CryoAC in the framework of the ATHENA science.

\newpage
\thispagestyle{empty}
\newgeometry{top=1.5cm, bottom=1.5cm}

\begin{table}[H]
\centering
\includegraphics[width=1\textwidth]{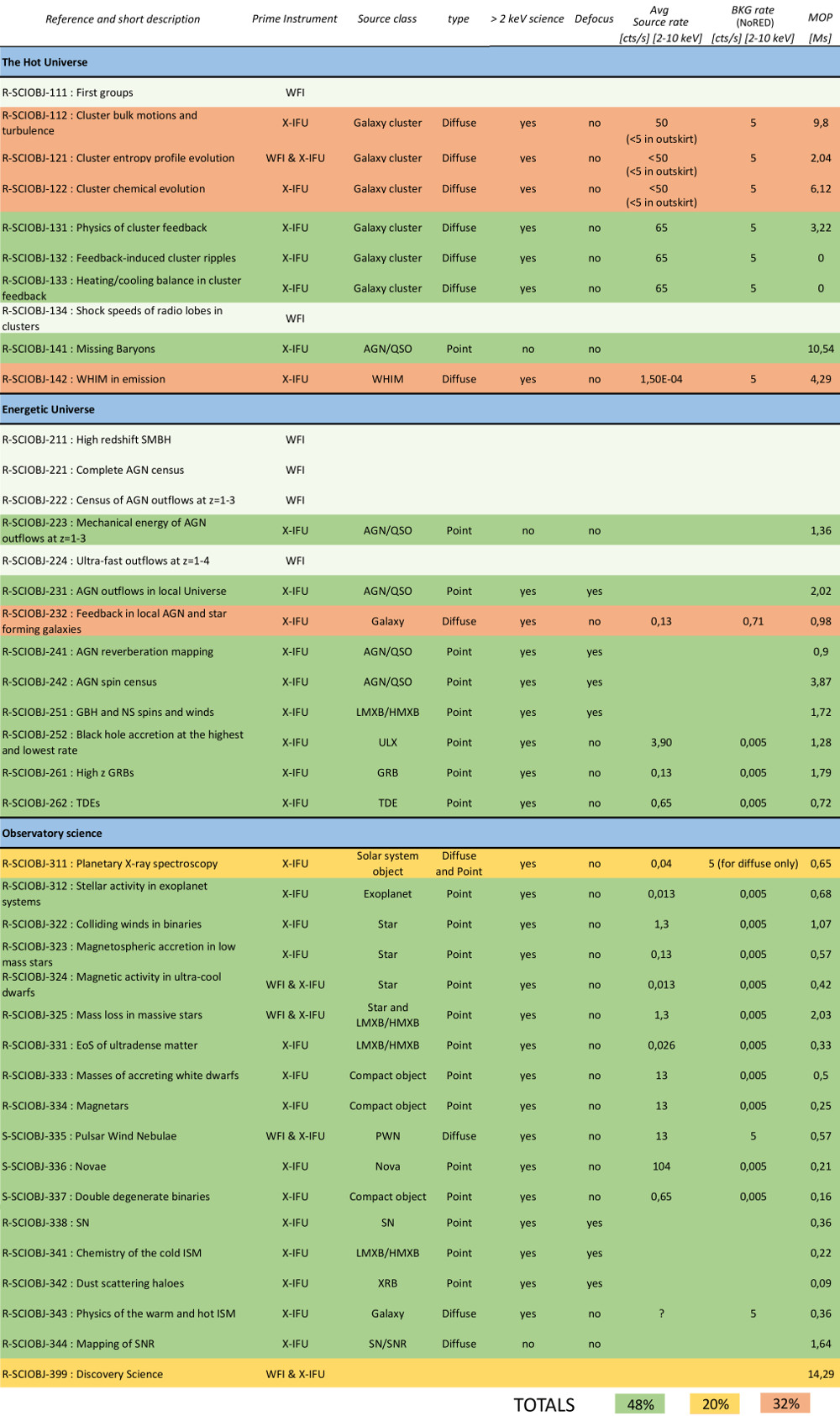}
\caption{Rough study of the ATHENA MOP, where the color code shows when the particle background rate is dominant with respect to the source one, assuming that no background reduction technique is applied to the X-IFU (red: background domination, green: source domination, yellow: sources partially or possibly background dominated). The percentages on the bottom refers to the observing time.}
\label{xibkg}
\end{table}

\restoregeometry

\newpage
\subsection{Background impact on highly obscured AGNs observation}

In order to show the importance of the background reduction also in possible X-IFU observations outside the current MOP, in this subsection I report as case study the simulation of the observation of a distant Compton Thick AGN. As shown in par. \ref{energeticuniverselab}, the detection of highly obscured AGNs is one of the main science requirements of the ATHENA mission. The corresponding science objective (\textit{R-SCIOBJ-221: Complete AGN census}) will be addressed mainly by the WFI intrument, thanks to its advanced survey capabilities, but in the following I will show that also the X-IFU could play an important role in this context.

I have simulated the observation of XID-202, the first higly obscured AGN ($N_H = 0.88 \cdot 10^{24}$ cm$^{-2}$) detected at cosmological distance (z=3.70) in the ultra-deep ($>$ 3 Ms) XMM-Newton survey in the Chandra deep field South \cite{xid202}. XID202 has a flux of 3.9$\cdot$10$^{-15}$ erg cm$^{-2}$ s$^{-1}$ in the 0.5-10 keV band, ad a \textit{rest frame} luminosity  $\sim$ 6 $\cdot$ 10$^{44}$ erg s$^{-1}$ in the 2-10 keV band. To perform the simulation I have used the software XSPEC \cite{xspec}. XSPEC is an X-ray spectral-fitting program, where it is possible to define the model of an astrophysical source and to simulate its observation with a given instrument and a given background level. The main feature of the X-ray spectrum of a Compton Thick AGN is the presence of a strong $K_{\alpha}$ iron emission line, which can allow the source detection and its redshift determination. The X-ray continuum is typically an absorbed power law, and it is useful to characterize some AGN fundamental properties (i.e. the obscuring gas column density). To model the AGN spectrum in XSPEC I have therefore used the following model:

\begin{equation}
\text{XID-202 Model} = zwabs \cdot (zpowerlw + zgauss)
\label{model}
\end{equation}
where \textit{zpowerl}, \textit{zgauss} e \textit{zwabs} are XSPEC models that respectively represent: a redshifted power-law (to model the continuum AGN emission), a redshifted gaussian line (to model the K$\alpha$ iron line) and the photo-electric absorption due to a distant absorber (to model the absorbtion due to dust and gas). The parameters of the model are reported in Tab. \ref{table:xspecpar}.

\begin{table}[H]
\centering
\begin{tabular}{cccc}
\toprule
Component & Parameter & Value & Unit\\
\midrule
zwabs & nH & $88\cdot 10^{22}$ & cm$^{-2}$\\
zwabs & z & 3.7 &\\
zpowerlw & $\Gamma$ & 1.82 &\\
zpowerlw & z & 3.7 &\\
zpowerlw & $N$ & $1.9\cdot10^{-5}$ & ph/(keV cm$^{-2}$ s$^{-1}$) \\
zgauss & $E_L$ & 6.38 & keV \\
zgauss & $\sigma$ & 0 & keV \\
zgauss & z & 3.7 &\\
zgauss & I & $4.0\cdot10^{-6}$ & ph/(cm$^{-2}$ s$^{-1}$) \\
\bottomrule
\end{tabular}
\caption{XID 202 model parameters. From: \cite{xidlotti}.}
\label{table:xspecpar}
\end{table}

The informations about the optics effective area, the detector quantum efficiency and its energy resolution are included in the X-IFU response matrices that can be found in the official site of the instrument \cite{xifusite}. I have used the latest response matrices, which have been produced for the so-called Athena cost-constrained configuration, with an effective area of $\sim$ 1.4 m$^2$ at 1 keV.

\begin{figure}[H]
\centering
\includegraphics[width=1\textwidth]{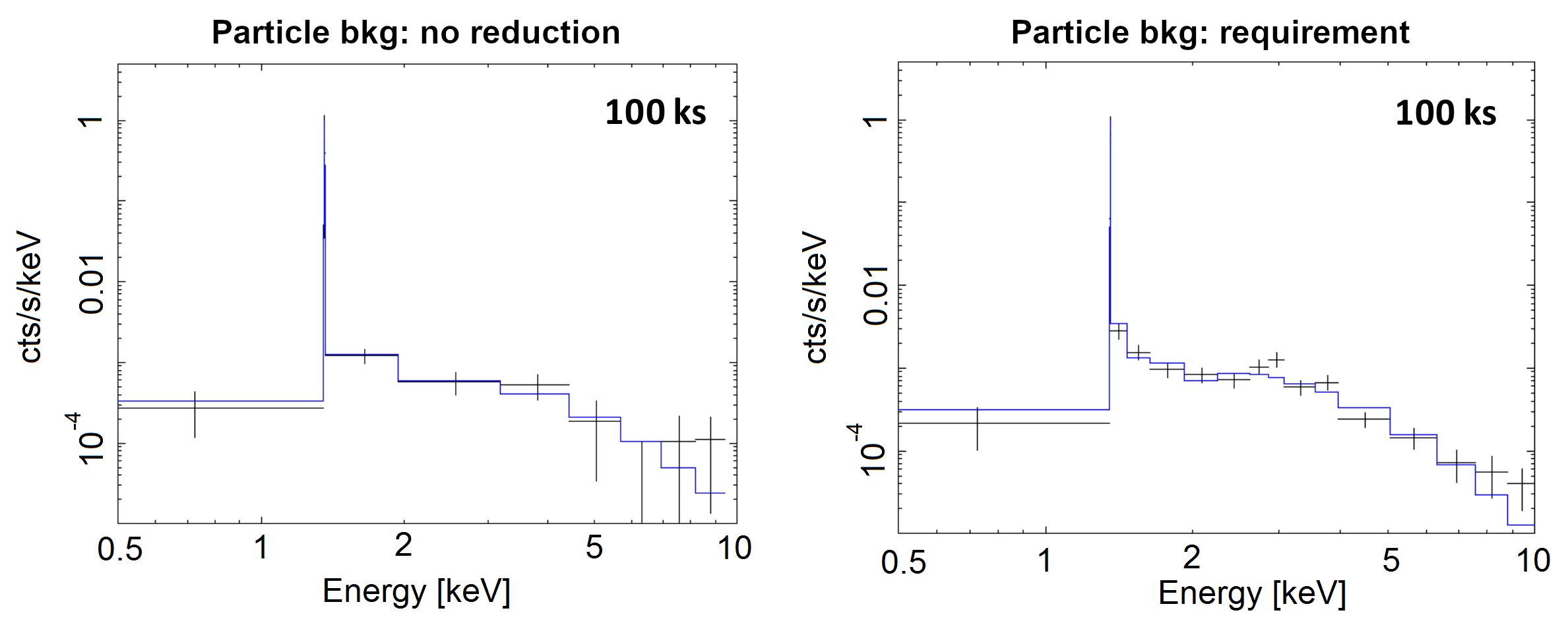}
\caption{Simulated X-IFU 100ks observation of the XID 202 highly obscured AGN, with two different particle background levels. \textit{Left:} simulation with the non-reduced background level. \textit{Rigth:} simulation with the requirement background level.}
\label{xid202}
\end{figure}

In Fig. \ref{xid202} the spectra obtained simulating an X-IFU 100 ks observation with two different background level are shown: without any background reduction technique (left plot, particle background set at 0.23 cts/cm$^{-2}$ s$^{-1}$ keV$^{-1}$), and with the requirement background level (right plot, particle background set at 5$\times$10$^{-3}$ cts/cm$^{-2}$ s$^{-1}$ keV$^{-1}$). The number of counts due to the source is $\sim$ 650, of which $\sim$ 290 in the iron line. The background counts are instead  $\sim$ 600 without reduction (500 background particles and 100 X-ray background photons), and $\sim$ 110 in the requirement configuration (10 background particles and 100 X-ray background photons). The simulated spectra have been fitted (blu lines) in order to assess the accuracy by which the source parameter could be recovered. Before fitting the spectra I have used the \textit{grppha} routine of the FTOOLS software package \cite{ftools} to rebin the data until each bin has a minimum of 30 counts. The best-fit parameters and the Signal to Noise Ratio (SNR) of the simulated observations are reported in Tab.\ref{table:xid202} 

\begin{table}[H]
\centering
\begin{tabular}{ccccc}
\toprule
Background & nH ($10^{22} \; cm^{-2})$ & $\Gamma$  & SNR\\
\midrule \\
Without reduction & $60^{+71}_{-42}$ & $1.3^{+1.4}_{-0.9}$ & 18\\\\
Requirement & $96^{+36}_{-21}$ & $1.8^{+0.5}_{-0.4}$ & 24\\\\
\bottomrule
\end{tabular}
\caption{The errors refer to a 90\% confidence level.}
\label{table:xid202}
\end{table}

The strong iron line is correctly identified in both the cases, and the source redshift z=3.7 is always recovered with negligible relative errors. The other spectal parameters (column density N$_{H}$ and power-law index $\Gamma$) are instead significantly better determined in the \virg requirement'' configuration. Note that the SNR goes from 18 without background reduction, to 24 with the requirement background. For comparison the same object observed with XMM for 3 Ms has a SNR $\sim$ 14. 

Given the count rate of the source and the different background levels, it is possible to determine the observation time $t$ needed to detect the source assuming a detection threshold of 5$\sigma$, by solving:

\begin{equation}
\dfrac{s \cdot t}{\sqrt{(s + b) \cdot t}} > 5
\end{equation}
where $s$ and $b$ are the source count rate and the background count rates, respectively. We obtain the following detection times:  $t > 7.5$ ks without background reductions and $t > 4.5$ ks in the requirement configuration. 

\bigskip
I have then repeated the simulation scaling the AGN flux of a factor 10. The \virg new'' AGN has a flux of 4$\cdot$10$^{-16}$ erg cm$^{-2}$ s$^{-1}$ in the 0.5-10 keV band, and a \textit{rest frame} luminosity of $\sim$ 6 $\cdot$ 10$^{43}$ erg s$^{-1}$ in the band 2-10 keV.

Given 100 ks of observation, the source counts are now 67, while those due to the background are unchanged (600 without reduction and 110 in the requirement configuration). In this case the source is detected only in the requirement configuration (SNR = 5), whereas without background reduction techniques we have SNR = 2.6, too low for a detection (with this background level it would be needed an observation time of $\sim$ 370 ks to have a detection). This is a demonstration of how the background level influences the minumum detectable flux of an istrument.

To conclude this case study, let us evaluate how many CT AGN we can expect to detect with ATHENA X-IFU in the requirement configuration. Using as reference the models reported in \cite{gilli}, we expect to found about 60 CT AGN per deg$^{2}$ with 2 $<$ z $<$ 7 and flux higher than $4\cdot 10^{-16}$ erg cm$^{-2}$ s$^{-1}$ \cite{gillisite}. Assuming now that at least the 50\% of the ATHENA mission lifetime (4 years) will be spent in high Galactic latitude surveys with t $>$ 100 ks with X-IFU, assuming an 80\% observation efficiency we expect to serendipitously find $\sim$ 170 CT AGN. This is a good result, that will be achievable only thanks to the CryoAC and the other background reduction techniques.

\chapter{The X-IFU background: estimates and reduction techniques}
\chaptermark{Background estimates and reduction techniques}
\label{beckmark}

The X-IFU performance would be strongly degraded by the particle background expected in the ATHENA L2 halo orbit, thus advanced reduction techniques - including the Cryogenic AntiCoincidence detector (CryoAC) - have been adopted to reduce this contribution by a factor $\sim$ 50 down to the requirement of 0.005 cts cm$^{-2}$ s$^{-1}$ keV$^{-1}$ (between 2 and 10 keV). This is needed to enable many core science objectives of the mission, like the study of the hot plasma in galaxy cluster outskirts and the characterization of highly obscured distant AGNs.

Given the importance of this topic, a great effort is being made to proper characterize the L2 environment and to accurately estimate the expected residual background for the X-IFU. This is not a trivial problem, due to the lack of experimental data relevant to the background of X-ray detectors in L2, and also to the complexity of the geometry and the particles interaction processes in the satellite, thus requiring the use of advanced Monte Carlo simulations to address the problem. In this context, ESA is supporting R$\&$D projects entirely dedicated to the ATHENA background issue, like AREMBES \cite{arembes} and EXACRAD \cite{exacrad}. The results of the these activites finally drive the design of the X-IFU FPA and the CryoAC.

In this chapter, after a brief introduction about the background issue for X-rays detectors, I will review the latest estimates of the main components that constitute the X-IFU background, introducing the reduction techniques that have been adopted in the instrument design. Finally, I will present the CryoAC conceptual design, showing how the top levels requirements translate into the anticoincidence detector specifications.

\section{An introduction to the background issue}

Observations in the X-ray waveband are often limited by the background, due to the low fluxes of the sources compared to other energy bands. Consider that the Crab nebula, which is used as calibrator, it is considered a bright X-ray source with a flux of $\sim$ 3 photons cm$^2$ s$^{-1}$ in the 1 - 10 keV band. The background can be therefore easily comparable, and in several cases also higher than the signal itself. In this case, the Signal to Noise Ratio (SNR) of an observation is given not only by the statistical fluctuations of the source count rate, but depends also on the background level. The background adds poissonian fluctuations to the total number of counts, thus reducing the SNR and the detection capabilities of the instrument. The SNR can in fact be expressed as \cite{fraser}:

\begin{equation}
\text{SNR} = \dfrac{S}{\sqrt{C}} = \dfrac{S}{\sqrt{S + B}}
\label{sn}
\end{equation} 

\noindent where S and B are respectively the source and the background net counts, and C=S+B are the total counts detected by the instrument.

For any satellite operating in the X-ray band, the background is composed of a diffuse X-ray component (X-ray Background) and an internal particle component (Particle Background, also called NXB - Non X-ray Background). The first includes any photon coming from astrophysical sources beside the source of interest and depositing energy within the detector energy band. The second is generated by particles traveling through the spacecraft and releasing energy inside the detector, also creating swarms of secondary particles along the way. Referring to the source flux as $F$ [ph cm$^{-2}$ s$^{-1}$ keV$^{-1}$], to the diffuse X-ray background as $b_x$ [ph cm$^{-2}$ s$^{-1}$ keV$^{-1}$ sr$^{-1}$] and to the internal particle background as $B_p$ [cts cm$^{-2}$ s$^{-1}$ keV$^{-1}$], it is possible to re-write eq. (\ref{sn}) as:

\begin{equation}
\text{SNR} = \dfrac{F \cdot A_{eff} \sqrt{t \cdot \Delta E}}{\sqrt{F \cdot A_{eff} + b_x \cdot \Omega \cdot A_{eff} + B_p \cdot A_d}}
\label{eqn:snr1b}
\end{equation}

\noindent where t [s] is the observation time, $\Delta E$ [keV] is the instrument energy bandwidth, $\Omega$ [sr] is the angular size of the source, A$_{eff}$ [cm$^2$ cts/ph] is the effective area of the instrument (including the quantum efficienty of the detector) and A$_d$ [cm$^2$] is the physical area of the detector in which the signal is collected. 

Note that the collection area $A_d$ is related to the angular size of the source by the square focal lenght $f_L$ of the telescope: $A_d = f_L^2 \cdot \Omega$. The eq. (\ref{eqn:snr1b}) can be therefore written as:

\begin{equation}
\text{SNR} = \dfrac{F \sqrt{A_{eff} \cdot t \cdot \Delta E}}{\sqrt{F + (b_x + B_p \cdot f_L^2/A_{eff})\cdot \Omega}}
\label{eqn:snr2}
\end{equation}

\noindent 
where $B_P \cdot f_L^2/A_{eff}$ is the particle background per sky solid angle. Given the values of $b_x$ and $B_P$, the importance of the internal particle background component with respect to the diffuse X-ray one should be evaluated taking into account the $f_L^2/A_{eff}$ ratio, which can be therefore used as figure of merit in an X-ray mission design.

From the relations (\ref{eqn:snr1b}) and (\ref{eqn:snr2}) it is possible to obtain the minimum flux F$_{min}$ detectable by an X-ray detector for a desired significance level n$_\sigma$. In the case of a background-dominated observation (S $<<$ B), it is:

\begin{equation}
F_{min} = \frac{n_\sigma}{A_{eff}} \sqrt{\dfrac{b_x \cdot \Omega \cdot A_{eff} + B_p \cdot A_d}{t \cdot \Delta E}} = n_{\sigma} \sqrt{\dfrac{(b_x + B_p \cdot f_L^2/A_{eff})\cdot \Omega}{A_{eff}\cdot t \cdot \Delta E}}
\label{eqn:fluxmin}
\end{equation}

The relation (\ref{eqn:fluxmin}) shows the necessity to reduce the background level in order to increase the capabilities of an X-ray intrument, thus enabling the detection of faint and diffuse sources. The X-ray Background can be reduced combining high energy resolution (to resolve and isolate its lines component) with high angular resolution (to resolve the individual point sources that contributes to this signal). The particle background can be instead reduced using specific reduction techniques, such as magnetic diverters, active anticoincidence devices and passive particle shields.

\newpage
\section{The X-IFU background}

\begin{figure}[htbp]
\centering
\includegraphics[width=0.7\textwidth]{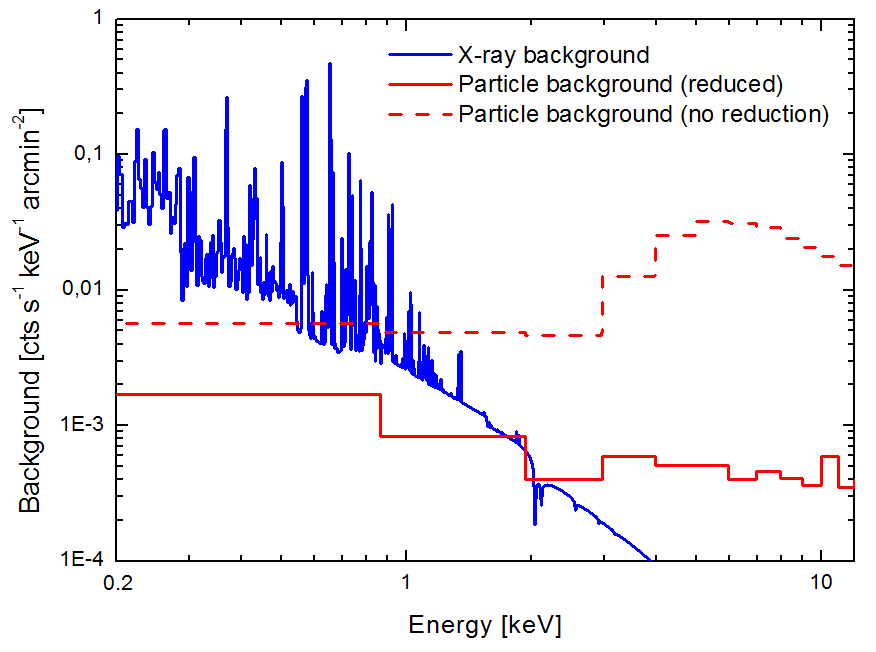}
\caption{The expected diffuse X-ray background (blu line) and internal particle background per solid angle (red line) for the X-IFU baseline configuration. The dashed red line shows the particle background in the case that no reduction technique is applied to the instrument.}
\label{bx}
\end{figure}

In Fig. \ref{bx} the expected diffuse X-ray background (blu line) and internal particle background per solid angle (red line) in the X-IFU baseline configuration are shown. The level of the particle component is also shown in the case that no background reduction technique is applied to the instrument (dashed red line). Note that in the baseline configuration the particle contribution is dominant with respect to the X-ray one for energies above 2 keV, while without reduction techiques the particle background would be dominant down to an energy $<$ 1 keV.

\noindent Referring to what is reported in \cite{lottiahead}, the main contributions to the two background components are listed below.

\bigskip

\noindent The \textbf{Diffuse X-ray Background} contains two main components:

\begin{itemize}

\item \textbf{Galactic X-ray Background} - At energies below $\sim$ 1 keV the line emission from the hot diffuse gas in our galaxy is dominant. A first important contribution is given by $\sim$ 2$\cdot$10$^6$ K gas in the Galactic halo. A second component is generated by the Local Hot Bubble (LHB): a $\sim$ 100 pc radius cavity around the solar system filled by hot rarefied interstellar gas with T $\sim$ 10$^6$ K. A last contribution is due to the solar wind charge exchange (SWCX) mechanism, that produces soft x-ray emission when heavy ions inside the solar wind (mostly oxygen) interact with the neutral gas in the solar system. 

\item \textbf{Cosmic X-ray Background (CXB)} - Above $\sim$ 1 keV an absorbed powerlaw of extragalactic origin dominates the X-ray Background spectrum. It is produced by the unresolved emission of distant X-ray sources like quasars, distant galaxy clusters and AGNs. As the observing time spent on the same field increases, this emission can be resolved into the single point sources of which it consists. For the X-IFU it is estimated that about the 80$\%$ of the CXB should be resolved \cite{xidlotti}.

\end{itemize}

\noindent The \textbf{Internal Particle Background} is induced by two different particle populations:

\begin{itemize}
\item \textbf{Soft solar protons} - The Sun contributes to the particle background via a strongly variable component of emitted soft protons (E $<$ $\sim$ 100 keV) that can be funnelled by the optics toward the focal plane. This component can be easily suppressed with the use of a magnetic diverter aimed to deflect the charged particles outside the FoV of the instruments.

\item \textbf{Galactic Cosmic Rays (GCR)} - Most of the particle background is generated by high energy GCR particles (E $>$ $\sim$ 100 MeV) crossing the spacecraft and depositing inside the TES array a fraction of their energy. These particles create secondaries along their way, which can in turn impact the detector inducing further background counts. The GCR-induced background component is mainly reduced with the presence of the Cryogenic Anti-Coincidence detector (CryoAC), and with the use of a passive shielding to reduce the flux of secondary particles.
\end{itemize}

\noindent In the next sections it will be shown in more detail how the different X-IFU background components are estimated/modelled.

\subsection{Diffuse X-ray Background}

The Diffuse X-ray Background is described following the model reported in \cite{xidlotti}, whose parameters are extracted from the observation of a 1 sr sky region centered on l = 90$^o$, b = +60$^o$ (galactic coordinates) performed by an array of microcalorimeters flown on a sounding rocket. This sky area avoids emission features, such as the Scorpion-Centaurus superbubble, and should be representative of typical high Galactic latitude pointings \cite{xidlotti}. Inside XSPEC, the model is given as the sum of a thermal emission \textit{apec} model\footnote{It is an emission spectrum from collisionally ionized diffuse gas that includes line emissions from several elements} to represent the hot gas in the galactic halo, and a \textit{powerlaw} model to represent the unresolved CXB component, both multiplied by an \textit{awabs} model to account for the interstellar medium absorption at low energies.  In addition, an unabsorbed thermal \textit{apec} component is used to describe the contributions from the Local Hot Bubble (LHB) and the Solar Wind Charge eXchange (SWCX):

\begin{equation}
\text{XSPEC model: } apec + wabs*(apec + powerlaw)
\end{equation}

The model parameters are reported in Table \ref{tab:diffusebkg} (from \cite{xidlotti}) and assume different normalizations of the extragalactic \textit{powerlaw} component: while in the observation of point sources the unresolved CXB emission must indeed be completely taken into account, in the observation of extended sources it can be mostly resolved. It is conservatively assumed that in the observation of diffuse sources the CXB component can be resolved to up to 80\% \cite{xidlotti}. 

\begin{figure}[H]
\centering
\includegraphics[width=0.8\textwidth]{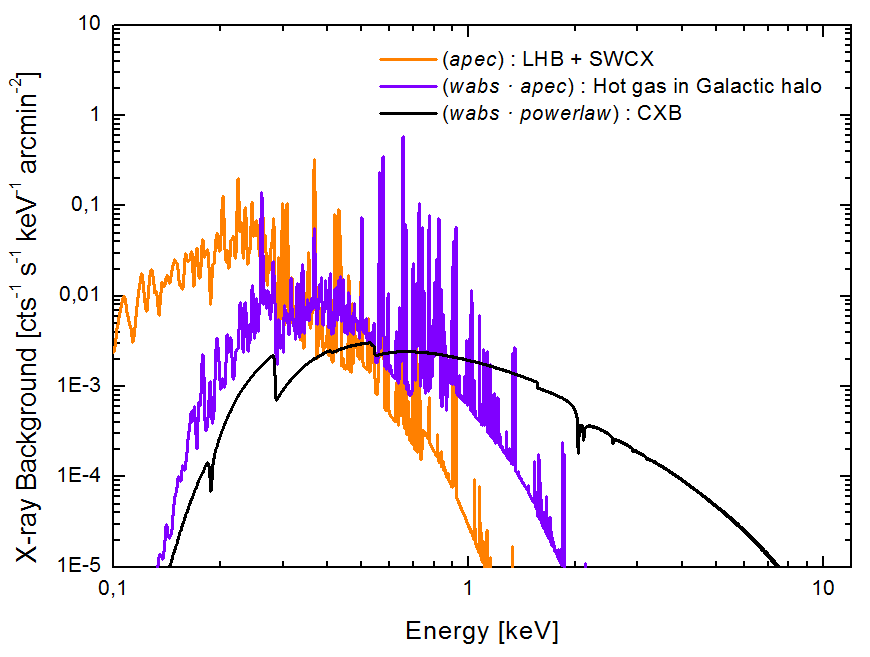}
\caption{Different contributions to the X-IFU diffuse X-ray background modelled inside XSPEC with the parameters reported in tab. \ref{tab:diffusebkg}.}
\label{xbkg}
\end{figure}

The different contribution to the X-ray background are finally shown in Fig. \ref{xbkg}, where the corresponding XSPEC models have been convolved with the X-IFU effective area.

$$  $$

\begin{table}[H]
\centering
\caption{Parameters for the XSPEC diffuse background model, from: \cite{xidlotti}.}
\label{tab:diffusebkg}
\begin{tabular}{ccccc}
\hline\noalign{\smallskip}
Component & Model & Parameter & Unit & Value \\
\noalign{\smallskip}\hline\noalign{\smallskip}
LHB + SWCX&\textit{apec} & $kT$ & keV & 0.099 \\
LHB + SWCX&\textit{apec} & abundance & & 1 \\
LHB + SWCX&\textit{apec} & redshift & & 0.0 \\
LHB + SWCX&\textit{apec} & norm (1 arcmin$^2$ source)& & 1.7 $\times$ $10^{-6}$ \\
ISM absorbtion&\textit{wabs} & $n_H$ & $10^{22}$ & 0.018 \\
Hot halo gas&\textit{apec} & $kT$ & keV & 0.225 \\
Hot halo gas&\textit{apec} & abundance & & 1 \\
Hot halo gas&\textit{apec} & redshift & & 0.0 \\
Hot halo gas&\textit{apec} & norm (1 arcmin$^2$ source)& & 7.3 $\times$ $10^{-7}$ \\
CXB &\textit{powerlaw} & photon index & & 1.52 \\
CXB &\textit{powerlaw} & norm (point surces) & & 1.0$\times$ $10^{-6}$ \\
CXB &\textit{powerlaw} & norm (1 arcmin$^2$ source) & & 2.0 $\times$ $10^{-7}$ \\
\noalign{\smallskip}\hline
\end{tabular}
\end{table}

\bigskip

$$ $$

\bigskip

\newpage
\subsection{Internal Particle Background: Soft solar protons}

The soft protons contribution to the X-IFU particle background has been been studied in detail in \cite{lottiahead} and \cite{lottispexp}. This proton flux can be easily damped with the use of a magnetic diverter (already used to deflect electrons in the XMM-Newton \cite{XMM} and Swift \cite{Swift} missions), so these studies first present an estimation of this component in absence of such a device, and then derive its requirements to reduce the soft protons induced background below the level required to enable the mission science.

\begin{figure}[H]
\centering
\includegraphics[width=0.9\textwidth]{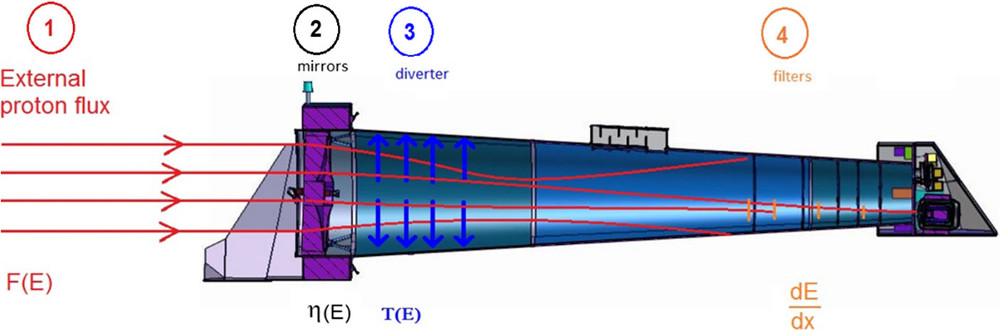}
\caption{Schematics of the steps followed by the soft protons through the telescope: 1) we start with the external soft protons environment, 2) the protons interact with the mirrors, with funnelling efficiency $\eta$, 3) the protons are deflected by the magnetic diverter with transmission efficiency $T$, 4) the protons experience energy loss $dE/dx$ inside the thermal filters before reaching the detector. From \cite{lottiahead}.}
\label{softprotons}
\end{figure}

The estimation of the soft proton background contribution is performed dividing the problem into steps, according to the path followed by the protons towards the ATHENA telescope (see Fig. \ref{softprotons}): at first it is determined the external fluxes impacting on the optics, then using ray-tracing simulations it is derived the mirrors funnelling efficiency, then it is computed the energy lost crossing the thermal filters and finally, put everything together, it is evaluated the flux expected on the instrument. 

\bigskip

\begin{figure}[H]
\centering
\includegraphics[width=0.6\textwidth]{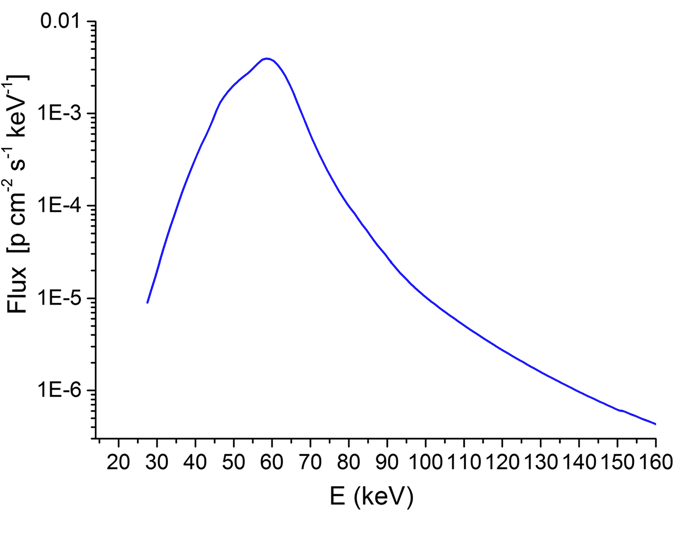}
\caption{Spectrum of the primary energies of the protons that reach the focal plane depositing in-band energy on the X-IFU without a Magnetic Diverter. The worst case scenario for the external proton flux has been assumed. See \cite{lottispexp} for details.}
\label{softprotonfluxes}
\end{figure}

Assuming the worst-case scenario for the external proton flux, in absence of a magnetic diverter it is expected for the X-IFU a soft proton induced background of 4$\times$10$^{-2}$ cts cm$^{-2}$ s$^{-1}$ keV$^{-1}$ \cite{lottispexp}. The major contribution is given by protons with intermediate energies: $\sim$ 80$\%$ of the background is given by protons with primary energy 45 keV $<$ E $<$ 65 keV. The spectra of the initial energies of the soft protons that reach the focal plane depositing in-band energy on the detector is reported in Fig. \ref{softprotonfluxes}.

Since the requirement for the soft protons induced background is 5 $\times$ $10^{−4}$ cts cm$^{−2}$ s$^{−1}$ keV$^{-1}$ (which is $\sim$ 10\% of the total Particle Background requirement), it is mandatory to adopt solutions to reduce or discriminate about the 90$\%$ of the proton flux at the focal plane level. As reported in \cite{lottispexp}, this can be done with a magnetic diverter able to deflect every particle with an energy below $\sim$ 70 keV and thus requiring, depending on the geometry of the system, a magnetic field in the 0.1 - 0.8 T range.

\subsection{Internal particle background: Galactic Cosmic Rays (GCR)}
\label{NXBsim}
No X-Ray mission has been operated in the L2 environment so far, thus it is not possible to estimate the expected particle background level relying on existing data\footnote{The Standard Radiation Environment Monitor (SREM \cite{srem}) on board the two L2 ESA missions Herschel and Planck have collected useful data, but they are not sufficient to properly characterize the high energy particle L2 environment as required for a X-ray mission \cite{sreml2}. Also the GCR impact on the Planck HFI low temperature detectors has been deeply studied \cite{planck}, but it is not possible to obtain complete information about the L2 environment from these data without known the particles path and their interactions inside the satellite.}. Concerning the GCR-induced component (which is the dominant one) the problem is also too complex to be treated by an analytical approach, and consequently the only way to estimate the background is by detailed Monte Carlo simulations, which are performed using the Geant4 toolkit \cite{geant4}. In order to create a reliable simulation three elements are required: 

\begin{itemize}
\item A model of the particle environment expected in the L2 environment, which is estimated extrapolating the data acquired from other satellites;
\item A representative mass model of the satellite, which shall be especially accurate for as regards the detector and its sourroundings. Note that the CAD drawing is too detailed for simulation, so it is is mandatory to develop a representative \virg reduced'' model;
\item The most appropriate physics list, i.e. a reliable set of physics models to describe the interaction of the particle environment with the mass model;
\end{itemize}

\noindent A detailed description of the Geant4 simulation developed to estimate the X-IFU background is reported in \cite{lottispexp}. Here I will quickly review the main simulation elements, finally reporting the latest simulation results. 

$$ $$

\subsubsection{L2 environment}

\begin{figure}[H]
\centering
\includegraphics[width=1.0\textwidth]{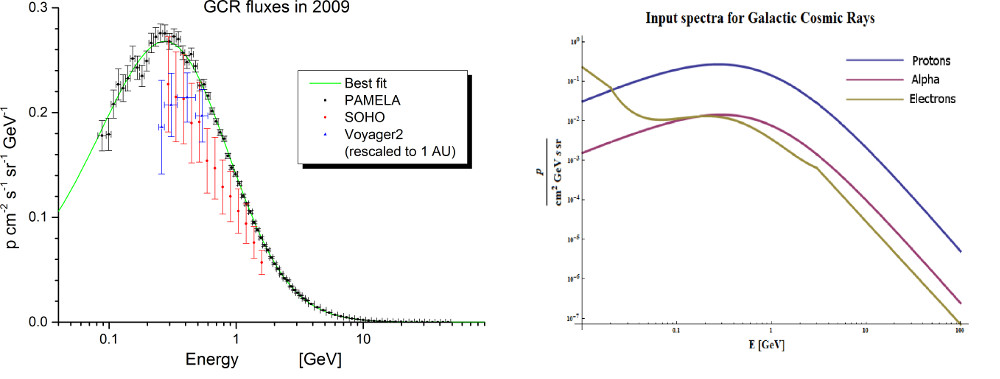}
\caption{(\textit{Left)}: The best fit curve (in green) of the GCR proton spectra obtained by the datasets from PAMELA, SOHO and Voyager2 . \textit{(Right)}: The new reference input spectra for GCR protons, alpha particles and electrons used in the simulations. From \cite{lottispie2018}}
\label{inputspectra}
\end{figure}

The dominant high energy particle contribution in the L2 environment is due to GCR protons \cite{lottispie2018}. The absolute level of the GCR protons spectrum and its uncertainty during the ATHENA mission lifetime (2031-2034) have been recently assessed by a dedicated study \cite{minervini}.  For this purpose data from the Voyager2, SOHO and PAMELA satellites and neutron monitor (NM) measurements have been analyzed and compared with the CREME96 and CREME2009 models previously used as input for the simulations. The comparison has shown that the spectra foreseen by CREME are not fully in accordance with the experimental data, and can underestimate the measured GCR flux by up to $\sim$ 3 times \cite{minervini}. As a consequence, a new reference GCR proton spectrum has been developed. This spectrum is reported in Fig. \ref{inputspectra}, where for sake of completeness are also shown the adopted alpha and electrons GCR components.

\bigskip

\subsubsection{Mass model}

Developing an accurate mass model of the instrument, especially in the detector proximity, it is crucial to proper estimate the particle background level, since it strongly depends on the materials, their placement, their shapes, and on the total mass shielding the detector from radiation.

\begin{figure}[H]
\centering
\includegraphics[width=0.8\textwidth]{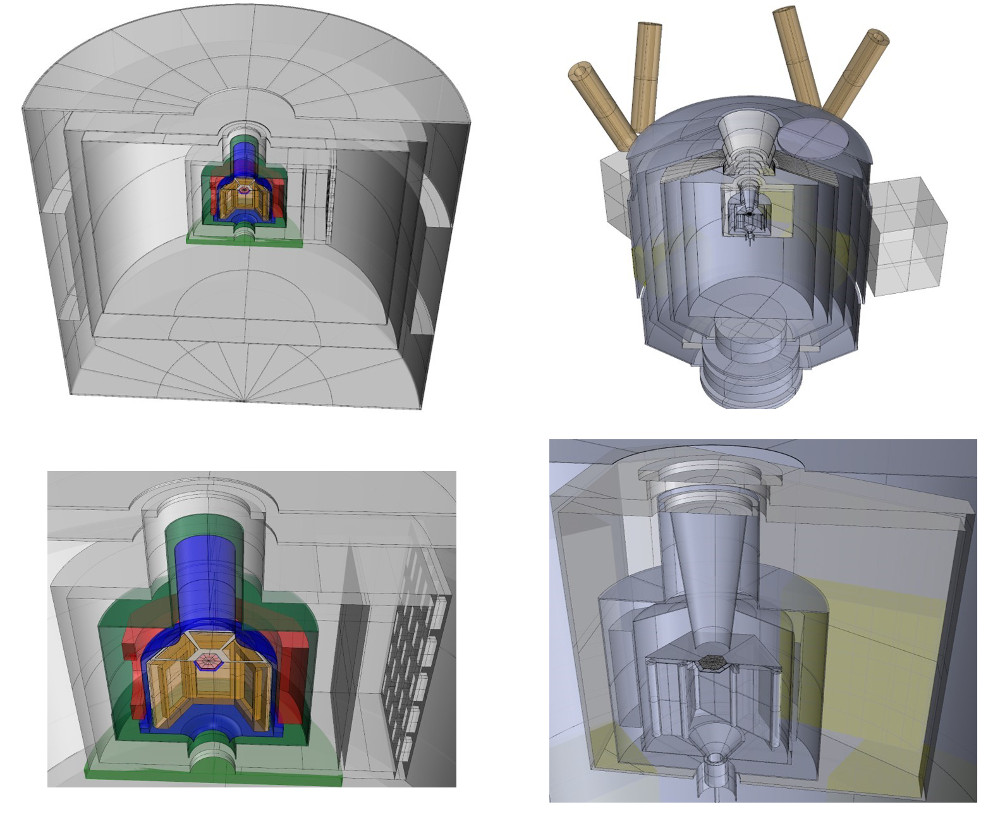}
\caption{The old and new cryostat mass models (top left and right, respectively) and FPA (bottom left and right).}
\label{massmodel}
\end{figure}

The X-IFU mass model used in the Geant4 simulation has been recently updated (on June 2017) starting from the updated CAD models of the cryostat (developed by CNES) and the FPA (developed by SRON) \cite{lottispie2018}. I have directly contributed to this activity, being in charge of the simplification of the CAD models before their insertion in the Geant4 toolkit. The comparison between the old and the new mass models is shown in Fig. \ref{massmodel}. With respect to the previous mass model, beside the revision of the structures surrounding the detector, we have updated the TES absorbers thicknesses from 4 $\mu$m to 4.2 $\mu$m for Bismuth, and from 1 $\mu$m to 1.7 $\mu$m for Gold, inserted a more realistic model of the thermal filters, and the aperture cylinder sustaining them.

The mass model have been then inserted into Geant4, where the different solids have been finally assigned to different \virg regions'', each with different settings of the cut for the generation of secondary particles:

\begin{itemize}
\item The detector, the supports, and the surfaces directly seen by the detector were assigned to the \virg inner region'' with the lowest possible cut values (few tens of nm, high detail level);
\item The remaining solids in the FPA were assigned to an \virg intermediate region'' with higher cut values (few $\mu$m);
\item The cryostat and the masses outside the FPA were assigned to the \virg external region'' were the cut (few cm) allowed the creation only of high energy secondary particles.
\end{itemize}

\subsubsection{Physics list}

In the Geant4 toolkit, there are several models representing the same physical processes, with different internal settings and applicability conditions. The so-called Physics List file defines which models to use during the simulation, and also their internal settings. The Geant4 consortium provides several pre-assembled Physics Lists, optimized for different purposes. 

The X-IFU background simulations initially used a pre-assembled Physics List that, according to software developers, was the most suited for space-based applications, namely the G4EmStandardPhysics-option4 (Opt4). Recently, in the context of the AREMBES ESA programme \cite{arembes}, it has been carried out an activity aimed to pinpoint the most suited physical models and settings to simulate the ATHENA detectors behavior in space. A custom Physics List dedicated to ATHENA, namely the \virg Space Physics List'' (SPL), has been finally defined comparing the performances of different Geant4 models in reproducing the available experimental data for the physical processes involved. It has been officially endorsed by ESA for the X-IFU, and it is now adopted as reference settings for the simulations. 

\subsubsection{Simulation results}
\label{bkgresults}

The latest background simulations \cite{lottispie2018} have been performed using the new GCR proton spectrum, the updated X-IFU mass model and the Space Physics List. Different FPA configurations have been tested to assess the expected particle background level: 

\begin{itemize}
\item FPA without any solution to reduce the particle background;
\item FPA with the insertion of the Cryogenic Anticoincidence detector (CryoAC);
\item FPA with the CryoAC and a passive kapton shield sourrounding the TES array;
\item FPA with the CryoAC and the passive shield in an improved design that foresees a bilayer with 20 $\mu$m of Bismuth and 250 of $\mu$m Kapton.
\end{itemize}

\begin{figure}[H]
\centering
\includegraphics[width=0.7\textwidth]{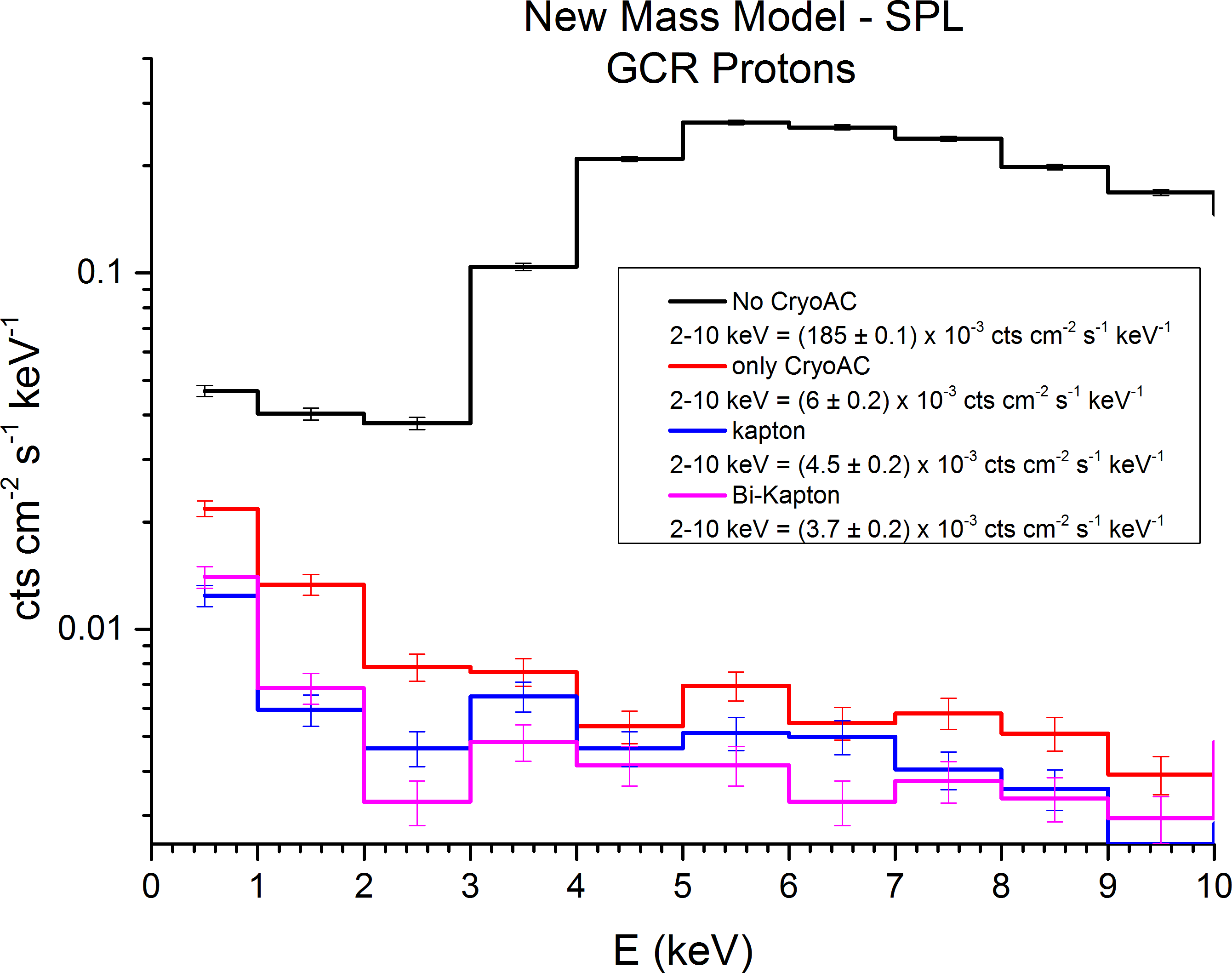}
\caption{Unrejected X-IFU background levels in different configurations of the instrument. From: \cite{lottispie2018}}
\label{pbkg}
\end{figure}

\noindent The background levels resulting from these different configurations using the Galactic Cosmic Ray (GCR) proton spectrum as input flux (note that it constitutes $\sim$ 90\% of the total particle input flux) are reported in Fig. \ref{pbkg}. 

\bigskip

Without any reduction technique (black line in the plot) the X-IFU would experience a particle background level of 0.185 [cts cm$^2$ s$^{-1}$ keV$^{-1}$] in the 2-10 keV energy band, 37 times above the requirement of 5 $\times$ 10$^{-3}$ [cts cm$^2$ s$^{-1}$ keV$^{-1}$].

The insertion of the CryoAC detector (red line) allows to reduce the background of a factor $\sim 30$ down to the level of 6 $\times$ 10$^{-3}$ [cts cm$^2$ s$^{-1}$ keV$^{-1}$], 20\% above the required level. This reduction is achieved effectively removing all the particles (primaries and secondaries) that are able to cross the main detector and reach the anticoincidence detector, such as MIPs, whose typical shape is no longer present in the spectrum.

At this point, the unrejected background is mostly induced by secondary electrons created in the structures surrounding the TES array, which are absorbed into it or bounces on its surface. This contribution is effectively damped by the Kapton shield (blue line), due to its low secondary electrons generation yield. Its insertion allows to reduce the background level by 25\%, reaching the level of 4.5 $\times$ 10$^{-3}$ [cts cm$^2$ s$^{-1}$ keV$^{-1}$]. 

The improved shield design, including the Bismuth layer, is finally able to further reduce the secondary electrons flux towards the detector, by shielding the 16 keV fluorescence line emitted by the Niobium (the inner magnetic shield in the FPA). In this configuration it is reached the \virg optimal'' background level of 3.7 $\times$ 10$^{-3}$ [cts cm$^2$ s$^{-1}$ keV$^{-1}$]. At this point, backscattered electrons have repeatedly proven to be the main component of the residual background: the $\sim$ 85\% of the residual background is induced by secondary electrons, and $\sim$ 80\% of these electrons are backscattered.

\bigskip

\begin{figure}[H]
\centering
\includegraphics[width=0.7\textwidth]{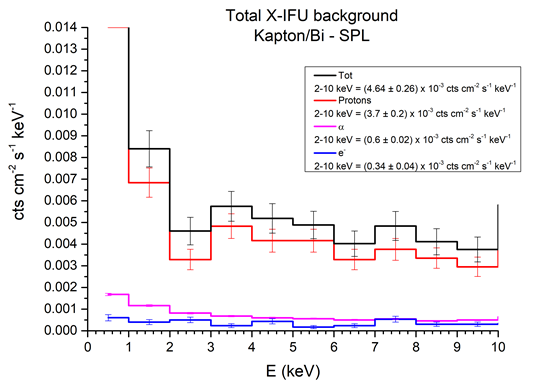}
\caption{The total X-IFU unrejected background level (GCR Protons, alpha particles and e$^-$) in the FPA configuration including the CryoAC and the Kapton/Bismuth passive shield. From: \cite{lottispie2018}}
\label{pbkg2}
\end{figure}

Once the optimal FPA configuration has been defined using only GCR protons as input, the simulation has been repeated accounting also for the remaining components of the L2 environment (electrons and alpha particles), obtaining the complete X-IFU background evaluation shown in Fig. \ref{pbkg2}, corresponding to a particle level of 4.64 $\times$ 10$^{-3}$ [cts cm$^2$ s$^{-1}$ keV$^{-1}$], slightly below the scientific requirement.

\bigskip

\newpage

\newpage
\section{The Cryogenic Anticoincidence Detector (CryoAC)}

As shown in the previous section, most of the particle background reduction in the X-IFU ($\sim$ 80 \%) is achieved thanks to the Cryogenic AntiCoincidence detector (CryoAC), whose presence is therefore mandatory to reach the background scientific requirement. The aim of this section is to provide the Conceptual design of the CryoAC detector, showing how the top-level requirements translates into the detector specification.

\subsection{Requirements}

The top-level particle background requirement for the X-IFU (ref. SCI-BKG-R-020) is formulated inside the \textit{Athena Science Requirements Document} \cite{athenascireq} as follow:

\bigskip

\noindent \textbf{Athena shall achieve a not focused non-X-ray background for High spectral resolution observations of $<$ 5 $\times$ 10$^{-3}$ counts s$^{-1}$ cm$^{-2}$ keV$^{-1}$ in 80\% (TBC) of the observing time between 2 keV and 10 keV.}

\bigskip

\noindent From the Geant4 simulations descends that to reach this requirement it is mandatory to achieve a total rejection efficiency for primary particles of $\sim$99.83\% and a total rejection efficiency for rejectable\footnote{A fraction of the secondaries flux is induced by \virg unrejectable'' particles (i.e., electrons backscattering on the detector surface releasing just a fraction of their energy, low energy particles that are completely absorbed inside the main array switching on only one pixel). It is not possible to veto these events, so a requirement on the rejection efficiency should be placed only on the remaining \virg rejectable'' component of the secondary particles flux.} secondary particles of $\sim$99.73\%.

The total rejection efficiency of the X-IFU system depends on the  distance and sizes of the CryoAC and the TES array (\virg geometrical rejection efficiency'', i.e. the ratio between the number of particles triggering the CryoAC after or before hitting the TES array and the number of particles crossing the TES array), and on the capability of the TES array to discriminate background events on its own, relying on the energy deposited and on the pixel pattern turned on by the impacting particles (\virg detector rejection efficiency'').

\begin{figure}[H]
\centering
\includegraphics[width=0.9\textwidth]{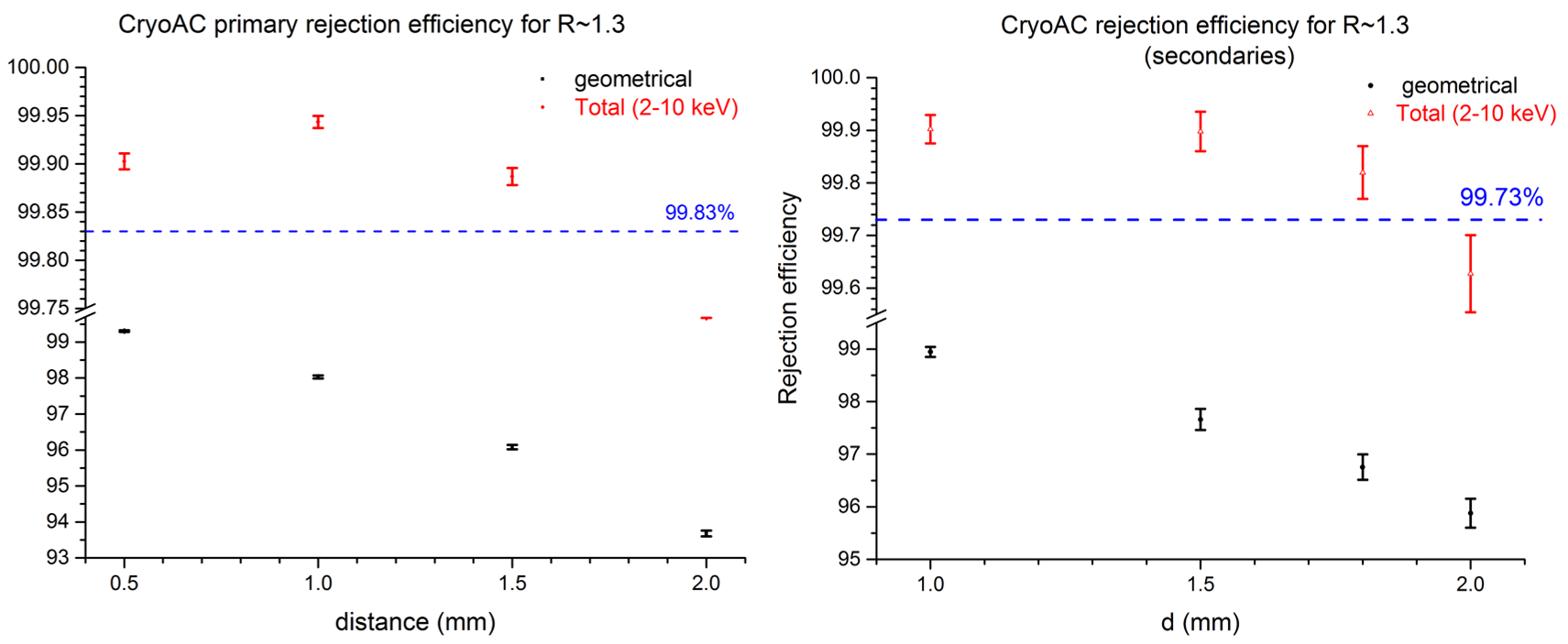}
\caption{Geometrical and total rejection efficiency as function of the distance between the TES array and the CryoAC, for primary particles (left), and secondaries (right).}
\label{cryoacgeomefff2}
\end{figure}

About the geometrical configuration of the system, the simulations show that placing the CryoAC at a distance of 1 mm from the TES array, and assuming a ratio of 1.3 between the linear size of the CryoAC and the TES array, the  geometrical rejection efficiency amounts to 98\% for primaries and to 98.94\% for secondaries (see Fig. \ref{cryoacgeomefff2}). These values allow to reach a total rejection efficiency inside the requirement for both particle species.

\bigskip

Note that, being the CryoAC placed less than 1 mm below the TES array, it must be able to operate at the same bath temperature T$_B$ = 50-55 mK TBC (this is the reason why the anticoincidence is a cryogenic detector), and it is also subjected to a strict power dissipation requirement from the FPA (P$_{CryoAC}$ $<$ 40 nW TBC).

\bigskip

The last requirement concerning the CryoAC is the deadtime requirement for the X-IFU (ref. XIFU-PAYLOAD-R-0260), which is formulated inside the \textit{XIFU performance requirements document} \cite{xifureq} as follow:

\bigskip

\noindent \textbf{The X-IFU dead time shall be lower than 2\% (TBC). Rationale: 1\% for background, 1\% for MXS (Modulated X-ray Sources)}

\bigskip

\noindent The \virg background'' deadtime represents the time after each CryoAC trigger during which the detector is not able to trigger another event. During this time the TES array signals must be discarded since it is not possible to perform the veto procedure (otherwise the TES array will collect background). It is equivalent to the CryoAC intrinsic deadtime, which is the time the detector needs to recover an operating point after a particle event.

\bigskip

The final CryoAC requirements descending from the top-level requirements here reported are summarized in Tab. \ref{table:cryoacrequie} and Tab. \ref{table:cryoacrequie2}

\bigskip

\begin{table}[H]
\footnotesize
\centering
\begin{tabular}{lll}
\toprule
\textbf{Parameter} & \textbf{Value} & \textbf{Comments} \\
\midrule
CryoAC geometrical rejection efficiency for primaries & $>$ 98 $ \% $ & To reach a total rejection efficiency \\
& &   of 99.83 $\%$ for primaries \\
& &   and of 99.73 $\%$ for secondaries \\
& &   \\
CryoAC intrinsic Deadtime & $ \leq $ 1$ \% $ & Total X-IFU deadtime $ < $2$ \% $\\
\midrule
\end{tabular}
\caption{CryoAC \virg scientific'' requirements.}
\label{table:cryoacrequie}
\end{table}

\begin{table}[H]
\footnotesize
\centering
\begin{tabular}{lll}
\toprule
\textbf{Parameter} & \textbf{Value} & \textbf{Comments} \\
\midrule
Thermal bath temperature T$_0$ & 50-55 mK & (TBC) \\
& & \\
Power dissipation at  T$_0$ & 40 nW & (TBC) \\
\midrule
\end{tabular}
\caption{CryoAC \virg FPA'' requirements.}
\label{table:cryoacrequie2}
\end{table}

\subsection{Concept Design and specifications}

To satisfy the requirements listed in the previuos section it has been defined a CryoAC Concept Design, which represents the current baseline for the detector. It is based on an absorber made of a thin single crystal (500 $\mu$m thick) of silicon, where the energy deposited by particles is sensed by TES detectors. This configuration offers important advantages. Being indeed the technology similar to the TES array, the integration, interfaces and SQUID readout issues can be substantially accommodated by using the same technology solutions. 

\begin{figure}[H]
\centering
\includegraphics[width=1\textwidth]{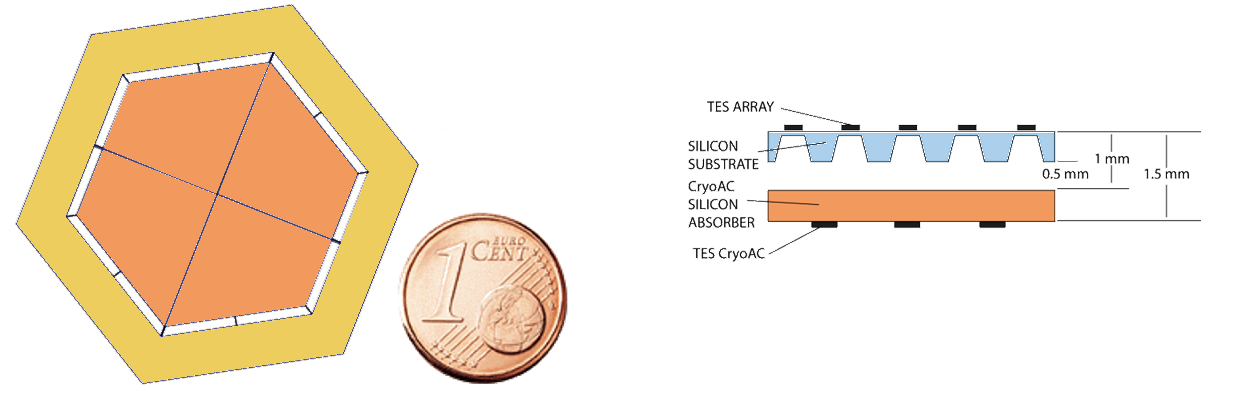}
\caption{\textit{Left:} The CryoAC detector design: 4 trapezoidal silicon pixels connected to a gold plated silicon rim by 4 bridges per pixel. \textit{Right:} Sketch of the distance between the TES array and the CryoAC}
\label{cryoacdesign}
\end{figure}

The detector is placed at a distance $<$ 1 mm below the TES-array absorbers (Fig. \ref{cryoacdesign} - Right), and the active part covers a full area of 4.9 cm$^2$, larger than the TES array one (2.3 cm$^2$). It is divided into 4 independent pixels that are connected to a gold plated silicon rim by narrow silicon beams, realizing a reliable and reproducible thermal conductance towards the thermal bath (represented by the rim itself). In order to obtain a fast response despite the large pixel size (1.23 cm$^2$ area per pixel), the detector is designed to work in the so-called athermal regime, collecting the first out-of-equilibrium family of phonons generated when a particle deposit energy into the silicon absorber. The detector design is shown in Fig. \ref{cryoacdesign} - Left.

Concerning the front end electronics, each CryoAC pixel will be read out by a different SQUID operated in a standard Flux Locked Loop (FLL) configuration. It is foreseen to adopt the same SQUID technology selected for the TES array, but having requirements tailored to the CryoAC needs. In order to keep the system redundant, each \virg pixel + SQUID'' will be served by a dedicated section inserted in the CryoAC Warm Front End Electronics. The baseline foresees that the veto operation will be performed on ground, given the expected modest telemetry rate.

\bigskip

\bigskip

\noindent The whole CryoAC detector sub-system will be composed of:

\begin{itemize}
\item The 4 pixels, each one based on a suspended Si absorber (1.23 cm$^2$ area, 500 $\mu$m thick);

\item The gold-plated surrounding silicon rim as interface to connect by narrow silicon beams the suspended pixels to a metallic supporting frame (this rim realizes the real CryoAC thermal bath);

\item About one hundred TES sensors per pixel, for increasing the phonon collecting area, in parallel readout to have redundancy so avoiding pixel loss due to possible TES broken/malfunctioning in case of series readout. The TES are made of Ir/Au bi-layer to properly work the detector at the 50-55 mK bath temperature;

\item Nb anti-inductive wiring of the TES in order to reduce the magnetic field of the bias current at less than 1 $\mu$T at 1mm;

\item A Metallic supporting structure of the detector, so providing the mechanical and thermal interface to the cold FPA plate;

\item An heater (TBC) to be deposited on the Si rim or the absorbers to optimize the CryoAC thermal bath temperature, in order to eventually reducing the TES bias current and thus the DC magnetic coupling towards the TES array.

\item  A Cold Front End Electronics constituted by 4 SQUIDs in a standard FLL configuration (one for each pixel) and 1 temperature sensor

\item Dedicated Warm Front End and Back End Electronics for analogue and digital processing of the signal
\end{itemize}

\noindent The detector design specifications able to satisfy the requirements are finally summarized in Tab. \ref{table:cryoacreq}

\begin{table}[H]
\centering
\includegraphics[width=1\textwidth]{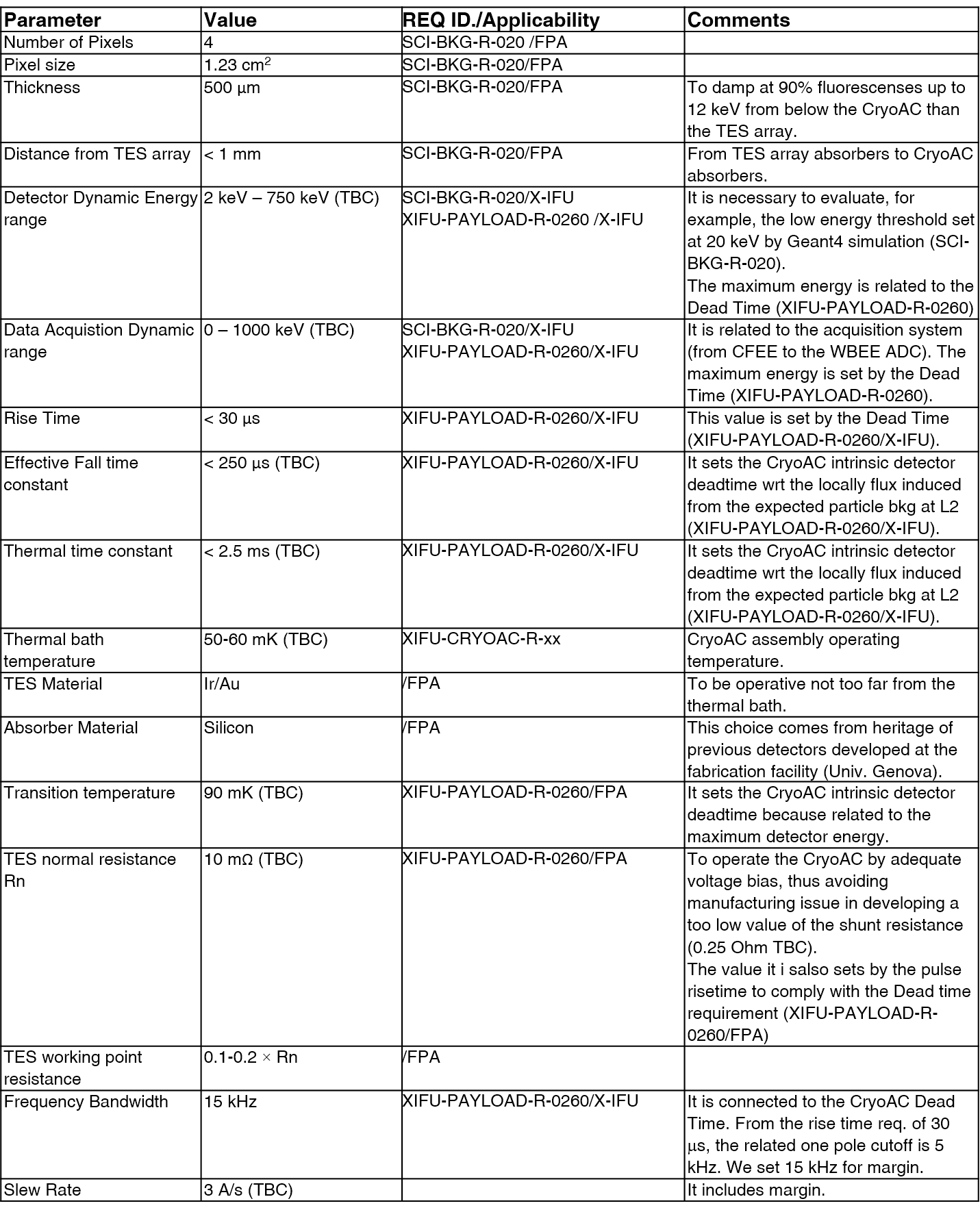}
\caption{CryoAC detector specifications.}
\label{table:cryoacreq}
\end{table}

\chapter{Going AHEAD beyond the X-IFU baseline}

The aim of this chapter is to present two studies that have been performed to understand how the baseline design and the specifications of the CryoAC can be improved in order to enhance its performance and provide additional capabilities to the X-IFU. These activities have been carried out in the context of AHEAD \cite{ahead} (integrated Activities in the High Energy Astrophysics Domain, grant agreement n. 654215), a project in the framework of the European Horizon 2020 program that supports also studies aimed at identifying possible solution beyond the ATHENA baseline. 

In the first study, we have investigated the possibility of exploiting the CryoAC not only as anticoincidence, but also as an hard X-ray detector, in order to enlarge the scientific capabilities of the X-IFU beyond the energy bandwidth of the TES array (0.2 - 12 keV). I have actively partecipated in all the parts of this study,  theoretically evaluating the detector performance and simulating the observation of possible astrophysical targets. 

The aim of the second study has been instead to propose a new solution for the X-IFU Focal Plane Assembly (FPA) design, foreseeing additional vertical CryoAC pixels on the sides of the TES array, in order to form a sort \virg anticoincidence box'' around the detector and to be able to further decrease the residual particle background of the instrument. In this case I have contributed to the study developing a realistic mechanical assembly to integrate the side CryoAC pixels in the FPA, thus providing a more representative environment to test this solution by Geant4 simulations.

\section{An assessment of the CryoAC capabilities in the hard X-ray band}

In this section I will report the feasibility study carried out to explore the observational capabilities of the CryoAC detector in the hard X-ray band (E $>$ 10 keV). The aim of the study is to understand if the present detector design can be improved in order to enlarge the scientific capability of the X-IFU instrument over a wider energy bandwidth. I remark that this is beyond the CryoAC baseline, being this instrument not aimed to perform X-ray spectroscopy but conceived to operate as anticoincidence particle detector. We will first evaluate the detector limit fluxes in the hard X-ray band, an then we will examine its scientific performance taking into account astronomical sources in a study case. The outcome of this scientific assessment will be the detector requirements enabling the CryoAC high energy observational capabilities, which will be reported in the conclusions.

\subsection{Evaluation of the CryoAC instrumental sensitivity}
\label{sec:3}

To assess the observational capabilities of an X-ray detector the fundamental quantity to evaluate is its \textit{minimum detectable flux}. This parameter characterizes the detector performances in terms of signal to noise ratio, determining the instrumental sensitivity and then the suitable astronomical targets. We recall that the minimum detectable flux $F_{min}$ [ph/cm$^2$/s/keV] within an observing time $t$ [s] and a $\Delta E$ energy band [keV] can be expressed as \cite{fraser}:
\begin{equation}
F_{min} = \frac{n_\sigma}{A_{eff} Q} \sqrt{\frac{B_p A_d + Q b_p \Omega A_{eff}}{t \Delta E}}
\label{eqn:fmin1}
\end{equation}
where $n_\sigma$ is the desired confidence level, $A_{eff}$ is the X-ray optics effective area [cm$^2$], Q is the detector response function [cts/ph],  $A_d$ its geometric area [cm$^2$], $B_P$ is the internal particle background level [cts/cm$^2$/s/keV], $b_p$ is the flux of the diffuse component of the background [ph/cm$^2$/s/keV/sr] and $\Omega$ is the detector aperture [sr]. Note that differently to the previous chapter (eq. \ref{eqn:fluxmin}), we have here explicited the contribution of the detector response function, not including it in the definition of the effective area. In the following sections we will evaluate all the parameters reported above, considering the baseline CryoAC design.

\subsubsection{ATHENA mirror effective area}
\label{sec:3-2}

\begin{figure}[H]
\centering
\includegraphics[width=0.6\textwidth]{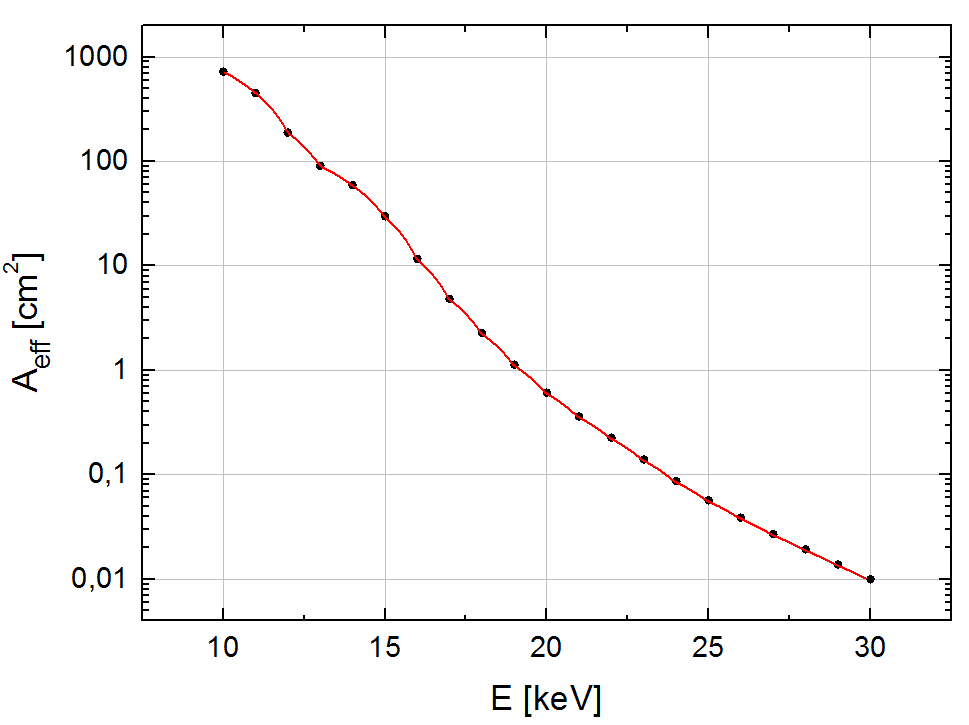}
\caption{On-axis effective area of the ATHENA telescope. The red line is an interpolation of the data courtesy of R. Willingale (black points).}
\label{fig:aeff} 
\end{figure}

The on-axis effective area of the ATHENA mirror in the range 10-30 keV is shown in Fig.~\ref{fig:aeff}. The data has been provided by R. Willingale (University of Leicester - Department of Physics and Astronomy, private communication) and refers to optics with a mirror module radius R$_{max}$ = 1469 mm, 2.3 mm of rib spacing and a surface roughness of 5 \AA. For information about the ATHENA optics and their development see \cite{athenaoptics}.

\noindent Fig.~\ref{fig:aeff} shows that above 10 keV the mirror area rapidly drops, hitting the value of 1 cm$^2$ around an energy of 19 keV. Based upon this data, we can already assess that within the ATHENA context the target hard X-ray band is limited in the range from 10 keV to $\sim$ 20 keV. 

\subsubsection{CryoAC response function}
\label{sec:3-1}

We define the CryoAC response function as the probability that an X-ray photon focused by the ATHENA optics will be detected by the CryoAC. This function includes several contributions as the trasmissivity of the X-IFU thermal filters, the trasmissivity of the TES array and the quantum efficiency of the CryoAC (i.e. the photoelectric absorption efficiency in the absorber).

To evaluate the CryoAC response we build a Monte Carlo simulation using the Geant4 toolkit \cite{geant4}. The simulation is based on a simplified model of the X-IFU Focal Plane Assembly (FPA), which includes the five thermal filters (with a total thickness of 280 nm Polyimide and 210 nm Al), the TES array consisting of 3840 pixels of 249 $\mu$m pitch (absorber composition is 1 $\mu$m of Au covered by 4 $\mu$m of Bi) and the CryoAC detector in the baseline design (500 $\mu$m thick Silicon absorber).

\begin{figure}[H]
\centering
\includegraphics[width=0.49\textwidth]{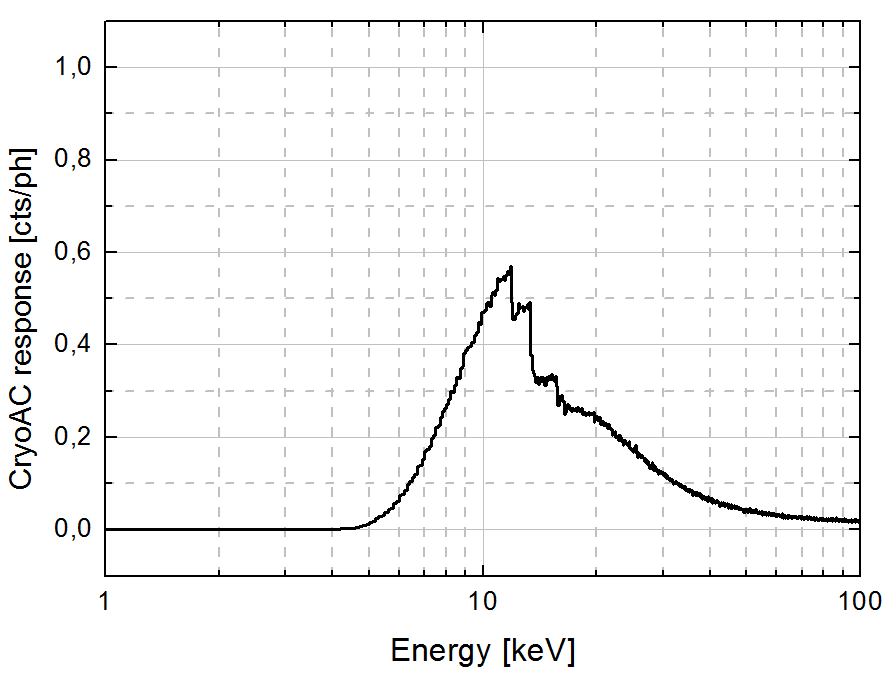}
\includegraphics[width=0.49\textwidth]{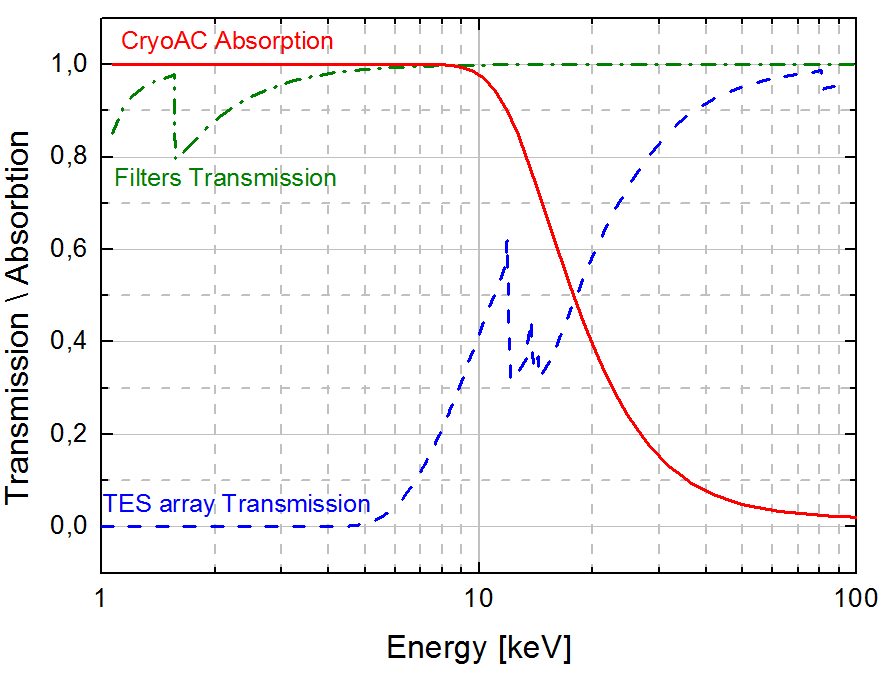}
\caption{(\textit{Left}): CryoAC response function estimated by means of a Geant4 simulation. (\textit{Right}): Different contributions to the CryoAC response function estimated using the attenuation coefficients tabulated by NIST \cite{nist}.}
\label{fig:Q} 
\end{figure}

The result of the simulation is shown in Fig.~\ref{fig:Q} - \textit{left}. The response function rapidly grows at low energy, showing a maximum around 12 keV (Q$_{max}$ = 0.57 cts/ph), and decreases at higher energies, dropping below the Q = 0.10 cts/ph level above 30 keV. To understand the contribution of the different elements in the FPA to the response function, in Fig.~\ref{fig:Q} - \textit{right} are reported the respective transmission/absorbtion curves estimated using the X-ray attenuation coefficients tabulated by NIST \cite{nist}.

Note that in the reference energy range from 10-20 keV, the current CryoAC design ensure a good absorbtion efficiency ($\sim$ 40\% at 20 keV), and so we do not consider useful to put effort into an upgrade of the absorber characteristics (i.e. the thickness or the material) at this stage.

\subsubsection{In orbit background}
\label{sec:3-3}

The diffuse component of the background has been estimated starting from the hard X-ray spectrum of the Cosmic X-ray Background (CXB) measured by IBIS INTEGRAL. It has been modeled using the analytical description proposed by Turler et al. in \cite{turler}. To obtain the background level we have folded this spectral flux with the optics effective area $A_{eff}$, the detector response function $Q$ and the aperture of a single CyoAC pixel $\Omega = 0.80 \cdot 10^{-6}$ sr (corresponding to a FoV of 9.4 arcmin$^2$).

The internal particle background has been instead estimated by means of the Monte Carlo simulations already developed to study the X-IFU in-orbit background (see sect. \ref{NXBsim}). In this case we have taken into account the use of the X-IFU TES array as ``reverse'' anticoincidence device. This is the opposite of what happens in observations with the X-IFU array, where it is the CryoAC that discriminates main detector events that happen in both detectors. In this context the main detector can act as an anticoincidence device for the CryoAC, reducing the particle background to some extent despite not being designed/optimized for the scope. The residual particle background level on a CryoAC pixel is roughly constant in the 10-30 keV energy band, counting about 0.012 cts/s/keV.

\begin{figure}[H]
\centering
\includegraphics[width=0.6\textwidth]{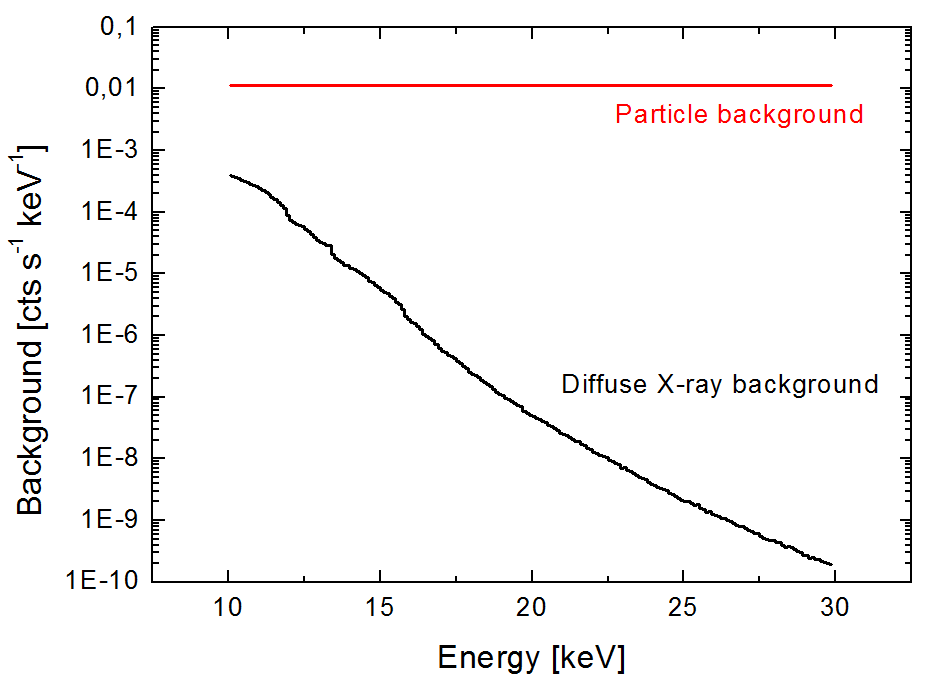}
\caption{Spectra of the particles and the diffuse X-rays backgrounds expected on a single CryoAC pixel in L2 orbit.}
\label{fig:background} 
\end{figure}

The estimated background spectra for a single CryoAC pixel in the band 10-30 keV are shown in Fig.~\ref{fig:background}. Note that above 10 keV the diffuse X-Ray component of the background is always negligible with respect to the particles one, and so it can be neglected.

\subsubsection{Limit fluxes and continuum sensitivity}
\label{sec:3-4}

\begin{figure}[H]
\centering
\includegraphics[width=0.49\textwidth]{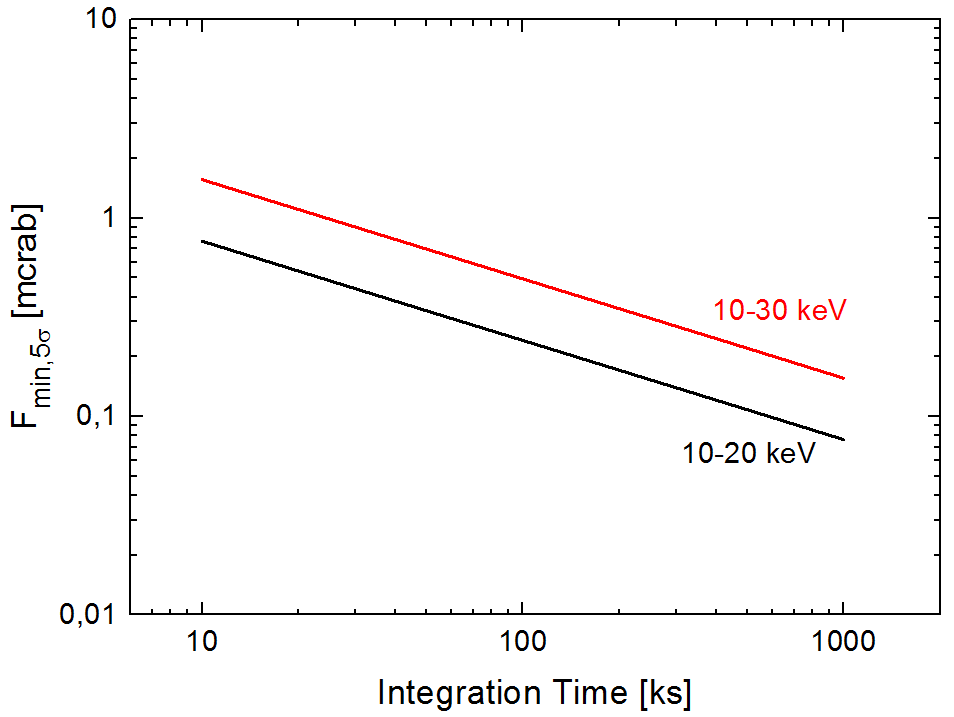}
\includegraphics[width=0.49\textwidth]{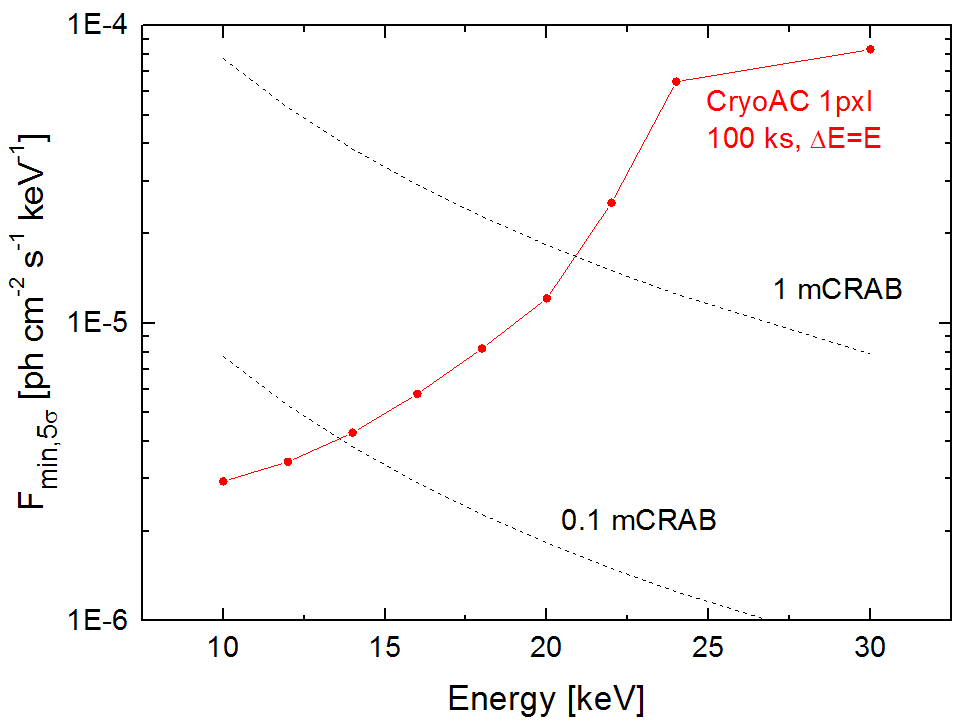}
\caption{(\textit{Left}): CryoAC minimum detectable flux as function of the observation time. (\textit{Right}):CryoAC continuum sensitivity for 100 ks of exposure time and an energy bandwidth $\Delta$E=E. Crab fluxes are overplotted as reference.}
\label{fig:fmin} 
\end{figure}

Including in eq. (\ref{eqn:fmin1}) the energy-dependence of the effective area $A_{eff}$ and the response function $Q$, we have then estimated the minimum detectable flux for a single CryoAC pixel and a 5$\sigma$ confidence level (n$_\sigma$ = 5). In Fig.~\ref{fig:fmin} \textit{left} the limit flux is reported as a function of the observation time in two different energy ranges, whereas in Fig.~\ref{fig:fmin} - \textit{right} it is shown as a function of energy, considering in eq. (\ref{eqn:fmin1}) an energy range $\Delta$E=E and a 100 ks exposure for each point. These two plots characterize the sensitivity of the CryoAC pixels to a continuum emission, showing that the limit fluxes with the current detector baseline are fractions of mCrab in the band 10-20 keV. 

To better evaluate the sensitivity of the detector, in Tab.~\ref{tab:fluxes} are reported the limit fluxes of the CryoAC compared with the reference ones for NUSTAR and ATHENA X-IFU (assuming t = 100 ks and n$_\sigma$ = 5).

\begin{table}[H]
\centering
\caption{Limit fluxes for the CryoAC compared with the reference ones for NUSTAR and ATHENA X-IFU (t = 100 ks, n$_\sigma$ = 5)}
\label{tab:fluxes}
\small
\begin{tabular}{lllll}
\hline\noalign{\smallskip}
Instrument & Energy range & $F_{min}$ & $F_{min}$ & Notes and refs. \\
 & [keV] & [erg/cm$^2$/s] & [mCrab] &  \\
\noalign{\smallskip}\hline\noalign{\smallskip}
CryoAC & 10-20 & 1.6$\cdot 10^{-12}$ & 0.2 & 1 pixel \\
CryoAC & 10-30 & 6.3$\cdot 10^{-12}$ & 0.5 & 1 pixel \\
NUSTAR & 10-30 & 5.0$\cdot 10^{-14}$ & 0.4$\cdot 10^{-2}$ & \cite{nustar}\\
ATHENA X-IFU & 2-10 & 3.2$\cdot 10^{-16}$ & 1.6$\cdot 10^{-5}$ & Point source \cite{xidlotti} \\ 
\noalign{\smallskip}\hline
\end{tabular}
\end{table}
\normalsize
The result of the first part of this scientific assessment is that the CryoAC can operate as hard X-ray detector without changes in the baseline design, supplying a moderate sensitivity in the 10-20 keV  band (Tab.~\ref{tab:fluxes}). We remark that this narrow energy band is limited by the drop of the optics effective area at high energies, and not by the detector features.

\subsection{Scientific Simulations}
\label{sec:4}

Once assessed the continuum sensitivity of the CryoAC, in this section we will present some scientific simulations of astronomical targets that we have performed to understand the spectroscopic capabilities of the detector. This will allow us to define an optimal reference value for the CryoAC required energy resolution. We point out that at present no requirement has been set for this parameter.

\subsubsection{Crab observation}
\label{sec:4-1}

\begin{figure}[H]
\centering
\includegraphics[width=0.7\textwidth]{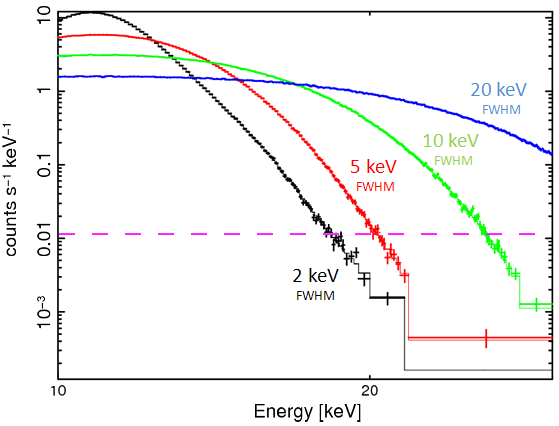}
\caption{Crab spectra for a 100ks observation with the CryoAC and different values of the energy resolution. The violet dashed line represents the background level.}
\label{fig:crab} 
\end{figure}

To have a first reference, we have simulated with XSPEC \cite{xspec} a 100 ks observation of the Crab Nebula. The Crab spectrum has been modeled as a power law with photon index $\alpha$ = 2.08 and normalization K = 9.3 ph/cm$^2$/s at 1 keV \cite{crab}. Starting from the quantities evaluated in the previous sections, we have generated several CryoAC response matrices, varying the detector energy resolution from 2 to 20 keV (FWHM). Note that the energy resolution has been assumed constant in the range 10-30 keV. The Crab spectra simulated accounting for the different response matrices are shown in Fig.~\ref{fig:crab}. 

We have fitted the simulated dataset in order to assess the accuracy by which the source parameter could be recovered. 
The results of the fit are shown in Table \ref{tab:crab}, where we have reported the 90\% confidence range of the spectral parameter obtained with the different energy resolutions. As expected the spectral resolution deeply influences the observation, significantly conditioning the errors on the recovered parameters. Based upon these results, we can preliminary conclude that an energy resolution of few keV (FWHM) in the 10-20 keV range is necessary to extend the scientific capability of the CryoAC.

\begin{table}[H]
\centering
\caption{Spectral parameter of the Crab Nebula (90\% confidence range) obtained simulating a 100 ks observation with the CryoAC for different values of the energy resolution.}
\label{tab:crab}
\begin{tabular}{ccccc}
\hline\noalign{\smallskip}
Energy resolution& $\alpha$ & $\Delta \alpha /\alpha$ & K & $\Delta K / K$ \\
(FWHM) &  &  & [ph/cm$^2$/s/keV] &  \\
\noalign{\smallskip}\hline\noalign{\smallskip}
2 keV & 2.07 - 2.10 & 1.4\% & 9.1 - 9.8 & 7.2\%\\
5 keV & 2.06 - 2.12 & 2.8\% & 8.8 - 10.2 & 15\%\\
10 keV & 2.05 - 2.18 & 6.8\% & 8.6 - 12.0 & 36\%\\
20 keV & 2.10 - 2.51 & 20\% & 9.8 - 26.0 & 174\%\\
\noalign{\smallskip}\hline
\end{tabular}
\end{table}

\subsubsection{A case study: HMXBs observation}
\label{sec:4-2}

We want now to evaluate the CryoAC scientific capabilities and the role of its energy resolution in combination with the X-IFU TES array. Given the sensitivity estimated in Sect.~\ref{sec:3-4}, a target source with a flux of some tens of mCrab observed with an exposure time around 100 ks is needed to have enough counts on the CryoAC for spectral analysis. Bright High-Mass X-ray Binaries (HMXBs) can be considered as an ideal candidate in this context. These sources can indeed present a high energy cut-off in the CryoAC energy band (10-20 keV), representing an interesting target for joint TES array/CryoAC observations.

\begin{figure}[H]
\centering
\includegraphics[width=0.7\textwidth]{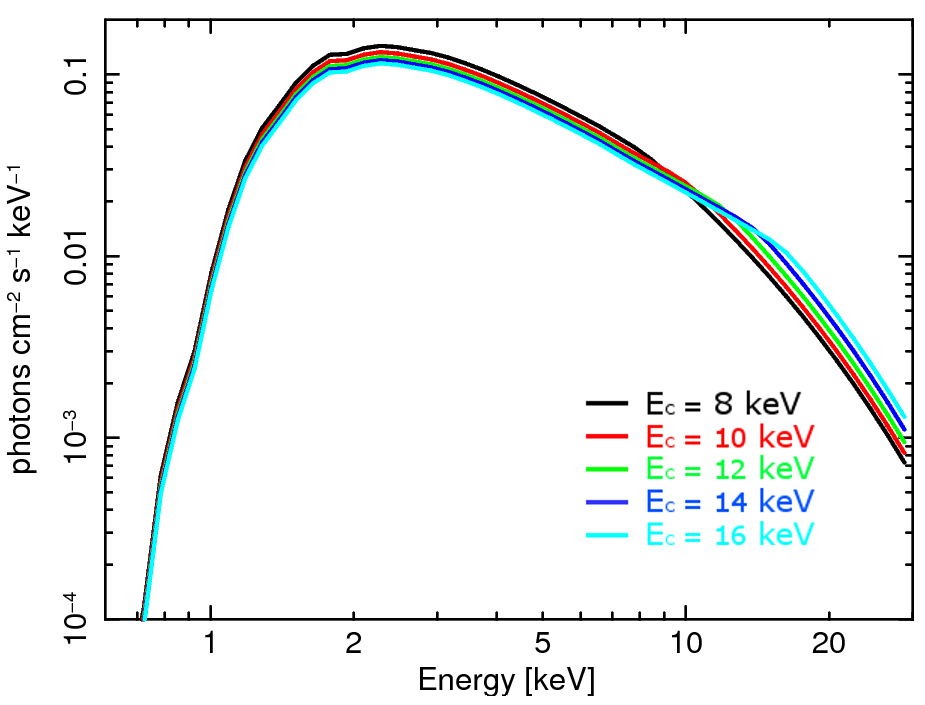}
\caption{The HMXB spectral models used in the simulations.}
\label{fig:hmxb} 
\end{figure}

As case study, we have therefore simulated the observation of High-Mass X-ray Binary spectra, exploring if the CryoAC could effectively extend the X-IFU energy band. For the HMXBs spectra we have used a simplified model with an absorbed power law spectrum and a high energy cut-off (\textit{wabs*highecut*powerlaw} in XSPEC), generating several models with different cutoff energies in the range 8-16 keV. The models are shown in Fig.~\ref{fig:hmxb} and their parameters are given in Table~\ref{tab:hmxbxtab}.

\begin{table}[H]
\centering
\caption{HMXB models parameters.}
\label{tab:hmxbxtab}  
\begin{tabular}{lll}
\hline\noalign{\smallskip}
Component & Parameter & Value \\
\noalign{\smallskip}\hline\noalign{\smallskip}
wabs & N$_H$ & 2$\cdot$10$^{22}$ cm$^2$\\
highecut & E$_C$ & 8, 10, 12, 14, 16 keV\\
highecut & E$_F$ & 10 keV\\
powerlaw & $\Gamma$ & 1.5\\
 & Flux 0.2 - 20 keV & 100 mCrab\\
\noalign{\smallskip}\hline
\end{tabular}
\end{table}

\begin{figure}[H]
\centering
\includegraphics[width=0.7\textwidth]{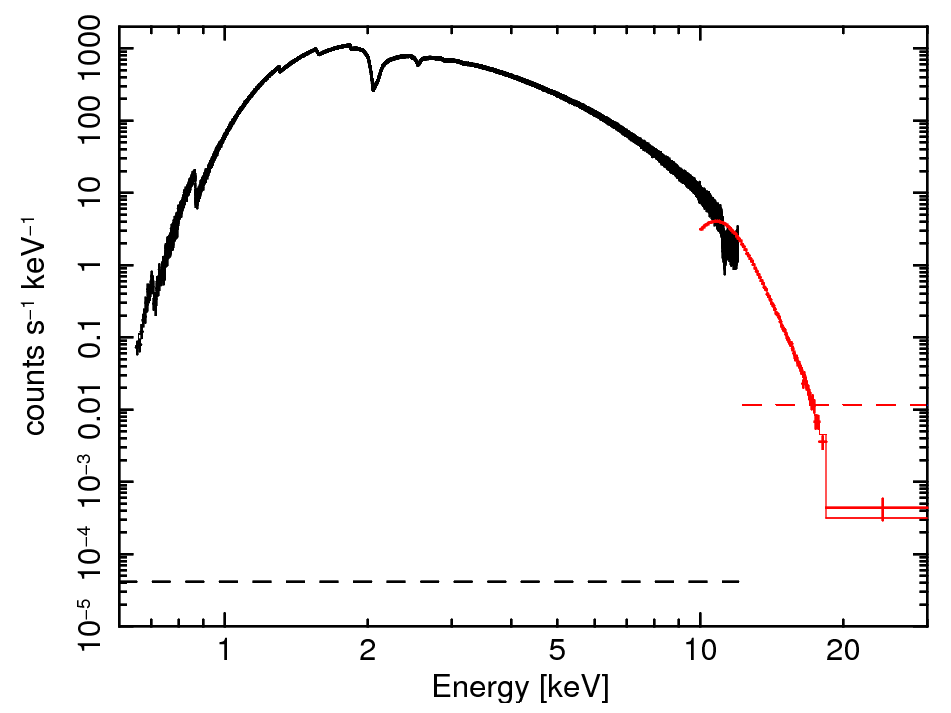}
\caption{HMXB spectrum with a cut-off at E$_C$ = 12 keV observed with both the X-IFU TES array (black data) and the CryoAC (red data) in 50 ks exposure time. Dashed lines represent the respectives background levels.}
\label{fig:hmxb2} 
\end{figure}

\noindent For each model we have simulated three different kind of observation: 
\begin{itemize}
\item 50 ks exposure with the X-IFU TES Array only
\item 50 ks exposure with both the X-IFU TES Array and the CryoAC (energy resolution $\Delta$E = 2keV FWHM)
\item 50 ks exposure with both the X-IFU TES Array and the CryoAC (energy resolution $\Delta$E = 5keV FWHM)
\end{itemize}
An example of a simulated spectrum is shown in Fig.~\ref{fig:hmxb2}.

\begin{table}[H]
\centering
\caption{High energy cut-off parameters obtained fitting the simulated spectra in the different configurations. The errors refer to the 90\% confidence level.}
\label{tab:hmxbtab2}      
\footnotesize 
\begin{tabular}{llll}
\hline\noalign{\smallskip}
Model & TES array &  TES array &  TES array \\
 &  & + CryoAC &  + CryoAC \\
& & ($\Delta$E = 2 keV) & ($\Delta$E = 5 keV) \\
\noalign{\smallskip}\hline\noalign{\smallskip}

&  &  &  \\
E$_C$ = 8 keV & E$_C$ = $8.01^{+0.02}_{-0.02}$ keV & E$_C$ = $8.00^{+0.02}_{-0.01}$ keV & E$_C$ = $8.01^{+0.02}_{-0.01}$ keV \\
& E$_F$ = $9.60^{+0.11}_{-0.11}$ keV & E$_F$ = $9.63^{+0.11}_{-0.11}$ keV & E$_F$ = $9.61^{+0.11}_{-0.11}$ keV \\
&  &  &  \\

&  &  &  \\
E$_C$ = 10 keV & E$_C$ = $9.97^{+0.04}_{-0.03}$ keV & E$_C$ = $9.93^{+0.03}_{-0.03}$ keV & E$_C$ = $9.94^{+0.03}_{-0.03}$ keV \\
& E$_F$ = $8.88^{+0.17}_{-0.43}$ keV & E$_F$ = $9.47^{+0.25}_{-0.24}$ keV & E$_F$ = $9.31^{+0.35}_{-0.35}$ keV \\
&  &  &  \\

&  &  &  \\
E$_C$ = 12 keV & E$_C$ = $9.2^{+0.3}_{-0.5}$ keV & E$_C$ = $11.6^{+0.2}_{-0.2}$ keV & E$_C$ = $11.0^{+0.2}_{-0.2}$ keV \\
& E$_F$ = $113^{+54}_{-30}$ keV & E$_F$ = $12.0^{+1.5}_{-1.5}$ keV & E$_F$ = $19.1^{+4.0}_{-3.0}$ keV \\
&  &  &  \\

&  &  &  \\
E$_C$ = 14 keV & E$_C$ = $9.3^{+0.4}_{-0.5}$ keV & E$_C$ = $13.4^{+0.5}_{-0.5}$ keV & E$_C$ = $14.7^{+2.0}_{-1.5}$ keV \\
& E$_F$ = $155^{+80}_{-39}$ keV & E$_F$ = $15.3^{+7.1}_{-4.7}$ keV & E$_F$ = $4.3^{+11.2}_{-4.3}$ keV \\
&  &  &  \\

&  &  &  \\
E$_C$ = 16 keV & E$_C$ = $8.1^{+0.8}_{-1.0}$ keV & E$_C$ = $14.3^{+0.9}_{-1.2}$ keV & E$_C$ = $8.0^{+0.6}_{-1.0}$ keV \\
& E$_F$ = $248^{+170}_{-93}$ keV & E$_F$ = $23^{+39}_{-12}$ keV & E$_F$ = $261^{+190}_{-79}$ keV \\
&  &  &  \\
\noalign{\smallskip}\hline
\end{tabular}
\end{table}
\normalsize
In Tab.~\ref{tab:hmxbtab2} are reported the high energy cut-off parameters obtained fitting the simulated spectra. The use of the CryoAC as hard X-ray detector allows to improve the characterization of the cut-off and the folding energies of the sources for E$_C$ $>$ 10 keV, whereas the TES array is unable to properly constrain them. Furthermore, note that also in this case the energy resolution of the CryoAC plays a fundamental role in the parameter characterization, significantly influencing the instrument performances.  

Finally, in order to evaluate the relevance of this case study in the context of the ATHENA mission, we have analyzed the last ATHENA Mock Observing Plan searching for bright HMXBs.  Within the planned X-IFU observations we have found: 
\begin{itemize}
\item 25 HMXB with average source intensity F$_{AVG} >$ 100 mCrab and planned exposure time t = 20 ks
\item 10 HMXB with 10 mCrab $< $ F$_{AVG} <$ 100 mCrab and planned exposure time in the range t = 20 - 200 ks
\end{itemize}
We can conclude that with a CryoAC energy resolution of 2 keV (FWHM) could be possible to characterize these source in the enlarged band up to 20 keV.

\subsection{Conclusions}
\label{sec:5}

We have performed this study to understand if the CryoAC detector onboard the ATHENA X-IFU can be optimized in order to extend the instrument scientific capability in the hard X-ray band.

We have found that in the baseline configuration the CryoAC could operate as hard X-ray detector in the narrow band 10-20 keV, with a limit flux (5$\sigma$, 100 ks, 1 pixel) of 1.6$\cdot$10$^{-12}$ erg/cm$^2$/s ($\sim$ 0.2 mCrab). The energy band is limited by the drop of the optics effective area at high energies, and not by the detector features.

Furthermore, we have found that an optimization of the CryoAC energy resolution could have a scientific return in the observation of bright sources with a spectral cut-off in this band. In this context we present the observation of HMXBs as case study, finding about 35 sources of this type in the ATHENA Mock Observing Plan. The required value that we have identified for the CryoAC energy resolution is $\Delta$E = 2 keV (FWHM) in the 10-20 keV band.

\newpage
\section{Side CryoAC: a design concept}

The main contribution to the X-IFU unrejected background is due to secondary electrons with energies $<$1 MeV that bounce off the detector surface, releasing a small fraction of their energy (back-scattering process). In the baseline X-IFU design, which foresees a planar CryoAC placed below the TES array, such a particle can not of course be detected by the anticoincidence. 

We have therefore studied the effectiveness of inserting additional CryoAC pixels in the FPA, enclosing the TES array in a sort of \virg anticoincidence box''.  This should allow to reject particles hitting the TES array from above the detector itself (inside a solid angle dependent on the side CryoAC  pixels features), thus reducing the unrejected background level and ultimately increasing the instrument sensitivity. 

\subsection{A first rough effectiveness evaluation}

\begin{figure}[htbp]
\centering
\includegraphics[width=0.4\linewidth]{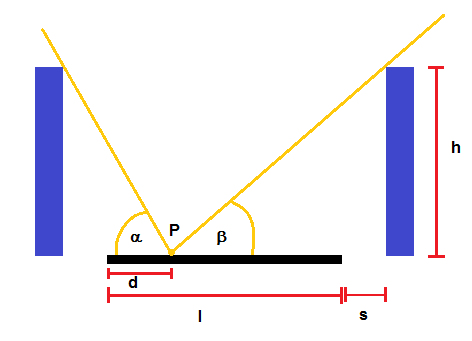}
\includegraphics[width=0.4\linewidth]{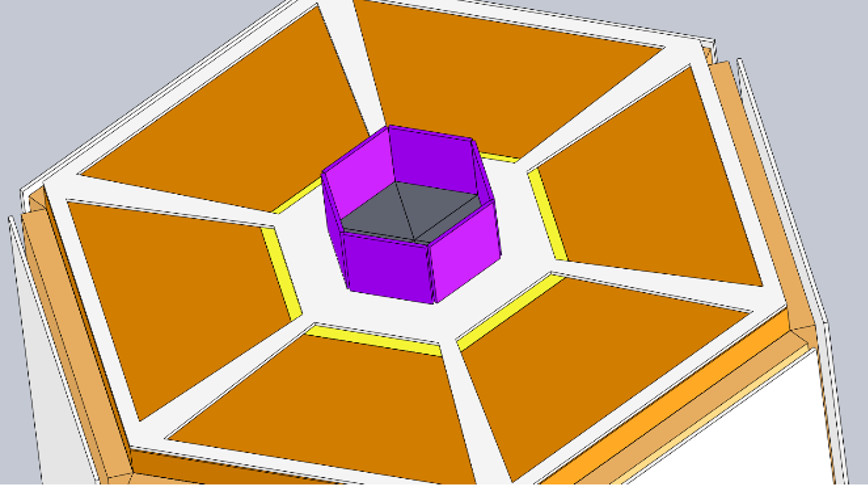}
\caption{\textit{Left:} Schematics of the CryoAC side pixels (in blue) placed around the TES array detector (in black). \textit{ Right:} Simplified CAD model of the X-IFU FPA with the insertion of the side CryoAC pixels (violet). The TES array is the grey hexagon in the center. }
\label{side_cryoac_sim}
\end{figure}

In a first order approximation, assuming cylindrical symmetry, the \virg side CryoAC'' concept can be represented by the scheme in Fig. \ref{side_cryoac_sim} - left, where $l$ is the TES array detector linear size, $h$ is the height of the side CryoAC pixels, and $s$ is the distance between the two systems. Given a point $P$ at a distance $d$ from the TES array edge, the solid angle covered by the side CryoAC pixels will be:

\begin{equation}
\Omega = 2 \pi cos[(\pi-\Theta)/2]
\label{solidangle}
\end{equation}

\noindent where

\begin{equation}
\Theta = \alpha +\beta = arctan[h/(d+s)] + arctan[h/(l-d+s)], 
\end{equation}

From these simple analytical evaluations, we found that with $l = 2.1$ cm (i.e. the hexagonal TES array diagonal), $s=0$, and the side CryoAC pixels square-shaped with an area of 1 cm$^2$ ($h=1$ cm), the anticoincidence box would cover $\sim$70$\%$ of the solid angle seen by the detector, so expecting roughly a factor 3 of reduction in the unrejected background. Note that the 1 cm$^2$  CryoAC pixels area has been chosen since we are already capable of producing devices of that size, and thus it represents a solid solution.

This kind of solution has been then implemented inside the Geant4 X-IFU background simulator, surrounding the TES array by 6 square shaped side CryoAC pixels (see Fig. \ref{side_cryoac_sim} - right). In this new configuration of the FPA, we found a residual background level of $1.5 \times 10^{−3}$ cts cm$^{−2}$ s$^{-1}$ keV$^{-1}$ in the 2-10 keV energy band (obtained with only GCR as input), essentialy a factor 3 below the value achieved with the baseline design (\virg CryoAC + Kapton-Bi'' , $3.7 \times 10^{−3}$ cts cm$^{−2}$ s$^{-1}$ keV$^{-1}$ in the 2-10 keV energy band) and reported in Sect. \ref{bkgresults}, thus confirming the first considerations of solid angle. 

\subsection{A more realistic concept design}

\begin{figure}[H]
\centering
\includegraphics[width=0.9\linewidth]{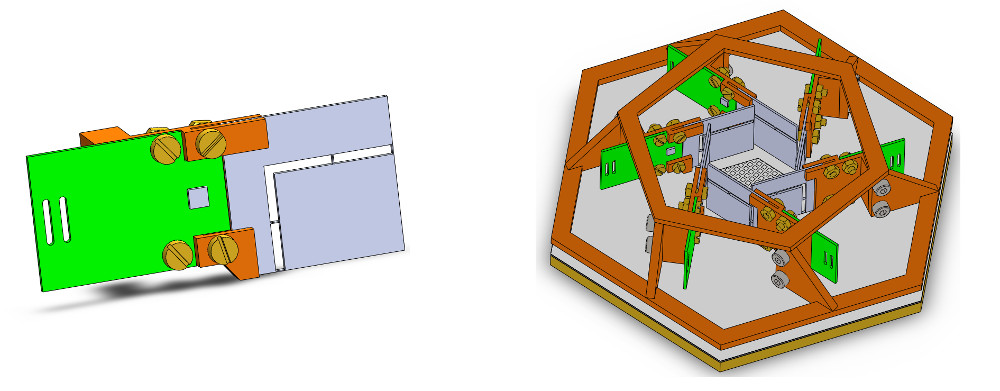}
\includegraphics[width=0.9\linewidth]{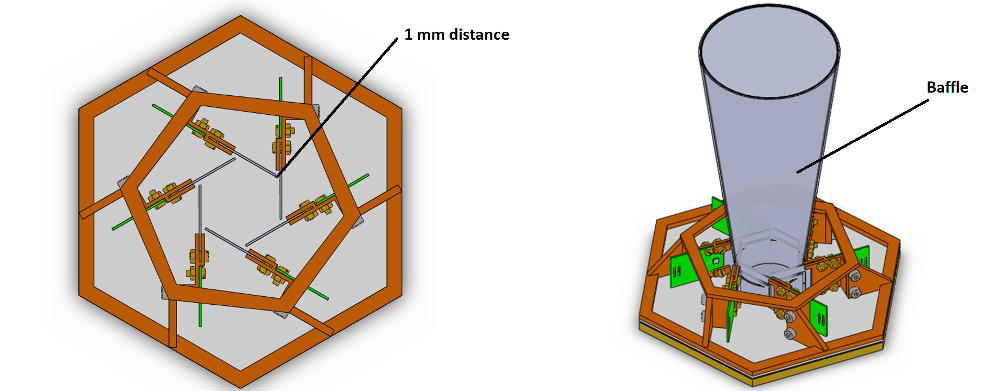}
\caption{\textit{Top Left:} The single side CryoAC pixel (light grey) attached to a supporting structure (brown) that sustains also the SQUID PCB electronic board (green). \textit{Top Right:} The complete side CryoAC assembly, composed by 6 pixels and a copper supporting structure coupled to the FPA plate. \textit{Bottom Left:} The side CryoAC assembly, top view. \textit{Bottom Right:} Interference between the CryoAC and the baffle hosting the Kapton-Bi electron shield in the FPA in its current design.}
\label{side_cryoac_pixel}
\end{figure}

The results obtained with this first simulation led us to further investigate the side CryoAC solution. With respect to the previous assessment, where the side CryoAC pixels were suspended with no supporting structure, we have then created a realistic mechanical assembly for the integration in the FPA, thus providing a more representative model to be probed by Geant4 simulation (Fig. \ref{side_cryoac_pixel}). This design concept has been discussed with SRON, which is responsible for the FPA design, in order to check its coupling with the others FPA structures and to preliminary verify its feasibility.

\bigskip

The single side CryoAC pixel assembly (Fig. \ref{side_cryoac_pixel} - Top Left) follow the CryoAC Demonstration Model design, that will be shown in the Chapt. \ref{DMchapt} of this thesis. It is composed by:

\begin{itemize}
\item The proper CryoAC pixel (light grey in the figure), where the 1 cm$^2$ silicon absorber is connected to a rim by four narrow silicon beams, which defines the thermal conductance between the absorber and the thermal bath (i.e. the rim).

\item A copper structure (brown in the figure) that clamps the silicon rim, supporting the CryoAC pixel. The proper thermalization of the pixel will be realized by gold wire bondings (not shown in the figure) connecting the rim with the copper areas clamping it.

\item A PCB board (green in the figure) with the pixel cold front end electronics, based on a SQUID chip (the little grey square in the figure). The eyelets in the PCB are designed to host the loom with the superconductive bias and readout wiring.
\end{itemize}

\noindent The six side CryoAC pixels can be individually integrated and tested, and then mounted on a copper structure coupled to the FPA plate (Fig. \ref{side_cryoac_pixel} - Top Right and Bottom Left). The structure supports the pixels keeping them at a distance of 2 mm from the plate, as requested by SRON. The minimum distance foreseen between adjacent pixels is 1 mm, and the total  weight of the system (supporting structure + pixels) is about 84 g. 

\bigskip

\begin{figure}[H]
\centering
\includegraphics[width=1.0\linewidth]{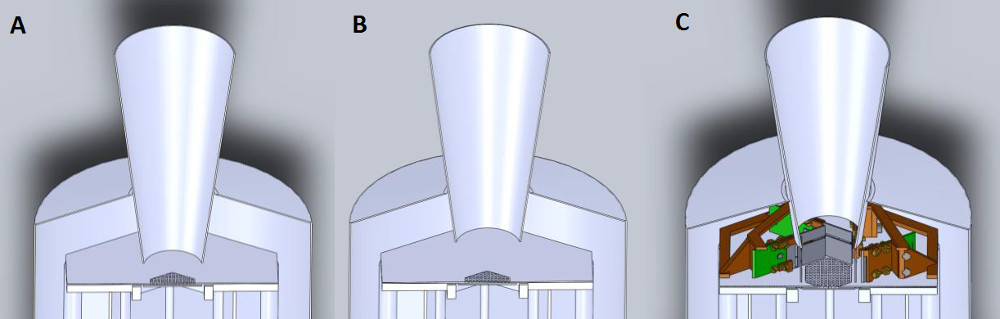}
\caption{\textit{Left}: The original FPA design (A). \textit{Center}: the modified FPA with the reduced baffle (B). \textit{Right} The modified FPA with the reduced baffle and the side CryoAC assembly inserted (C).}
\label{side_cryoac_mm}
\end{figure}

With the current design of the X-IFU FPA, this side CryoAC assembly would interfer with the baffle hosting the passive Kapton-Bi electron shield (see Fig. \ref{side_cryoac_pixel} - Bottom Right). In order to accomodate the system, it is therefore mandatory reduce the lenght of the baffle (and the shield) by 6.5 mm (margin included). To disentangle the contribution to the background level of this modifications from the impact of the additional CryoAC pixels, we have then ran three different Geant4 simulations, with three different mass models (see Fig. \ref{side_cryoac_mm}):

\begin{itemize}
\item[\textbf{A}] - The original X-IFU FPA mass model without any modification;
\item[\textbf{B}] - The mass model modified by cutting the baffle and the passive electron shield, but without inserting the side CryoAC assembly;
\item[\textbf{C}] - The final mass model modified both cutting the baffle and inserting the side CryoAC assembly 
\end{itemize}

\begin{figure}[H]
\centering
\includegraphics[width=0.6\linewidth]{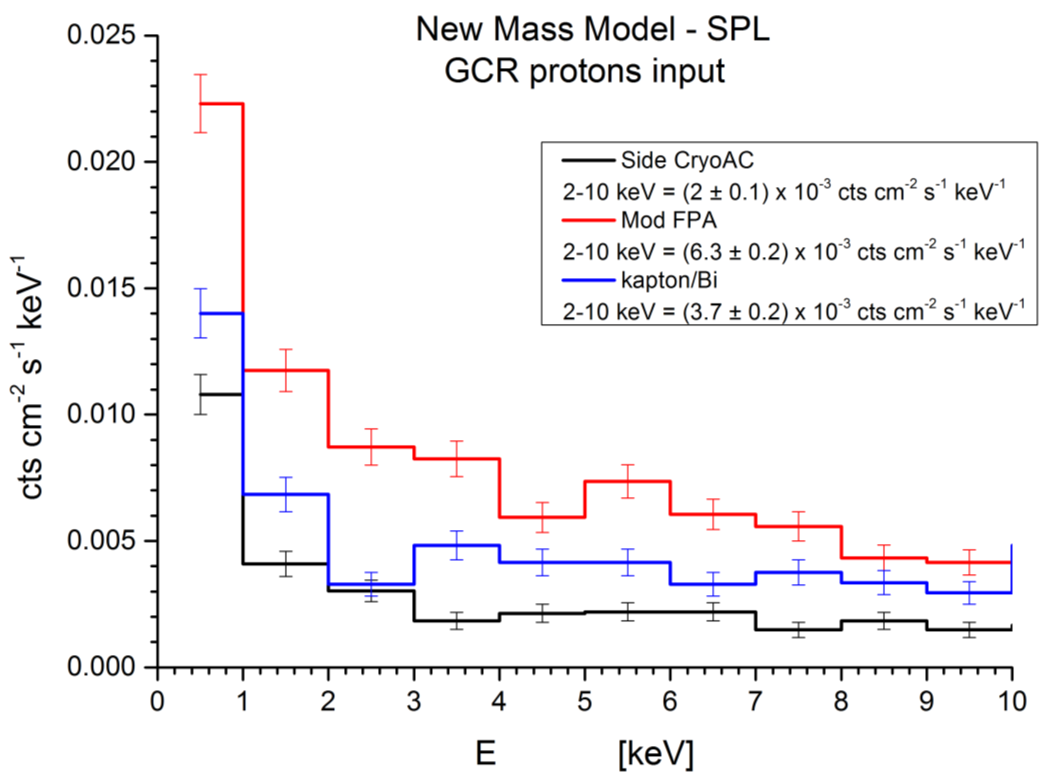}
\caption{Unrejected background spectra in the baseline configuration (blue line), with the modified FPA (red line), and with the lateral CryoAC (black line).}
\label{side_cryoac_res}
\end{figure}

\noindent The results of these simulations are shown in Fig. \ref{side_cryoac_res}. Referring to the three different mass models, we have:

\begin{itemize}

\item[\textbf{A}] The baseline reference background level is 3.7$\cdot$10$^{-3}$ cts/cm2/s/keV in the 2 - 10 keV band (obtained with only GCR protons as input)

\item[\textbf{B}] The residual background increases to a level of 6.3$\cdot$10$^{-3}$ cts/cm2/s/keV in this modified version of the FPA (i.e. a factor $\sim$ 2 above the baseline). This is due to the fact that removing the lower part of the passive Kapton/Bi shield, quite close to the TES array, the detector is now directly exposed to the secondary electron generated in the inner FPA Niobium shield.

\item[\textbf{C}] With the insertion of the side CryoAC assembly the residual background is reduced to 2.0$\cdot$10$^{-3}$ cts/cm2/s/keV, a factor $\sim$ 3 below the modified FPA level (\textbf{B}), as expected from solid angle argument (see eq. \ref{solidangle}), and a factor of $\sim$ 2 below the baseline solution level (\textbf{A}).
\end{itemize}

\subsection{Conclusions}

With this preliminary concept study we have shown that the insertion of six additional CryoAC vertical pixels in the FPA could provide a reduction in the residual background level of about a factor 2 with respect to the X-IFU baseline design. This solution should be therefore considered for a trade-off study at the X-IFU system level, evaluating the benefits of the achievable background reduction with respect to the impact of this additional system to the whole instrument TRL.

\part{The X-IFU Cryogenic Anticoincidence Detector (CryoAC): the experimental path towards the Demonstration Model}

\chapter{CryoAC: The detector physics}

The aim of this chapter is to provide a basic introduction to the physics underlying the CryoAC detector. First, I will present TES microcalorimeters, solving their standard electrothermal model and introducing the main noise sources and the SQUID readout. Then, I will focus on the athermal phonon mediated TES microcalorimeters (which the CryoAC belongs to), developing an electrothermal model to show their response function. Finally, I will show some useful equation to evaluate the main thermal parameters of these microcalorimeters.

\section{An introduction to TES microcalorimeters}

\begin{figure}[H]
\centering
\includegraphics[width=0.8\linewidth]{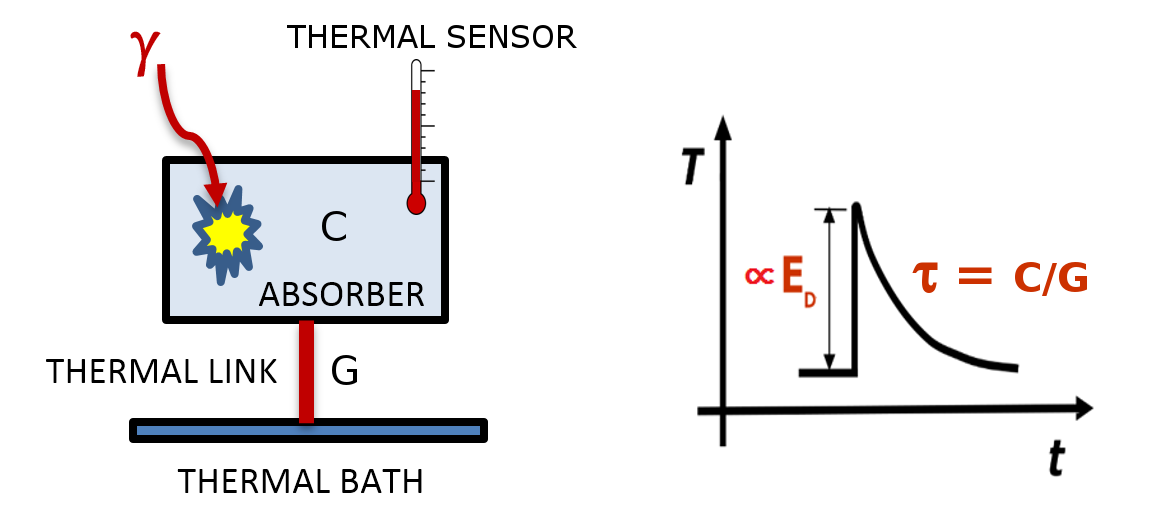}
\caption{Schematic illustration of a simple calorimeter detector. (Left): An absorber of heat capacity C is coupled to a thermal bath by a thermal conductance G. A higly sensitive thermal sensor monitors the absorber temperature. (Right): Thermal response of the calorimeter to a deposit of energy into the absorber.}
\label{TES_microcalorimeter_1}
\end{figure}

A calorimeter detector is a device able to convert in heat pulse the energy deposited by a particle into an absorber, measuring it via precise thermometry. The measurable physical quantity is the temperature raise of the absorber $\Delta T$, which is related to the deposited energy $E_d$ by the fundamental relation:

\begin{equation}
\Delta T = \dfrac{E_d}{C}
\label{eq:calorimeter}
\end{equation}
where C is the heat capacity of the absorber. A calorimeter detector is composed by three main components (Fig. \ref{TES_microcalorimeter_1}-Left): the absorber, a highly sensitive thermal sensor to measure its temperature variation, and a thermal link towards a thermal bath at constant temperature. This last part plays the important role of restoring the absorber base temperature after a heat pulse, with a characteristic thermal time of:

\begin{equation}
\tau_{th} = \frac{C}{G}
\end{equation}
where G is the thermal conductance of the thermal link between absorber and bath. The thermal response of a calorimeter to a particle energy deposition shows therefore a typical pulse shape, with decay time $\tau_{th}$ and height proportional to $E_d$ (Fig. \ref{TES_microcalorimeter_1}-right). 

The energy resolution achievable by a calorimeter is intrinsically limited by the thermal fluctuation due to the random exchange of phonons between the absorber and the thermal bath (phonon noise), and it can be approximated by \cite{moseley}:
\begin{equation}
\Delta E_{th} = 2.36 \sqrt{k_B T^2 C}
\label{eq:calorimeter2}
\end{equation}
where $k_B$ is the Boltzmann constant. From eq. (\ref{eq:calorimeter}) and (\ref{eq:calorimeter2}) it is clear that very low heat capacity and temperature are needed to measure small amounts of energy (like the ones of incoming X-ray photons) with high spectral resolution. Since the heat capacity is proportional to the absorber volume (and to a power of its temperature), this implies the need to work with small devices operated at cryogenic temperature, and hence named \textit{cryogenic} \textit{micro}-calorimeters. Consider that for a typical X-ray application C $\sim$ some pJ/K and T $\sim$ 100 mK, in order to have at 6 keV a temperature raise of fractions of mK and a resolution of few eV. 
\bigskip

The thermal sensors employed in cyogenic microcalorimetry are of different types. Usually they are based on materials whose electrical resistivity is strongly temperature-dependent, like superconductors or doped semiconductors (resistive microcalorimeters), but there are also sensors based on different principles, like the ones using paramagnetic materials with a temperature-dependent magnetization (Metallic Magnetic Calorimeters - MMC \cite{mmc}), or the ones exploiting the temperature dependence of the kinetic inductance effect (Thermal Kinetic Inductance Detectors - TKID \cite{tkid}). 

\begin{figure}[H]
\centering
\includegraphics[width=0.6\linewidth]{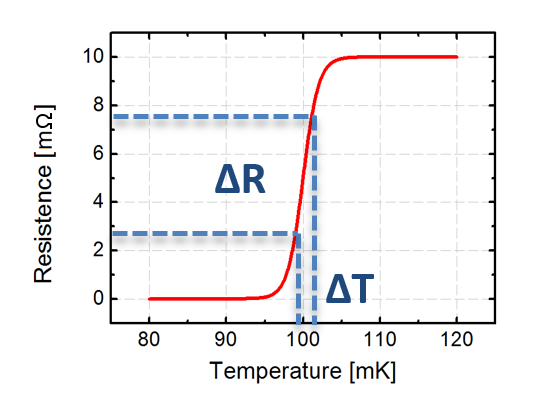}
\caption{Superconducting transition of a TES with a critical temperature of $\sim$ 100 mK and a normal resistance of 10 m$\Omega$. Inside the sharp transition the resistance significantly varies in fraction of mK, suggesting the use as a sensitive thermometer.}
\label{TES_microcalorimeter_2}
\end{figure}

Here we will deal with the Transition Edge Sensors (TES), which consist of a superconducting film operated in the narrow temperature region between the normal and superconducting state, where the electrical resistance varies between zero and its normal value (Fig. \ref{TES_microcalorimeter_2}).

\noindent The temperature sensitivity of a TES is described by the adimensional parameter $\alpha$:

\begin{equation}
\alpha = \frac{d \; log(R)}{d \; log(T)} = \frac{T}{R}\frac{dR}{dT}
\end{equation} 
In the TES used for X-ray spectroscopy $\alpha$ is usually of the order of 100, about two order of magnitude higher than the typical semiconductor thermistor thermometer.
\bigskip

The TES are active devices, since they need to be biased in order to be operated inside the superconductive transition, at a temperature higher than the thermal bath. They are usually voltage-biased, balancing the thermal power flowing towards the bath with the joule power dissipated into the device (P$_J$ = V$^2$/R$_{TES}$). Note that in this way, when a particle deposits energy into the absorber and the TES temperature raise, the bias joule power drops down due to the raise of the TES resistance, helping to cool the device back to its initial state. This effect is known as \textit{electrothermal feedback (ETF)}, and it is useful for several reasons:

\begin{itemize}
\item The ETF increases the stability of the TES bias point;
\item The ETF reduces the pulse decay time, allowing the acquisition at higher count-rate and thus enabling the observation of bright sources and the use of large effective-area optics;
\item The ETF reduces the sensibility of the TES performance to the fluctuations in the bath temperature, easing the requirements for temperature stability.;
\item The ETF reduces the detector sensitivity to the TES parameter variation (i.e. T$_C$, transition width, magnetic field environment, ...), making it easily to operate large arrays of TES devices.
\end{itemize}

\begin{figure}[H]
\centering
\includegraphics[width=0.6\linewidth]{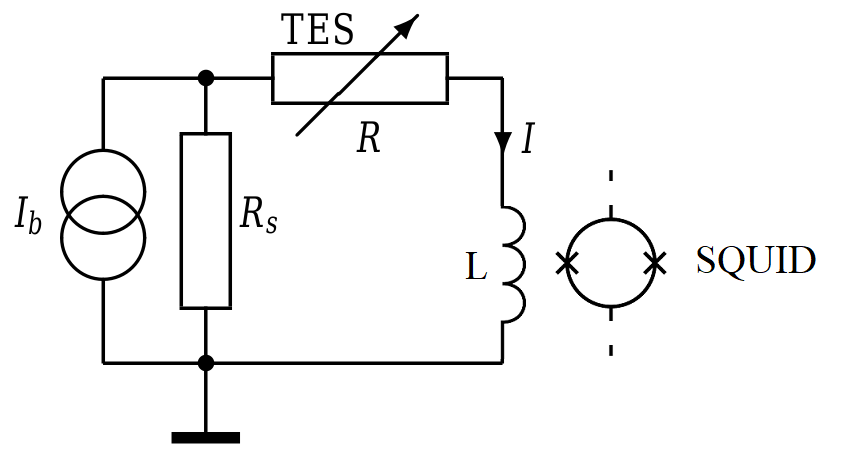}
\caption{Typical circuit diagram for operating a TES under voltage bias with a SQUID readout.} 
\label{rcircuit}
\end{figure}

\noindent The TES voltage bias is typically implemented applying a bias current $I_b$ to a small shunt resistor $R_s$ in parallel with the TES (Fig. \ref{rcircuit}). In this configuration, a change in the TES resistance manifests as a change in the current $I$ passing through the TES. To measure this current variation, a cryogenic low noise amplifier with a low impedance it is needed, in order to match with the low resistance of the TES. An optimal choise is to use a Superconductive Quantum Interference Device (SQUID), an extremely sensitive magnetometer measuring the magnetic flux threading it. To do this the TES is usually operated in series with a superconductive input coil L, and the change in the TES current is measured as a change in the magnetic flux inductively coupled to the SQUID. 

In the next section we will present the \virg standard'' differential equations describing the electrothermal evolution of a TES microcalorimeter, following \cite{irwin}.

\subsection{The \virg standard" electrothermal model}
\label{standardmodel}
\begin{figure}[H]
\centering
\includegraphics[width=0.7\linewidth]{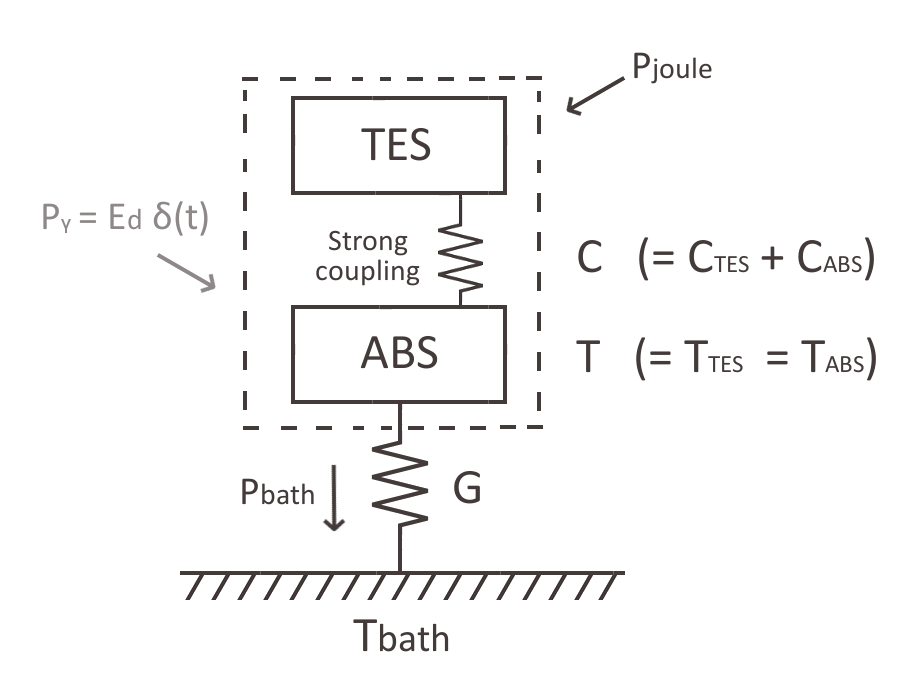}
\caption{Thermal circuit representing a simple TES microcalorimeter. TES and Absorber are so strongly coupled to be considered as a single system with heat capacity C and temperature T (dashed box). This system is linked to the thermal bath at $T_{bath}$ through the thermal conductance G. The input joule power $P_{joule}$ is due to the TES bias, while $P_{\gamma}$ is the signal power dissipated in the system when a particle deposits energy into the absorber. $P_{bath}$ is the power flowing towards the bath.}
\label{model_standard}
\end{figure}

\begin{figure}[H]
\centering
\includegraphics[width=0.9\linewidth]{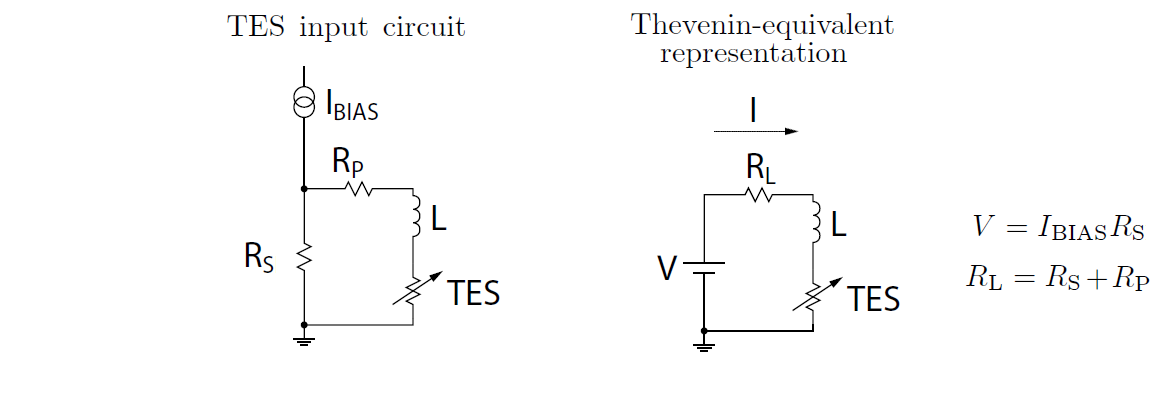}
\caption{The TES electrical circuit and its Thevenin-equivalent representation. \textit{(Left)} A bias current $I_{BIAS}$ is applied to the shunt resistor $R_S$ in parallel with the TES, the total inductance L (including the input coil and the stray wire contribution) and a parasitic resistance $R_P$. \textit{(Right)} In the Thevenin-equivalent circuit a bias voltage $V = I_{BIAS} R_S$ is applied to a load resistor $R_L = R_S + R_P $, the TES and the input coil. From: \cite{irwin}.}
\label{tes_circuit}
\end{figure}

The thermal and electrical circuits representing a simple TES microcalorimeter are shown in Fig. \ref{model_standard} and \ref{tes_circuit}. Note that in the thermal circuit the TES and the absorber are so strongly coupled to be considered as a single system with heat capacity $C = C_{TES} + C_ {ABS}$ and global temperature $T = T_{TES} = T_{ABS}$. 

The system is made of two coupled differential equations describing the electrical and thermal circuits: the thermal equation determines the temperature T, and the electrical equation determines the current I passing through the TES. We have:

\begin{equation}
  \begin{cases}
  C \dfrac{dT}{dt} + P_{bath} = P_{joule} + P_{\gamma}\\
  \\
  L \dfrac{dI}{dt} + I R_L + I R_{TES} = V
  \end{cases}
  \label{system1}
\end{equation}

\noindent where the bias joule power is:

\begin{equation}
P_{joule} = I^2 R_{TES}
\end{equation}

\noindent and the power flow to the thermal bath can be generically described by a power-law \cite{irwin}:

\begin{equation}
P_{bath} = k(T^n - T_{bath}^n)
\end{equation}

\noindent where k and n are determined by the nature of the thermal link to the bath. Finally, the signal power $P_{\gamma}$ can be expressed as:

\begin{equation}
P_\gamma = E_d \delta (t)
\end{equation}

\noindent where $E_d$ is the energy deposited by a particle into the absorber and $\delta (t)$ is the Dirac delta-function.

\bigskip
In absence of signal ($P_{\gamma} = 0$), the system will be in a steady-state characterized by the values of temperature, current and TES resistance $T_0$, $I_0$ and $R_0$, representing the TES bias point. Searching for solution in the small-signal limit, the system \ref{system1} can be linearized around this steady-state. We can approximate:

\begin{equation}
  \begin{aligned}
  T &\approx T_0 + \delta T \\
  I &\approx I_0 + \delta I \\
  R_{TES} &\approx R_0 + \alpha_0 (R_0 / T_0) \delta T + \beta_0 (R_0 / I_0) \delta I \\
  P_{bath} &\approx P_{bath,0} + G \delta T \\
  P_{joule} &\approx P_{J,0} + \alpha_0 (P_{J,0}/T_0) \delta T + (\beta_0 + 2) (P_{J,0}/I_0) \delta I
  \end{aligned}
  \label{approxlin}
\end{equation}

\noindent where $\alpha_0$ and $\beta_0$ are the temperature and the current sensitivity of the TES, defined as:

\begin{equation}
\alpha_0 = \left( \frac{T}{R} \frac{\partial R}{\partial T} \right)_{I_0, T_0}
\;\;\; , \;\;\;
\beta_0 = \left( \frac{I}{R} \frac{\partial R}{\partial I} \right)_{I_0, T_0}
\end{equation}

\noindent and G is the thermal conductance of the thermal link to the bath evaluated at $T_0$:
\begin{equation}
G = \left(\frac{dP_{bath}}{dT}\right)_{T_0} = k n T_0^{(n-1)}
\end{equation}

\bigskip
\noindent Substituting \ref{approxlin} into \ref{system1} and dropping the second-order terms we obtain: 

\bigskip
\begin{equation}
\begin{cases}

C \dfrac{d \delta T}{dt} + G \delta T = \alpha_0 \dfrac{P_{J,0}}{T_0} \delta T + (\beta_0 + 2) \dfrac{P_{J,0}}{I_0} \delta I + P_{\gamma} \\
\\

L \dfrac{d \delta I}{dt} + R_L \delta I + \alpha_0 \dfrac{I_0 R_0}{T_0} \delta T + \beta_0 R_0 \delta I + R_0 \delta I = 0
\end{cases}
\end{equation}
\bigskip

\noindent Note that in the thermal equation there are now two additional source terms due to the dependece of the bias joule power $P_{joule}$ on $\delta I$ an $\delta T$. This is the mathematical manifestation of the previous introduced electrothermal feedback phenomenon.

\bigskip
Doing some algebra we then obtain:

\begin{equation}
  \begin{cases}
  \dfrac{d(\delta T)}{dt} + \dfrac{(1-\mathscr{L})}{\tau_{th}} \delta T - \dfrac{I_0 R_0 (2+ \beta_0)}{C} \delta I = \dfrac{P_\gamma}{C} \\
  \\
 \dfrac{d (\delta I)}{dt} + \dfrac{G \mathscr{L}}{I_0 L} \delta T + \dfrac{\delta I}{\tau_{el}}  = 0
  \end{cases}
  \label{afteralgebra}
\end{equation}

\noindent where $\mathscr{L}$ is named \virg Loop gain' and it is defined as: 

\begin{equation}
\mathscr{L} = \dfrac{\alpha_0 P_{J,0}}{G T_0}
\end{equation}

\noindent $\tau_{th}$ is the thermal characteristic time of the microcalorimeter:
\begin{equation}
\tau_{th} = C/G
\end{equation}

\noindent and $\tau_{el}$ is the electrical characteristic time of the TES circuit:
\begin{equation}
\tau_{el} = \dfrac{L}{R_L + R_0 (1 + \beta_0)}
\end{equation}

\bigskip
The system \ref{afteralgebra} can be easily solved in the limit $L \sim 0$, i.e. when $\tau_{el}$ is much less than all the others  characteristic times of the system. In this approximation we have:

\begin{equation}
(G \mathscr{L} / I_0) \delta T \approx - [R_L + R_0(1+ \beta_0)]\delta I
\end{equation}

\noindent and so we obtain:

\begin{equation}
  \begin{cases}
  \dfrac{d(\delta T)}{dt} + \dfrac{\delta T}{\tau_{eff}} =  \dfrac{P_\gamma}{C}\\
  \\
 \dfrac{d (\delta I)}{dt} + \dfrac{G \mathscr{L}}{I_0 L} \delta T + \dfrac{\delta I}{\tau_{el}}  = 0
  \end{cases}
  \label{finalsystem}
\end{equation}

\noindent where $\tau_{eff}$ is the effective thermal time constant of the system, lower than $\tau_{th}$ due to the electrothermal feedback:

\begin{equation}
\tau_{eff} = \tau_{th} \dfrac{(1 + \beta_0 + R_L/R_0)}{1+\beta_0+R_L/R_0+(1-R_L/R_0)\mathscr{L}}
\end{equation}

\noindent Finally, let us define the parameter:
\begin{equation}
\mathscr{L^*} = \dfrac{\mathscr{L}}{1+\beta_0+R_L/R_0 + (1-R_L/R_0)\mathscr{L}}
\end{equation}
which tends to the unity for high $\mathscr{L}$.

\bigskip 
The solution of the system \ref{finalsystem} is given by:

\begin{equation}
\delta T = \dfrac{E_d}{C} e^{-t/\tau_{eff}}
\end{equation}

\begin{equation}
\boxed{ \delta I = - \dfrac{E_d \mathscr{L^*}}{I_0 R_0 \tau_{eff}}  \dfrac{(e^{-t/\tau_{eff}} - e^{-t/\tau_{el}} )}{(1-\tau_{el}/\tau_{eff})} }
\label{solution1}
\end{equation}

\begin{figure}[H]
\centering
\includegraphics[width=0.6\linewidth]{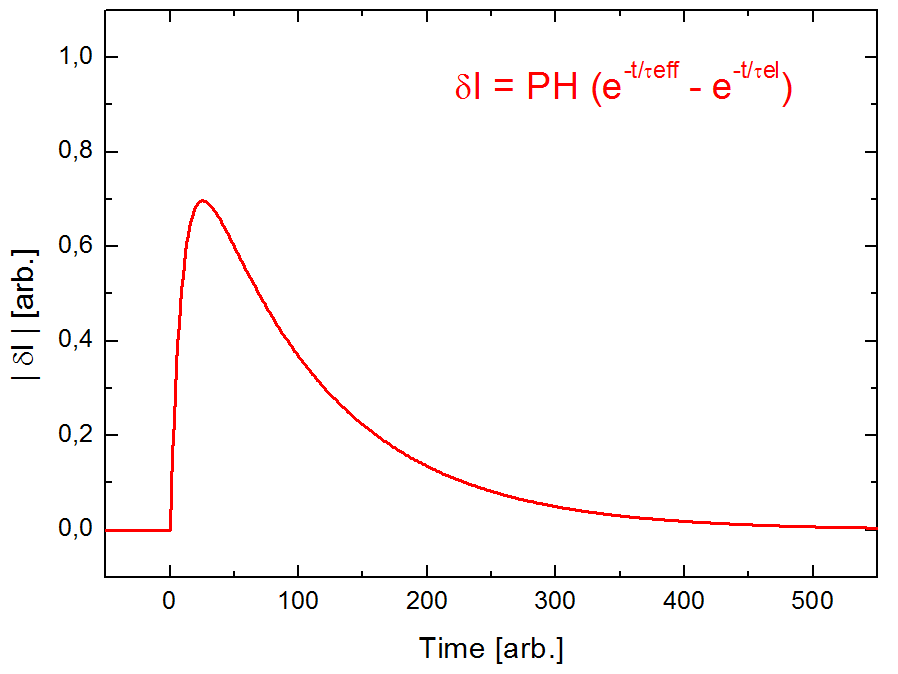}
\caption{The TES response to an instantaneous deposit of energy into the absorber is a pulse with Pulse Height proportional to $E_d$, rise-time $\tau_{el}$ and decay time $\tau_{eff}$.}
\label{singlepulse}
\end{figure}

\noindent In the low-signals and low-inductance limits, the TES response to an  instantaneous deposit of energy into the absorber is therefore a pulse (Fig. \ref{singlepulse}) with rise-time $\tau_{el}$, decay time $\tau_{eff}$ and Pulse Height:

\begin{equation}
PH = -\dfrac{E_d \mathscr{L^*}}{I_0 R_0 \tau_{eff} (1-\tau_{el}/\tau_{eff})}
\end{equation}

\subsection{Noise}

In this section I will introduce the main noise sources in a TES microcalorimeter, showing how they are observed as noise currents at the system output (following the formalism reported in \cite{bergmann}).

\subsubsection{Phonon noise}
The phonon noise is due to the fluctuation of the thermal power flowing towards the thermal bath due to the random exchange of phonons between the absorber and the bath through \virg G''. This fluctuation is given by:

\begin{equation}
\delta P_{ph} = \sqrt{4 \gamma k_B T_0^2 G}
\end{equation}

\noindent where $\gamma \sim n/(2n +1)$ is a parameter accounting for the thermal gradients over the thermal link. The phonon noise is observed at the output of the system with a current spectral density [$A/\sqrt{Hz}$]:

\begin{equation}
\delta I_{ph}(f) = \frac{\sqrt{4 \gamma k_B T_0^2 G}}{I_0 R_0}  \mathscr{L^*} \frac{1}{\sqrt{1 + (2 \pi f \tau_{eff})^2}} \frac{1}{\sqrt{1 + (2 \pi f \tau_{el})^2}}
\end{equation}

\subsubsection{TES Johnson noise}
The Johnson noise is due to the random transport of charge over a resistor. It manifests as a voltage fluctuation across the TES:

\begin{equation}
\delta V_{jo} = \sqrt{4 k_B T_0 R_0}
\end{equation}

\noindent The related Noise Equivalent Power (NEP) and noise current spectral density are:

\begin{equation}
\delta P_{jo}(f) = \sqrt{4 k_B T_0 R_0} \frac{GT_0}{I_0 R_0 \alpha_0}\left( \frac{1 + \beta_0 + \frac{R_L}{R_0}}{1 + \frac{R_L}{R_0}}\right) \sqrt{1 + (2 \pi f \tau_{th})^2} 
\end{equation}

\begin{equation}
\delta I_{jo}(f) = \frac{\sqrt{4 k_B T_0 R_0}}{R_0 + R_L} \left( \frac{1+\beta_0 + \frac{R_L}{R_0}}{1 + \beta_0 + \frac{R_L}{R_0} + (1-\frac{R_L}{R_0})\mathscr{L}} \right) \frac{\sqrt{1 + (2 \pi f \tau_{th})^2}}{\sqrt{1 + (2 \pi f \tau_{eff})^2}} \frac{1}{\sqrt{1 + (2 \pi f \tau_{el})^2}}
\end{equation}

\subsubsection{Shunt Johnson noise}
The Johnson noise manifests also as a voltage fluctuation across the shunt resistor:

\begin{equation}
\delta V_{sh} = \sqrt{4 k_B T_S R_S}
\end{equation}

\noindent where $T_s$ is the shunt temperature. In this case the NEP and the output current spectral density will be:

\begin{equation}
\delta P_{sh}(f) = \sqrt{4 k_B T_S R_S} \frac{GT_0}{I_0 R_0 \alpha_0}\left(\frac{1 + \beta_0 + \frac{R_L}{R_0}}{1 + \frac{R_L}{R_0}}\right) \sqrt{(1-\mathscr{L})^2 + (2 \pi f \tau_{th})^2}
\end{equation}

\begin{equation}
\delta I_{sh}(f) = \frac{\sqrt{4 k_B T_S R_S}}{R_0 + R_L} \left(\frac{1+\beta_0 + \frac{R_L}{R_0}}{1 + \beta_0 + \frac{R_L}{R_0} + (1-\frac{R_L}{R_0})\mathscr{L}}\right) \frac{\sqrt{(1-\mathscr{L})^2 + (2 \pi f \tau_{th})^2}}{\sqrt{1 + (2 \pi f \tau_{eff})^2}} \frac{1}{\sqrt{1 + (2 \pi f \tau_{el})^2}}
\end{equation}

\subsubsection{Energy resolution}

The theoretical energy resolution of a TES microcalorimeter can be evaluated starting from the NEP calculated in the previous paragraphs: 

\begin{equation}
\Delta E \;\;\; \text{(FWHM)}= \frac{2.36}{\sqrt{\int_0^\infty \dfrac{4 df}{\delta P_{ph}^2 + \delta P_{jo}^2 + \delta P_{sh}^2}}}
\label{deltaENEP}
\end{equation}

\noindent An approximated solution of eq. (\ref{deltaENEP}), valid in the limit of strong electrothermal feedback ($\mathscr{L} \gg 1$), is given by:

\begin{equation}
\Delta E \;\;\; \text{(FWHM)} = 2.36 \sqrt{\xi \cdot k_B \cdot T_0^2 \cdot C} 
\end{equation}

\noindent where: 

\begin{equation}
\xi = \frac{2\sqrt{2 n}}{\alpha_0}
\end{equation}

\noindent In a real calorimeter, the energy resolution is therefore dependent not only on the heat capacity of the system and the operating temperature, but also on the TES sensitivity.

\subsubsection{Excess noise}

At the time of this writing, TES x-ray microcalorimeters have been demonstrated with impressive energy resolutions of $\Delta E \sim 1.6$ eV at 5.9 keV (single pixel) and $\Delta E \sim 2.5$ eV at 5.9 keV (array of $\sim$ 40 pixels) [Data shown during the 17th international workshop on Low Temperature Detectors (LTD17) \cite{ltd17} ]. However, the performance still has not achieved the limits predicted by theory. This is due to additional noise sources in the TES itself that are usually classified as \virg Excess Noise''. The excess noise could be divided in two main macro-categories \cite{irwin}:

\begin{itemize}
\item \textit{Internal Thermal Fuctuation Noise (ITFN):} it is a phononic noise due to the non-uniformity of the TES temperature, and it can be explained in terms of internal thermal fluctuations between distributed heat capacities inside the TES.

\item \textit{Excess electrical noise:} it is a noise with the same frequency dependence as Johnson noise, which has been observed by multiple research groups using many different TES geometries and materials. There is not an universally accepted explanation of this noise, but it have been found strong correlations between it and the TES sensitivity parameters $\alpha$ and $\beta$, the applied magnetic field and the geometry of the detector.

\end{itemize}

\subsection{SQUID readout}
\label{squidpar}

To conclude this introduction to TES microcalorimeter, in this section we will present the SQUID and its typical readout scheme. 

\begin{figure}[H]
\centering
\includegraphics[width=0.5\linewidth]{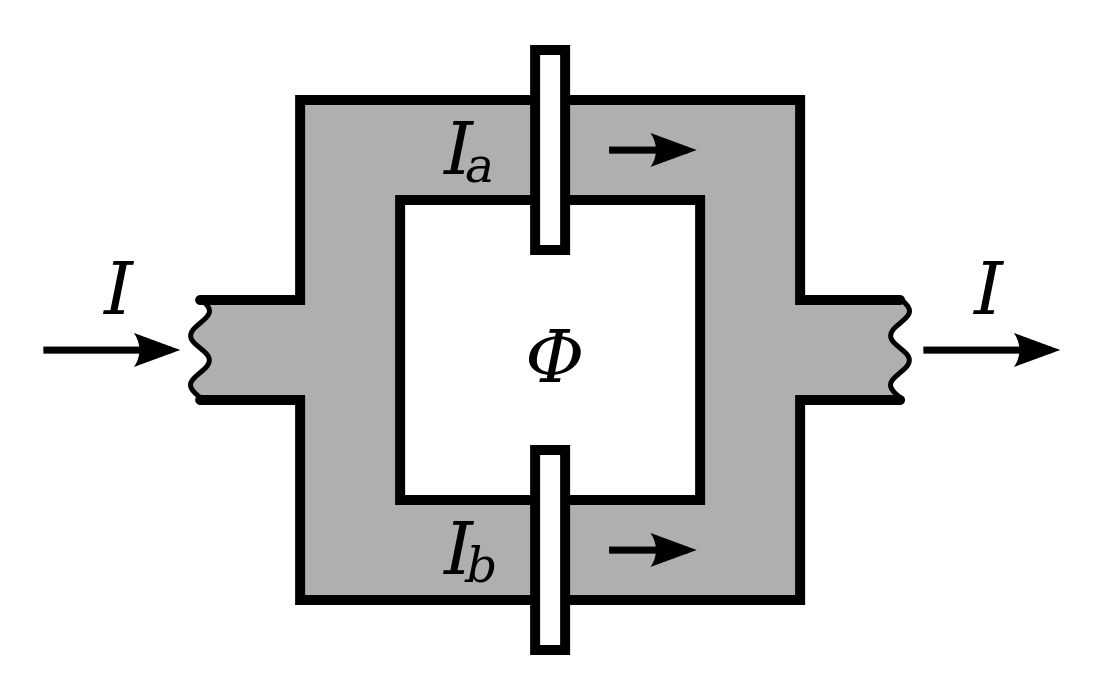}
\caption{Diagram of a SQUID. The bias current I splits into the two branches. As soon as the current in either branch exceeds the critical current of the JJ, a voltage appears across SQUID. This voltage is a periodic function of the external magnetic flux threading the SQUID loop ($\Phi$). From: \cite{squidwiki}.} 
\label{SQUID1}
\end{figure}

\bigskip
The SQUID (Superconductive Quantum Interference Device) is a very sensitive magnetometer able to measure changes in the magnetic flux less than $10^{-15}$ T$\cdot$m$^2$. It is based on a superconducting loop interrupted by two diametrical opposed and identical thin insulating barriers (Josephson Junctions - JJ), as shown in Fig. \ref{SQUID1}.

\begin{figure}[H]
\centering
\includegraphics[width=0.5\linewidth]{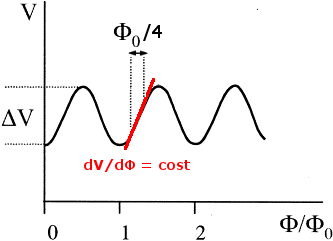}
\caption{SQUID voltage response to an external magnetic flux (V-$\Phi$ characteristic SQUID curve). The red line highlights the narrow linearity range of the response function ($\Delta \Phi_{IN} < \sim \Phi_0/4$)  }
\label{SQUID2}
\end{figure}

\noindent When biased by a current exceeding the critical current of the JJ, the SQUID shows a periodic voltage response to the magnetic flux which passes through it (Fig. \ref{SQUID2}).

\noindent The periodicity  is equal to the flux quantum $\Phi_0$, defined as:

\begin{equation}
\Phi_0 = \frac{h}{2e} \simeq 2.067 \cdot 10^{-15} \;\;\; T \cdot m^2
\end{equation}
where $h$ is the Planck constant and $e$ is the electron charge.

\begin{figure}[H]
\centering
\includegraphics[width=0.6\linewidth]{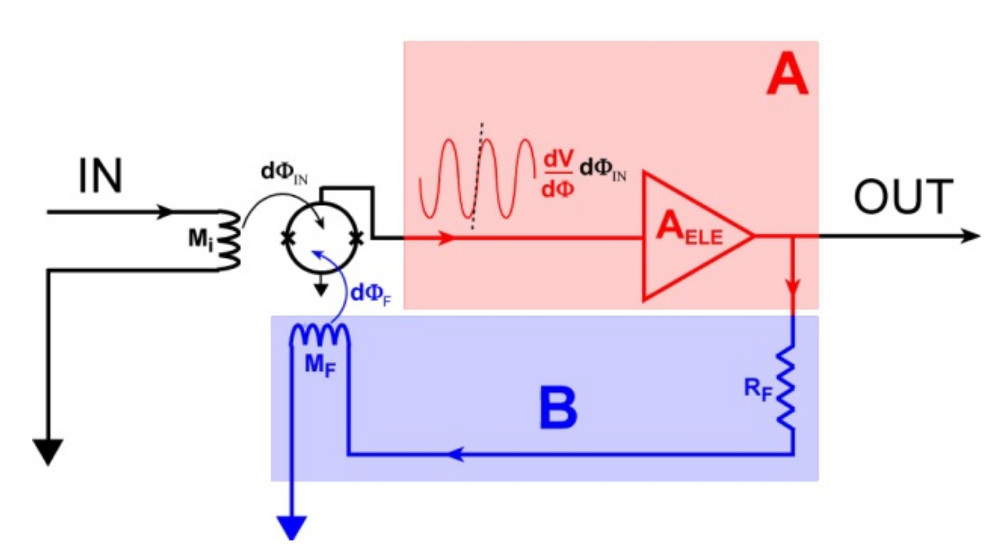}
\caption{Circuit diagram of a typical SQUID Flux Locked Loop (FLL) configuration.}
\label{SQUID_feedback}
\end{figure}

\bigskip
To linearize the response to external magnetic fluxes, SQUIDs are usually operated in the so-called Flux Locked Loop (FLL) configuration (Fig. \ref{SQUID_feedback}), following a typical feedback system. In this configuration the SQUID voltage output is amplified and applied to a feedback resistor, in order to generate a current going into a feedback coil that counteracts the SQUID input magnetic flux. This feedback allows to reduce the magnetic flux excursions around the SQUID operating point, keeping it in the narrow linearity range of the V-$\Phi$ characteristic (Fig. \ref{SQUID1}). 

\noindent By solving the typical closed feedback system, the gain of a SQUID operating in a FLL configuration (i.e. the ratio between the SQUID voltage output $d V_{OUT,FLL}$ and the input flux variation $d\Phi_{IN}$) is given by: 

\begin{equation}
G_{SQUID,FLL} = \frac{d V_{OUT, FLL}}{d \Phi_{IN}} = \frac{A}{1+A \cdot B} \sim \frac{1}{B} = \frac{R_{F}}{M_{F}} \;\;\;\;\; [V/\Phi_0]
\end{equation}
where A and B generically represent the Amplifier and the Feedback blocks of the diagram, and we have assumed A$\cdot$B $\gg$ 1. Note that in this case A = A$_{ele}$ $\cdot$ dV/d$\Phi$ and B = $M_F$/$R_F$, where $R_{F}$ is the feedback resistor and $1/M_F$ is the mutual inductance between the feedback coil and the SQUID (normally expressed in [A/$\Phi_0$]).

\bigskip
In a TES system, the total gain given by the FLL SQUID readout (i.e. the ratio between the SQUID voltage output and the change in the current passing through the TES $\delta I$) is therefore:

\begin{equation}
G_{TOT} = \frac{d V_{OUT, FLL}}{\delta I} = \frac{G_{SQUID,FLL}}{1/M_{IN}} = \frac{R_F}{M_F} \frac{1}{M_{IN}} \;\;\;\;\;\; [V/A]
\end{equation}
where $1/M_{IN}$ is the mutual inductance between the input coil and the SQUID.

\subsection{Multiplexing}

\begin{figure}[H]
\centering
\includegraphics[width=0.9\linewidth]{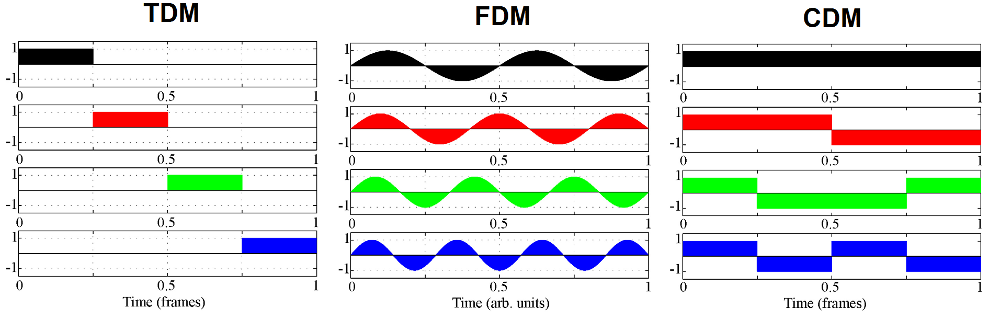}
\caption{TDM (Time Division Multiplexing), FDM (Frequency Division Multiplexing) and CDM (Code Division Multiplexing) modulation functions for the example of a four-pixel multiplexer readout. From: \cite{cdm}.}
\label{multiplexing}
\end{figure}

For the sake of completeness, in this small paragraph I want to mention the issue of multiplexing. In order to develop kilo-pixel order TES arrays, it has been necessary to develop cryogenic multi-channel readout techniques to reduce the number of wires required between temperature stages, due to the limited cooling power at cold. These \textit{multiplexing} techniques enables several pixels to be readout by the same SQUID. The most important are the Time Division Multiplexing (TDM), the Frequency Division Multiplexing (FDM) and the Code Division Multiplexing (CDM), whose modulation functions are shown in Fig. \ref{multiplexing}.

The TDM simply consits of a time scan,  at regular intervals, of the pixel chain. The pixels are DC-biased, and a second SQUID stage acts as \virg supercondictive switch'', enabling this time scanning. For the FDM, instead, each pixel is AC biased with a specific carrier frequency, so that at each frequency of the output signal can be associated only one pixel. The energy spectrum is obtained by de-modulation techniques at the same bias modulation frequencies. This is the baseline scheme that will be used for the X-IFU TES array readout. Finally, in the CDM the polarities of the pixels bias are switched in the pattern of an Walsh matrix, which can then be used to reconstruct the pixels signals.

\newpage
\section{Athermal phonon mediated TES microcalorimeters}
\label{athermalsection}

\begin{figure}[H]
\centering
\includegraphics[width=0.9\linewidth]{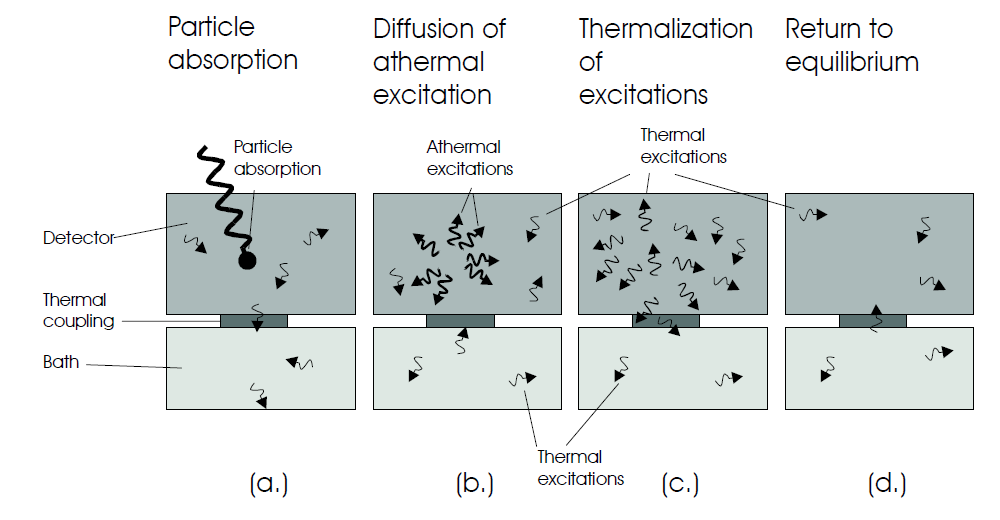}
\caption{The absorption of a particle in a low temperature detector based on a crystalline absorber. The process is divided in the following steps: (a.) the detector is at equilibrium until it absorbs a particle. (b.) the absorption event creates a number of athermal excitations in the detector. (c.) the athermal excitation inelastic scattering multiplies thermal excitations. (d.) the detector cools back to equilibrium, ready to measure a new particle. From: \cite{lindeman}.}
\label{lindeman_scheme}
\end{figure}

So far, we have considered TES microcalorimeters as purely thermal detectors, describing their behaviour with the assumption of thermodynamic equilibrium. Actually, in some cases it is useful to design these detectors in order to operate in an athermal regime, exploiting the TES capability to collect also out-of-equilibrium signal. This is the case of the so-called \virg athermal phonon mediated TES microcalorimeter'', in which the TES are used to detect the athermal excitation that propagates in a crystalline absorber after a particle interaction (Fig. \ref{lindeman_scheme}). The main advantage of working in the athermal regime is the possibility to operate with fast time constants (pulse decay time typically less than 1 ms) even with very large absorbers (up to tens of cm$^3$ \cite{pyle}). This feature is exploited for example in rare event searches experiments, like the ones to search for dark matter \cite{cresst} \cite{cdms} or double beta decay \cite{cuore}.

Here we deal with this kind of detector because the CryoAC itself is designed to efficiently collect the athermal excitation generated in its silicon absorber. This in order to exhibit a fast component of the pulse with a sharp rise front, that can be efficiently used to trigger the background particles and set the veto flag.

\bigskip
To explain the principle function of these detectors we will start by recalling the notion of phonon. In a solid, atoms are elastically bounded to their neighbors, forming a lattice structure. A phonon is the quantum mechanical description of an elementary vibrational motion in which the lattice oscillates at a single frequency. In other words, phonons are to vibrations in a lattice what photons are to electromagnetic radiation in vacuum. A phonon can be represented by its wavevector $k$, frequency $\nu$, and polarization (i.e longitudinal, transverse, ...). If the atoms in the unit cell of the lattice move in phase, the phonon is named \virg acoustic'' (long wavelength limit), whereas when the atoms move out of phase it is named \virg optical'' (short wavelength limit). At thermal equilibrium, the average number of phonons with a given frequency it is given by the Planck distribution:

\begin{equation}
<n(\nu)> = \frac{1}{e^{h \nu/(k_B T)} -1}
\end{equation}

\noindent showing a peak at $h \nu \sim 2.8 \cdot k_B T$ ($\nu_{Peak} \sim $ 6 GHz at 100 mK)

\bigskip
When a particle interaction occurs in a low temperature crystal, energy is deposited in the form of electron-hole pairs and high energy optical phonons. The optical phonons decay very rapidly (ns) into acoustical phonons of about half of the Debye frequency (13.5 THz in Si), which propagate from the interaction site across the crystal, decaying into phonon of lesser frequency due to lattice anharmonicity. These phonons decay until they become ballistic, meaning that their mean free path become longer than the dimensions of the crystal. At this point they start to bounce around the crystal as particles in a box. Initially, these ballistic phonons are not in thermal equilibrium with the bulk of phonons in the crystal, and thus they are called \virg athermal''. Finally, after a time scale going from tens of $\mu$s to several ms (depending on the crystal characteristics and dimensions), they completely thermalize, causing a rise in the overal crystal lattice temperature.  

\begin{figure}[H]
\centering
\includegraphics[width=0.9\linewidth]{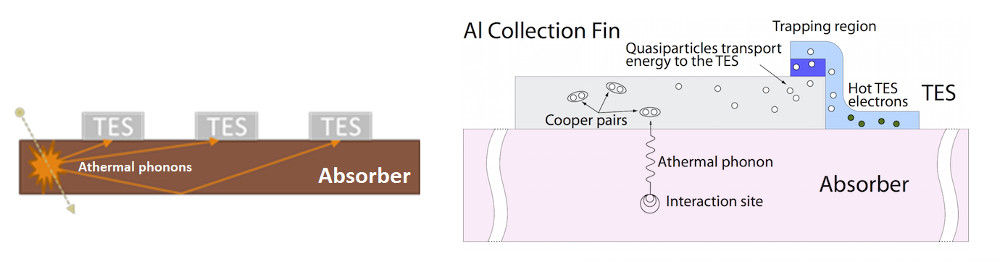}
\caption{Schemes of the detection principle in an athermal phonon mediated microcalorimeter. \textit{Left:} The ballistic athermal phonons generated when a particle deposit energy into the absorber can enter the TES deposited on the crystal surface, quickly heating the TES electron system. \textit{Right:} Aluminum collectors can be used to further improve the athermal collection efficiency. The phonons hitting the Al collectors can be absorbed into them, breaking cooper pairs and generating electron-like quasiparticles that diffuse towards the TES. From: \cite{figueroa}}
\label{athermals}
\end{figure}

\bigskip
The athermal phonons mediated TES microcalorimeters exploit the fact that the ballistic athermal phonons can enter the TES deposited on the crystal surface, being efficiently absorbed by the free electrons of the metal film (Fig. \ref{athermals}-Left). This interaction quickly heats the TES electron system, generating a fast athermal signal that happens before the usual thermal one (related to the absorber temperature rise). Note that for a detector based upon detecting athermals, it is imperative to collect as much of the phonon signal before it thermalizes. This requires that much of the surface area of the absorber (crystal) must be instrumented, so that when an athermal phonon reaches the surface the probability of being absorbed in a TES is improved. For this reason, these detectors are frequently featured also with a layer of aluminum collectors directly connected to the TES. The athermal phonons hitting the aluminum collectors can be absorbed into them, breaking cooper pairs and generating electron-like quasiparticles. These quasiparticles then diffuse through the aluminum towards the TES, where they deposit their energy contributing to the fast athermal signal (Fig. \ref{athermals}-Right).

\subsection{A model for athermal phonon mediated TES microcalorimeters}

\begin{figure}[H]
\centering
\includegraphics[width=0.8\linewidth]{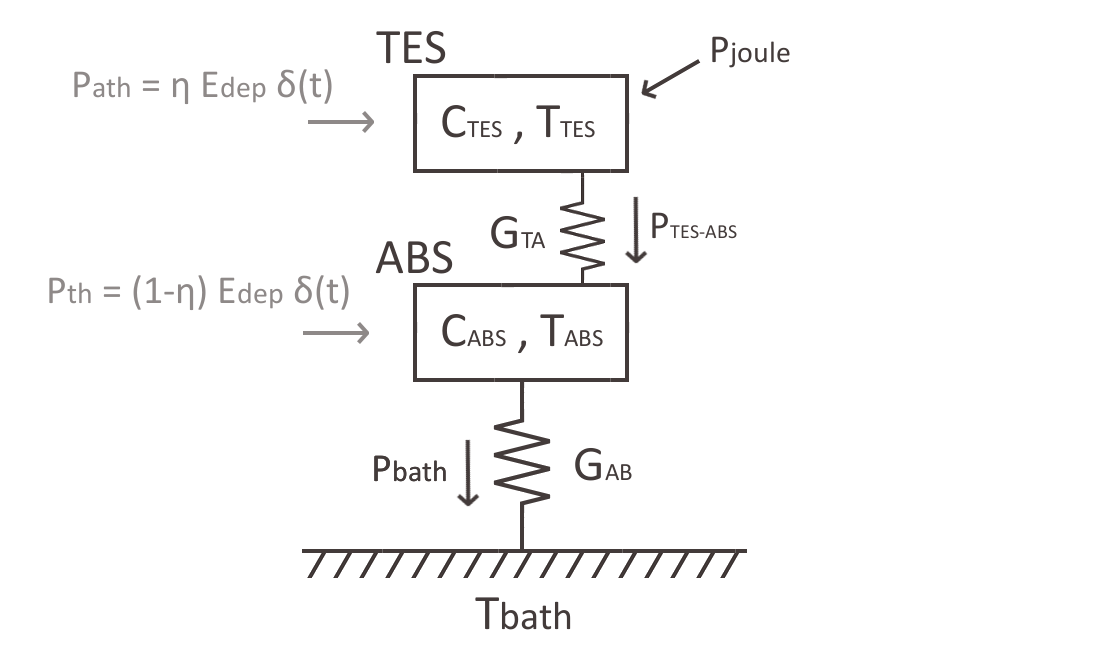}
\caption{Thermal circuit representing a simple athermal phonon mediated TES microcalorimeter. The TES is connected to the crystalline absorber through the thermal conductance $G_{TA}$, and the absorber is linked to the thermal bath at $T_{bath}$ through the thermal conductance $G_{AB}$. The input joule power $P_{joule}$ is due to the TES bias, while $P_{ath}$ and $P_{th}$ are the signal power injected in the TES and in the absorber when a particle deposits energy into the system. $P_{bath}$ is the power flowing towards the bath, whereas $P_{TES-ABS}$ is the power flowing from the TES to the absorber.}
\label{model_decoupled}
\end{figure}

\begin{figure}[H]
\centering
\includegraphics[width=0.9\linewidth]{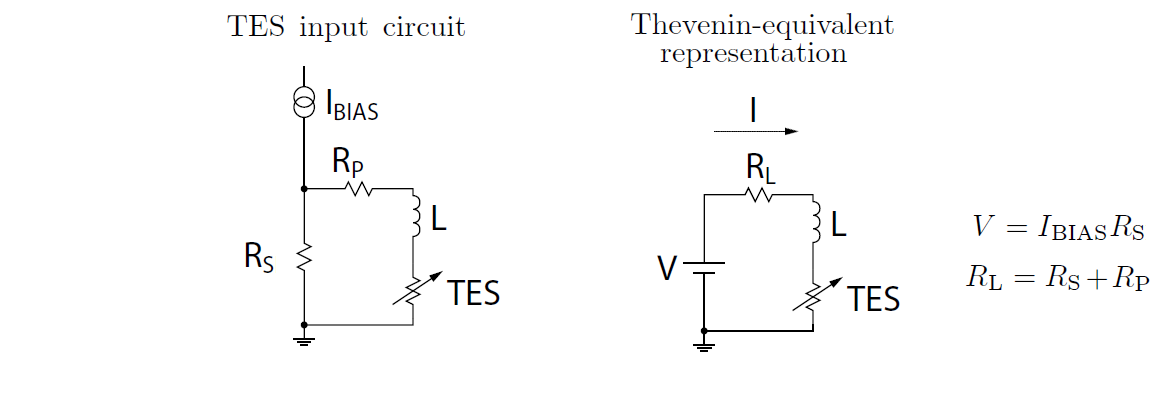}
\caption{The TES electrical circuit and its Thevenin-equivalent representation are the same as those shown in Sect. \ref{standardmodel} \textit{(Left)} A bias current $I_{BIAS}$ is applied to the shunt resistor $R_S$ in parallel with the TES, the input coil L and a parasitic resistance $R_P$. \textit{(Right)} In the Thevenin-equivalent circuit a bias voltage $V = I_{BIAS} R_S$ is applied to a load resistor $R_L = R_S + R_P $, the TES and the input coil. From: \cite{irwin}.}
\label{tes_circuit2}
\end{figure}

In this section we will develop a simple electrothermal model for athermal phonon mediated TES microcalorimeters, starting from the thermal and electrical circuits shown in Fig. \ref{model_decoupled} and \ref{tes_circuit2}. Note that in this case, differently from the \virg standard'' model in Sect. \ref{standardmodel}, we can not assume a strong coupling between the TES and the absorber. Indeed, at very low temperature, the thermal impedance between the electron and the phonon systems in the TES can be important, and the TES free electrons will have generally a different temperature from the phonons in the TES itself (which can be neglected in the model due to their very small heat capacity) and the phonons in the crystalline absorber.

In this case the system is described by three coupled differential equations, governing the evolution of three state variables: the TES electrons temperature $T_{TES}$, the absorber phonons temperature $T_{ABS}$ and the currrent I passing through the TES. We have:

\begin{equation}
  \begin{cases}
  C_{TES} \dfrac{dT_{TES}}{dt} + P_{TES-ABS} = P_{joule} + P_{ATH}\\
  \\
  C_{ABS} \dfrac{dT_{ABS}}{dt} + P_{bath} - P_{TES-ABS} = P_{TH}\\
  \\
  L \dfrac{dI}{dt} + I R_L + I R_{TES} = V
  \end{cases}
  \label{systemath}
\end{equation}

\noindent where the bias joule power is:

\begin{equation}
P_{joule} = I^2 R_{TES}
\end{equation}

\noindent and the power flows from the TES to the absorber and from the absorber to the thermal bath can be generically described by power-laws: 

\begin{equation}
P_{TES-ABS} = k_{TA} (T_{TES}^{n_{TA}} - T_{ABS}^{n_{TA}})
\end{equation}
\begin{equation}
P_{bath} = k_{AB} (T_{ABS}^{n_{AB}} - T_{bath}^{n_{AB}})
\end{equation}

\noindent where the couples of parameters $k_{TA}$,$n_{TA}$  and $k_{AB}$,$n_{AB}$ are determined by the nature of the thermal links. Finally, the signal power is divided into two contributions: the first \virg athermal'' injected directly into the TES system by the athermal phonons that enter the metal film

\begin{equation}
P_{ATH} = \eta E_d \delta (t)
\end{equation}

\noindent and the second \virg thermal'' injected into the absorber by the rest of the phonons that thermalize in the crystal

\begin{equation}
P_{TH} = (1-\eta) E_d \delta (t)
\end{equation}

\noindent where $E_d$ is the total energy deposited by a particle, $\delta (t)$ is the Dirac delta-function and $\eta$ is a parameter representing the TES athermal phonons collecting efficiency.  

\bigskip
In absence of signal ($P_{ATH}$ = 0, $P_{TH}$ = 0) the system will be in a steady-state characterized by the values of temperatures, TES current and TES resistance $T_{TES,0}$, $T_{ABS,0}$, $I_0$ and $R_0$. Similarly to what done in
 sect. \ref{standardmodel}, we can linearize the system \ref{systemath} around this steady state, approximating: 
 
\begin{equation}
  \begin{cases}
  T_{TES} \approx T_{TES,0} + \delta T_{TES} \\
  T_{ABS} \approx T_{ABS,0} + \delta T_{ABS} \\
  I_{TES} \approx I_0 + \delta I \\
  R_{TES} \approx R_0 + \alpha_0 (R_0 / T_0) \delta T + \beta_0 (R_0 / I_0) \delta I \\
  P_{joule} \approx P_{J,0} + \alpha (P_{J,0}/T_{TES,0}) \delta T_{TES} + (\beta + 2) (P_{J,0}/I_0) \delta I \\
  P_{bath} \approx P_{bath,0} + G_{AB} \delta T_{ABS} \\
  P_{TES-ABS} \approx P_{TES-ABS, 0} + G_{TA, Ttes0} \delta T_{TES} - G_{TA, Tabs0} \delta T_{ABS}
   \end{cases}
   \label{linearization}
\end{equation}

\noindent where $\alpha_0$ and $\beta_0$ are the temperature and current sensitivity of the TES:
\begin{equation}
\alpha_0 = \left( \frac{T}{R} \frac{\partial R}{\partial T} \right)_{I_0, T_0}
\;\;\; , \;\;\;
\beta_0 = \left( \frac{I}{R} \frac{\partial R}{\partial I} \right)_{I_0, T_0}
\end{equation}

\noindent $G_{AB}$ is the thermal conductances between the absorber and the thermal bath evaluated at $T_{ABS,0}$:

\begin{equation}
G_{AB} = \left(\frac{dP_{bath}}{dT_{ABS}}\right)_{T_{ABS,0}} = k_{AB} n_{AB} T_{ABS,0}^{(n_{AB}-1)}
\end{equation}

\noindent and $G_{TA, Ttes0}$ and $G_{TA, Tabs0}$ represent the thermal conductance between the TES and the absorber evaluated at $T_{TES,0}$ and $T_{ABS,0}$, respectively:

\begin{equation}
G_{TA, Ttes0} = \left(\frac{\partial P_{TES-ABS}}{\partial T_{TES}}\right)_{T_{TES,0}} = k_{TA} n_{TA} T_{TES,0}^{(n_{TA}-1)}
\end{equation}

\begin{equation}
G_{TA, Tabs0} = - \left(\frac{\partial P_{TES-ABS}}{\partial T_{ABS}}\right)_{T_{ABS,0}} = k_{TA} n_{TA} T_{ABS,0}^{(n_{TA}-1)}
\end{equation}

\bigskip
\noindent Substituting \ref{linearization} into \ref{systemath} and dropping the second-order terms we obtain:

\begin{equation}
  \begin{cases}
  C_{TES} \dfrac{d \delta T_{TES}}{dt} + G_{TA, Ttes0} \delta T_{TES} - G_{TA, Tabs0} \delta T_{ABS} = \\ \\
\;\;\;\;\;\;\;\;\;\;\;\;\;\;\;\;\;\;\;\;\;\;\;\;\;\;\;\;\;\;\;\;\;\;\;
= \alpha_0 \dfrac{P_{J,0}}{T_{TES,0}} \delta T_{TES} + (\beta_0 + 2) \dfrac{P_{J,0}}{I_0} \delta I + P_{ATH}\\
  \\
  C_{ABS} \dfrac{d \delta T_{ABS}}{dt} + G_{AB} \delta T_{ABS} - G_{TA, Ttes0} \delta T_{TES} + G_{TA, Tabs0} \delta T_{ABS} =  P_{TH} \\
  \\
  L \dfrac{d \delta I}{dt} + R_L \delta I + \alpha_0 \dfrac{I_0 R_0}{T_{TES,0}} \delta T_{TES} + \beta_0 R_0 \delta I + R_0 \delta I = 0
  \end{cases}
\end{equation}
\bigskip

\noindent Doing some algebra we then obtain: 

\begin{equation}
  \begin{cases}
 \dfrac{d \delta T_{TES}}{dt} + \dfrac{(1 - \mathscr{L})}{\tau_{tes}} \delta T_{TES} - \dfrac{\delta T_{ABS}}{\tau_{ta}} - \dfrac{(2+\beta_0)I_0 R_0}{C_{TES}}\delta I = \dfrac{P_{ATH}}{C_{TES}} \\
 \\
 
 \dfrac{d \delta T_{TABS}}{dt} - \dfrac{\delta T_{TES}}{\tau_{at}} + \delta T_{ABS} \left( \dfrac{1}{\tau_{abs}} + \dfrac{1}{\tau_{at}} \right) = \dfrac{P_{TH}}{C_{ABS}} \\
\\
 \dfrac{d (\delta I)}{dt} + \dfrac{G_{TA,Ttes0} \mathscr{L}}{I_0 L} \delta T_{TES} + \dfrac{\delta I}{\tau_{el}}  = 0 
 
  \end{cases}
  \label{systemath2}
\end{equation}

\noindent where $\mathscr{L}$ is the \virg Loop gain'':
\begin{equation}
\mathscr{L} = \dfrac{\alpha_0 P_{J,0}}{G_{TA,Ttes0} T_{TES,0}}
\end{equation}

\noindent $\tau_{tes}$, $\tau_{ta}$, $\tau_{at}$ and $\tau_{abs}$ are characteristic times of the system:

\begin{equation}
\tau_{tes} = C_{TES}/G_{TA,Ttes0}
\end{equation}

\begin{equation}
\tau_{ta} = C_{TES}/G_{TA,Tabs0}
\end{equation}

\begin{equation}
\tau_{at} = C_{ABS}/G_{TA,Tabs0}
\end{equation}

\begin{equation}
\tau_{abs} = C_{ABS}/G_{AB}
\label{tthermal}
\end{equation}

\noindent and $\tau_{el}$ is the electrical characteristic time of the TES circuit:

\begin{equation}
\tau_{el} = \dfrac{L}{R_L + R_0 (1 + \beta_0)}
\end{equation}

\bigskip
The system \ref{systemath2} can be solved in the limit $L \sim 0 $ (i.e. assuming $\tau_{el}$ much less than all the others characteristic times of the system) and assuming a strong decoupling between the TES and the absorber (i.e.  $G_{TA,Tabs0}$ and $G_{TA,Ttes0}$ much less than all the others thermal conductances of the system). Under these approximations, we obtain:

\begin{equation}
\begin{cases}
\dfrac{d \delta T_{TES}}{dt} + \dfrac{\delta T_{TES}}{\tau_{eff}} = \dfrac{\delta T_{ABS}}{\tau_{ta}} + \dfrac{P_{ATH}}{C_{TES}} \\
 \\
 
 \dfrac{d \delta T_{TABS}}{dt} + \dfrac{\delta T_{ABS}}{\tau_{abs}} = \dfrac{P_{TH}}{C_{ABS}} \\
\\
 \dfrac{d (\delta I)}{dt} + \dfrac{G_{TA,Ttes0} \mathscr{L}}{I_0 L} \delta T_{TES} + \dfrac{\delta I}{\tau_{el}}  = 0 
\end{cases}
\label{finalsystemath}
\end{equation}

\noindent where $\tau_{eff}$ is defined as:
\begin{equation}
\tau_{eff} = \tau_{tes} \dfrac{(1 + \beta_0 + R_L/R_0)}{1+\beta_0+R_L/R_0+(1-R_L/R_0)\mathscr{L}}
\label{teffathermal}
\end{equation}

\noindent Finally, let us define the adimensional parameter:
\begin{equation}
\mathscr{L^*} = \dfrac{\mathscr{L}}{1+\beta_0+R_L/R_0 + (1-R_L/R_0)\mathscr{L}}
\end{equation}
which tends to the unity for high $\mathscr{L}$.

\bigskip
The solution of the system \ref{finalsystemath} is given by:
\begin{equation}
\delta T_{ABS} = \dfrac{(1-\eta)E_d}{C_{ABS}} e^{-t/\tau_{abs}}
\end{equation}

\begin{equation}
\begin{split}
\delta T_{TES} = & \dfrac{\eta E_d}{C_{TES}} e^{-t/\tau_{eff}} + \dfrac{(1-\eta)E_d}{C_{ABS}} \left(\dfrac{T_{ABS}}{T_{TES}}\right)^{(n_{TA}-1)} \dfrac{\tau_{eff}}{\tau_{tes}}\dfrac{(e^{-t/\tau_{abs}} - e^{-t/\tau_{eff}})}{(1- \tau_{eff}/\tau_{abs})}
\end{split}
\end{equation}

\begin{empheq}[box=\fbox]{align}
\delta I \sim - \dfrac{E_d \mathscr{L^*}}{I_0 R_0 \tau_{eff}} & \left[ \dfrac{\eta(e^{-t/\tau_{eff}}  - e^{-t/\tau_{el}}) }{(1-\tau_{el}/\tau_{eff})} + \right. \nonumber \\
& \left. + (1-\eta) \dfrac{C_{TES}}{C_{ABS}} \left(\dfrac{T_{ABS}}{T_{TES}}\right)^{(n_{TA}-1)} \dfrac{(e^{-t/\tau_{abs}} - e^{-t/\tau_{eff}})}{(1-\tau_{eff}/\tau_{abs})} \dfrac{\tau_{eff}}{\tau_{tes}} \right]
\end{empheq}

\begin{figure}[H]
\centering
\includegraphics[width=0.6\linewidth]{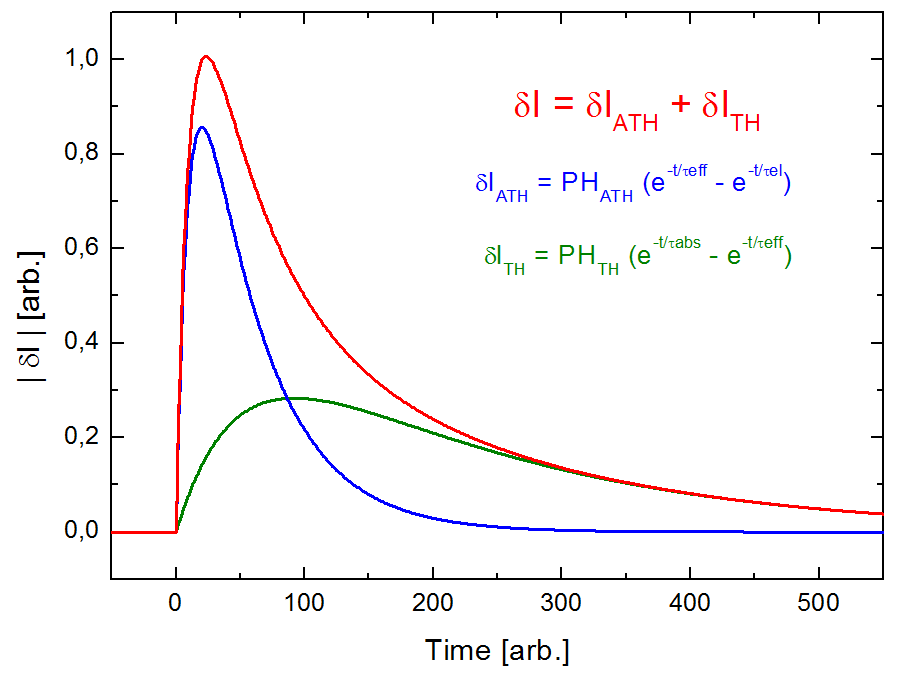}
\caption{The response of an athermal phonon mediated microcalorimeter to a particle event is given by the sum of two different components: an athermal pulse and a thermal pulse.}
\label{doublepulse}
\end{figure}

\noindent In the low-signal and low-inductance limits, assuming a strong decoupling between the TES electrons and the absorber phonons, the response of an athermal phonons mediated microcalorimeter to a particle event is therefore given by the sum of two different pulse components (Fig. \ref{doublepulse}):  

\begin{equation}
\delta I \sim \delta I_{ath} + \delta I_{th} = PH_{ath} (e^{-t/\tau_{eff}} - e^{-t/\tau_{el}}) + PH_{th} (e^{-t/\tau_{abs}} - e^{-t/\tau_{eff}})
\label{endmodel}
\end{equation}

\noindent where:

\begin{equation}
PH_{ath} = -\dfrac{\eta E_d \mathscr{L^*}}{I_0 R_0 \tau_{eff} (1-\tau_{el}/\tau_{eff})}
\end{equation}

\begin{equation}
PH_{th} = -\dfrac{(1-\eta) E_d \mathscr{L^*}}{I_0 R_0 \tau_{eff} (1-\tau_{eff}/\tau_{abs})}\frac{\tau_{eff}}{\tau_{tes}}\frac{C_{TES}}{C_{ABS}}\left(\dfrac{T_{ABS}}{T_{TES}}\right)^{(n_{TA}-1)}
\end{equation}

\subsection{Energy resolution}
The intrinsic limit to the energy resolution of an athermal phonon mediated microcalorimeter is degraded by the number fluctuation of the electron-hole pairs in the absorber \cite{leman}. Indeed, when a particle deposit energy into the crystalline absorber, a part of the energy (about 30$\%$ in Silicon) is spent to generate electron-hole pairs. The number fluctuation of these pairs follows the Fano statistics:

\begin{equation}
\sigma_{e-h \; number} = \sqrt{\frac{F \cdot E_d}{\varepsilon_0}}
\end{equation}

\noindent where $E_d$ is the deposited energy, $F$ is the Fano factor (0.116 for the Silicon) and $\varepsilon_0$ is the average electron-hole production energy (3.81 eV in Silicon).

\bigskip
This number fluctuation is associated to an energy fluctuation that is added at the usual phononic term in the evaluation of the intrinsic energy resolution of the microcalorimeter, giving:

\begin{equation}
\Delta E = 2.36 \sqrt{\left(\sqrt{\xi \cdot k_B \cdot T_0^2 \cdot C}\right)^2 + \left( \frac{E_{gap}}{\varepsilon_{det}} \sqrt{\frac{F \cdot E_{d}}{\varepsilon_0}}\right)^2} \;\;\;\;\; (FWHM) 
\end{equation}
where $\varepsilon_{det}$ is a detector efficiency parameter and $E_{gap}$ is the gap energy (for Silicon at 100 mK is $\sim$ 1.2 eV).

\newpage
\section{Thermal parameters evaluation}

In this last section of the chapter I want to review some useful formula to evaluate the thermal parameters (i.e. heat capacities and thermal conductances) present in the models described in the previous sections.

\subsection{Heat capacities}
\label{hcest}
The heat capacity is the amount of heat required to raise the temperature of an object or substance with one degree. Usually, heat capacity is divided by the amount of substance, mass, or volume, so defining molar heat capacity (heat capacity per mole), specific heat capacity (or specific heat as heat capacity per unit mass) and volume (volumetric) specific heat (as specific heat per unit volume) \cite{lazanu}.

\subsubsection{Normal metals}
At low temperature the heat capacity of normal metals is almost entirely due to the electron heat capacity and thus is linear in temperature. A convenient form for the heat capacity is \cite{irwin}:

\begin{equation}
C_{\text{Normal metal}}(T) = \dfrac{\rho}{A} \gamma_e V T
\end{equation}
where $T$ is the temperature, $V$ is the sample volume, $\gamma_e$ is the molar electrons specific heat, $\rho$ is the mass density, and $A$ is the atomic weight.

\subsubsection{Superconductors}
The BCS theory predicts that at $T_C$ the heat capacity of a superconductor is 2.43 times the normal-metal value \cite{irwin}:

\begin{equation}
C_{\text{Superconductor}}(T) = 2.43 \dfrac{\rho}{A} \gamma_e V T
\end{equation}
Note that the TES is operated within the superconducting transition, so the heat capacity varies within a factor of 2.43 as the resistance change.

\subsubsection{Crystalline materials}
In a crystal the heat capacity is the sum of contribution from the lattice (phonons) and electronic systems. In the frame of the Debye theory, in the low temperature limit (T $\ll$ $\Theta_D$, where $\Theta_D$ is the Debye temperature) the lattice contribution is represented by a $T^3$ dependence. The electronic heat capacity is instead linear in temperature, and it is proportional to the concentration of free electrons. \cite{lazanu}

\begin{equation}
C_{\text{Crystal}}(T) = C_{\text{Phonons}}(T) + C_{\text{Electrons}}(T) = c_p T^3 + c_e T
\end{equation}
with:
\begin{equation}
c_p = \dfrac{12 \pi^{4}}{5} R \dfrac{\rho}{A} V \dfrac{1}{\Theta_D^3}
\end{equation}

\begin{equation}
c_e = \dfrac{\rho}{A} \gamma_{e} V
\end{equation}
where $R$ is the universal gas constant and $\gamma_e$ refers to the free electrons molar specific heat.

\subsection{Thermal Conductances}
\label{Gestim}

The power flowing through a thermal link to a thermal bath is generally described assuming a power-law dependence \cite{irwin}:
\begin{equation}
P = k (T^n - T_B^n)
\end{equation}
where the values of $k$ and $n$ are determined by the nature of the thermal link. The differential thermal conductance of the thermal link is then defined as:
\begin{equation}
G = \dfrac{dP}{dT} = n k T^{(n-1)}
\end{equation}

\subsubsection{Normal metals}
The thermal conductance across normal metals at low temperature is dominated by the Wiedmann-Franz thermal conductance of the normal electrons \cite{irwin}:
\begin{equation}
G_{\text{Normal metal}} = \dfrac{L_0}{R} T
\end{equation}
where $L_0$ in the Lorentz number ($L_0 \approx$ 24.4 nW $\Omega$ $K^{-2}$) and $R$ is the electrical resistance of the metal. Another way to express the same conductance (useful to evaluate the thermal conductance of a wirebond) is \cite{pyle}: 
\begin{equation}
G_{\text{Normal metal}} = \dfrac{v_f d_e}{3} \dfrac{\rho}{A}\gamma_e \dfrac{S}{L} T
\end{equation}
where $v_f$ is the Fermi velocity, $d_e$ the electron diffusion length, $S$ the metal cross sectional area and $L$ its lenght. The power-law index for a normal metal thermal link is therefore:
\begin{equation}
n_{\text{Normal metal}} = 2
\end{equation} 

\subsubsection{Superconductors}
Near $T_C$, the thermal conductance across a superconductor is close to the normal metal value. However, well below $T_C$, the electron thermal conductance becomes exponentially small, as the electrons are bound in Cooper pairs that do not scatter to conduct heat. At $T < \sim 10$ $T_C$ the supeconductor thermal conductance can generally be assumed to be zero. \cite{irwin}

\begin{equation}
\begin{cases}
G_{\text{Superconductor}} \approx G_{\text{Normal metal}} & \text{for } T \sim T_C \\
G_{\text{Superconductor}} \approx 0 & \text{for } T \ll T_C 
\end{cases}
\end{equation}

\subsubsection{Crystalline materials}
The thermal conductance across a crystal is due both to the transport of heat by phonons and free electrons. At low temperature the first contribution goes as $T^3$, while the second goes as $T$ \cite{lazanu}:
\begin{equation}
G_{\text{Crystal}} = G_{\text{Phonons}}(T) + G_{\text{Electrons}}(T) = g_p \dfrac{S}{L} T^3 + g_e \dfrac{S}{L} T
\end{equation}
where $g_p$ and $g_e$ are the phonons and the free electrons thermal coupling parameters. Note that in this case also the power flow is the sum of the two different contributions, represented as power laws with indexes:
\begin{equation}
n_{\text{Crystal phonons}} = 4
\end{equation}

\begin{equation}
n_{\text{Crystal free electrons}} = 2
\end{equation}

\subsubsection{Boundary interfaces}
At subkelvin temperatures, the boundary thermal conductance between two carefully bonded solids (from body \virg 1'' to body \virg 2'') - usually called Kapitza conductance - is well described by the acoustic impedance mismatch between the two media. It can be derived as \cite{probst}: 

\begin{equation}
G_{\text{Kapitza}} = \dfrac{C_1}{2V_1} \langle v_\bot \alpha \rangle \propto T^3
\end{equation}
where $C_1/V_1$ is the phononic heat capacity per unit volume of the first material, $v_\bot$ is the phonon group velocity normal to the interface, $\alpha$ is the transmission probability and $\langle v_\bot \alpha \rangle$ is averaged over modes and wave vectors of the incident phonons. Note that the cubic temperature dependence is due to the phononic heat capacity temperature dependence. In this case the power-law index for the power flow is therefore:
\begin{equation}
n_{\text{Kapitza}} = 4
\end{equation}  

\subsubsection{Phonons and electrons thermal coupling in metals}
In metals at very low temperature, the thermal impedance between the electron and the phonon systems can be important, and the two systems will have generally different temperatures. The thermal conductance related this coupling is usually described with a temperature dependence with exponent between 4 and 5 \cite{irwin} \cite{probst}: [Probst]:
\begin{equation}
G_{\text{Electrons-Phonons}} = g_{ep} T^{\;4 \text{ or } 5}
\end{equation} 
where $g_{ep}$ is a thermal coupling parameter. In this case the power-law index for the power flow is therefore:

\begin{equation}
n_{\text{Electrons-Phonons}} = \text{ 5 or 6}
\end{equation}  

\chapter{The Cryogenic setup and the test activity}

The aim of this chapter is to present the cryogenic setup where the CryoAC prototypes are integrated and tested, and to show the typical measurements that are performed to characterize these samples and to verify their performance. 

First, I will introduce the two refrigeration systems of the IAPS High Energy (HE) cryolab, an Adiabatic Demagnetization Refrigerator (ADR) and a $^3$He/$^4$He Dilution Refrigerator (DR). The first cryostat has been operating for several years, and it has been used to perform numerous measurements presented in this thesis. The latter has been installed during the second year of my Phd, and I have been in charge of the design and the installation of the cryogenic setup needed to perform the CryoAC samples test activity.

Finally, I will present the typical test plan of a CryoAC prototype, showing the main measurements that we perform to characterize and test the sample.

\section{The Adiabatic Demagnetization Refrigerator}
\label{ADRsection}

\begin{figure}[H]
\centering
\includegraphics[width=0.8\textwidth]{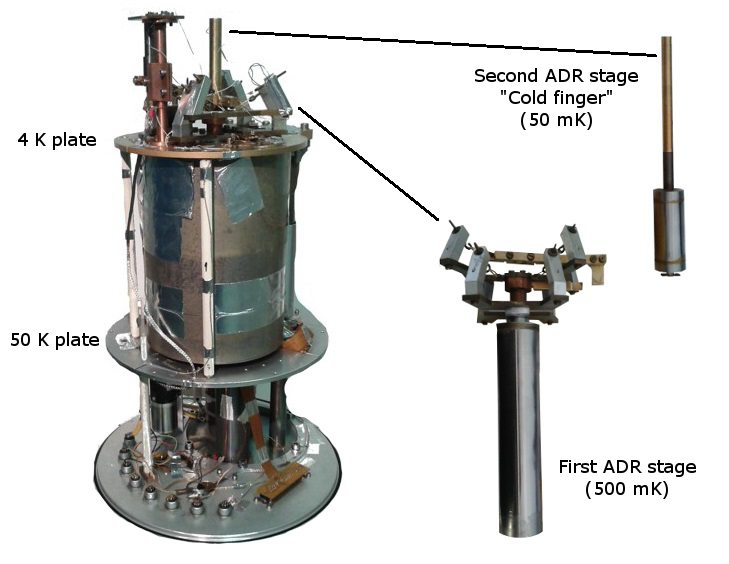}
\caption{The Vericold ADR cryostat operated in the IAPS HE cryolab. \textit{Left:} Global view of the refrigerator. \textit{Right:} Detail of the cold ADR stages, unmounted from the cryostat assembly. On the bottom there is the first stage ($\sim$ 500 mK), connected to the pill containing the GGG paramagnetic salt. On the top is shown the cold finger (base temperature $\sim$ 35 mK) connected to the pill containing the FAA paramagnetic salt.}
\label{ADR_pictures}
\end{figure}

The first cryostat operated in the IAPS HE cryolab has been a dry system (i.e. it does not need any cryogen like liquid nitrogen or liquid helium) composed by a Pulse Tube Refrigerator (PTR) and an Adiabatic Demagnetization Refrigerator (ADR). It is a commercial system assembled by the Vericold company. The PTR is a two-stages mechanical cooler providing 5 W of cooling power at 50 K on its first stage and 350 mW at 4 K on its second stage. It is used to pre-cool the system in order to operate the ADR unit, which is also constituted by two stages: the first reaching $\sim$ 500 mK and the second (the \virg cold finger'') having a base temperature of $\sim$ 35 mK. A view of the system is shown in Fig. \ref{ADR_pictures}, where the different stages of the cryostat are highlighted. Information about the PTR operation can be easily found in literature \cite{pulsetube}. Here I will introduce the ADR working principle, showing how to theoretically estimate the cooling energy and the hold time of the refrigerator. Then, I will show a typical test setup used to perform the CryoAC samples test activity.

\subsection{Working principle}
\begin{figure}[H]
\centering
\includegraphics[width=0.9\textwidth]{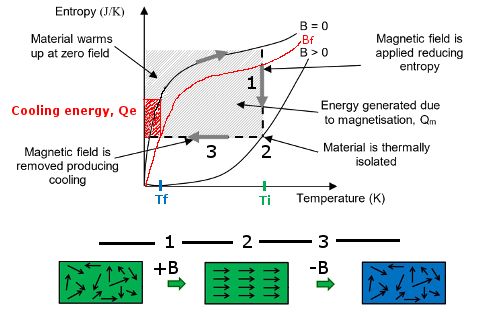}
\caption{The ADR refrigeration cycle. On the \textit{Top} is plotted the entropy of the paramagnet as a function of its temperature, with and without the external magnetic field. The magnetic refrigeration cycle is highlighted in the plot, and schematically represented on the \textit{Bottom}. See the text for details.}
\label{ADR_cycle}
\end{figure}

An ADR provides cyclic cooling by transfering entropy between the magnetic regions in a paramagnetic material and its random thermal vibrations. More specifically, in the 10's mK range, the paramagnetic material is cooled down by controlling the alignment of the electronic dipole moments of its magnetic ions by the use of an external magnetic field. This process is known as \virg Magnetic Refrigeration''. The magnetic refrigeration cycle is shown in Fig. \ref{ADR_cycle}, and it involves three main steps:
 
\begin{itemize}
\item[(1)] An external magnetic field B is applied to the paramagnetic material, thus its dipole moments align with the field reducing the magnetic entropy of the system. In this phase the paramagnet is in strong thermal contact with its surroundings, so the magnetization energy $Q_m$ generated by the process is transferred towards them in the form of heat, which is absorbed by a pre-cooler (i.e. the PTR). In the ideal case the magnetization is isothermal and $Q_m$ is equal to $T_i$ dS, where $T_i$ is the paramagnet temperature and dS is the change in its magnetic entropy;
\item[(2)] the paramagnetic material is then thermally isolated from its surroundings; 
\item[(3)] the magnetic field is finally reduced, and the dipole moments in the paramagnet return disordered, absorbing entropy from the thermal vibration of the material and producing cooling (note that the paramagnet is now isolated, so this process is ideally adiabatic). Once reached the desired final temperature $T_f$ (in correspondence of a magnetic field $B_f < B$ ), a \textit{cooling energy} $Q_E < Q_m$ is available to cool down masses connected to the paramagnet and keep them cold for a given \textit{hold time}, depending to the applied power load. Once the cooling energy has been totally consumed, the system needs to be recycled, reconnecting the paramagnet to the surroundings and restarting from the point 1.
\end{itemize} 
 
In our system the first ADR stage use as paramagnetic material the Gandolinium Gallium Garnet (GGG) salt, and the second stage the Ferric Ammonium Alum (FAA) salt. A magnetic field up to 6 T is generated by a superconductive electromagnet, and the cold stages, suspended by a kevlar strand system, are then isolated from the rest of the cryostat by using a mechanical heat switch. 

\subsection{Cooling energy and hold time evaluation}
In this section, I want to show how to evaluate the minimum temperature, the cooling power and the hold time of an ADR stage. In particular, I will refer to the second stage of our cryostat (i.e. the cold finger). The entropy of the FAA paramagnetic salt can be evaluated as \cite{pobell}: 

\begin{equation}
\begin{aligned}
S = n_{mol} \cdot R \cdot ( \;\; (x/2) \left[ coth(x/2) - (2J+1)coth \left( x(2J+1)/2 \right) \right] + \\
+ ln \left[ sinh \left( x(2J+1)/2 \right) /sinh(x/2) \right] \;\; )
\end{aligned}
\label{entropy}
\end{equation}

\noindent where

\begin{equation}
x = \frac{\mu_B \cdot g}{k_B} \cdot \frac{B}{T}
\label{btratio}
\end{equation}

\noindent and the other parameters are summarized in Tab. \ref{tab:parADR}. Here T is the temperature of the salt and B is the total magnetic field, which is the sum of the internal field of the salt ($B_{int}$) and the applied external field.

\begin{table}[H]
\centering
\caption{Parameters needed to evaluate the entropy of the FAA salt in our Vericold cryostat.}
\label{tab:parADR}
\begin{tabular}{lll}
\hline\noalign{\smallskip}
Parameter & Description & Value \\
\noalign{\smallskip}\hline\noalign{\smallskip}
$\mu_B$ & Bohr magneton & $\sim$ 9.274 $\cdot$ $10^{-24}$ [J/T]\\
R & Ideal gas constant & $\sim$  8.314 [J/(K$\cdot$mol)] \\
k$_B$ & Boltzmann constant & $\sim$ 1.38 $\cdot$ $10^{-23}$ [J/K] \\
g & Landé factor & 2 (for FAA salt) \\
J & Total angular momentum & 5/2 (for FAA salt) \\
n$_{mol}$& Number of moles & 0.15 in our ADR \\
$B_{int}$ & Internal magnetic field & 50 mT (for FAA salt) \\
\noalign{\smallskip}\hline
\end{tabular}
\end{table}

As previously stated, the demagnetization phase of the ADR cycle is ideally adiabatic (S = const), thus from eq. (\ref{entropy}) and (\ref{btratio}) we have that during the cooling the $B/T$ ratio is kept. The achievable minimum temperature is therefore given by:

\begin{equation}
\frac{B_{min}}{T_{min}} = \frac{B_{max}}{T_{i}} \Rightarrow T_{min} = \frac{B_{int}}{B_{max}}T_{PTR} = \frac{50\;mT}{6\;T} 4.3\;K = 36\;mK 
\end{equation}

\noindent where $B_{max}$ is the maximum magnetic field generated by the superconductive electromagnet, $B_{min}=B_{int}$ is the internal magnetic field of the FAA salt and $T_{i}=T_{PTR}$ is the temperature of the second PTR stage, at which the ADR cycle starts. Analogously, the final total magnetic field needed to bring the cold finger at a given final temperature (50 mK in this example), is given by:

\begin{equation}
\frac{B_{f}}{T_f} = \frac{B_{max}}{T_{max}} \Rightarrow B_{f} = \frac{B_{max}}{T_{PTR}} T_f \Rightarrow B_{50\;mK} = \frac{6\;T}{4.3\;K} 50\;mK = 70\;mT
\end{equation}

\begin{figure}[H]
\centering
\includegraphics[width=0.6\textwidth]{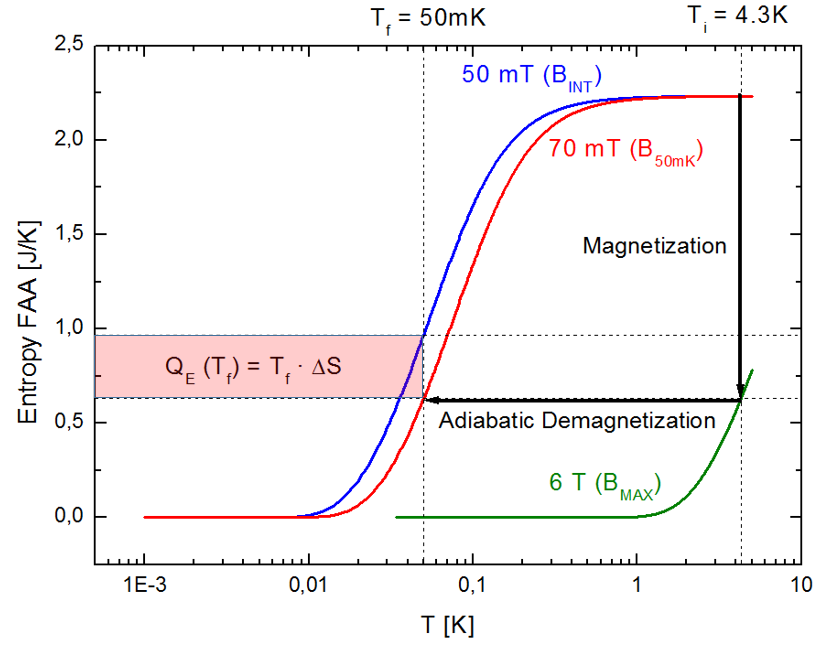}
\caption{Evaluation of the ADR second stage cooling energy at 50 mK.}
\label{coolingenergy}
\end{figure}

\noindent Once assessed this value, the available cooling energy at $T_f = $50 mK can be easily evaluated from the plot in Fig. \ref{coolingenergy}, evaluating the difference between the entropy of the FAA salt at $B_{50\;mK}$ and $B_{int}$ at that temperature:

\begin{equation}
Q_E (50\;mK) = 50\;mK \cdot \left[S_{FAA}(B_{int}, 50\;mK) - S_{FAA}(B_{50 mK}, 50\;mK)\right] = 17\;mJ
\end{equation}

\begin{figure}[H]
\centering
\includegraphics[width=0.6\textwidth]{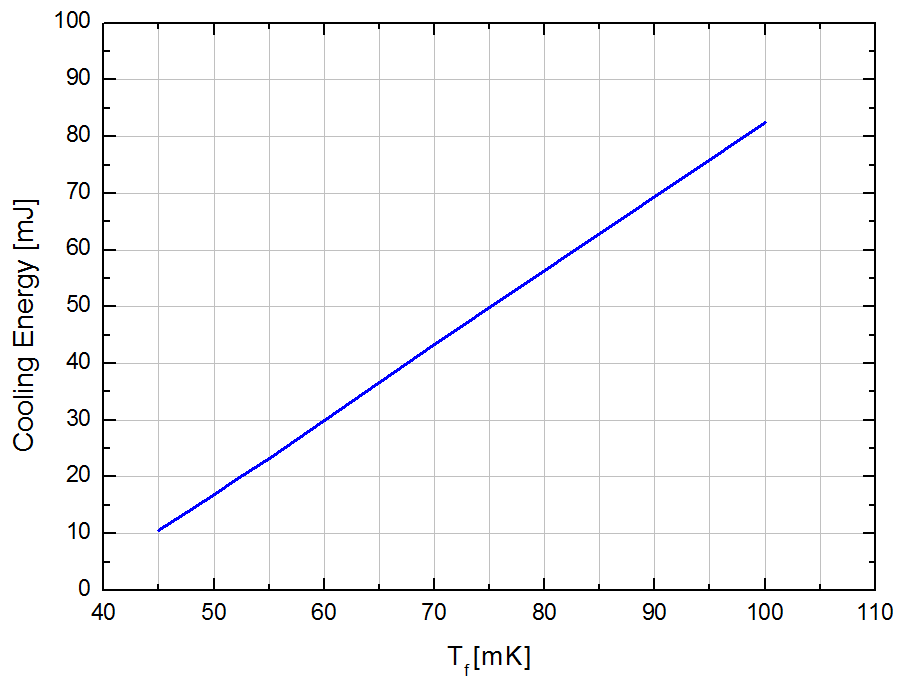}
\caption{Cooling energy available on the FAA ADR stage as a function of its final temperature.}
\label{coolingenergy2}
\end{figure}

In Fig. \ref{coolingenergy2} the cooling energy available on the FAA ADR stage is reported as a function of the salt final temperature, showing a linear trend for $T_f$ in the range from 45 to 100 mK.

\begin{figure}[htbp]
\centering
\includegraphics[width=0.6\textwidth]{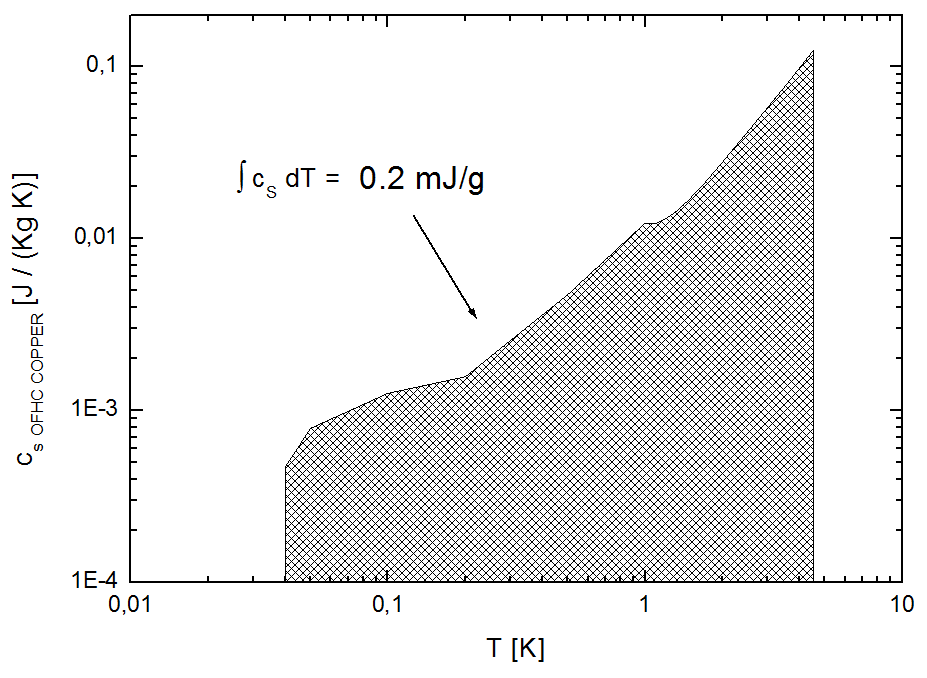}
\caption{Specific heat of the OFHC copper as a function of its temperature. The highlighted area represents the energy per unit of mass needed to cool down the material from 4.3 K to 50 mK. Data from \cite{nist2}.}
\label{Cscopper}
\end{figure}

In order to asses the hold time of the system, we have to consider that part of the cooling energy will be used to cool down the masses connected to the FAA salt pill, like the cold finger and the sample holder. The amount of this energy ($Q_{cool}$) is given by:

\begin{equation}
Q_{cool} = \int_{T_i}^{T_f} c_s(T) M dT
\end{equation}

\noindent where $c_s(T)$ [J/(Kg K)] is the specific heat of the material to be cooled and M is its mass. For reference, consider that for OFHC copper the energy needed to cool down a mass M from 4.3 K to $\sim$ 50 mK is approximately (see the plot in Fig. \ref{Cscopper}): 

\begin{equation}
Q_{cool\;,\;OFHC\;copper\;4.3\;K \rightarrow 50\;mK} \sim 0.2 \; [mJ/g] \cdot M
\end{equation}

Finally, it is possible evaluate the hold time $t_{H}$ by simply dividing the residual available cooling energy by the power load $P_L$ applied to the stage:

\begin{equation}
t_H (T_f, P) = \frac{(Q_E (T_f) - Q_{cool})}{P_L}
\end{equation}

\begin{figure}[H]
\centering
\includegraphics[width=0.45\textwidth]{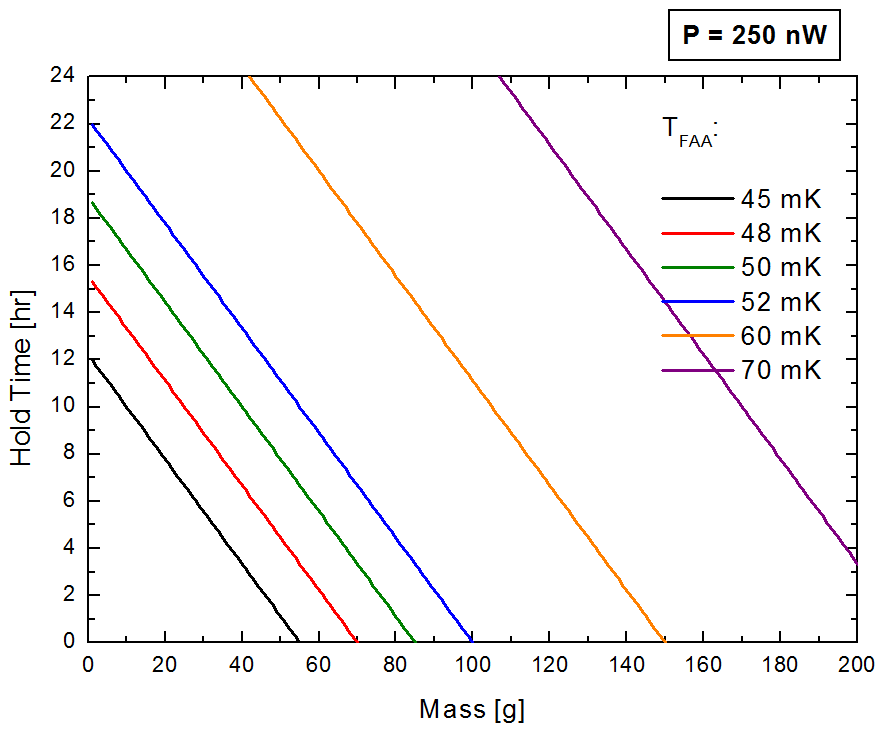}
\includegraphics[width=0.45\textwidth]{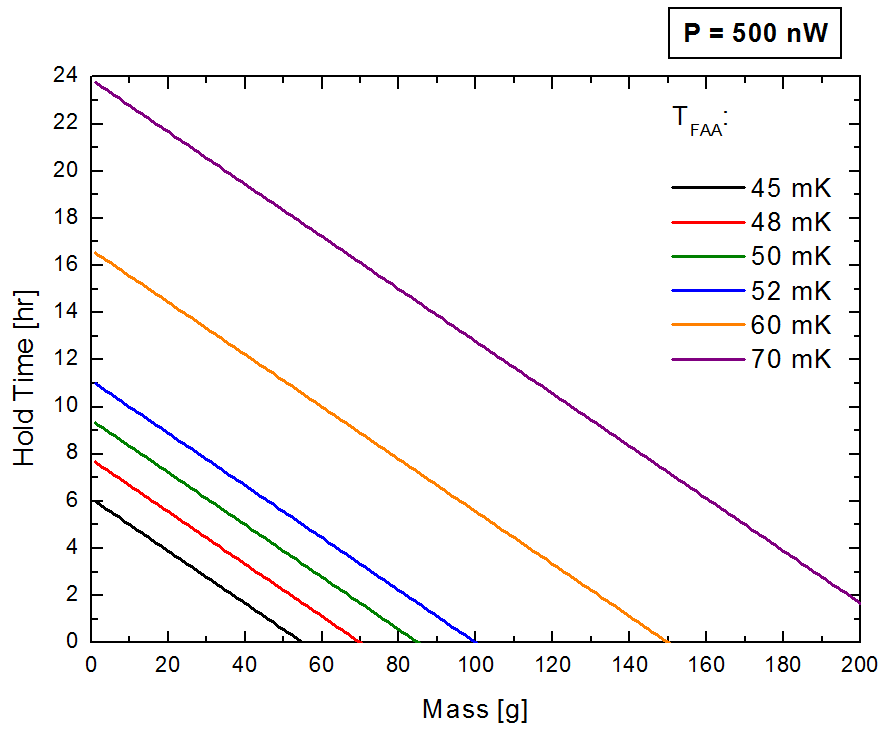}
\includegraphics[width=0.45\textwidth]{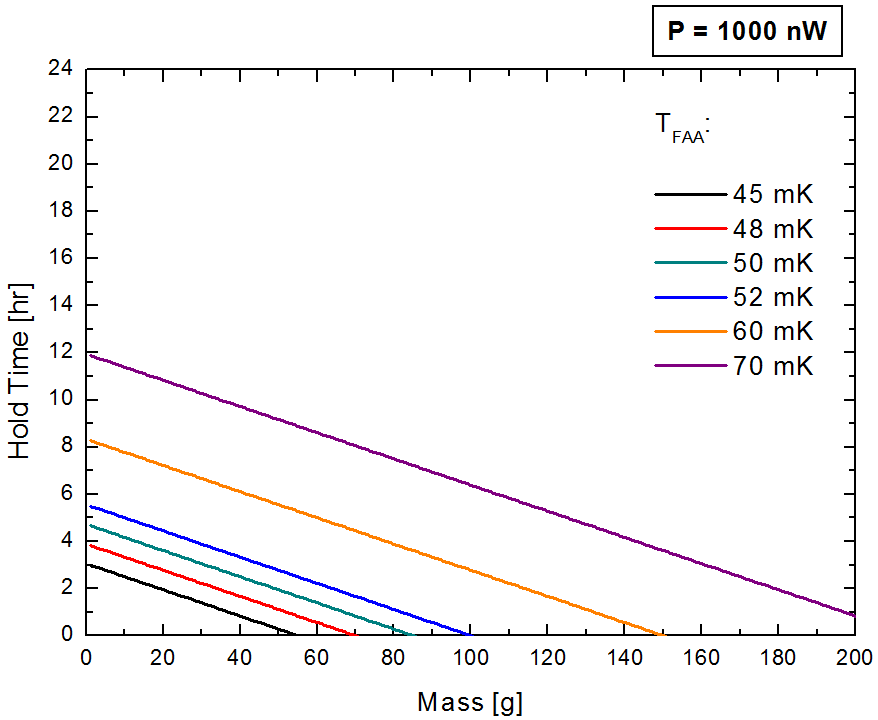}
\caption{Evaluation of the Hold Time of the ADR cold stage as a function of the copper mass connected to the FAA salt pill, the desired final temperature and the applied power load.}
\label{HoldTime}
\end{figure}

\noindent In Fig. \ref{HoldTime} it is shown the evaluation of the hold time for the cold stage of our system as a function of the mass of copper anchored to the FAA salt pill, for different final temperatures and applied power loads.

\bigskip

\bigskip

\noindent The main parameters evaluated in this section are summarized in Tab. \ref{ADRsummary}
\begin{table}[H]
\centering
\caption{Theoretical evaluation of the main performance of the ADR cryostat}
\label{ADRsummary}
\begin{tabular}{lc}
\hline\noalign{\smallskip}
Parameter & Evaluation\\
\noalign{\smallskip}\hline\noalign{\smallskip}
Initial temperature & 4.3 K\\
Base temperature & 36 mK\\
Cooling energy at 50 mK & 17 mJ \\
Cooling energy at 100 mK & 83 mJ \\
Hold time at 50 mK (P = 250 nW and M = 50g) & 7.5 hr \\
Hold time at 50 mK (P = 1000 nW and M = 50g) & 2.8 hr \\
\noalign{\smallskip}\hline
\end{tabular}
\end{table}

\subsection{The ADR test setup}

\begin{figure}[htbp]
\centering
\includegraphics[width=0.8\textwidth]{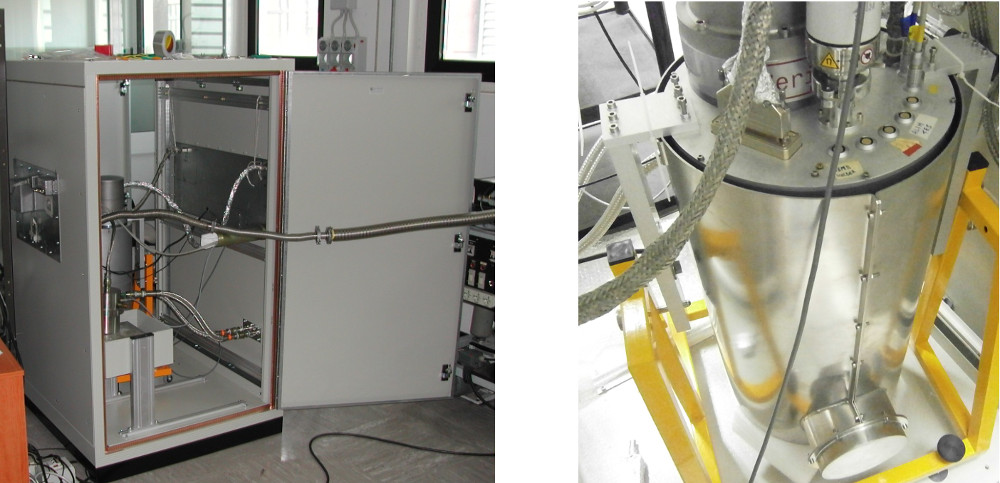}
\caption{\textit{Left : }Photo of the Vericold cryostat inside the Faraday cage for RF shielding. On the left wall of the cage are visible the I/O low pass filters. \textit{Right :} The cryostat enclosed in a mu-metal shield.}
\label{ADRpicts}
\end{figure}

During its operations, the ADR cryostat is located within a Faraday cage to minimize EMI towards the CryoAC samples and the SQUID, and it is surrounded by a mu-metal shield to damp the DC magnetic field (Fig. \ref{ADRpicts}). Since TES and SQUID are quite susceptible to EMI propagating through bias, signals and thermometers lines, all the cage I/O are low pass filtered (cut off frequency $\sim$ 10 MHz).

\begin{figure}[htbp]
\centering
\includegraphics[width=0.6\textwidth]{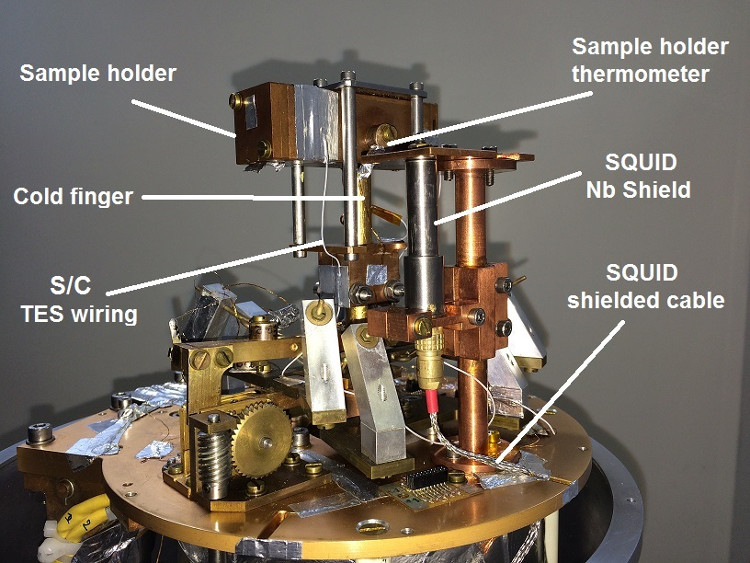}
\caption{Setup of the ADR cold stages for the CryoAC test activity.}
\label{ADRsetup}
\end{figure}

\noindent A typical setup of the cryostat cold stages for the CryoAC prototypes test activity is shown in Fig. \ref{ADRsetup}. The detector is mounted inside an OFHC copper sample holder anchored to the cold finger. The holder is also suitable to host a radioactive source in order to illuminate the detector. The temperature of this block is monitored by a Ruthenium Oxide thermometer. A commercial Magnicon SQUID \cite{magnicon} for the TES readout is placed inside a Nb capsule acting as superconductive shield, which is anchored to the 4 K cryostat plate. I remark that this picture refers to the initial phase of my PhD, and that part of my job has been then focused to improve the cryogenic test setup. The main results of this activity will be presented in the chapter \ref{testsetupch} of this thesis.

\bigskip

\newpage
\section{The Dilution Refrigerator}
\label{DRsection}
\begin{figure}[H]
\centering
\includegraphics[width=0.6\textwidth]{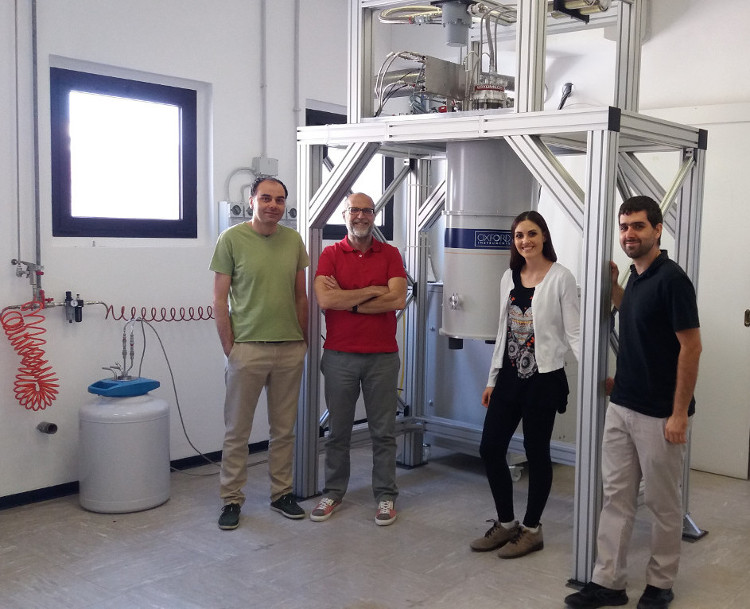}
$\;\;\;$
\includegraphics[width=0.275\textwidth]{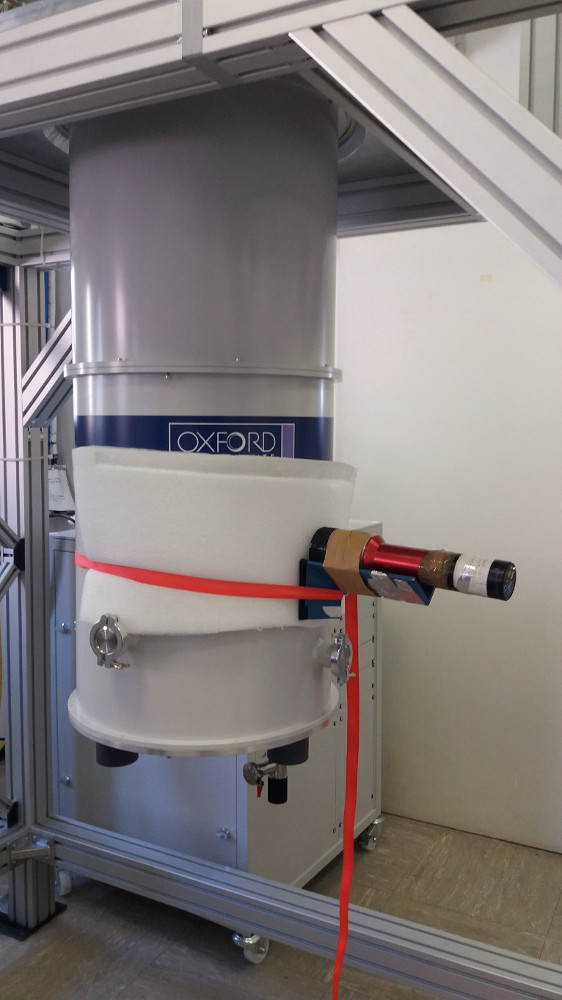}
\caption{\textit{Left:} The new cryogenic system installed in the IAPS HE Cryolab. \textit{Right:} The gamma detector of the nuclear orientation thermometer used to verify the system performance.}
\label{dilpicts}
\end{figure}

In July 2017, we have installed in the IAPS HE cryolab a new cryogenic system produced by the Oxford Cryogenics (Fig. \ref{dilpicts}-Left). It is a dry system constituted by a pre-cooling double stage PTR (providing 40 W of cooling power at 45 K and 1.5 W at 4.2 K) and a $^3$He/$^4$He dilution unit. The purchase of this new system has been motivated by the need to make up for the lack of reliability of the old ADR refrigerator (which after years of operation often needs to be refurbished), and by the necessity to have available better thermal performance towards the development of the CryoAC fligth model.

The performance of the new system have been verified at the end of the installation by using a nuclear orientation thermometer\footnote{This kind of device is capable of providing absolute temperature measurements in the millikelvin range. Their working principle is based on the temperature dependence of the gamma emission anisotropy of a radioactive source (in our case $^{60}$Co).}, which has been used to calibrate on-site the thermometers on the cold cryostat stage (Fig. \ref{dilpicts}-Right). The main measured cryostat performance are reported in Tab. \ref{tab:dilution}. 

\begin{table}[H]
\centering
\caption{Main performance of the new cryogenic system.}
\label{tab:dilution}
\begin{tabular}{ll}
\hline\noalign{\smallskip}
Parameter & Value \\
\noalign{\smallskip}\hline\noalign{\smallskip}
Base temperature &  6.6 mK\\
Cooling power at 20 mK &  $>$ 12 $\mu$W\\
Cooling power at 100 mK &  $>$ 450 $\mu$W\\
Cooling power at 115 mK &  $>$ 600 $\mu$W\\
Cooling time (from room to base temperature) & 25.5 hrs \\
Thermal stability at 50 mK & $\pm$ 0.05 mK (RMS)\\
\noalign{\smallskip}\hline
\label{tabdil}
\end{tabular}
\end{table}

\noindent Note that a dilution refrigerator, unlike an ADR, provides a continuous cooling, being able to run uninterrupted at base temperature for as long as several months. The performance of this refrigerator are therefore evaluated not in terms of \virg cooling energy'', but in terms of \virg cooling power'', which represents the maximum power load that the cryostat can handle  mantaining the cold stage at a given temperature. 

In the next paragraphs I will first introduce the principle of working of a $^3$He/$^4$He Dilution Refrigerator. Then, I will show the integration in the new system of the setup needed to perform the CryoAC samples test activity.

\subsection{Working principle}

A dilution refrigerator provides continuous cooling to temperature as low as some mK, obtaining its cooling power from the heat of mixing of two isotopes of helium: $^3$He and $^4$He. To explain the cooling process it is therefore necessary to introduce the properties of the $^3$He-$^4$He mixture.

\begin{figure}[H]
\centering
\includegraphics[width=0.5\textwidth]{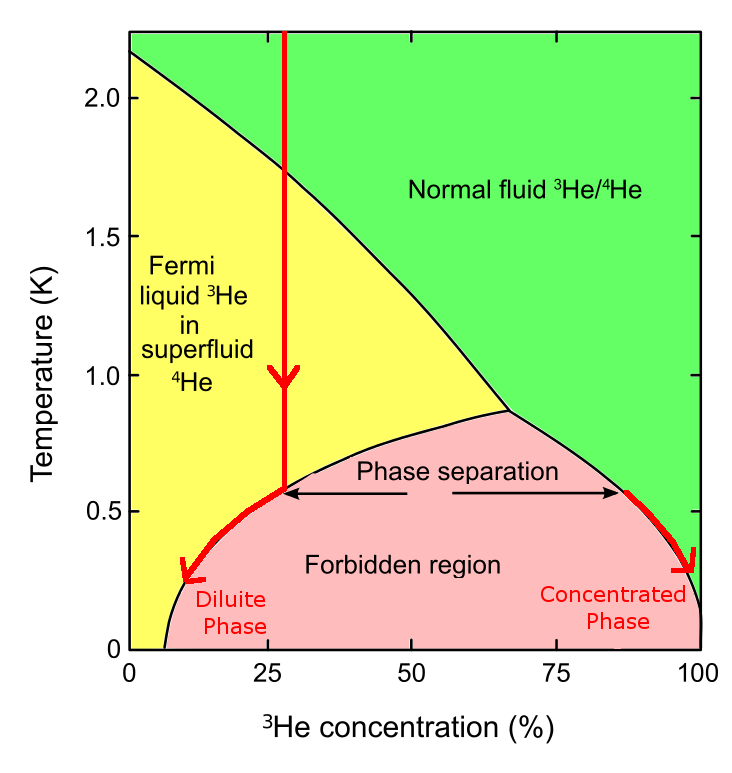}
\caption{Phase diagram of liquid $^3$He-$^4$He showing the phase separation experienced by the mixture at low temperature. This process is at the base of the dilution refrigeration. See text for details. Plot from: \cite{drwiki}.}
\label{DR_principle}
\end{figure}

Pure $^4$He (nuclear spin l=0) obeys Boson statistics, and it becomes superfluid under 2.17 K due to the Bose–Einstein condensation. This mechanism is not available for pure $^3$He (nuclear spin l=1/2), which obeys Fermi statistics and becomes superfluid only at much lower temperature (T $<$ 2.5 mK). Consequently, the superfluid transition temperature of a $^3$He-$^4$He mixture strongly depends on the $^3$He concentration, as shown in the phase diagram in Fig. \ref{DR_principle}. The diagram also shows a forbidden region, where the single phase solution for the mixture becomes unstable. 

Following the red line in the plot, we have that cooling down a $^3$He-$^4$He mixture, it will first become superfluid, and then it will undergo a spontaneous phase separation, forming an $^3$He-poor superfluid phase (the dilute phase) and a $^3$He-rich normal phase (concentrated phase). At very low temperatures the concentrated phase will be essentially pure $^3$He, while the dilute phase will contain about 6.5$\%$ $^3$He and 93.5$\%$ $^4$He. This finite solubility of $^3$He in $^4$He at low temperature is the key of the dilution refrigeration. Indeed, removing $^3$He atoms from the diluite phase, $^3$He atoms from the concentrated phase will cross the phase boundary to occupy the vacant energy states in the diluite one. This process is endothermic, since the enthalpy of $^3$He in $^4$He is higher than for pure $^3$He, thus it will absorb heat from the environment (i.e. the mixing chamber). We can roughly compare this process to the vapour-liquid cooling where atoms pass from the liquid to the gas phase absorbing latent heat from the surroundings. 

\begin{figure}[H]
\centering
\includegraphics[width=0.95\textwidth]{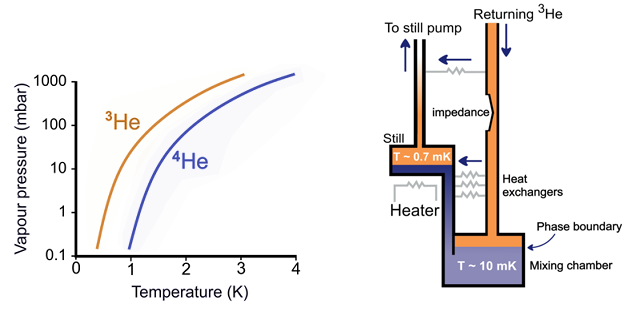}
\caption{\textit{(Left:)} Vapour pressure of the $^3$He and $^4$He isotopes as a function of the temperature. From: \cite{oxforddr}. \textit{(Right:)} Schematic of a dry closed-cycle dilution refrigerator. The pure $^3$He is shown in orange, while the pure $^4$He is in navy. The lighter blue tone represents the diluite phase of the mixture (mainly composed by $^4$He). The $^3$He is removed from the diluite phase by connecting the mixing chamber with the \virg still'' distiller, and it is then re-pumped to the top of the mixture. In the returning line an impedance ensures that the pressure is high enough ($\sim$ 0.3 bar) to liquefy the $^3$He, which is initially cooled by the outgoing $^3$He vapour from the still thanks to an heat exchanger located just before the impedance. Several different types of heat exchanger then precool the liquid $^3$He before it reaches the mixing chamber. From: \cite{oxforddr}.}
\label{DR_principle2}
\end{figure}

To remove only $^3$He from diluted $^4$He, in a dilution refrigerator the mixing chamber is connected to a distiller (the \textit{still}), which distils the $^3$He from the $^4$He due to the difference in their vapour pressures (Fig. \ref{DR_principle2}-Left). A room temperature gas handling system is then used to repump the $^3$He into a returning line towards the mixing chamber, in a closed lcycle (Fig. \ref{DR_principle2}-Right). Note that closed-cycle dilution refrigerators require gravity for their operation, thus they cannot be used for space applications. The gravity is indeed needed to separate the concentrated and dilute phases in the mixing chamber, and the liquid and vapour phases in the still.

The cooling power $\dot{Q}$ of a dilution unit is in first approximation given by the enthalpy difference $\Delta$H between the $^3$He in the dilute and the $^3$He in the concentrate phase multiplied by the $^3$He flow rate $\dot{n}_3$ [mol/s] \cite{oxforddr}:

\begin{equation}
\dot{Q}\;[W]= \dot{n}_3 \Delta H = 84 \; \dot{n}_3 \; T_{mxc}^2 
\end{equation}

\noindent where $T_{mxc}$ [K] is the temperature of the mixing chamber. In reality the returning $^3$He is always slightly warmer than the outgoing $^3$He due to the non-ideal heat exchangers. A more accurate cooling power expression therefore takes also into account the temperature $T_{ex}$ of the last heat exchanger before the mixing chamber:

\begin{equation}
\dot{Q}\;[W] = \dot{n}_3 (95 \; T_{mxc}^2 - 11 \; T_{ex}^2)
\label{cpdil}
\end{equation}

Note that the cooling power can be increased by apply power to the still, thus enhancing the $^3$He circulation rate. This is possible until the still temperature become too high and the vapour pressure of $^4$He become significant, reducing the dilution process efficiency due to the $^4$He circulation. In practice, a $^4$He fraction of ~10$\%$ in the circulated gas is acceptable, resulting in an optimal still temperature between 0.7 and 0.8 K

\subsection{Test Setup integration and residual cooling power measurement}

\begin{figure}[H]
\centering
\includegraphics[width=0.30\textwidth]{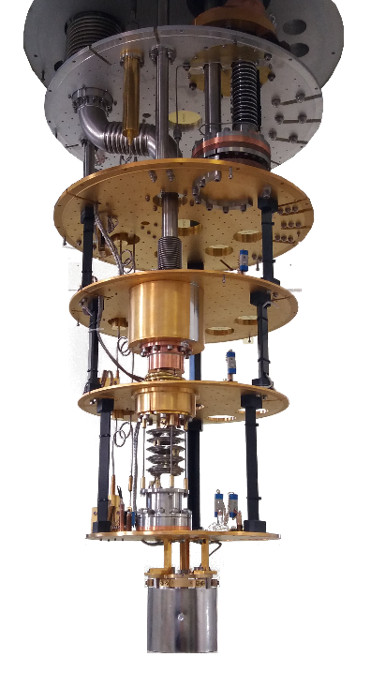}
\includegraphics[width=0.69\textwidth]{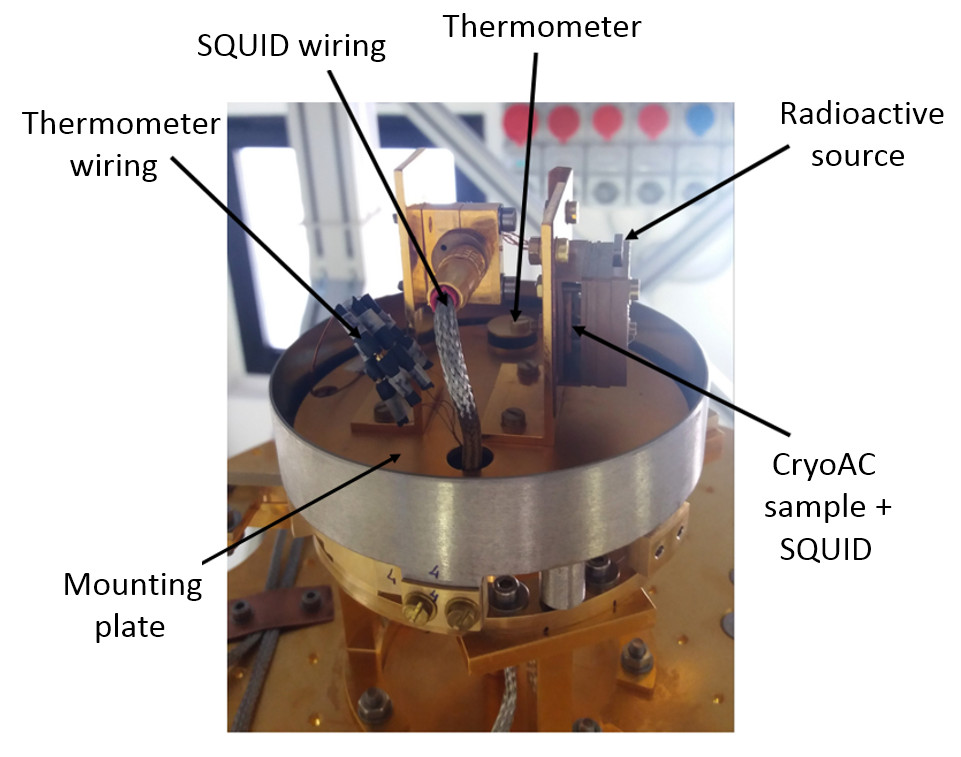}
\caption{\textit{Left:} The cryogenic magnetic shielding system integrated on the Mixing Chamber Plate of the new cryostat. \textit{Right:} Test setup mounted inside the magnetic shield.}
\label{dilinside}
\end{figure}

After the installation, we have integrated in the cryostat the test setup needed to test the CryoAC prototypes, in particular: 

\begin{itemize}
\item We have mounted on the Mixing chamber plate a cryogenic magnetic shielding system initially developed for the ADR (Fig. \ref{dilinside}-Left). The development of this system it is reported in the Chapt. \ref{testsetupch} of this thesis. The system has been integrated thanks to custom mechanical structures ad-hoc developed, including a gold-plated copper plate where are mounted the holders for the CryoAC samples and their cold front end electronics.
\item A Ruthenium Oxide thermometer and the S/C wiring for its readout. The thermometer has been mounted inside the magnetic shielding (Fig. \ref{dilinside}-Right) in order to monitor the sample holder temperature.
\item The wiring needed to bias and readout two independent SQUID system operating in FLL mode. One of these lines can be alternatively used to operate an heater to be placed inside the magnetic shield.    
\end{itemize}

\begin{figure}[H]
\centering
\includegraphics[width=0.6\textwidth]{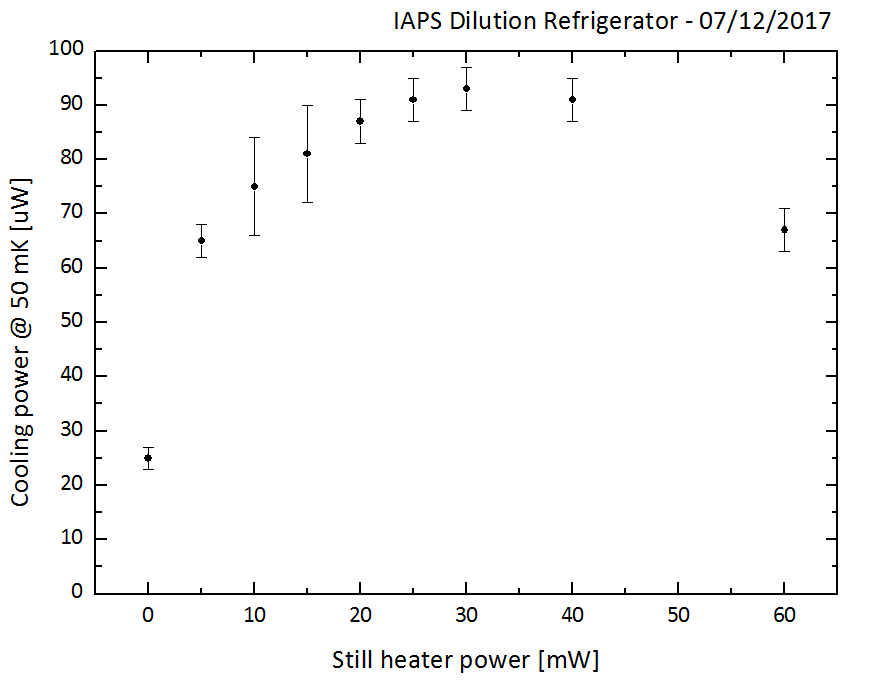}
\caption{Cooling power of the dilution unit at 50 mK as a function of the power applied to the still.}
\label{dilrefpower}
\end{figure}

At the end of the test setup installation, we have performed another performance measurement, evauating the residual cryostat cooling power at 50 mK as a function of the power applied to the Still (Fig. \ref{dilrefpower}). 

\noindent 
The measurement shows the trend expected by theory. Indeed, the cooling power initially increases with the power applied to the still heater (going from $\sim 25$ $\mu$W @ P$_{still} = 0$ mW to $\sim 90$ $\mu$W @ P$_{still} = 30$ mW) due to the increase of the $^3$He circulation rate (eq. \ref{cpdil}). Then, further increasing the still power, it starts to go down due to the enhancement of the $^4$He level in the circulating gas.

Also the approximately square dependence of the cooling power from the temperature (eq. \ref{cpdil}) results respected. Given a cooling power of 450 $\mu$ W @ 100 mK (Tab. \ref{tabdil}), we could indeed expect a cooling power of $\sim$ 110 $\mu$ W @ 50 mK (a factor 4 reduction). This is roughly consistent with the maximum measured value ( 93 $\pm$ 4 $\mu$W ) if we take into account also the additional power load on the mixing chamber due to the new installed wiring.
 
\newpage
\section{CryoAC samples typical test plan}

After presenting in the last sections the cryogenic test setup, I want here to report the typical test plan of a CryoAC prototype.

\bigskip
\noindent At first we perform the typical characterization measurements for TES detectors, that consist in the acquisition of the transition curve R-T, the evaluation of the detector electrical bandwidth and the analysis of its current-voltage characteristics (I-V curves). The R-T curve is used as benchmark of the quality of the TES, and it enables us to evaluate the critical temperature, the transition width and the thermometric responsivity of the sensor. From the electrical bandwidth measurement we are able to estimate the total impedance coupled with the TES in its bias/readout circuit, thus evaluating the  electrical characteristic time of the system. Finally, the I-V curves allow us to determine the optimal bias point of the detector and to evaluate its operation in the electrothermal feedback regime.

Afterwards, we test the detector response under X and Gamma ray illumination. We use the 6 keV line from a $^{55}$Fe radioactive source to check the detector energy threshold, and the 60 keV photons from a low activity $^{241}$Am source to study its response at higher energy. Tests at even higher energies, if necessary, can be performed by the detection of cosmic muons ($\sim$ 200 keV in 500 $\mu$m Silicon thick for a MIP). A deep analysis of the acquired pulses is then performed to extract the characteristic rise and decay times, obtain the energy spectra and, in general, fully frame how the detector is working.

\bigskip
All these measurements are performed by acquiring the detector output signal with a National Intruments PXI-DAQ system. The data processing is then performed by software that I have developed in the Labview 2014 environment during my PhD. 

\subsection{R vs T curve}

\begin{figure}[H]
\centering
\includegraphics[width=0.7\textwidth]{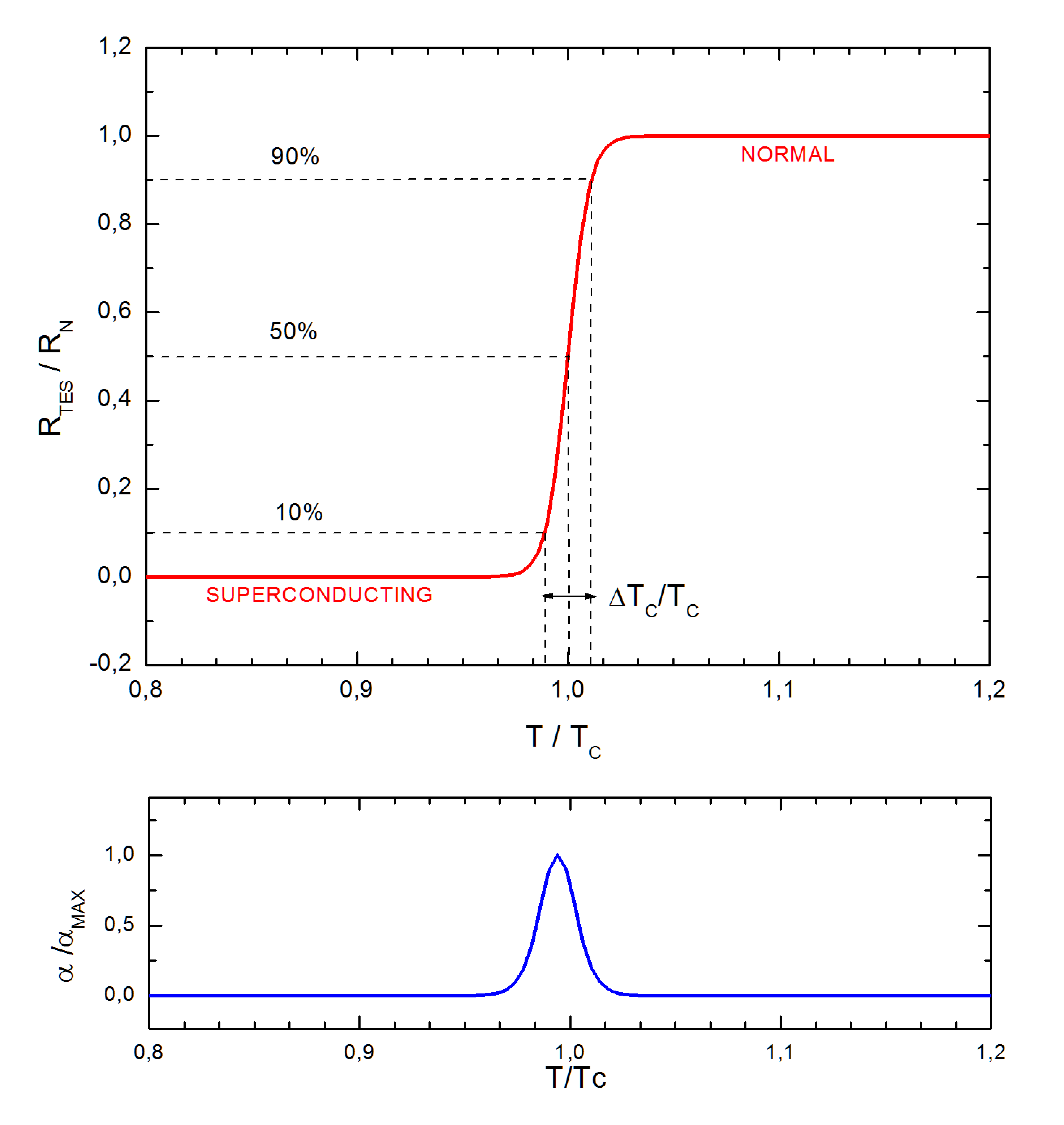}
\caption{\textit{Top:} Tipical transition curve for a TES sensor, normalized to the values of the normal resistance $R_N$ and  the temperature in the middle of the transition $T_C$  (sometimes $T_C$ is also referred to the beginning of the transition). $\Delta T_C$ is the 10\%-90\% transition width. \textit{Bottom:} Normalized thermometric responsivity $\alpha = \partial ln(R_{TES})/\partial ln(T)$ as a function of the temperature (evaluated from the upper transition).}
\label{RTmeas}
\end{figure}

The R-T curve represents the sensor resistance as a function of its temperature. We evaluate from this curve the main characteristic parameter of the sensor: the transition temperature $T_C$, the transition width $\Delta T_C$, the thermometric responsivity $\alpha$ and the normal resistance $R_N$ (Fig. \ref{RTmeas}).

\begin{figure}[H]
\centering
\includegraphics[width=0.6\textwidth]{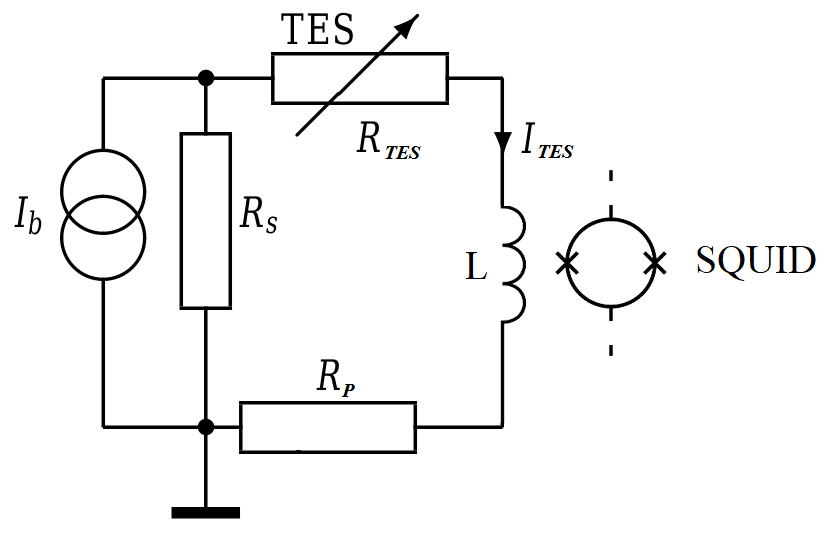}
\caption{TES bias and readout circuit. $I_b$ is the bias current, and $I_{TES}$ the current through the TES.}
\label{TEScirc}
\end{figure}

To measure the R-T curve we connect the sensor to its typical SQUID readout circuit (Fig. \ref{TEScirc}) and we use the signal modulation technique by applying a low-frequency sinusoidal bias current, thus increasing the signal-to-noise ratio. After solving the network, the sensor resistance can be determined by measuring the current through the TES with the SQUID operating in FLL mode, as follows:

\begin{equation}
R_{TES} = R_{S} \cdot \left( \frac{I_b}{I_{TES}} -1 \right) - R_P
\end{equation}

\noindent where $R_S$ is the shunt resistor and $R_P$ is the parasitic resitance in the TES branch of the circuit. The bias current is kept sufficiently small (few $\mu A_{PP}$) to minimize the sensor self-heating so avoiding systematic effects. The temperature of the sensor is varied by slow sweeping ($\sim$ 1 mK/min rate) the bath temperature by the cryostat software control.

\newpage
\subsection{Electrical bandwidth}

The electrical bandwidth of the sensor is evaluated by sending a sinusoidal current through the TES and measuring the attenuation of the output signal as a function of the input frequency $f$. Given the circuit in Fig. \ref{TEScirc}, we expect the typical attenuation of an RL network:

\begin{equation}
\dfrac{I_{TES}(f)}{I_{TES}(f=0)} = \left( 1 + (f/f_t)^2  \right)^{-1/n}
\end{equation}

\noindent where n = 2 is the order of the RL filter and the cutoff frequency $f_t$ is given by:

\begin{equation}
f_t = \frac{(R_{TES} + R_{P} + R_{S})}{2 \pi L}
\end{equation}

\noindent Note that L is the total impedance coupled with the TES, including both the contributions of the SQUID input coil and the wiring (stray inductance).

\begin{figure}[H]
\centering
\includegraphics[width=0.7\textwidth]{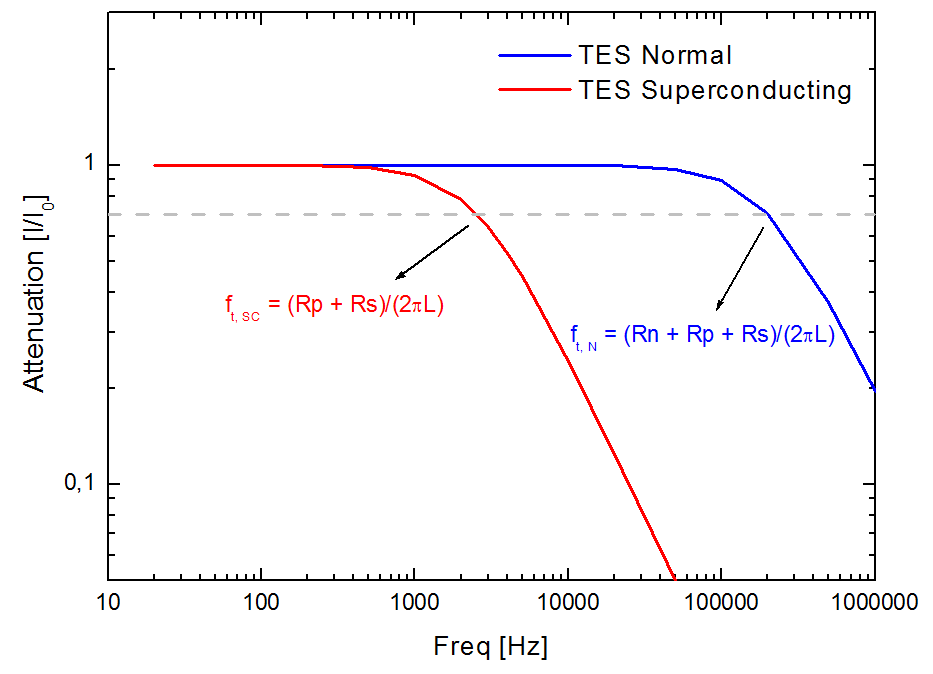}
\caption{Typical electrical band width measurement for a TES sensor. The red line refers to the TES in the superconducting state, the blue one to the TES normal. The increase of the cutting frequency is due to the increase of the resistance in the RL circuit.}
\label{band_meas}
\end{figure}

We usually measure the electrical bandwidth both when the TES is in the superconducting ($R_{TES}$ = 0) and in the normal state ($R_{TES}$ = $R_{N}$), in order to evaluate L from the cutting frequencies and check the ratios between $R_S$, $R_N$ and $R_P$ (previously evaluated by the R-T curve measurement). A typical measurement of the TES electrical bandwidth is shown in Fig. \ref{band_meas}.

\subsection{I-V characteristics}

\begin{figure}[H]
\centering
\includegraphics[width=0.8\textwidth]{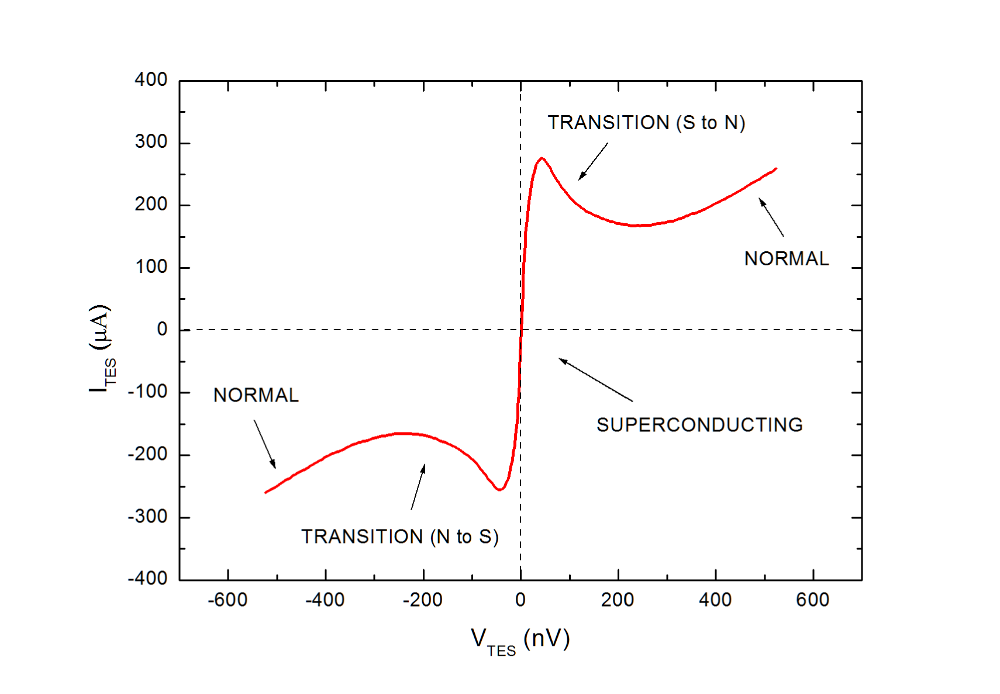}
\caption{Typical I-V curve for a TES sensor, measured at increasing values of voltage, from positive to negative values. The different phases of the superconducting transition are clearly visible.}
\label{IVmeas}
\end{figure}

The I-V curves represent the current passing through the TES as a function of the voltage difference across it. They are acquired for different thermal bath temperatures (in our tests usually from 50 mK to the TES critical temperature). The analysis of such a family of I-V curves allow us to find and characterize the optimal working point of the detector, estimating the electrothermal feedback loop gain L, and the thermal conductance between the sensor and the thermal bath. A typical I-V curve for a TES sensor is shown in Fig. \ref{IVmeas}.

The I-V curves is measured sweeping the voltage across the TES by changing $I_b$ (see Fig. \ref{TEScirc}) and measuring the current through it by the SQUID operating in FLL mode. For each I-V curve the bath temperature is kept constant by the cryostat thermoregulation technique. To sweep the voltage across the TES we are able to use two different techniques: we can vary a DC bias current at regular intervals or use a function generator to apply a very low frequency triangular wave (few mHz) as circuit bias current.

\subsection{Detector illumination}

After the characterization measurements we test the detector response under X and Gamma ray illumination. We perform two different tests: the first one using a $^{55}$Fe radioactive source to check if the detector threshold is below 6 keV, and the second one illuminating the sample by a low activity $^{241}$Am source, shielded in order to provide only 60 keV photons. This allow us to investigate the response of the silicon absorber bulk since that, unlike 6 keV photons, the 60 keV photons can be absorbed along the entire thickness of the absorber due to their large penetration depth in Silicon ($\sim$ 13 mm). Tests at even higher energies, if necessary, can be performed by the detection of cosmic muons ($\sim$ 200 keV in 500 $\mu$m Silicon thick for a MIP).

\bigskip

\begin{figure}[H]
\centering
\includegraphics[width=0.6\textwidth]{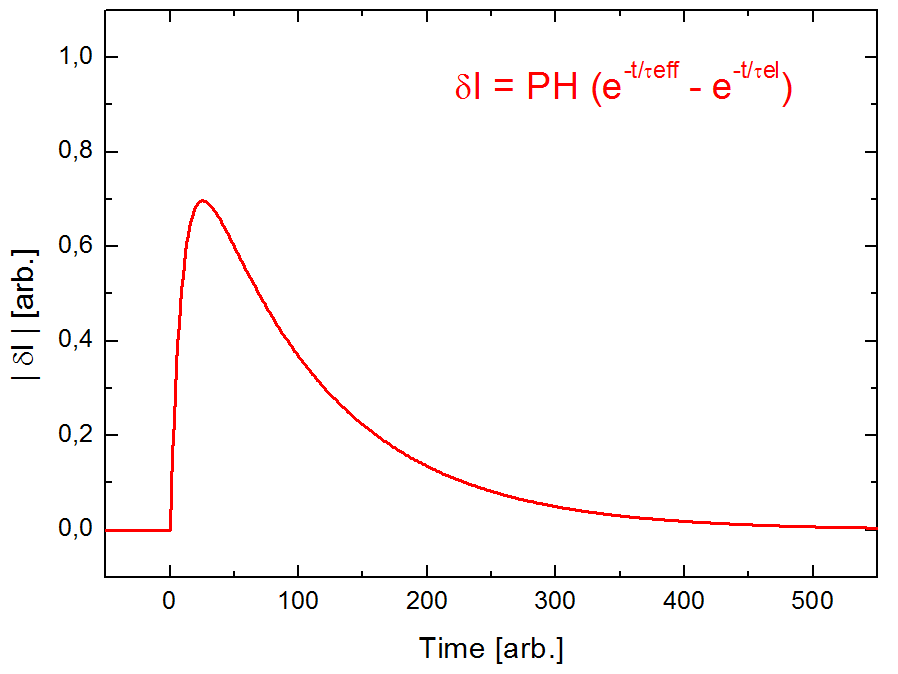}
\caption{Pulse acquired by an old CryoAC prototype (AC-S5) illuminated by the $^{241}$Am source \cite{acs5}. The pulse clearly shows the two expected components, highlighted by a double-pulse fit procedure. As shown in the section \ref{athermalsection} the fast one is due to the thermalization of athermal phonons directly in the TES network (athermal component), the slow one is instead linked to the rise of the absorber temperature (thermal component).}
\label{pulsetypical}
\end{figure}

\noindent The pulse acquisition is performed as follow:
\begin{itemize}

\item The bath temperature is set at a fixed $T_{bath}$ value and the sensor is voltage-biased applying a constant DC bias current in the TES circuit. The electrothermal characteristics of the working point has been already determined from the previous I-V curve analysis.

\item The SQUID output signal is amplified, and eventually band-pass filtered to increase the signal-to noise ratio, and then it is acquired by the ADC board.

\item The acquired pulses are then analyzed by an offline software that I have developed in the IDL environment.

\end{itemize}
In Fig. \ref{pulsetypical} it is shown a typical pulse acquired by a CryoAC sensor prototype.

We first perform a raw-analysis of the acquired pulses to have an idea on how the microcalorimeter is working, obtaining the energy spectra from the pulse height and/or the pulses in-time integral distributions, and extracting the characteristic rise and decay time of the pulses. 

Afterwards, we perform a double-pulse fit analysis to disentangle the foreseen athermal component of the pulses from the thermal one (see Fig. \ref{pulsetypical}), in order to investigate how the TES network works in collecting the athermal phonons. We remind that the athermal signal is the fast anticoincidence flag that we will use to reject the particle background, so it is important for us to study this component.

During my PhD I have also developed a further pulse analysis technique, based on the Principal Component Analysis (PCA), that will be introduced in the chapter \ref{pca} of this thesis.

\chapter{The CryoAC pre-DM samples}
\label{preDMchapter}

The technological maturity of an instrument (or a sub-system) with respect to a specific space application is classified according to a "Technology Readiness Level" (TRL) on a scale from 1 up to 9 (see Tab. \ref{tab:TRL} for a summary). For a detector, the research and development activities needed to increase the TRL involve the development of models and prototypes progressively more representative of the final Flight Model, in order to investigate the system critical technologies and verify its functional performances. 

When I joined the CryoAC development team at the beginning of my PhD, the fifth single-pixel anticoincidence prototype (AC-S5) had just been tested, achieving for the detector a TRL between 3 and 4. The activity of the team has then been focused to the development of the CryoAC Demonstration Model (DM), a single pixel prototype able to probe the detector critical technologies (i.e. the operation with a 50 mK thermal bath and the threshold energy at 20 keV with a representative pixel area of $\sim$ 1 cm$^2$), in order to reach in combination with an STM (Structural Thermal Model) TRL = 5 at subsystem level. To establish the final DM design, we have first developed and tested two further prototypes, namely ACS7 and ACS8, which enabled us to reach TRL $\sim$ 4 and to better understand the processes ruling the detector dynamic. 

In this chapter, after a brief review of the previous CryoAC prototypes, I will present these two pre-DM samples, reporting the main results obtained with them and showing how they helped us to fix the final DM design.

\begin{table}[H]
\centering
\caption{ISO Technology Readiness Level (TRL) summary adopted by ESA \cite{esatrl}}
\label{tab:TRL}
\begin{tabular}{ll}
\hline\noalign{\smallskip}
\footnotesize{TRL} & \footnotesize{Level Description}\\
\noalign{\smallskip}\hline\noalign{\smallskip}
\footnotesize{1} & \footnotesize{Basic principles observed and reported}\\
\footnotesize{2} & \footnotesize{Technology concept and/or application formulated}\\
\footnotesize{3} & \footnotesize{Analytical and experimental critical function and/or characteristic proof-of-concept}\\
\footnotesize{4} & \footnotesize{Component and/or breadboard functional verification in laboratory environment}\\
\footnotesize{5} & \footnotesize{Component and/or breadboard critical function verification in relevant environment}\\
\footnotesize{6} & \footnotesize{Model demonstrating the critical functions of the element in a relevant environment}\\
\footnotesize{7} & \footnotesize{Model demonstrating the element performance for the operational environment}\\
\footnotesize{8} & \footnotesize{Actual system completed and accepted for flight (\virg flight qualified'')}\\
\footnotesize{9} & \footnotesize{Actual system \virg flight proven'' through successful mission operations}\\
\noalign{\smallskip}\hline
\end{tabular}
\end{table}

\section{From the beginning to the pre-DM samples}

\begin{figure}[H]
\centering
\includegraphics[width=0.85\textwidth]{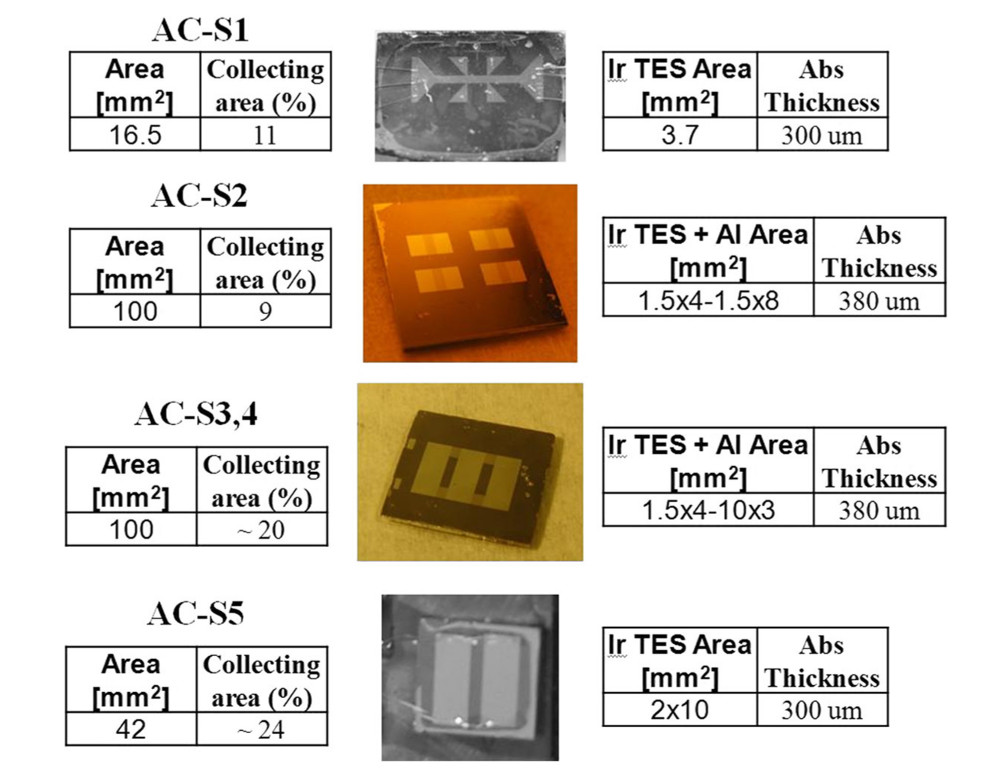}
\caption{Review of the CryoAC samples produced and tested from 2009 up to mid-2015. The tables show some geometric characteristics. The collecting area is the ratio between the sum of Ir and Al area to the full absorber area. From:  \cite{macculi2014}}
\label{CryoAC_samples}
\end{figure}

In the past years, along the development program of a cryogenic anticoincidence detector for X-ray missions (from IXO to ATHENA), the CryoAC collaboration has produced and tested several samples of microcalorimeters based on a silicon absorber sensed by iridium TES (Fig. \ref{CryoAC_samples}) \cite{macculi2014}. These samples have been produced and preliminary characterized at the Genova University (Phys. Dept.) and then deeply tested at IAPS.

\begin{figure}[H]
\centering
\includegraphics[width=0.45\textwidth]{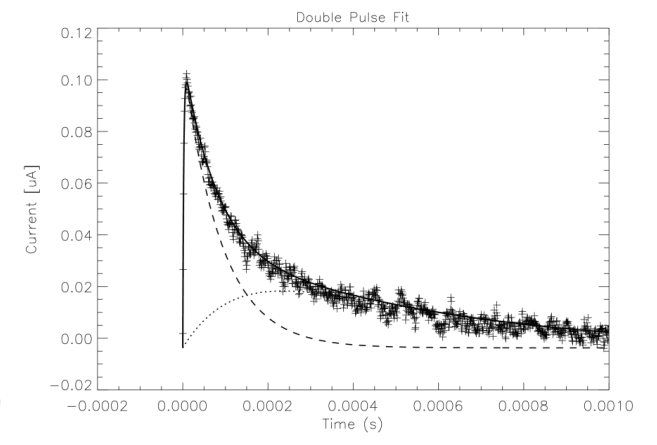}
\includegraphics[width=0.45\textwidth]{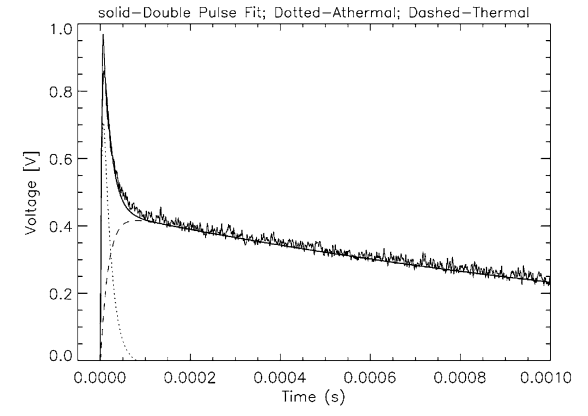}
\includegraphics[width=0.45\textwidth]{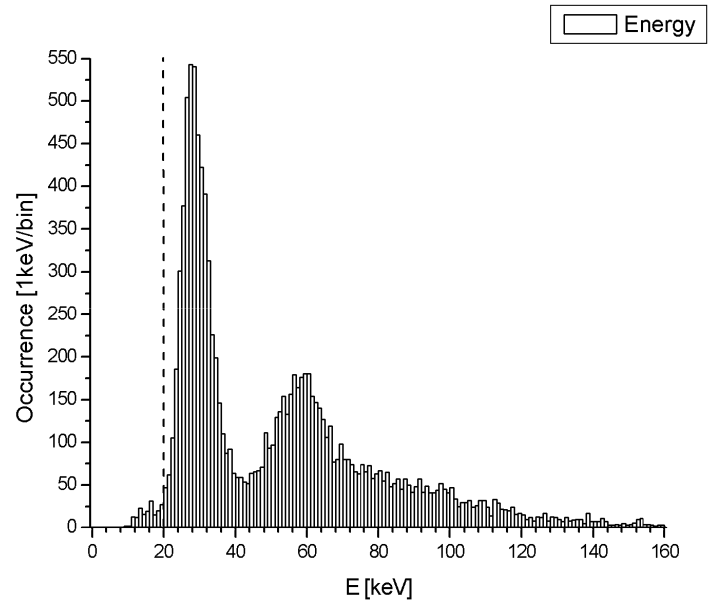}
\includegraphics[width=0.45\textwidth]{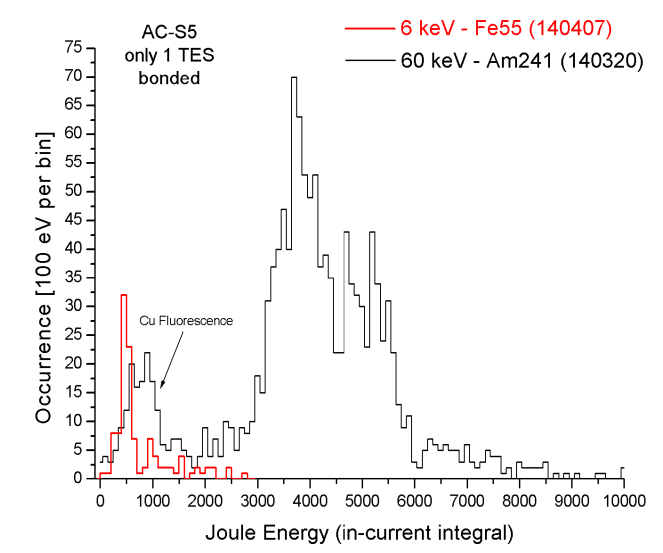}
\caption{Some of the main results obtained with the first CryoAC prototypes. \textit{Top-Left :} 6 keV pulse acquired by AC-S1. The fit highlights the two components of the pulse: athermal and thermal. From: \cite{macculi2010}. \textit{Top-Right :} 60 keV pulse acquired by AC-S2. Also in this case the two components of the pulse are clearly visible. From: \cite{macculi2012}. \textit{Bottom-Left :} Energy spectrum obtained illuminating AC-S3 (1 cm$^2$ area) by a shielded $^{241}$Am source. The low energy threshold of the detector (dashed line) is about 20 keV. From: \cite{macculi2010}. \textit{Bottom-Right :} Energy spectra obtained illuminating AC-S5 by a $^{55}Fe$ (red line) and a shielded $^{241}$Am (black line) sources. The 6 keV line from $^{55}Fe$ has been detected, thus the detector threshold is below 20 keV. From: \cite{acs5}. }
\label{historyplots}
\end{figure}

It started from a first simple sensor (AC-S1), where a single Ir film (3.7 mm$^2$ area) was deposited over the 16.5 mm$^2$ Si absorber. The following prototype (AC-S2) was the first with a representative 1 cm$^2$ absorber area ($\sim$ the area of a CryoAC single pixel), and the first exploiting aluminum collectors, which were directly connected to the 4 Ir TES obtaining a total absorber coverage up-scaled from the previous sample. From the first results obtained with these prototypes, it has been understood that the detector response was a combination of thermal and athermal pulses (see Fig. \ref{historyplots} - top plots), so realizing the necessity to increase the athermal phonons collection efficiency, in order to use the related signal as fast anticoincidence flag. 

The AC-S3 sample has been therefore developed increasing the area of the Al pads, to double the athermals collecting area. Then an identical sample has been produced (AC-S4), exploiting an improved Al film quality. These samples have been the first to show a low energy threshold of about 20 keV with 1 cm$^2$ absorber area (Fig. \ref{historyplots} - Bottom left). On the other hand, they also showed a delayed pulse rise front (up to hundreds of $\mu$s), due to the quasiparticle recombination in the Al pads (we will come back to this point on Sect. \ref{ACS3vsACS8}). 

Finally, another tentative solution has been pursued to increase the athermal phonons collection, increasing in the AC-S5 sample the Ir TES area rather than to add the Al collectors. This sample has worked fine (Fig. \ref{historyplots} - Bottom right), but its design is hard to replicate on larger area, due to the high thermal capacitance of the wide Ir film, suggesting the need of a different approach.

\newpage
\section{The AC-S7 and AC-S8 pre-DM prototypes}

\begin{figure}[H]
\centering
\includegraphics[width=0.7\linewidth]{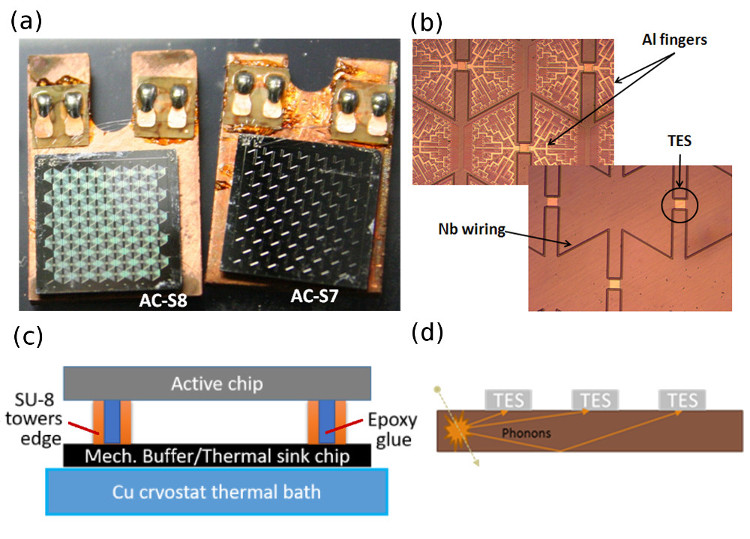}
\caption{\textit{(a):} The AC-S8 and AC-S7 samples. 
\textit{(b):} Details about the TES network, Nb wiring and Al fingers. 
\textit{(c):} Scheme of ''tower'' geometry with highlighted absorber, epoxy thermal link and buffer chip.
\textit{(d):} Scheme of the detection principle: the first burst of ballistic phonons could be directly collected by the TES network.}
\label{ACS7and8}
\end{figure}

The two ``pre-DM'' prototypes (Fig.~\ref{ACS7and8}a,b) have been produced at the Genova University (Phys. Dept.). Their structure is based on a 1 cm$^2$ Silicon absorber (380 $\mu$m thick) sensed by a network of 65 Ir TES in parallel connected through Nb lines, and readout as a single object. To ensure a reproducible thermal conductance towards the thermal bath, the absorber is connected to a Silicon buffer (in strong thermal contact with the bath) via 4 SU-8 hollow towers filled with epoxy glue (Fig.~\ref{ACS7and8}c). The shape of the towers allows us to control their thermal conductance, which is expected to be some 10$^{-8}$ W/K @ 100 mK. The main parameters of the samples are reported in Tab.~\ref{tab:ACS7-8}. 

The AC-S7 TES network was designed to ensure a large and uniform surface coverages for an efficient athermal phonons collection,  while constraining down the heat capacity contribution of the metal film due to the quite small TES size. To investigate the possibility to further improve the athermals collecting efficiency, the AC-S8 sample is featured also with a network of Al-fingers directly connected to the TES. I remind that the athermal phonons can be absorbed in the Al film, breaking Cooper pairs and generating electron-like quasiparticles. The quasiparticles diffuse through the fingers towards the TES, where they deposit their energy contributing to the fast athermal signal (see Sect. \ref{athermalsection}). The "tree-like" design of the Al finger network has been realized to minimize the quasiparticle path towards the TES, thus reducing their recombination probability in the Al.

\begin{table}[H]
\centering
\caption{Main parameters of the AC-S7 and AC-S8 samples.}
\label{tab:ACS7-8}
\begin{tabular}{ll}
\hline\noalign{\smallskip}
Parameter & Value \\
\noalign{\smallskip}\hline\noalign{\smallskip}
Absorber Silicon size &  10x10 mm$^2$, 380 $\mu$m thick \\
TES (x65) Iridium size &  100x100 $\mu$m$^2$, $\sim $200 nm thick \\
Niobium wiring & $\sim$ 870 nm thick\\ 
Aluminum fingers (only on AC-S8) &  $\sim$ 420 nm thick \\
\noalign{\smallskip}\hline
\end{tabular}
\end{table}

\subsection{Fabrication process}

\begin{figure}[H]
\centering
\includegraphics[width=0.8\linewidth]{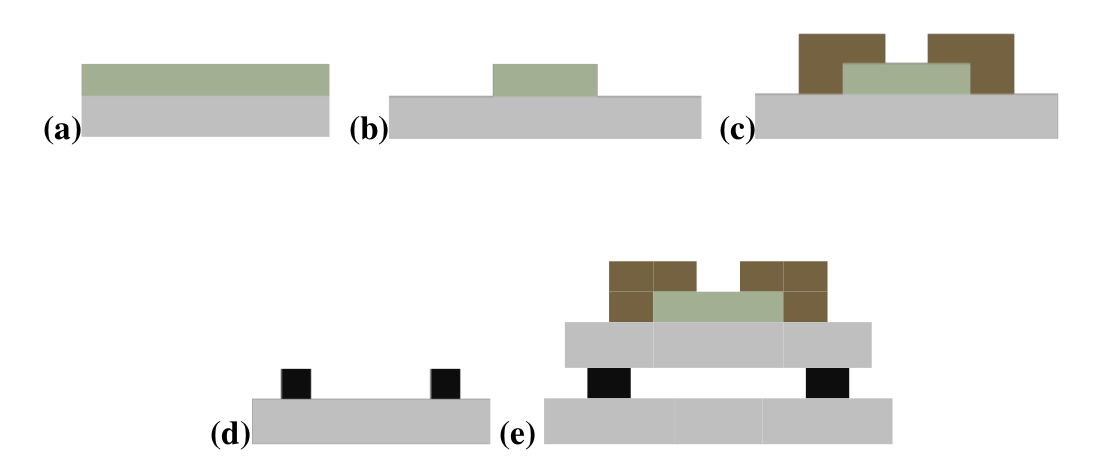}
\caption{Fabrication process: (a) iridium film deposition by pulsed laser deposition; (b) film patterning by positive photolotography and dry etching; (c) niobium wiring deposition and patterning by negative photolitography, RF-sputtering and lift-off process; (d) SU-8/epoxy tower building on buffer chip; (e) active and buffer chip coupling.}
\label{fabrication}
\end{figure}

\begin{figure}[H]
\centering
\includegraphics[width=0.6\linewidth]{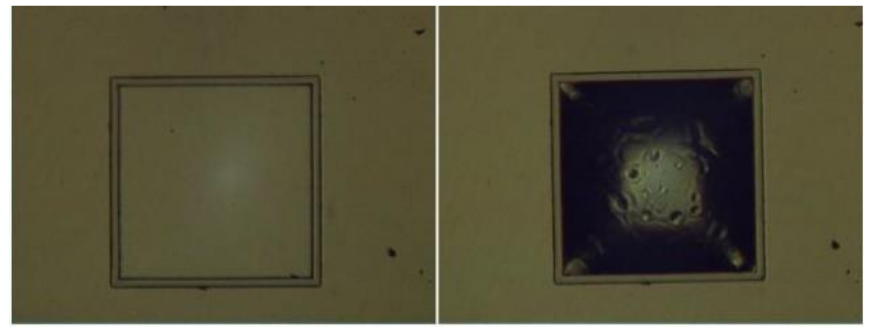}
\caption{Pictures of the empty SU-8 tower perimeter (left) and tower filled with EPO-TEK 301-2 epoxy glue (right).}
\label{towers}
\end{figure}

The samples have been produced from a commercially available silicon wafer (resistivity 10 $\Omega$/cm). They are composed by two different chips mounted in a \virg tower'' configuration in order to control the thermal conductance towards the thermal bath. The upper chip is the \virg active'' part of the detector, which is fabricated in three steps: in the first one, the iridium film is grown on the 1 cm$^2$ silicon chip. The growth is performed by pulsed laser deposition (PLD) technique, in a dedicated chamber with a base pressure of about 5$\times$10$^{-9}$ mbar. To ensure a good uniformity over the chip, the sample is rotated during the deposition in order to compensate any difference in growth rate. Its thickness, measured with an optical profilometer, is 200 nm (Fig. \ref{fabrication}(a)). In the second step, positive photolithography and dry etching are used to pattern the iridium film and to obtain 65 identical square TES on the silicon substrate. The dry etch is performed by ion milling using argon ion gun (Fig. \ref{fabrication}(b)). The last step is the wiring deposition. Negative photolithography, niobium deposition by RF-sputtering at base pressure of 5$\times$10$^{-8}$ mbar, and lift-off process are used to grow niobium strip and connect all the 65 TES in their parallel configuration (Fig. \ref{fabrication}(c)). After the active chip fabrication is completed, the buffer chip is prepared by using SU-8 permanent photoresist to build the tower walls (Fig. \ref{fabrication}(d)). The EPO-TEK 301-2 epoxy glue is then used to fill them (see the picture in Fig. \ref{towers}) and, finally, the active chip is fixed on the buffer (Fig. \ref{fabrication}(e)).

\newpage
\section{AC-S7 characterization and test}

\begin{figure}[H]
\centering
\includegraphics[width=0.45\linewidth]{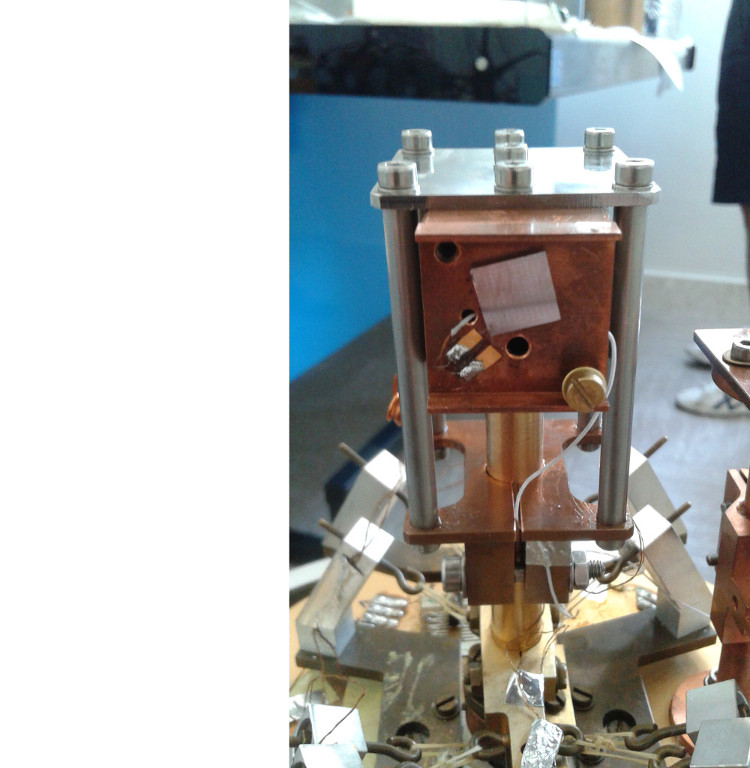}
$\;$
\includegraphics[width=0.52\linewidth]{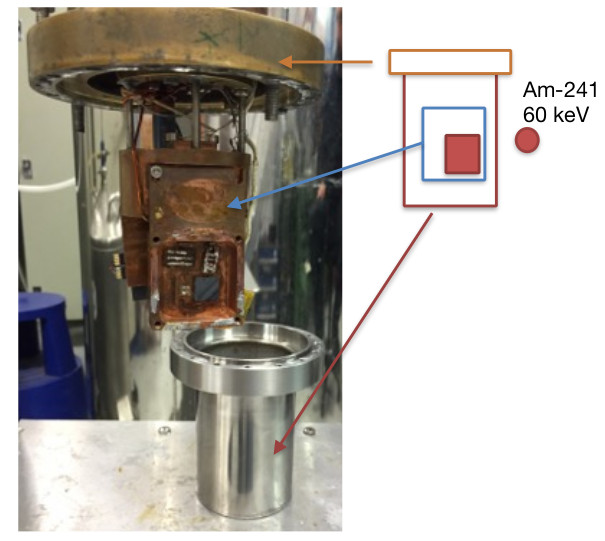}
\caption{\textit{Left:} AC-S7 glued on the sample holder anchored to the cold finger of the IAPS ADR cryostat. \textit{Right:} AC-S7 glued on the mixing chamber block of the UniGe dilution refrigerator. In this setup, the sample has been illuminated by an $^{241}$Am source placed outside the IVC steel shield.}
\label{ACS7setup}
\end{figure}

The AC-S7 prototype has been first characterized in the ADR setup at the IAPS HE cryolab and then, following an ADR cryostat issue,  it has been re-transferred to the Genova University in order to be illuminated by an $^{241}$Am source (in the dilution refrigerator of the phys. dept. cryolab). The two different setups are shown in Fig. \ref{ACS7setup}.

\subsection{Transition curve and IV characteristics}
\begin{figure}[H]
\centering
\includegraphics[width=0.6\textwidth]{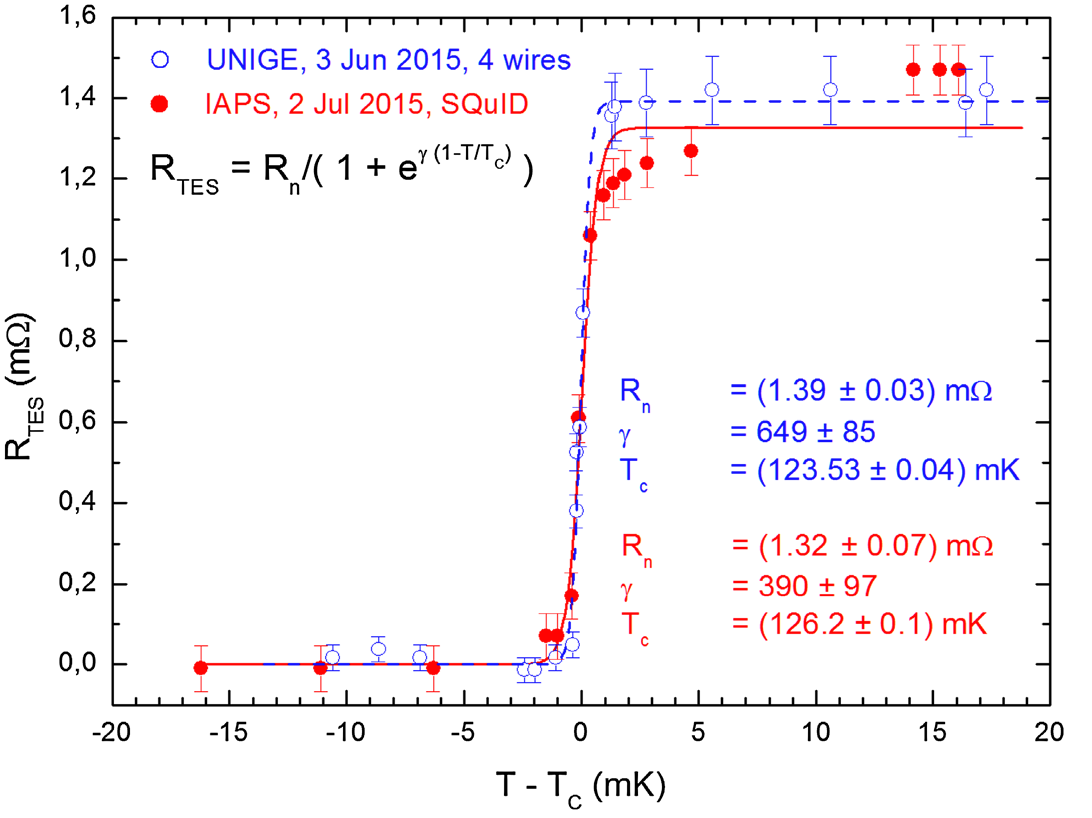}
\caption{Comparison between the two transition curves measured at UniGe and IAPS laboratories. The temperature data error is $<$ 100 $\mu$K$_{PP}$, not visible in this picture. The error bars related to the IAPS resistance data have been assessed by evaluating the contribution from the shunt resistor, the bias current, and the feedback voltage. The error bars related to the UniGe data come directly from the lock-in acquisition. $\gamma$ is a fitting parameter related to the $\alpha$ one evaluated at the critical temperature T$_C$: $\gamma = 2\cdot \alpha$. Please note that here R(T$_C$) = R$_N$/2.}
\label{ACS7-RT}
\end{figure}

The transition curves acquired in the two setups are shown in Fig. \ref{ACS7-RT}. They have been measured by two different readout techniques: 4-wire technique with Lock-in amplifier (blue-data,  UniGe dilution refrigerator setup), and SQUID FLL readout with modulation signal technique (red-data, IAPS ADR). The acquired data, after having removed the parasitic resistance, have been fitted with the function \cite{mooroka}:

\begin{equation}
R_{TES} = R_N / \left( 1 + e^{\gamma(1-T/T_C)} \right)
\end{equation}

\noindent where $R_N$ represents the TES normal resistance, $T_C$ is the transition critical temperature (at R$_N$/2) and $\gamma$ is a fitting parameter related to TES thermal responsivity evaluated at the critical temperature: $\gamma = 2\cdot \alpha|_{T_C}$. The plot shows a transition width from 10 to 90\% of about 2 mK and a temperature shift of the central value of 2.7 mK that is within the presently absolute temperature calibration uncertainties among the two cryogenic systems. The normal resistance is about 1.4 m$\Omega$ and it shows very small deviations within the errors. The gamma parameters are consistent well below 2$\sigma$ errors. We consider this comparison a good reproducibility test of the transition curve measurement and a demonstration that the Genova laboratory manufacturing processes of these large area samples are well consolidated and stable.

\begin{figure}[H]
\centering
\includegraphics[width=0.8\textwidth]{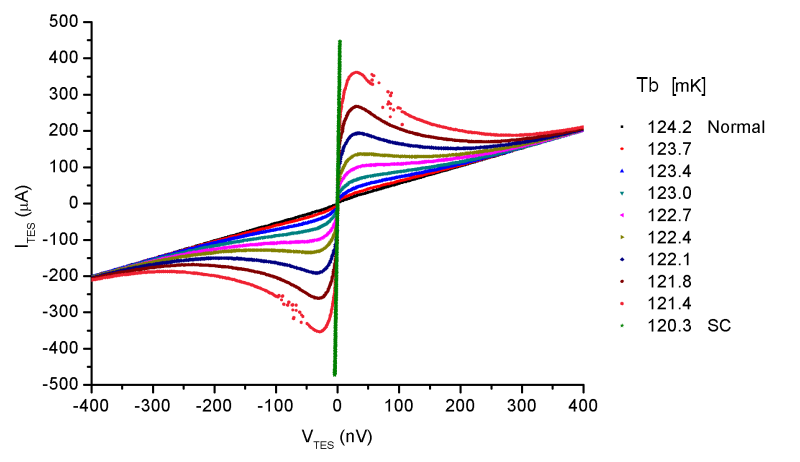}
\caption{A subset of characteristic I-V curves obtained at different bath temperatures in the interval 120 - 125 mK. The spikes on the red curve acquired at T$_B$ = 121.4 mK are typical of all curves in this V-T region.}
\label{ACS7-IV}
\end{figure}

In Fig. \ref{ACS7-IV} are reported some I-V curves acquired in the UniGe setup at different bath temperatures. Each curve has been acquired by sweeping the TES voltage bias at fixed thermal bath temperature (the uncertainty on temperature is 0.03 mK). The black line (T$_B$ = 124.2 mK) shows a curve with the TES in the normal state , whereas the green line (T$_B$ = 120.3 mK) shows only a superconductive behavior because the maximum current source available was not able to drive the device into the transition. In all the other curves the TES is driven in transition and we observe the expected IV negative slope in correspondence with the TES possible working points. Note that the measurement at T$_B$ = 121.4 mK (red curve) is characterized by some instability, as all the other curves acquired with the bath below 121.6 mK (not shown in the plot).

\subsection{Illumination by the 60 keV Line from 241Am}
\label{acs7ill}

\begin{figure}[H]
\centering
\includegraphics[width=1\textwidth]{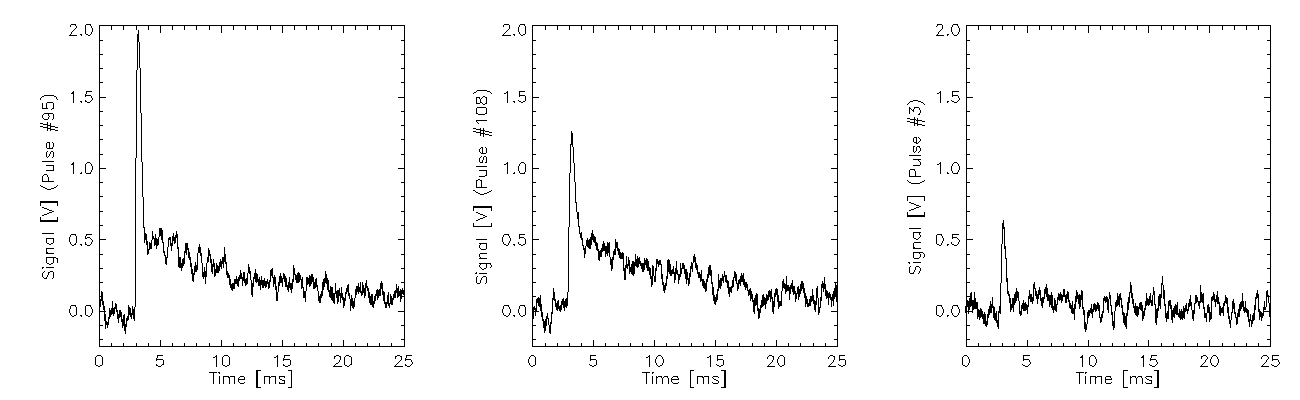}
\caption{Examples of triggered pulses with different shapes.}
\label{ACS7-rawpulses}
\end{figure}

The detector has been then biased with V$_{TES}$ = 50 nV at T$_B$ = 121.8 mK and illuminated by the $^{241}$Am source, which was shielded to filter out the products of the decay in Np and provide only the 60 keV line. For the TES readout has been adopted a Supracon SQUID (model VCblue). The SQUID output has been filtered by a low noise band-pass 1Hz - 10kHz and acquired with an ADC sampling rate of 100ks/s. The collected dataset - relative to about 2 hr of acquisition time - consists of 1270 pulses selected by a software trigger, each one containing 8000 samples. Some triggered pulses are shown in Fig. \ref{ACS7-rawpulses}.

The pulses clearly show the expected two different components: fast athermal and slow thermal. The first two pulses in Fig. \ref{ACS7-rawpulses} show a similar thermal component and different athermal ones. In the third pulse only the athermal component is visible, while the thermal one is submerged in the noise.

\subsubsection{Raw Analysis}

\begin{figure}[H]
\centering
\includegraphics[width=1\textwidth]{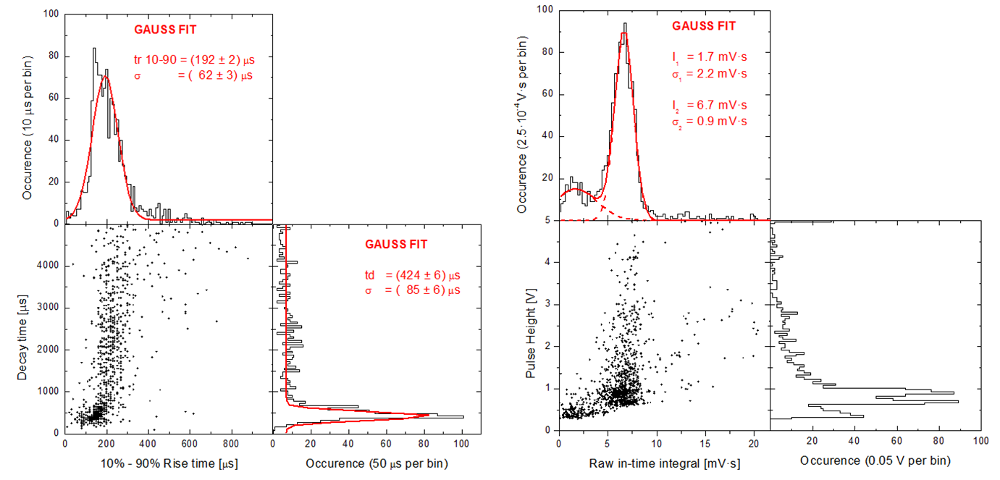}
\caption{Left: Characteristic decay time vs 10\%-90\% rise time from the raw analysis. Right: Pulse Height spectrum vs raw in-time integral of the pulses.}
\label{ACS7-rawanalysis}
\end{figure}

A very raw analysis has been initially performed to know pulses timings and spectrum, in order to roughly understand how the detector works. The timing results are shown in Fig. \ref{ACS7-rawanalysis} - left. The 10\%-90\% rise time constant is well defined around the value of $\tau_{R,10\%-90\%}$ = (192 $\pm$ 62) $\mu$s, while the distribution of the 1/e decay times peaks around $\tau_D$ = (424 $\pm$ 85) $\mu$s, showing a tail to higher values attributable to the roughness of this kind of analysis and to the noisy pulse shape. Note that the decay time here reported refers to the athermal component of the pulses, taking into account as reference the maximum value of the recorded data for each pulse.

In Fig. \ref{ACS7-rawanalysis} - right are shown the preliminary energy spectrum, represented by the distribution of the raw in-time integral of the pulses, and the Pulse Height (PH) distribution. The energy spectrum shows, as expected, a well-shaped principal line, corresponding to the 60 keV photopeak. The bump at lower energy is compatible with the correspondent Compton Edge of 60 keV photons ($E_{CE,60keV}$ = 11.3 keV), whose shape is due to the convolution with the instrumental response. We remember that in Silicon the cross sections for photoelectric absorption and Compton scattering at 60 keV are quite similar ($\sigma_{PE,Si,60keV}$ = 0.13 cm$^2$/g ; $\sigma_{C,Si,60keV}$ = 0.15 cm$^2$/g), and so we expect about half of the interacting photons in the Compton channel.

Lastly, note that the Pulse Height strongly depends on the athermal phonons collecting efficiency, and as a consequence the PH spectrum is less shaped than the raw in-time integral distribution. We will clarify this point later, by performing the double pulse fitting analysis.

\subsubsection{Double pulse fitting procedure}

\begin{figure}[H]
\centering
\includegraphics[width=1\textwidth]{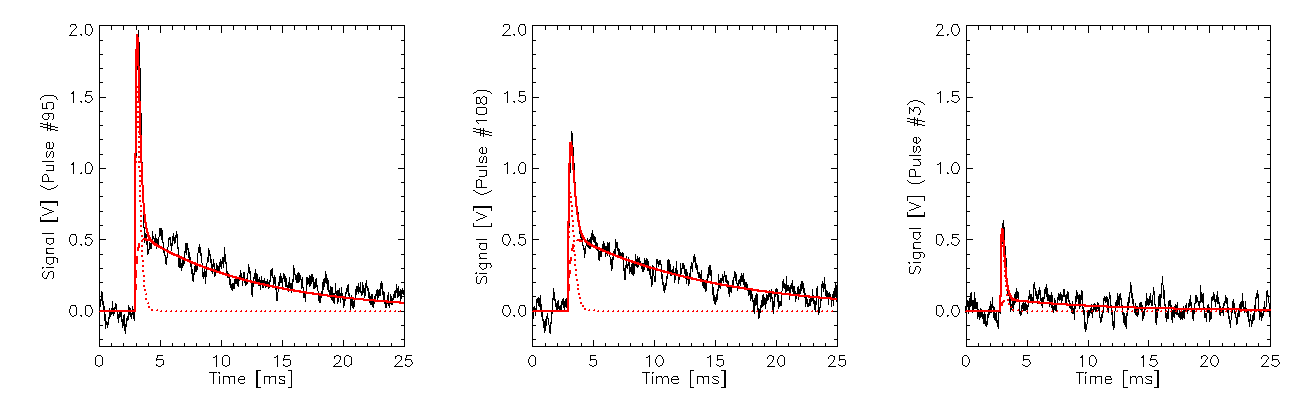}
\caption{Results of the double pulse fitting procedure over the triggered pulses (solid red line). Dotted red line and dashed red line respectively represent the athermal and the thermal component.}
\label{ACS7-fitpulses}
\end{figure}

\begin{figure}[H]
\centering
\includegraphics[width=1\textwidth]{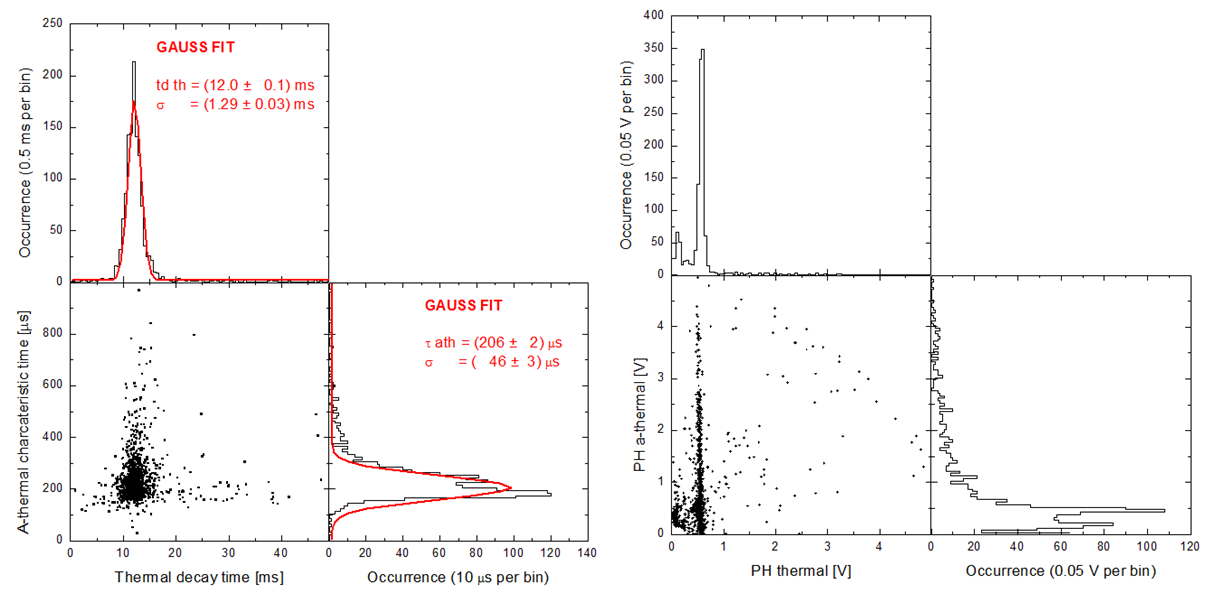}
\caption{ Left: Athermal characteristic time vs Thermal decay time from the double-pulse fitting procedure. Right: Athermal Pulse Height spectrum vs the thermal one.}
\label{ACS7-fitanalysis}
\end{figure}

The double pulse fitting procedure allow us to better investigate the detector dynamic, disentangling the athermal component of the pulses from the thermal one. The data shown in Fig. \ref{ACS7-rawpulses} and the preliminary raw analysis show that the typical acquired pulses present the athermal rise and decay time constants close to each other, probably both limited by the L/R filtering of the superconducting circuit. Hence, referring to \cite{irwin} for the description of a pulse with a single time constant, we performed the double fit analysis by means of the equation: 

\begin{equation}
\label{doublepulseoneath}
P(t) = PH_{ath} \frac{t}{\tau_{ath}} e^{(1-t/\tau_{ath})} + PH_{th}\left(e^{-t/\tau_{D, th}} - e^{-t/\tau_{R, th}}\right) \;\;\;,\;\;\; \tau_{R, th} = \tau_{ath}
\end{equation}

\noindent where PH$_{ath}$ is the athermal pulse height, $\tau_{ath}$ is the athermal peaking time, PH$_{th}$ the thermal pulse height, $\tau_{D,th}$ the thermal decay time and $\tau_{R,th}$ the thermal rise time. The two time constants $\tau_{R,th}$ and $\tau_{ath}$ are assumed to be equivalent since the thermal phonon family rises while the athermal one decays (see Chapter \ref{athermalsection}). An example of the fitting procedure is shown in Fig. \ref{ACS7-fitpulses}, whereas  the Fig. \ref{ACS7-fitanalysis} shows the fitting parameters distributions. 

The timing results are reported in Fig. \ref{ACS7-fitanalysis} - left. The thermal characteristic decay times show a well shaped Gaussian distribution with a central value of $\tau_{D,th}$ = 12.0 ms and a small dispersion. The athermal peaking times are well defined around the value of $\tau_{ath}$ = (206 $\pm$ 46) $\mu$s, fully compatible with the 10\%-90\% rise-time from the raw analysis. Note also that given the functional form in eq. (\ref{doublepulseoneath}), the athermal 1/e decay time corresponds to about twice the peaking time, and so it is consistent with the results of the previous analysis.

In Fig. \ref{ACS7-fitanalysis} - right are shown the PH distributions. The spectra are plotted on the same scale to appreciate the difference between the thermal and athermal regimes. From the thermal point of view AC-S7 works fine in spectroscopic mode, providing both the 60 keV line and the expected Compton bump. The athermal regime shows a lower spectroscopic capability, but the spectrum is still shaped, showing that the 65 TES network is working fine. The presence of two different lines in the athermal spectrum is currently under investigation but, roughly, it is probably an artifact due to the low signal-to-noise ratio for the lowest pulses. Note that the athermal spectrum shape has an impact on the raw ``global" PH spectrum shown in Fig. \ref{ACS7-rawanalysis}-left, which therefore shows a low spectroscopic capability.

\begin{figure}[H]
\centering
\includegraphics[width=0.6\textwidth]{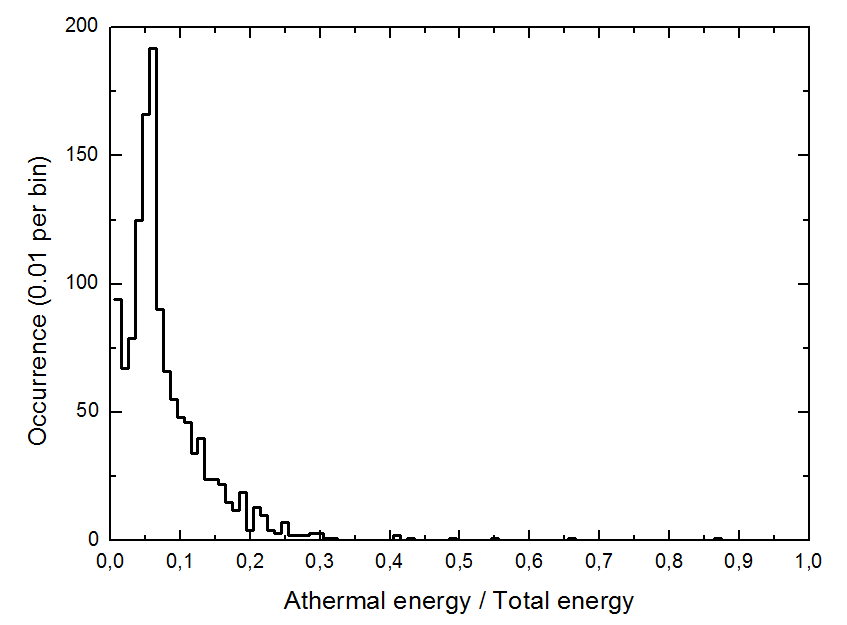}
\caption{Distribution of the ratio between the athermal and the total (athermal + thermal) energy of the pulses.}
\label{ACS7-epsilon}
\end{figure}

Finally, we show the distribution of the ratio between the athermal and the total (athermal + thermal) energy associated to each pulse (Fig. \ref{ACS7-epsilon}). The fraction of the athermal energy with respect to the total one is few \% (peak around 6\%). We will use this value later, comparing the athermal phonon collection efficiency of AC-S7 with the one of the AC-S8 samples (exploiting the additional Al finger layer).

\subsubsection{Energy spectrum}

\begin{figure}[H]
\centering
\includegraphics[width=0.7\textwidth]{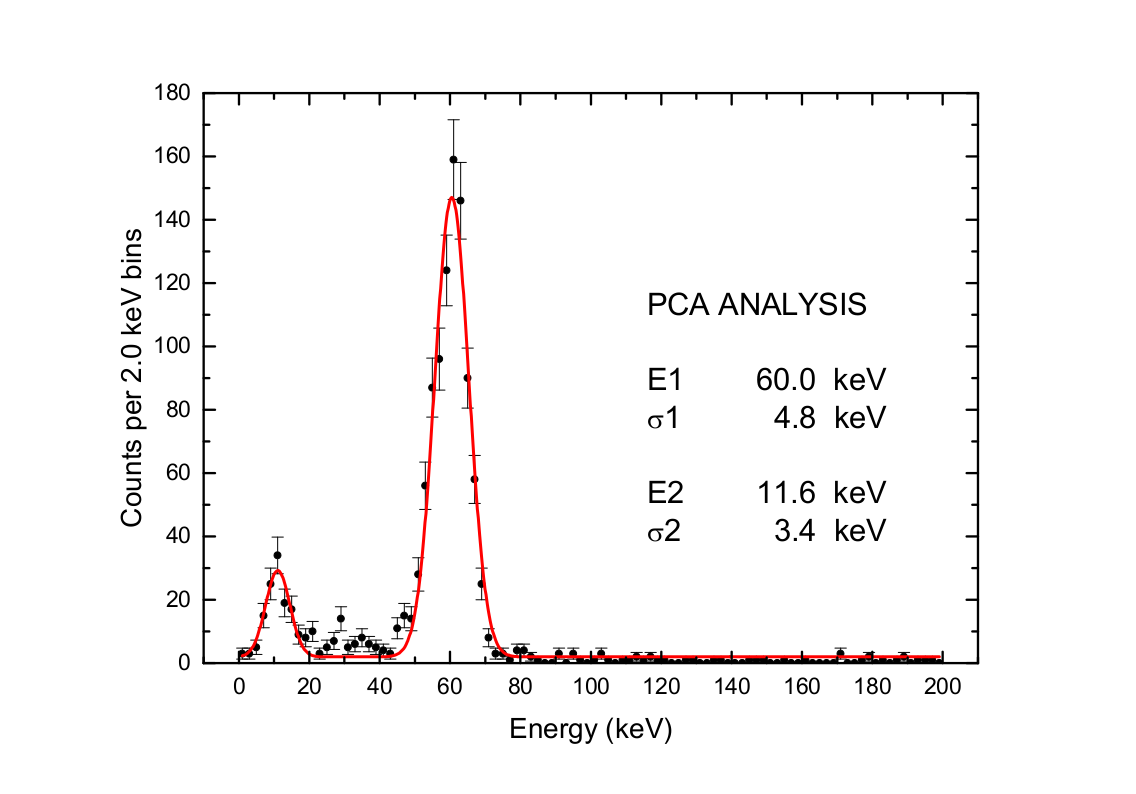}
\caption{Energy spectrum acquired by AC-S7 illuminated with the shielded $^{241}$Am source. The spectrum has been obtained performing a Principal Component Analysis (PCA) of the trigger pulses. More details are presented in Chapt. \ref{pca}}
\label{ACS7-pca}
\end{figure}

To conclude this section, in fig. \ref{ACS7-pca} it is shown the best energy spectrum extracted from the acquired data, which has been obtained performing a Principal Component Analysis (PCA) of the triggered  pulses. This method of data processing is illustred in detail in the Chapt. \ref{pca} of this thesis. With respect to the raw spectrum shown in Fig. \ref{ACS7-rawanalysis} - right, the PCA provides a more narrow 60 keV line ($\sigma$ = 4.8 keV). Also the bump at lower energy is better shaped, still compatible the hypothesis that we are observing the Compton edge expected at 11.3 keV. This allows us to better evaluate the energy threshold of the detector, which is below the 20 keV, compatible with our requirements.

\newpage
\section{AC-S8 characterization and test}

\begin{figure}[H]
\centering
\includegraphics[width=0.65\linewidth]{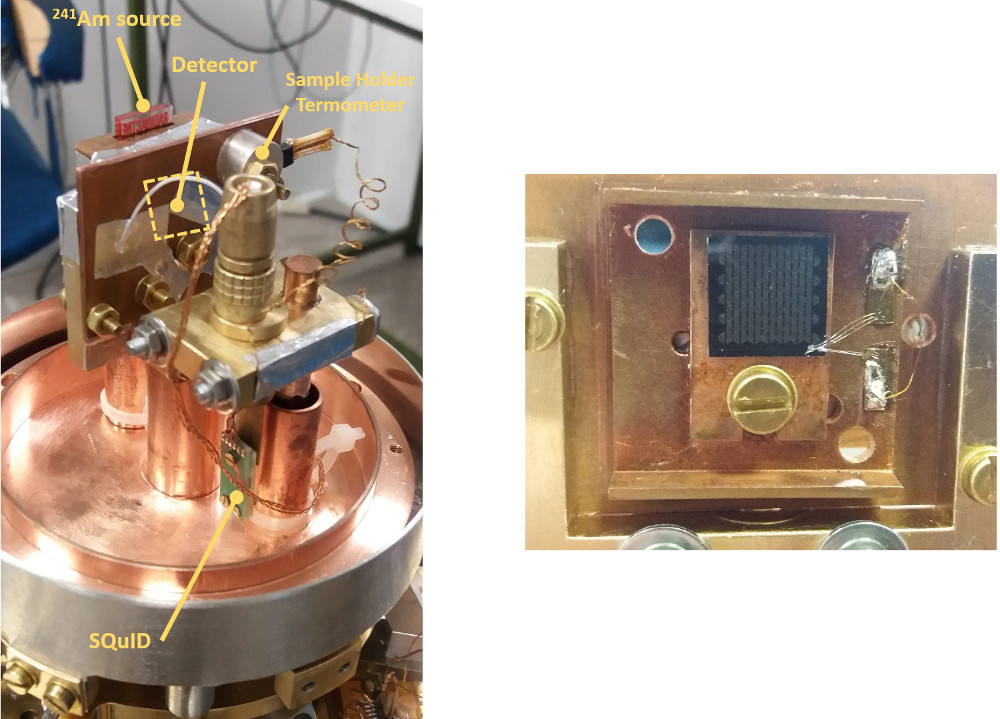}
\caption{ \textit{Left:} cryogenic setup developed to test the AC-S8 prototype in the IAPS ADR. \textit{Right:} detail of the AC-S8 sample mounted on the sample holder anchored to the ADR cold finger. The sample has been then covered by a copper sheet.}
\label{ACS8setup}
\end{figure}

The AC-S8 prototype has been tested in the ADR of the IAPS HE CryoLab, after the upgrading of the cryogenic setup with the installation of a magnetic shielding system at the 2.5 K stage (the development of this system is reported in detail in the Chapt. \ref{testsetupch} of this thesis). The detector has been coupled to a commercial SQUID array chip including a $R_S=$0.2 m$\Omega$ shunt resistor (Magnicon series array C6X16F), and operated with a Magnicon XXF-1 electronics. The integration of the sample in the cryogenic setup is shown in Fig. \ref{ACS8setup}.

\subsection{Transition and I-V characteristics}

\begin{figure}[H]
\centering
\includegraphics[width=0.43\linewidth]{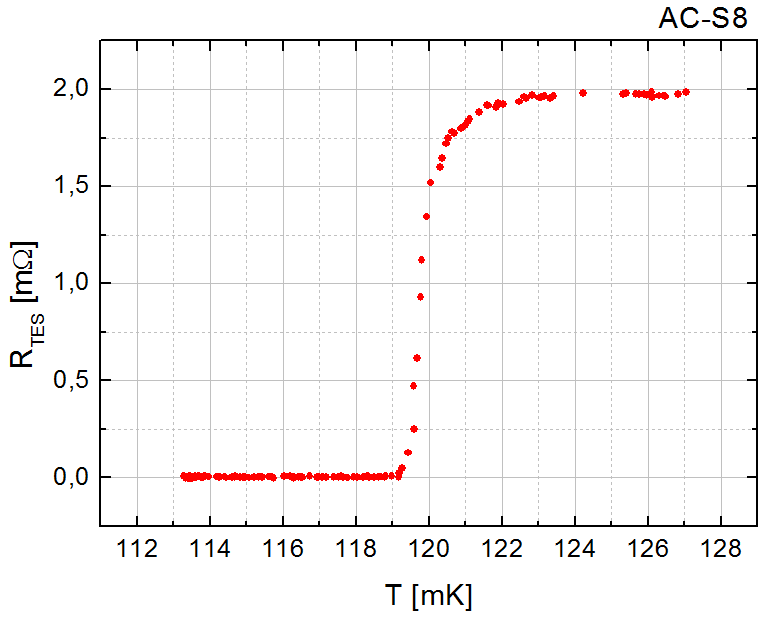}
\includegraphics[width=0.56\linewidth]{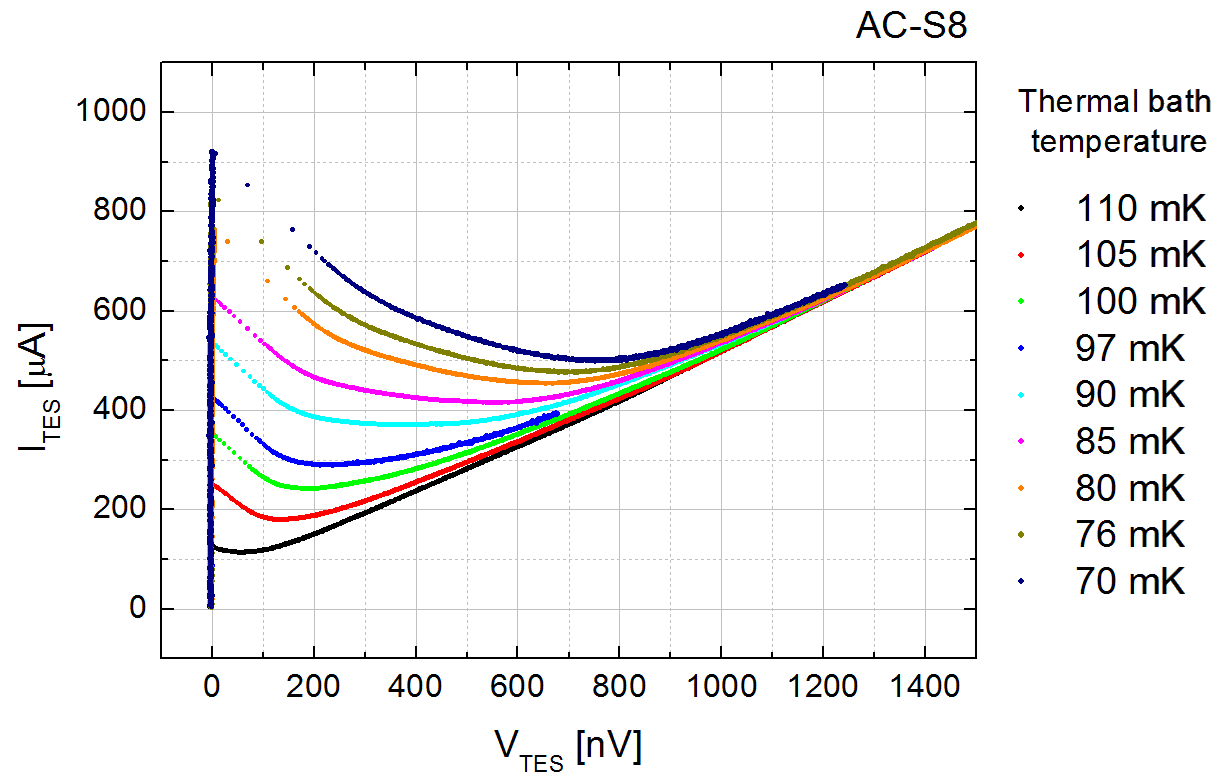}
\caption{Characterization measurements performed on the AC-S8 sample. \textit{Left:} Transition curve.
\textit{Right:} I-V curves.}
\label{RTIV}
\end{figure}

The transition curve of the sample is shown in Fig.~\ref{RTIV}-Left. The measurement has been performed biasing the sensor with a low frequency sine wave (f = 22Hz)  and sweeping the bath temperature. The bias current was kept small (few $\mu A_{PP}$) to minimize the TES self-heating and thus avoiding systematic effects. The plot shows a normal resistence R$_{N}$ = 2.0 m$\Omega$, a transition temperature T$_{C}$ = 120 mK (evaluated at R$_{TES}$/R$_{N}$ = 0.5) and a 10\%-90\% transition width $\Delta T_{C, 10\%-90\%} <$ 2 mK. The narrowness of the transition probes the homogeneity  of the TES network, which we remark is constituted by 65 identical Ir films connected in parallel, validating the sample manufacturing process also in presence of the additional Al finger network. 

In Fig.~\ref{RTIV}-Right are reported some characteristics I-V curves of the prototype. The curves have been acquired for decreasing voltage values (i.e. return branch, from normal to superconductive state) at fixed thermal bath temperatures. The wide explored temperature range (from T$_{B}$ = 70 mK to T$_{B}$ = 110 mK) shows that the detector can operate with the thermal bath quite far from the transition, without showing the instabilities that affected the AC-S7 sample (that so was probably due to setup issues).

\subsection{The pulse dynamic}

\begin{figure}[H]
\begin{center}
\includegraphics[width=0.6\linewidth]{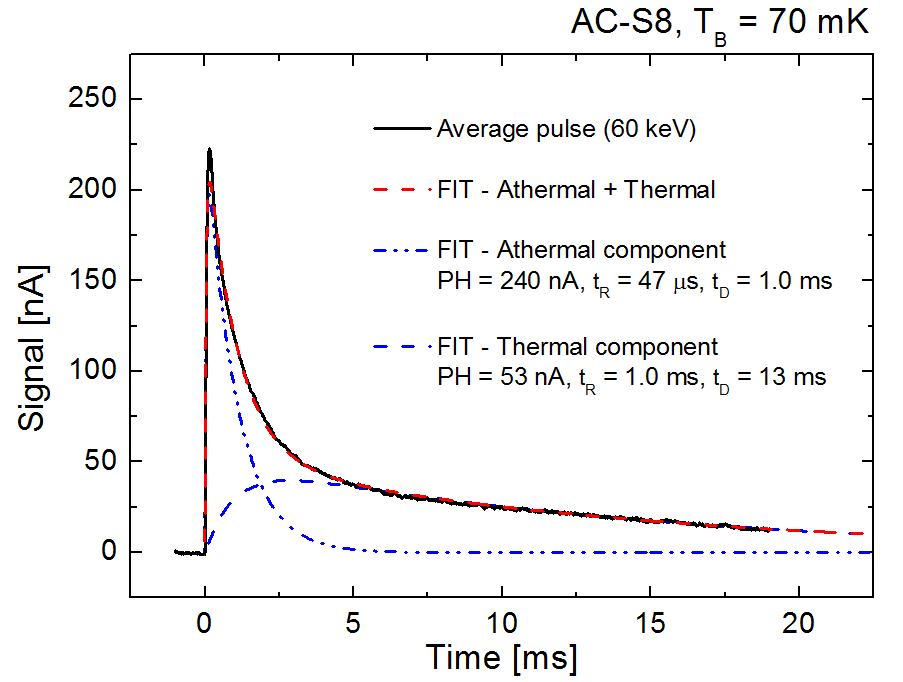}
\caption{Average 60 keV pulse acquired illuminating AC-S8 with a $^{241}$Am source. The result of the double pulse fitting procedure is overplotted and the best-fit parameters are reported.}
\label{avpulse}
\end{center}
\end{figure}

Also the AC-S8 detector has been illuminated with a low activity $^{241}$Am gamma source. The source was shielded with a copper sheet (0.5 mm thick) in order to provide only 60 keV photons, with an expected count rate of $\sim$ 3 cps. The bath temperature was set to 70 mK and the detector biased in the working point R$_{TES}$/R$_N$ = 0.3 (V$_{TES}$ = 340 nV, I$_{TES}$ = 615 $\mu A$).
We remark that this kind of detectors, due to the high number of TES, require high current (mA) in order to overcome the critical current so driving the TES from superconductive inside the transition region. Thus, with respect to our previous test procedures, we have driven the TES normal by increasing the bath temperature, injected the bias current, then decreased the thermal bath down to a good working point.
During 400 s of acquisition, 1244 pulses have been triggered, so having a count rate consistent with what expected. The average 60 keV pulse is shown in Fig.~\ref{avpulse}.  The result of a double pulse fitting procedure is overplotted, highlighting the expected fast athermal and the slow thermal components of the signal. The fitting function is: 

\begin{equation}
I(t) = PH_{ath} \cdot ( e^{-t/\tau_{D, ath}} - e^{-t/\tau_{R, ath}} ) + PH_{th} \cdot ( e^{-t/\tau_{D, th}} - e^{-t/\tau_{R, th}} )
\label{dpfunction}
\end{equation}
where for both the pulse components PH, $\tau_R$ and $\tau_D$ represent the Pulse Height and the characteristic rise and decay times. 

\noindent Note that, as expected, the decay time of the pulse athermal component results equal to the rise time of the thermal one ($\tau_{D,ath}$ = $\tau_{R,th}$ = 1.0 ms). This is a first positive check of the detector model developed in Sect. \ref{athermalsection}, which predicts that these times should be equal, both corresponding to the system characteristic time $\tau_{eff}$  (eq. \ref{endmodel}).

\subsubsection{From AC-S3 to AC-S8: comparing the pulse rising}
\label{ACS3vsACS8}
\begin{figure}[H]
\centering
\includegraphics[width=0.9\linewidth]{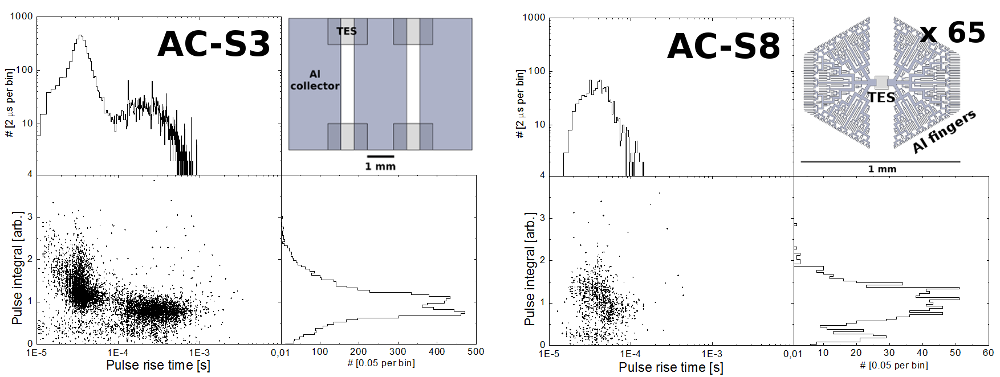}
\caption{\textit{Left:} Scatter plot of the ''Pulse integral'' vs ''Rise time'' of the 60 keV pulses acquired by AC-S3. Two different families of pulses are evident. \textit{Right:} The same plot referring to the pulses acquired by AC-S8. In this case the slow family of pulses is not present.}
\label{scatters}
\end{figure}

The measured average pulse athermal rise time $\tau_{R,ath}$ = 47 $\mu$s (Fig.~\ref{avpulse}) is compatible with the L/R $\sim$ 50 $\mu$s  of the TES circuit. This was evaluated taking into account the L = 90 nH total inductance and the $R_{P}$ = 1.0 m$\Omega$ parasitic resistance measured in series with the TES. The fast athermal rise time shows that the configuration of Al fingers does not slow down the rising of the pulse, as was instead observed in different prototypes developed in the past. Consider for comparison AC-S3, the old sample with 3 rectangular Al islands (2x5 mm$^2$) connecting 4 Ir/Au TES (each one 1x1.5 mm$^2$). In that case the wide area athermals collectors permitted the quasiparticle recombination during their diffusion in the aluminum, generating a late family of phonons \cite{macculi2012}. This phenomenon is well highlighted in Fig.~\ref{scatters}, where are shown the scatter plots \textit{Pulse integral} vs \textit{Rise time} of the 60 keV pulses acquired by AC-S3 and AC-S8. In the old sample two different families of pulses are evident (Fig.~\ref{scatters} - Left): the slowest family is due to the phonons generated by the quasiparticle recombination in the Al collectors. AC-S8 has been designed to prevent this issue, being based on a tree-like network of shorter Al fingers (max size 25x500 $\mu$m$^2$). In this way the quasiparticle path towards the TES is minimized, and the probability of recombination significantly reduced. As a result, the slow family of pulses is not present (Fig.~\ref{scatters} - Right).

\subsubsection{The arised setup issues}

The distribution of the in-time integral of the pulses in Fig.~\ref{scatters} - Right represents the preliminary energy spectrum acquired by AC-S8, with the main peak corresponding to the 60 keV events. The spectrum shows a very poor energy resolution ($\Delta$E/E $\sim$ 0.8 at 60 keV), making it difficult to provide an assessment of the detector low energy threshold. This is due to the fact that the detector has not been operated in good conditions. First, the high value of the parasitic resistance ($R_P$ = 1 m$\Omega$ = 0.5$R_N$) strongly affected the electrothermal feedback gain loop. This resistance was located between the input coil and the TES, at the contact between our superconducting wiring and the Nb pads of the SQUID packaging. Further, the shunt resistance on board the SQUID chip (at 500 mK) generated a high Johonson noise.

In the design of the DM (which is presented in the Chapt. \ref{DMchapt}) we have considered and solved these issues, placing the SQUID close to the detector (few mm, at the 50 mK stage), and realizing the connection between the input coil and the TES with ultrasonic aluminum wire bonding. Furthermore, we have designed the DM TES with a different aspect ratio, in order to have a higher normal resitance ($R_{N, DM} \sim$ 10 m$\Omega$). 

We remark that the setup limitations do not affect the results reported in the following sections, which have been obtained by means of average pulse analysis in order to increase the signal to noise ratio.

\subsubsection{From AC-S7 to AC-S8: the improvement in the athermals collection}
From the double-component fit in Fig.~\ref{avpulse}, it is possible to evaluate the ratio between the athermal and the total energy associated to the average pulse, which are proportional to the integral of the respective fit functions.
We found $\varepsilon_{AC-S8}$ = $E_{ath}/(E_{ath} + E_{th}) |_{AC-S8}$ = $\sim$ 26\%, more than a factor 4 above the AC-S7 value $\varepsilon_{AC-S7} \sim 6\%$. We interpret this as an evidence of the higher AC-S8 athermals collection efficiency due to the Al finger network. 

\begin{figure}[H]
\centering
\includegraphics[width=0.65\linewidth]{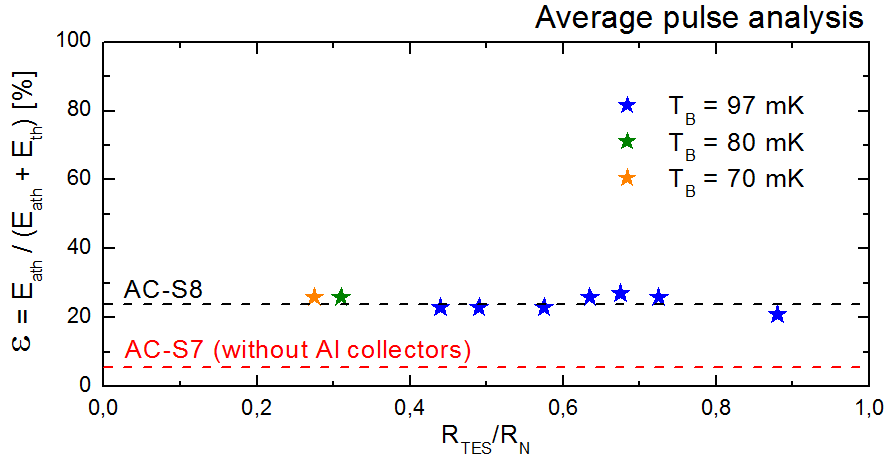}
\caption{Ratio between the athermal and the total (athermal + thermal) energy of the average 60 keV pulses acquired with AC-S8 in different working points. The mean value is compared with the AC-S7 one.}
\label{eps}
\end{figure}

To consolidate this result we have repeated the average pulse analysis biasing the detector at different transition levels and with different thermal bath temperatures. The results are reported in Fig.~\ref{eps}, which shows that the $\varepsilon$ ratio is roughly independent from the detector working point, confirming the efficiency increasing.

\subsubsection{Response of the detector in different bias points}

\begin{figure}[H]
\centering
\includegraphics[width=0.4\linewidth]{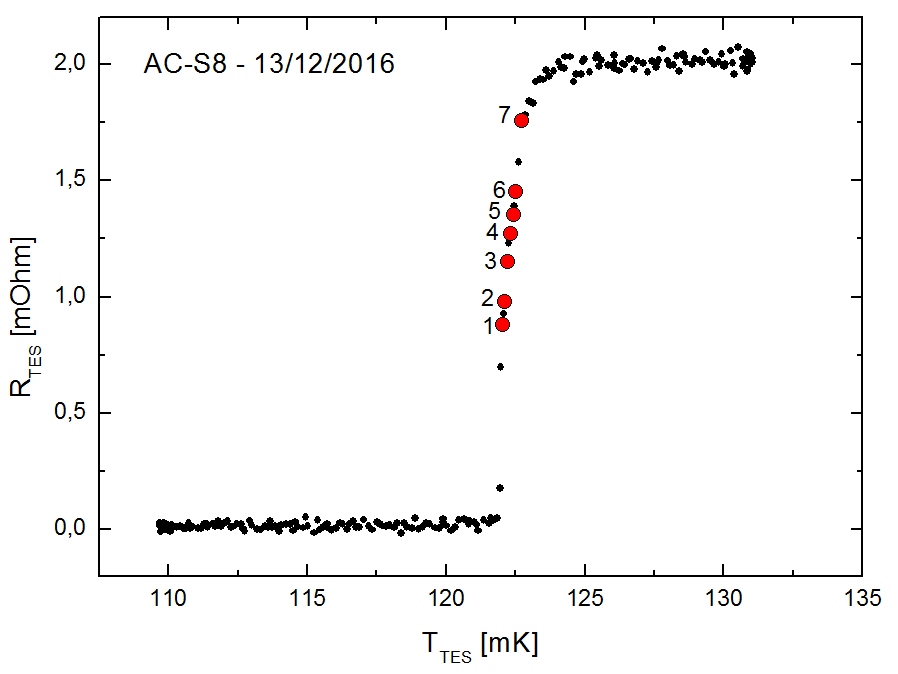}
\includegraphics[width=0.4\linewidth]{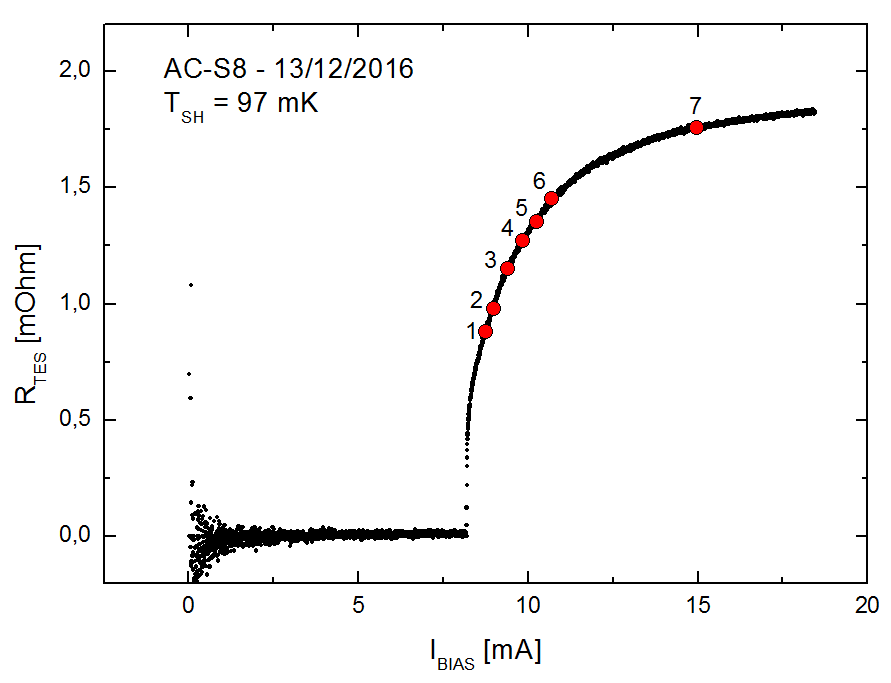}
\caption{Different points in which the detector has been biased, highlighted on the curves $R_{TES}$ vs $T_{TES}$ (\textit{Left}) and $R_{TES}$ vs $I_{BIAS}$ (\textit{Right}). The bath temperature was set to 97 mK. It has been not possible to bias the detector at resistance below the one of the point \virg 1'' ($R_{TES}$ = 0.9 m$\Omega$) because the system was unstable.}
\label{multibias}
\end{figure}

By analysing the data acquired biasing the detector at different transition level, we have also studied the response of the detector as a function of the bias point, once fixed the thermal bath temperature (in this case at $T_B$ = 97 mK). The seven different detector working points used for this analysis are shown in Fig. \ref{multibias}, where they are highlighted on both the curves $R_{TES}$ vs $T_{TES}$ and $R_{TES}$ vs $I_{BIAS}$.

\begin{figure}[H]
\centering
\includegraphics[width=0.9\linewidth]{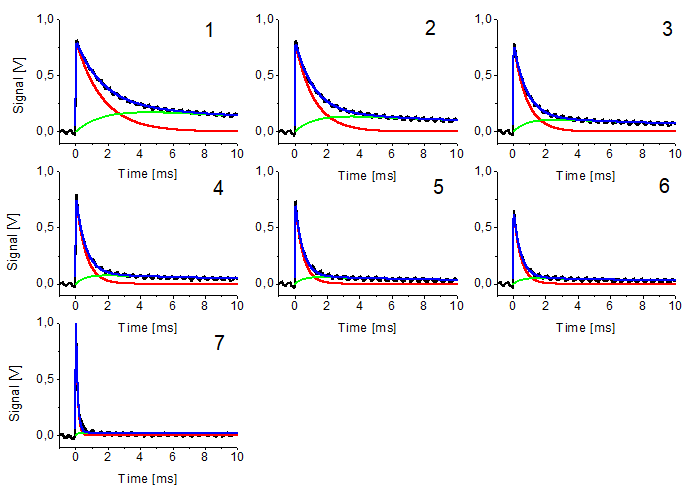}
\caption{Average 60 keV pulses acquired biasing AC-S8 in the seven points shown in Fig. \ref{multibias}, at $T_B$ = 97 mK. The result of the double pulse fitting procedure is overplotted.}
\label{multibias_avpulses}
\end{figure}

\noindent The average 60 keV pulses acquired in the different bias points are shown in Fig. \ref{multibias_avpulses}. Each pulse has been fitted by the double pulse function reported in eq. (\ref{dpfunction}), in order to disentangle the athermal (red lines) and the thermal (green lines) components.

\begin{figure}[H]
\centering
\includegraphics[width=0.48\linewidth]{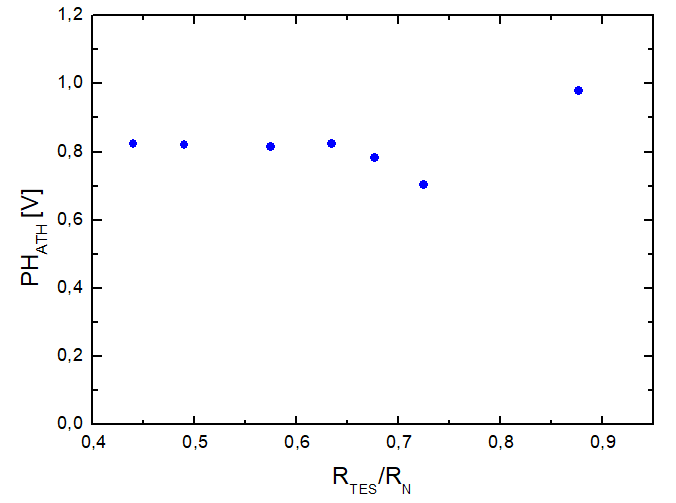}
\includegraphics[width=0.48\linewidth]{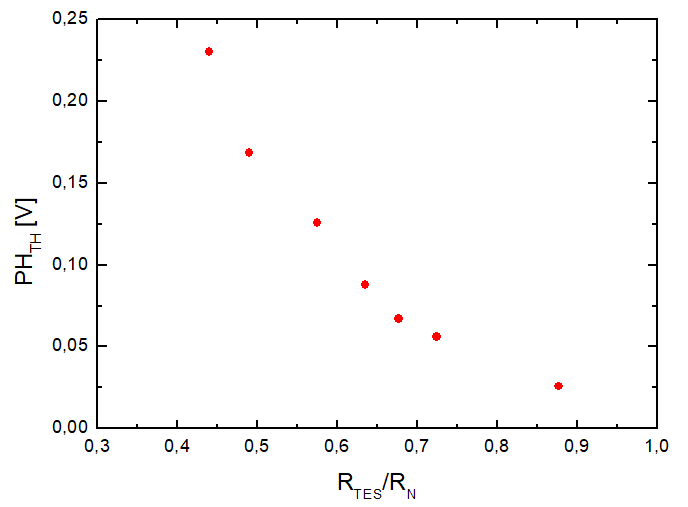}
\caption{Average 60 keV pulses athermal (Left) and thermal (Right) characteristic decay times as a function of the TES resistence in the bias point, at $T_B$ = 97 mK.}
\label{multibias_trends}
\end{figure}

The main result of this analysis is reported in Fig. \ref{multibias_trends}, where the athermal and the thermal characteristic decay times resulting from the fits are plotted as a function of the TES resistence in the bias point. As expected, due to the strong electron-phonon decoupling in the TES, only the athermal decay time is dependent on the detector bias point, decreasing with the increase of the TES resistance due to the enhancement of the loop gain (see eq. \ref{teffathermal}). Differently, the thermal decay time remains roughly constant, not being affected by the electrothermal feedback but depending only from the heat capacity of the absorber and the thermal conductance towards the thermal bath (see eq. \ref{tthermal}). This is another check of the validity of the model developed in Sect. \ref{athermalsection}, that predicts this detector behaviour.

\subsection{Operation at Tb = 50 mK}

\begin{figure}[H]
\centering
\includegraphics[width=0.9\linewidth]{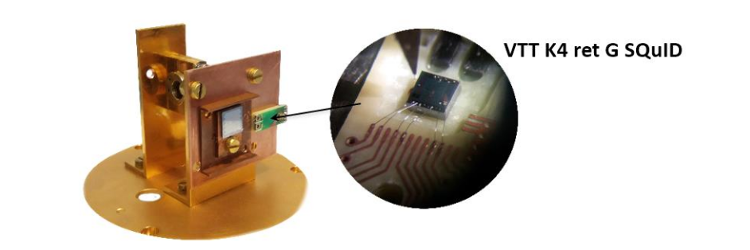}
\caption{Cryogenic setup developed to test the AC-S8 sample with the VTT K4 ret G SQUID (developed for the CryoAC DM), in the IAPS dilution refrigerator. The SQUID has been mounted inside a commercial Magnicon package, and it has been wire-bonded in order to be operated in FLL mode by the electronics Magnicon XXF-1.}
\label{ACS8_50mK}
\end{figure}

Once installed the Dilution Refrigerator in the IAPS HE cryolab, the AC-S8 sample has been integrated into a new setup in order to be tested in the new cryogenic system (Fig. \ref{ACS8_50mK}). Here the detector has been coupled to a SQUID (model K4 ret G) produced for the CryoAC DM by the VTT Technical Research Centre of Finland \cite{VTT}. 

\begin{figure}[H]
\centering
\includegraphics[width=0.9\linewidth]{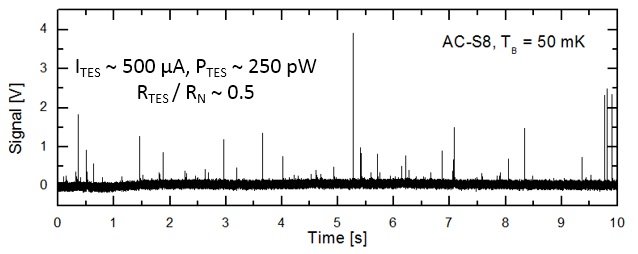}
\caption{Functional test of AC-S8 operated at a bath temperature of 50 mK (CryoAC DM requirement). The plot shows the raw output SQUID signal. The detector has been illuminated by the shielded $^{241}$Am source (60 keV photons).}
\label{ACS8_50mK_1}
\end{figure}

In this setup, also thanks to the new cryostat, it has been possible to operate for the first time the detector at a bath temperature of 50 mK (Fig. \ref{ACS8_50mK_1}). This is an important result because no CryoAC prototype has ever been operated at a so low bath temperature before. I remark that one of the requirement of the CryoAC DM is to demonstrate the detector functionality at $T_B$ = 50-55 mK, and this measurement represents an important step forward in this regard.

\begin{figure}[H]
\centering
\includegraphics[width=0.9\linewidth]{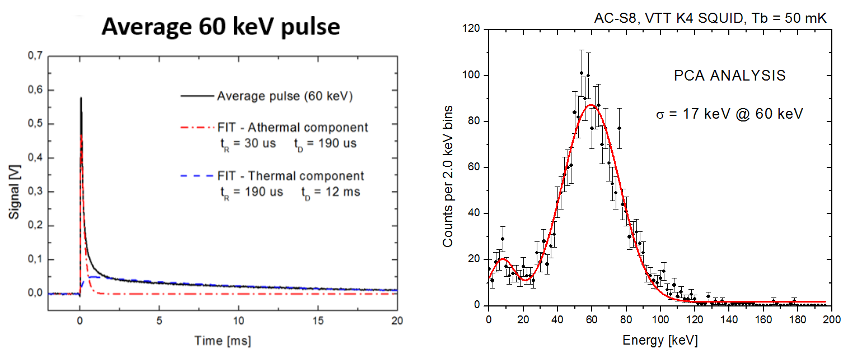}
\caption{Main results of the functional test of AC-S8 operated at a bath temperature of 50 mK. \textit{Left:} Average 60 keV pulse fitted by using a double component pulse function. The best-fit prameters are reported.\textit{Right:} Energy spectrum obtained illuminating the detector by the shielded $^{241}$Am source.}
\label{ACS8_50mK_2}
\end{figure}

\noindent In Fig. \ref{ACS8_50mK_2} the average 60 keV pulse and the raw energy spectrum obtained illuminating the detector by the shielded $^{241}$Am source are shown.

The average pulse is well shaped (Fig. \ref{ACS8_50mK_2} - Left), and it shows both the athermal and thermal components. The rise time is quite fast ($\tau_R$ = 30 $\mu$s), and it is roughly compatible with the CryoAC requirement ($\tau_{R, CryoAC}$ $\leq$ 30 $\mu$s). Also the athermal decay time is very fast ($\tau_{D,ATH}$ = 190 $\mu$s), as a results of high electrothermal feedback due to the low bath temperature, and it is whithin our requirement ($\tau_{eff, CryoAC}$ $\leq$ 250 $\mu$s). On the other hand, the thermal decay time of the detector is still too high ($\tau_{D,TH}$ = 12 ms, while the requirement is $\tau_{D, CryoAC}$ $\leq$ 2.5 ms), showing us the need to have a better control of the absorber heat capacity and thermal conductance towards the bath. 

The energy spectrum (Fig. \ref{ACS8_50mK_2} - Right) shows the expected 60 keV photopeak with a good gaussian profile, and a rise at lower energy roughly compatible with the expected Compton edge. Note that the measured energy resolution ($\sigma$ = 17 keV @ 60 keV) is not representative of the detector performance, since the detector working point has not be optimized (this has been a functional and not a performance test).

\newpage
\section{Conclusions}

In this chapter I have reported the main results of the test activities performed on the pre-DM samples AC-S7 and AC-S8, the first CryoAC prototypes based on large networks of TES ($\times$ 65) sensing a 1 cm$^2$ silicon absorber. Summarizing, we have:

\begin{itemize}
\item Measured for both AC-S7 and AC-S8 narrow superconductive transitions ($\Delta$Tc $\sim$ 2 mK), demonstrating the uniformity of TES large networks and thus validating their manufacturing processes.

\item Shown that the athermal contribution to the pulses energy increased of a factor $\sim$ 4  from AC-S7 (without Al collectors) to AC-S8 (with Al collectors), so obtaining a first evidence of the improvement in the athermal collection efficiency due to the use of the Al collectors. Furthermore we have demonstrated that the AC-S8 collectors design is able to prevent the quasi-particles recombination in the Al, assuring a fast pulse rising front.

\item Obtained the first validation checks for the detector model developed in Sect. \ref{athermalsection}, showing that the athermal decay time and the thermal rise time are equal, and that (due to the electron phonon decoupling in the TES) only the athermal decay time is influenced by the electrothermal feedback, while the thermal decay time is roughly indepentent from the detector working point (once fixed the bath temperature).

\item Shown with AC-S7 that this kind of detector can achieve a low energy threshold lower than 20 keV, and it exhibits an energy resolution of $\sim$11 keV @ 60 keV (FWHM).

\item Demonstrated by AC-S8 that this kind of detector can be operated with the thermal bath at 50 mK (CryoAC DM requirement), showing a pulse rise time and athermal decay time within the CryoAC requirements ($\tau_R$ = 30 $\mu$s, $\tau_{D,ATH}$ = 190 $\mu$s).
\end{itemize}

\begin{figure}[htbp]
\centering
\includegraphics[width=0.6\textwidth]{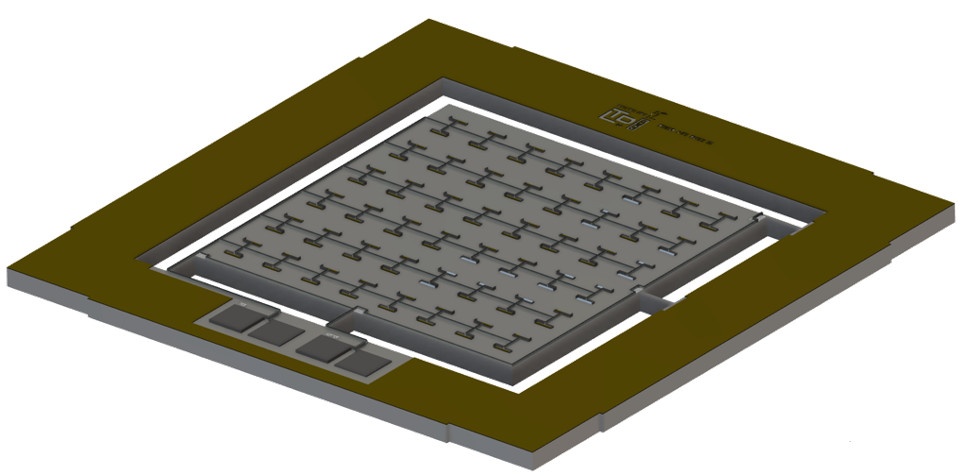}
\caption{Sketch of the CryoAC DM design.}
\label{introDM}
\end{figure}

\noindent On the other hand, we have also faced some problem during the detectors operation. This has allowed us to intervene on the CryoAC DM design, in order to fix these issues. In particular:

\begin{itemize}
\item The AC-S7 and AC-S8 samples have a very low normal resistance ($R_N \sim$ 2 m$\Omega$). This makes it difficult to operate the detectors in a good voltage bias, strongly affecting the electrothermal feedback gain loop and thus degrading the detector performance. We have therefore designed the CryoAC DM TES with a different aspect ratio, in order to have a higher normal resitance ($R_{N, DM} \sim$ 10 m$\Omega$). 

\item High bias current are needed to bias the detector with the bath at low temperature (hundred of $\mu$A to the TES network, corresponding to some mA to the whole TES bias circuit). To deal with this, in the CryoAC DM design we have foreseen an additional platinum heater deposited on the absorber. In this way, it will be possible to operate the detector also by smaller bias current, locally increasing the absorber temperature. The heater can also be used just to drive the TES into the normal state and bias it without overcoming the high critical current (several mA), and then turned off during the detector operations. In addition, the CryoAC DM TES are foreseen to be Ir:Au bilayers, in order to lower the transition critical temperature. 

\item The slow measured thermal decay time ($>$ 10 ms) showed us the need to better control the heat capacity of the absorber and the thermal conductance towards the thermal bath. In the CryoAC DM this latter point is addressed by realizing a suspended structure for the silicon absorber, and realizing the thermal conductance towards the bath by means of narrow silicon beams (Fig. \ref{introDM}). Furthermore, we have started an activity aimed at characterizing the silicon wafers from which the CryoAC DM is produced, in order to better evaluate its thermal properties.

\end{itemize}

\noindent The AC-S7 and AC-S8 prototype represent the last step towards the development of the CryoAC DM, which will be presented in the Chapt. \ref{DMchapt}.

\bigskip

\chapter{CryoAC pulse processing by Principal Component Analysis (PCA)}
\chaptermark{CryoAC pulse processing by PCA}
\label{pca}

The traditional method to process X-ray pulses acquired from TES microcalorimeters and generate their energy spectrum is the Optimal Filter, where the pulses are cross-correlated with a template generated starting from the average pulse and the acquired noise spectrum. This method maximizes the signal to noise ratio under the conditions that the noise and the pulse shape are stationary \cite{busch}. 

In the CryoAC detector this last condition is not satisfied. Our pulses, indeed, are the combination of two different signals: the fast athermal (due to the thermalization of athermal phonons directly in the TES), and the slow thermal (due to the temperature rise of the absorber bulk). The final pulse shape strongly depends on the quantity of athermal phonons collected by the TES network for a given particle event, and so it is not stationary.

To properly analyze the CryoAC data we therefore need to disentangle the athermal component of the pulses from the thermal one. As shown in the previous chapter, our typical analysis foresees a double-pulse fitting procedure over the triggered data. This analysis has allowed us to investigate the pulse dynamics in the last CryoAC prototypes, and to obtain energy spectra with a better resolution than the one achievable by the raw pulses in-time integral. On the other hand this procedure is strongly model-dependent, requiring a priori knowledge of the pulses fitting functions.

In this chapter, I present a different approach for the pulses processing, based upon Principal Component Analysis (PCA). This kind of analysis requires no prior knowledge of the dataset and it has been successfully used to process cryogenic microcalorimeters signals with severe pulse shape variation, recovering better energy spectra than achievable with traditional analysis \cite{busch} \cite{yan}. 

Initially, I will report the first implementation of this method on CryoAC pulses, which has been performed to process the data acquired by the AC-S7 prototype. Then, I will show the validation of the developed procedures  by applying this method also on simulated CryoAC signals.

\newpage
\section{PCA implementation for the AC-S7 pulse analysis}

The Principal Component Analysis is a statistical procedure that allows to represent a complex dataset in terms of a small set of orthogonal components that have the largest variance (Principal Components). It is widely used for data processing and dimensionality reduction in several research areas, from Neurosciences to Astronomy and Astrophysics. A simple introductive tutorial about PCA can be found in Ref. \cite{shlens}.

To apply this method to our pulses, I have followed the PCA implementation described in Ref. \cite{yan}. The goal is to represent each pulse as the linear combination of the Principal Components of the signal, which are selected from the eigenvectors of the time covariance matrix of the pulses. In this way it is possible to identify the significant characteristics pulse shape components, filtering out the noise. The Principal Components can be then combined to form a representation of the pulses energy, obtaining the energy spectra.

In the next sections it will be reported step-by-step how the PCA method has been implemented to analyze the pulses acquired by illuminating the AC-S7 prototype by the $^{241}$Am source (see Par. \ref{acs7ill} for informations about this acquisition).

\subsection{Pulse processing and Principal Component selection}
\label{PCAproclist}

The \virg AC-S7 - $^{241}$Am'' raw dataset consists of 1270 triggered pulse records, each one containing N$_S$ = 8000 samples (including 500 pre-trigger points) acquired with a sampling time $\Delta$t = 10 $\mu$s. The data have been processed as follow:

\begin{itemize}

\item Piles-up and non-saturated pulse records have been discarded, keeping N$_P$ = 1207 pulses (95\% of all triggered events).
\item The selected pulses have been collected in the matrix DATA (N$_S$ $\times$ N$_P$). Every column in the matrix contains a different event, so the matrix elements DATA[i,j] represents the i-th sample of the j-th pulse:

\smallskip

\begin{equation}
DATA = 
\begin{bmatrix}
    P_1(t=0) & P_2(t=0) & \dots  & P_{N_p}(t=0) \\
    P_1(t=\Delta t) & P_2(t=\Delta t) & \dots  & P_{N_p}(t=\Delta t) \\
    \vdots & \vdots & \ddots & \vdots \\
    P_1(t=T) & P_2(t=T) & \dots  & P_{N_p}(t=T) \\
\end{bmatrix}
\end{equation}

\smallskip

\item The data time covariance matrix COV (N$_{S}$ $\times$ N$_{S}$) has been defined as:
\begin{equation}
COV = DATA \times (DATA)^T
\end{equation}
\item According to the PCA procedure, the goal is to find a new a basis where the covariance matrix is diagonal, writing it in the form:
\begin{equation}
COV = EV \times \Lambda \times (EV)^{-1}
\end{equation}
where EV (N$_{S}$ $\times$ N$_{S}$) is the matrix of the COV eigenvectors and $\Lambda$ is the diagonal matrix of its eigenvalues (sorted from the highest to the lowest). Eigenvalues and eigenvectors of the covariance matrix have been therefore computed using the DSYEV Fortran90 routine of the Linear Algebra PACKage \cite{LAPACK}.
\item Finally, each pulse has been represented as the linear combination of the computed eigenvectors. To do this, the pulses have been rotated into the new basis EV:
\begin{equation}
R = (EV)^T \times DATA
\label{Rmatrix}
\end{equation}
The matrix R (N$_{S}$ $\times$ N$_{P}$) contains the pulses projections onto all the eigenvectors, and so it is a complete alternative representation of the initial dataset.

\end{itemize}

\begin{figure}[H]
\centering
\includegraphics[width=1\textwidth]{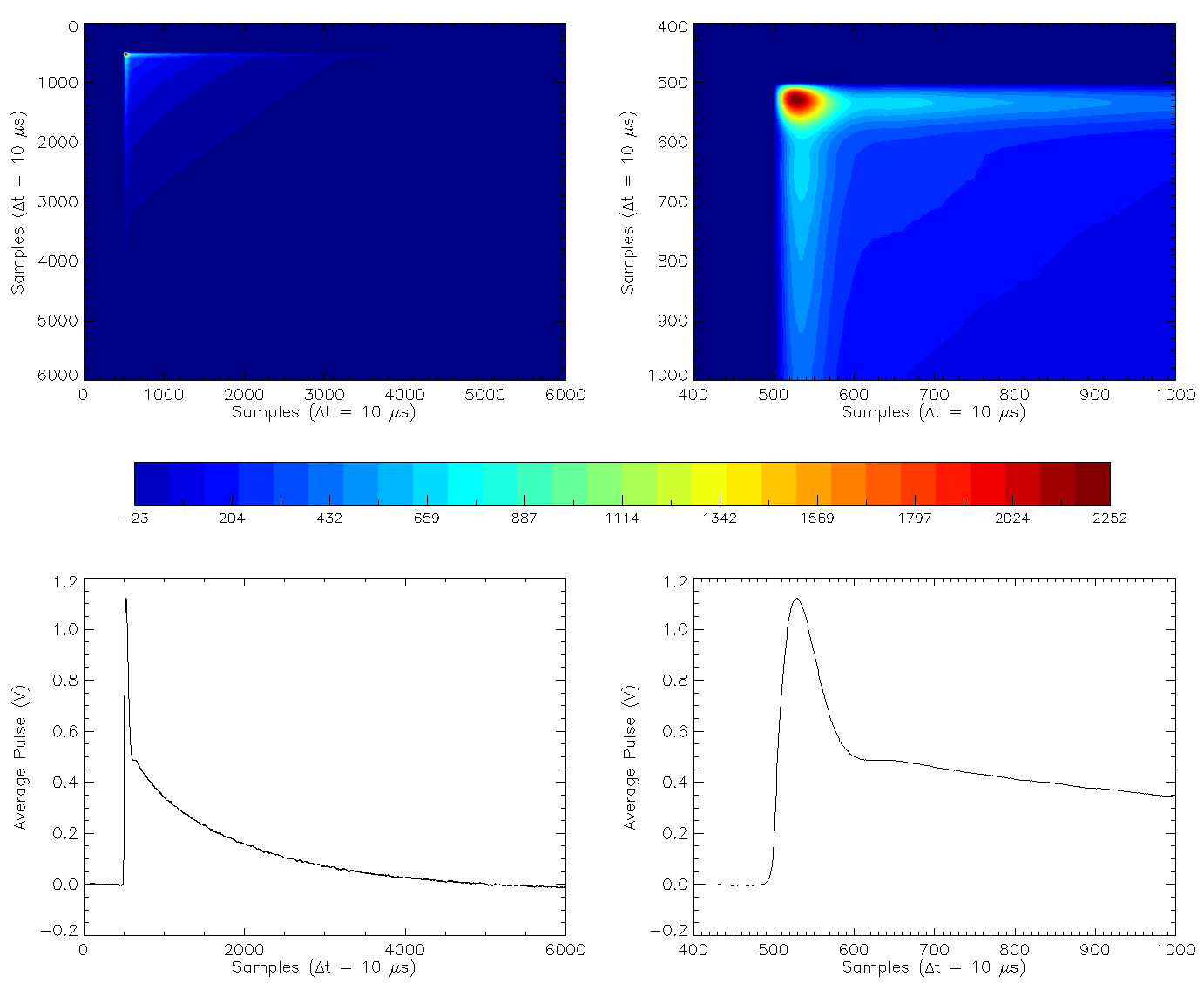}
\caption{\textit{Top:} Representation of the covariance matrix \virg COV'' created from the AC-S7 pulse dataset, in two differets sample (i.e. time) scales. The units of the color-coded legend are V$^2$. \textit{Bottom:} The dataset average pulse plotted in the two differets sample (i.e. time) scales.}
\label{PCA_cov}
\end{figure}

\begin{figure}[H]
\centering
\includegraphics[width=0.6\textwidth]{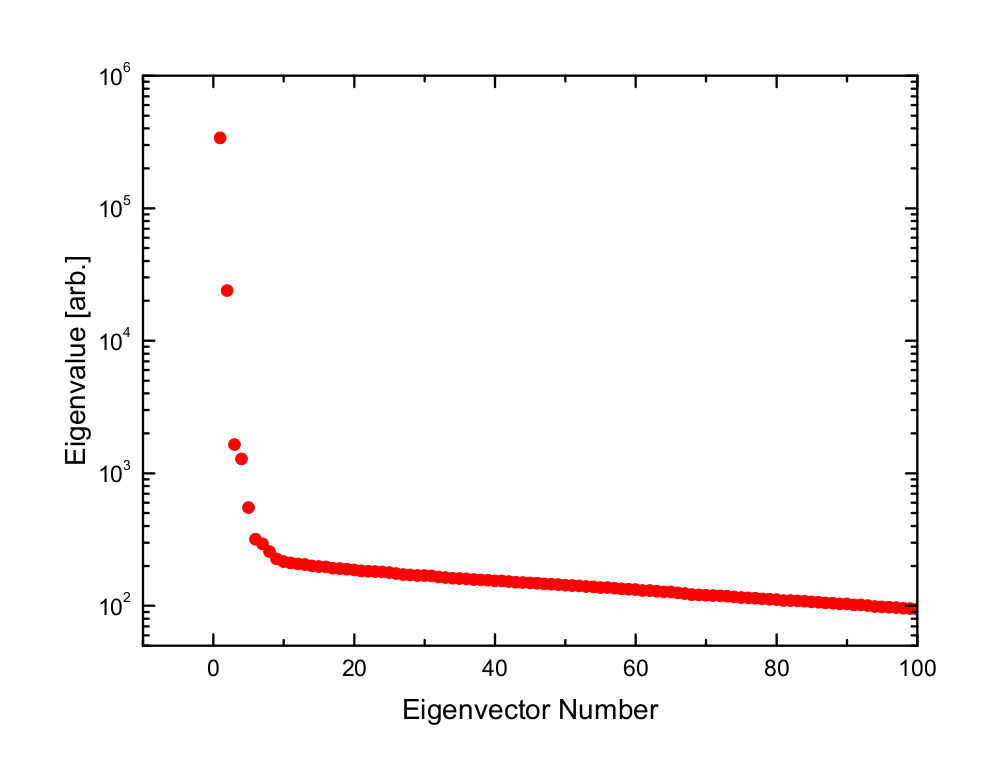}
\caption{Spectrum of the largest 100 eigenvalues of the covariance matrix.}
\label{PCA_evalues}
\end{figure}

\begin{figure}[H]
\centering
\includegraphics[width=0.95\textwidth]{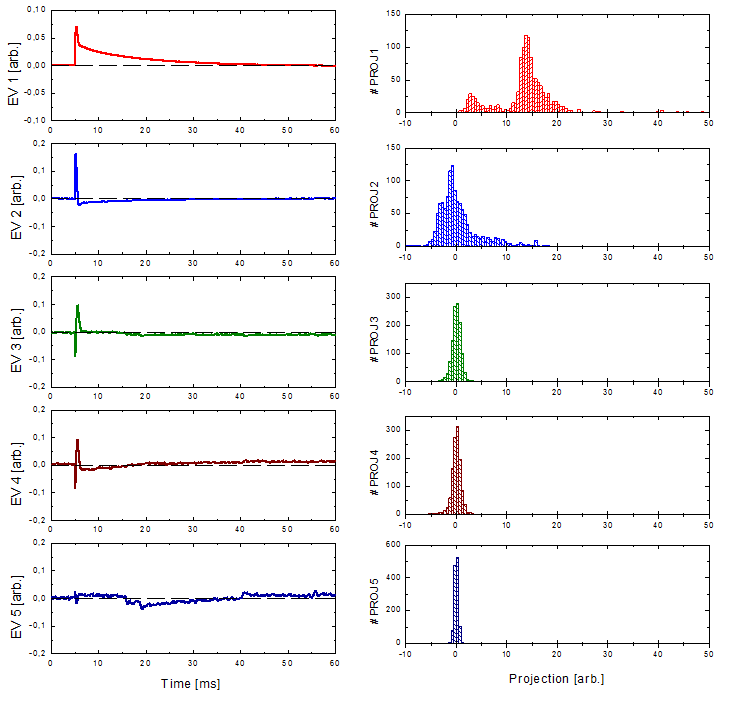}
\caption{Left: The first five eigenvectors of the pulses covariance matrix. Right: Histogram of the projections of the pulses onto the corresponding (left) eigenvector. }
\label{PCA_evectors}
\end{figure}

\noindent The covariance matrix of the \virg AC-S7 - $^{241}$Am'' dataset is represented in Fig. \ref{PCA_cov}, and in Fig. \ref{PCA_evalues} it is shown the spectrum of its largest 100 eigenvalues. In Fig. \ref{PCA_evectors} the eigenvectors corresponding to the five largest eigenvalues are plotted with the respective histograms of the pulses projections.
\bigskip

Once obtained the projections matrix R (eq. \ref{Rmatrix}), the main task is to determine which eigenvectors are responsible for the shape of the pulses and which are encoding only noise-structures. Following Ref. \cite{busch}, it is possible to identify the significant eigenvectors as the ones with the largest corresponding eigenvalues. The eigenvalues are indeed a measure of the information encoded in the pulses by the corresponding eigenvectors.

From fig. \ref{PCA_evalues}  and \ref{PCA_evectors}, it is clear that the majority of the information regarding the pulses shape is encoded by the first two eigenvectors. Eigenvectors 3-4 encode some minor variations in the pulses, while the others contains no shape information and corresponds to the noise in the dataset. 

It is interesting to note that the first eigenvector is essentially an average pulse. For the limit case of no pulse-shape variation, stationary noise and detector linear response, it would be the only representative eigenvector. Its projection histogram could then be used as representation of the energy spectrum, and the PCA should be equivalent to the conventional optimal filter.

\begin{figure}[H]
\centering
\includegraphics[width=1\textwidth]{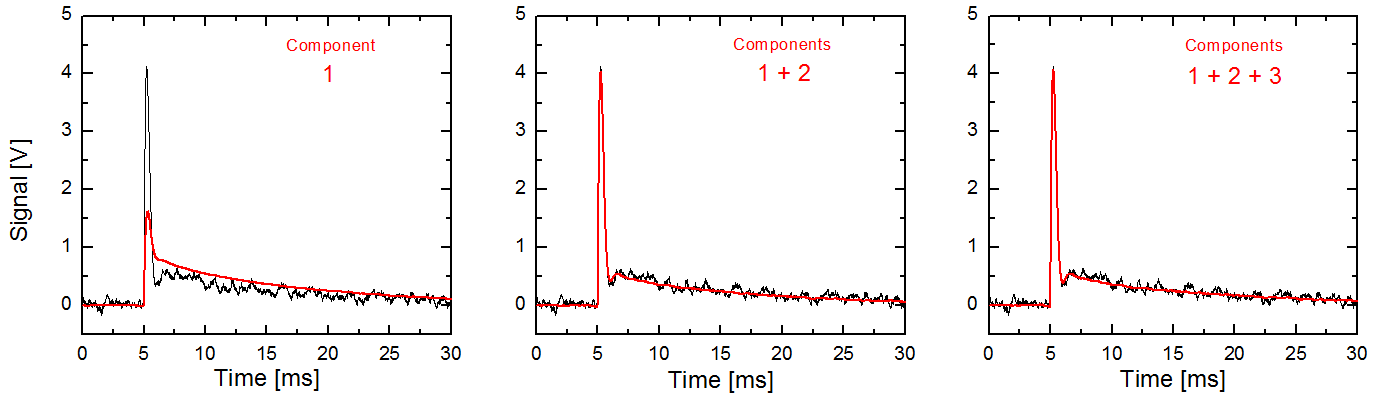}
\caption{An example raw pulse (black line) and its correspondent PCA reconstruction for an increasing number of considered components (red line).}
\label{PCA_pulses}
\end{figure}

In our case, we choose to select the first two eigenvectors to extract energy information from our dataset. We can see in Fig. \ref{PCA_pulses} that these two components are sufficient to well describe the pulse shape and so, in first approximation, we can neglect the contribution of the subsequent eigenvectors. 

\bigskip

\bigskip

\begin{figure}[H]
\centering
\includegraphics[width=0.9\textwidth]{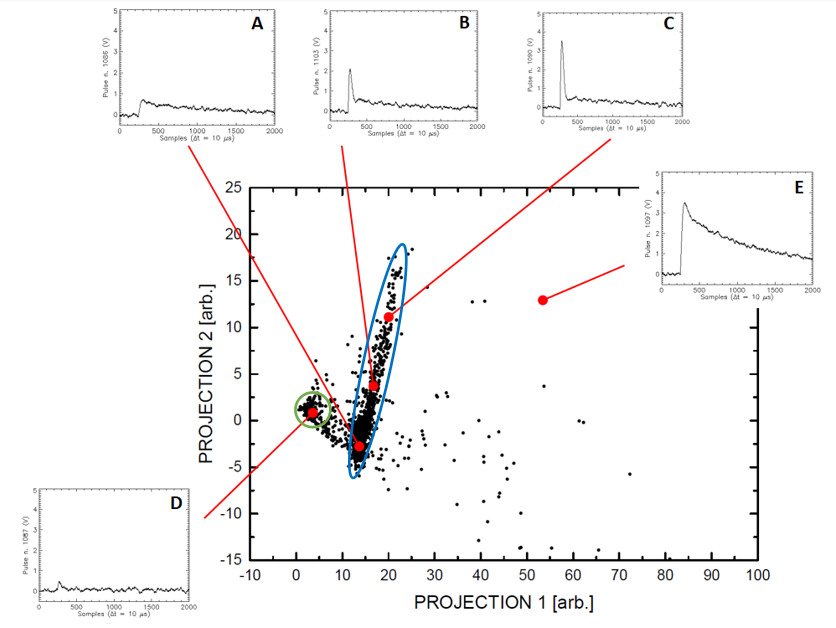}
\caption{2D scatter plot of the projection of each pulse onto the first two eigenvectors of the covariance matrix. The blu eclipse highlights the main cluster of points, corresponding to the 60 keV photopeak pulses, and the green circle a smaller cluster with low energy events (i.e. fluorescences and compton scattering into the absorber). The encapsulated plots represent raw pulse in the dataset placed in different positions of the projections space.}
\label{PCA_scatter}
\end{figure}

To obtain the energy spectra we have then examined the 2D scatter plot of the pulses projections onto the two selected eigenvectors (Fig. \ref{PCA_scatter}). Every black point in the plot represents a different pulse. We can associate the main cluster of points (blu region) to the 60 keV line that we mainly expect from the $^{241}$Am source. Note that the position of the pulses in the cluster is related to the height of the athermal component: the higher the athermal, the higher the position in the projections plane (see the incapsulated A, B and C pulses plots). The dataset includes another small cluster (green circle) with lower energy events (i.e. fluorescences and compton scattering events in the absorber, incapsulated D pulse), and also isolated points corresponding to higher energy events, probably background muons (incapsulated E pulse). 

\begin{figure}[H]
\centering
\includegraphics[width=0.9\textwidth]{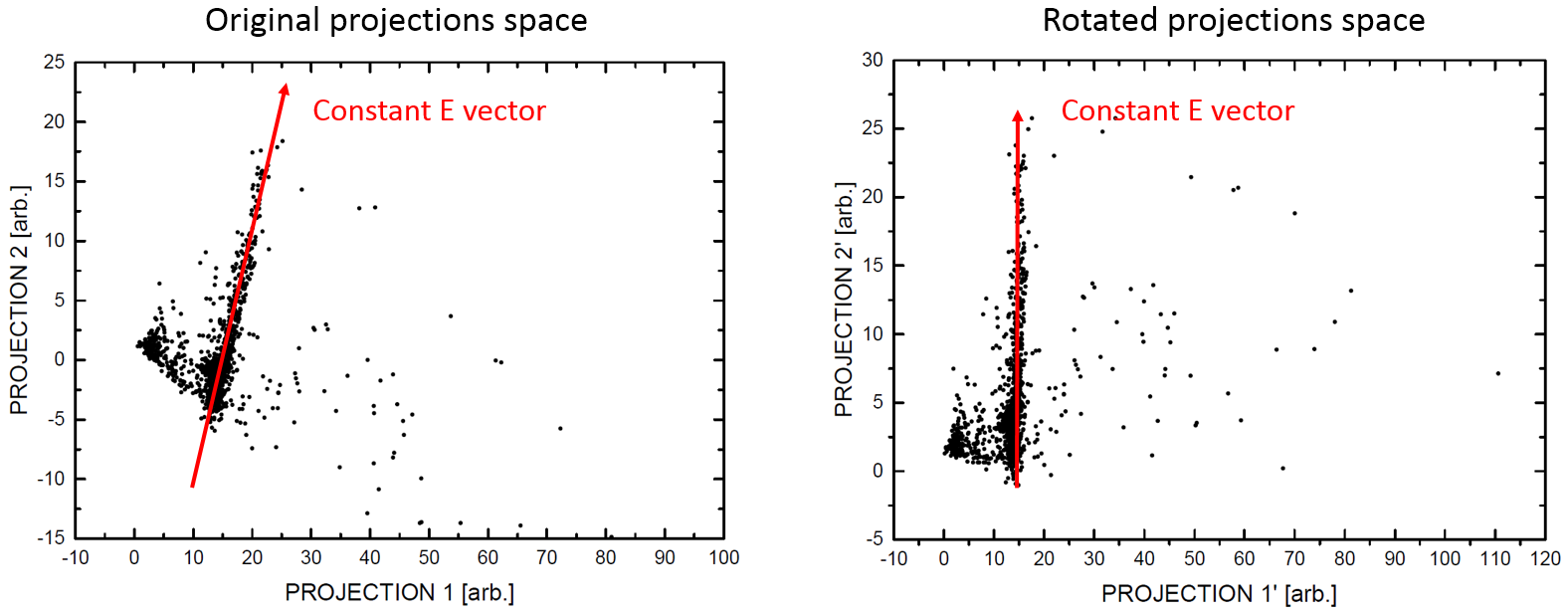}
\caption{\textit{(Left:)} 2D scatter plot of the pulses in the original projections space (same plot of Fig. \ref{PCA_scatter}). The red arrow indicates, as labeled, the costant energy direction, obtained by linear fitting the main \virg 60 keV photopeak'' cluster. \textit{(Right: )} New space obtained rotating the original one, in order to put the main cluster vertical.}
\label{PCA_rotation}
\end{figure}

Thanks to the 60 keV principal cluster we are able to determines the direction of constant energy in the 2D projections space. Following the prescription in \cite{yan}, this direction defines an axis that can be used as reference to rotate of an angle $\alpha$ the projections space, putting the main cluster vertical (Fig. \ref{PCA_rotation}). In the \virg rotated'' space, the x and y axis represent the projections of the pulses into the new vectors:

\begin{equation}
X' = EV\;1 \cdot cos(\alpha) - EV\;2 \cdot sin(\alpha)
\label{rot1}
\end{equation}

\begin{equation}
Y' = EV\;1 \cdot sin(\alpha) + EV\;1 \cdot cos(\alpha)
\label{rot2}
\end{equation}

\noindent where $EV\;1$ and $EV\;2$ are the eigenvector used to define the original projection space.

\bigskip

Finally, the projections onto the new x-axis (which is perpendicular to the constant E vector, and so it represents the $\Delta$E direction) can be used to generate the energy spectrum of the pulses. We noticed also that in the rotated space the area of the pulses is totally given by the new vector X' (i.e. the  Y' has a null in-time integral), as evidence that all the information about the pulses energy is encoded by the projections onto this new vector.

\subsection{Energy spectra comparison}

\begin{figure}[H]
\centering
\includegraphics[width=1\textwidth]{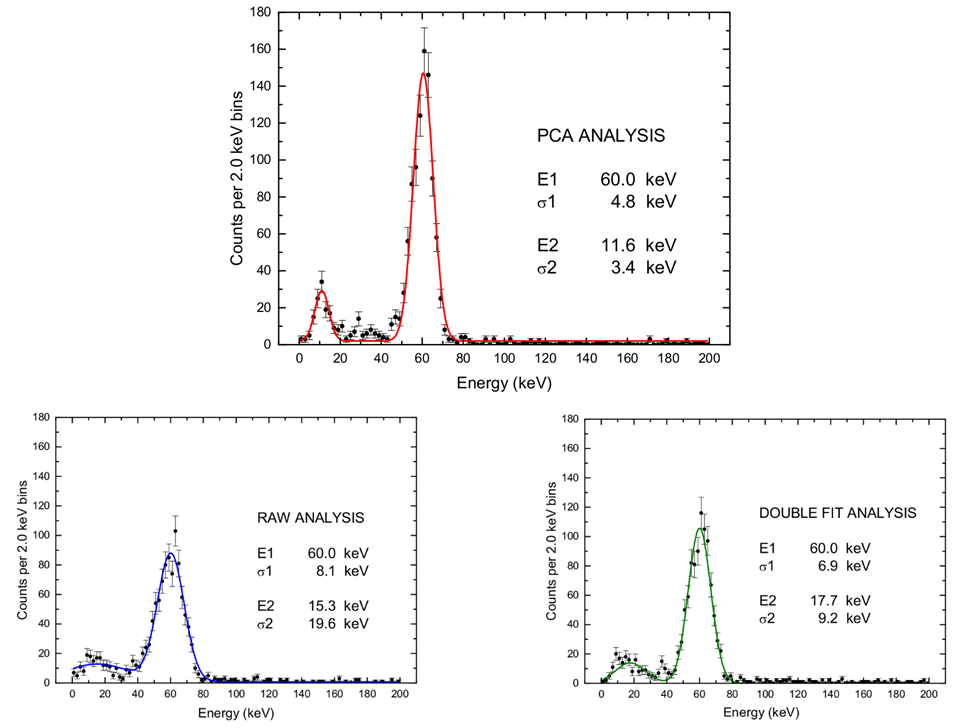}
\caption{\textit{(Top:)} Energy spectrum obtained from the AC-S7 dataset by applying the PCA technique. \textit{(Bottom:)} Energy spectra obtained from the same dataset by means of the raw pulse analysis \textit{(left)} and the double pulse fitting procedure \textit{(right}).}
\label{PCA_spectra}
\end{figure}

The energy spectrum obtained from the PCA is shown in Fig. \ref{PCA_spectra}, in comparison with the spectra from the raw and the double pulse fitting analysis (see par. \ref{acs7ill}).  All the spectra have been calibrated to the 60 keV line and fitted with a double peak Gaussian function.

The PCA technique provides a more narrow 60 keV line, improving the spectral resolution of a factor 1.4 compared to the fit procedure and of a factor 1.7 with respect to the raw analysis. Also the bump at lower energy is better shaped, enforcing the hypothesis that it corresponds to the Compton edge expected at 11.3 keV. Although the CryoAC is not aimed to perform spectroscopy, improving the energy resolution is a remarkable result because it enable us to better determine the low threshold energy of the detector. This parameter is important to reach high particle rejection efficiency (the lower the detector low energy threshold, the higher the particle flux the CryoAC is able to veto), and so it must be well characterized. Finally, we remark that this new analysis method is totally model independent, not requiring a priori knowledge of the pulses fitting functions.

\section{PCA validation with a simulated dataset}

\begin{figure}[H]
\centering
\includegraphics[width=0.45\textwidth]{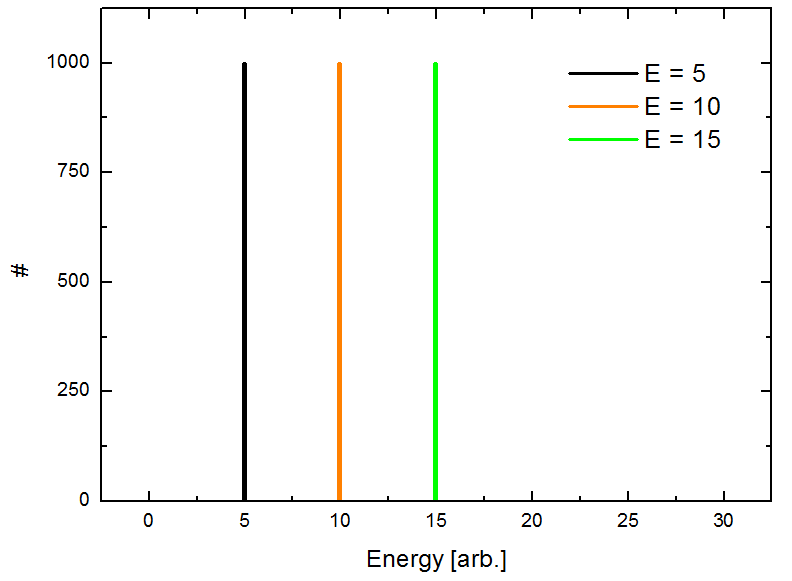}
\includegraphics[width=0.45\textwidth]{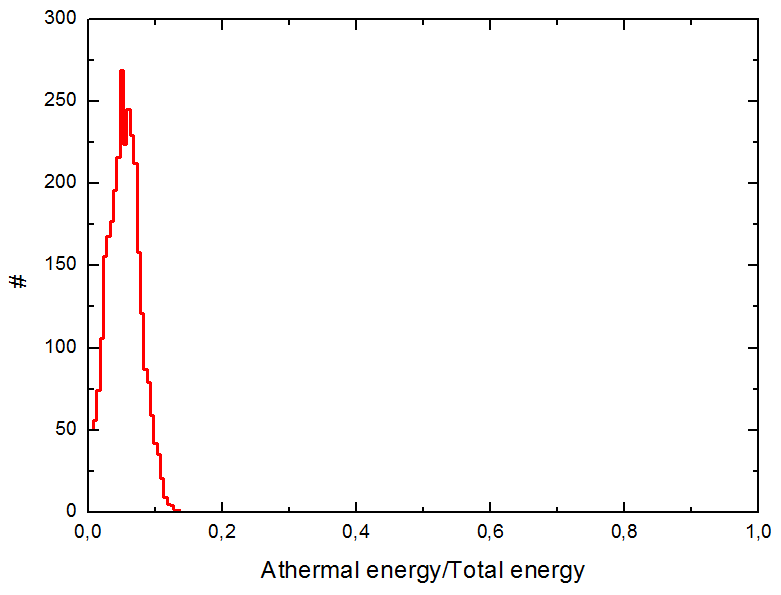}\\
\includegraphics[width=0.45\textwidth]{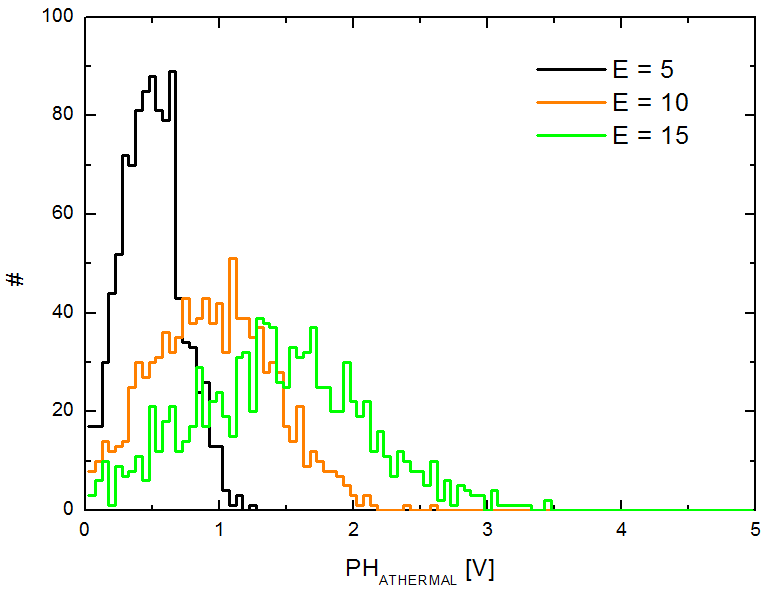}
\includegraphics[width=0.45\textwidth]{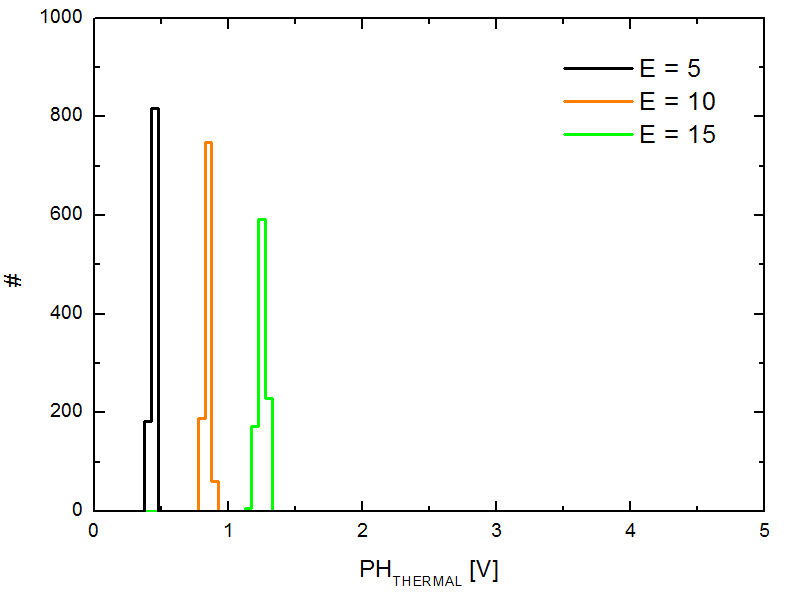}
\caption{\textit{(Top - left:)} Distribution of the simulated pulses as a function of their intrinsic energy, \textit{(Top - right:)} Distribution of the $\varepsilon$ ratio between the athermal and the total energy of the simulated pulses, \textit{(Bottom -left:)} Distribution of the generated athermal Pulse Heights $PH_{ath}$, \textit{(Bottom -right:)} Distribution of the generated thermal Pulse Heights $PH_{th}$}.
\label{PCA_gen}
\end{figure}

\begin{figure}[H]
\centering
\includegraphics[width=0.9\textwidth]{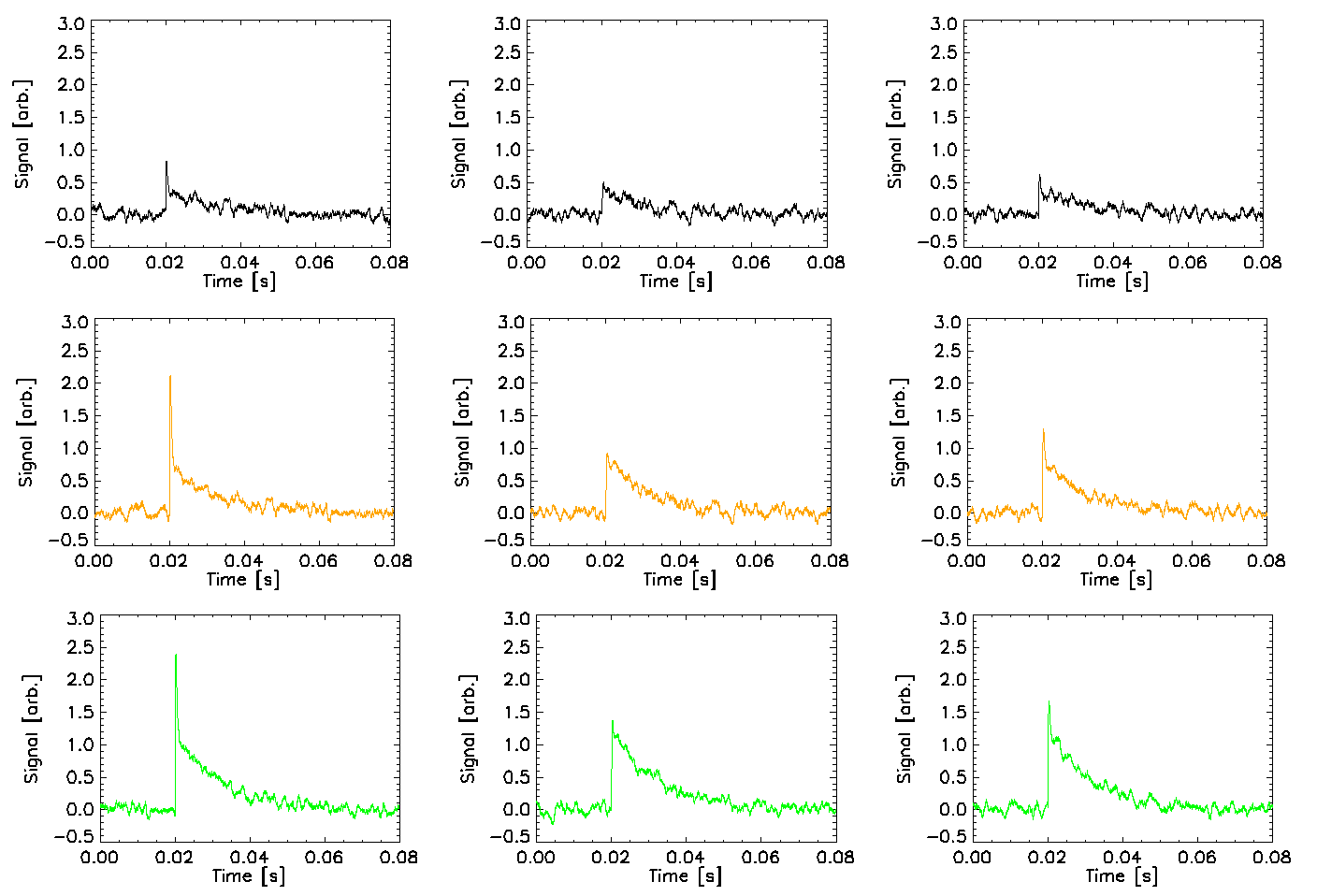}
\caption{Simulated pulses with an energy of 5 a.u. (black, top line), 10 a.u. (orange, middle line) and 15 a.u. (green, bottom line). Note that the amplitude of the fast athermal component is strongly variable, properly replicating the pulse shape variation really observed by AC-S7.}
\label{PCA_gen1}
\end{figure}

To validate the performed analysis, the developed PCA procedures have been applied also to a set of simulated pulses. The pulses have been generated at three different energies (i.e. 5, 10 ad 15 a.u.), starting from the function used to fit the AC-S7 pulses (see par. \ref{acs7ill}):

\begin{equation}
\label{doublepulseoneath2}
P(t) = PH_{ath} \frac{t}{\tau_{ath}} e^{(1-t/\tau_{ath})} + PH_{th}\left(e^{-t/\tau_{D, th}} - e^{-t/\tau_{R, th}}\right)
\end{equation}

\noindent where $\tau_{ath} = \tau_{R, th}$ = 200 $\mu$s and $\tau_{D, th}$ = 12 ms have been fixed. The value of the pulse in-time integral has been also fixed in order to obtain the desired pulse energies, and then $PH_{ath}$ and $PH_{th}$ have been randomly generated assuming the ratio $\varepsilon$ between the athermal and the total (athermal + thermal) energy of the pulse following a gaussian distribution with central value $\varepsilon_{c} = 5\%$ and dispersion $\sigma_\varepsilon = 2.5\%$ (discarding the $\varepsilon$ negative values). Note that all these values have been chosen to replicate the observed AC-S7 pulse shape variation (see par. \ref{acs7ill}). Finally, a gaussian white noise of fixed $\sigma$, corresponding to roughly 0.5 a.u. in the energy scale, has been added to the pulses (in the pulse signal scale $\sigma$ = 0.06 [arb.], while the pulses with energy 5 a.u. have an average Pulse Height of about 0.5 [arb.]). The simulated dataset consists of 3000 pulses, 1000 for each energy. The distributions of their intrinsic energy, $\varepsilon$ parameter and the Pulse Heights are reported in Fig. \ref{PCA_gen}, and some generated pulse is shown in Fig. \ref{PCA_gen1}.
\bigskip

\begin{figure}[H]
\centering
\includegraphics[width=0.8\textwidth]{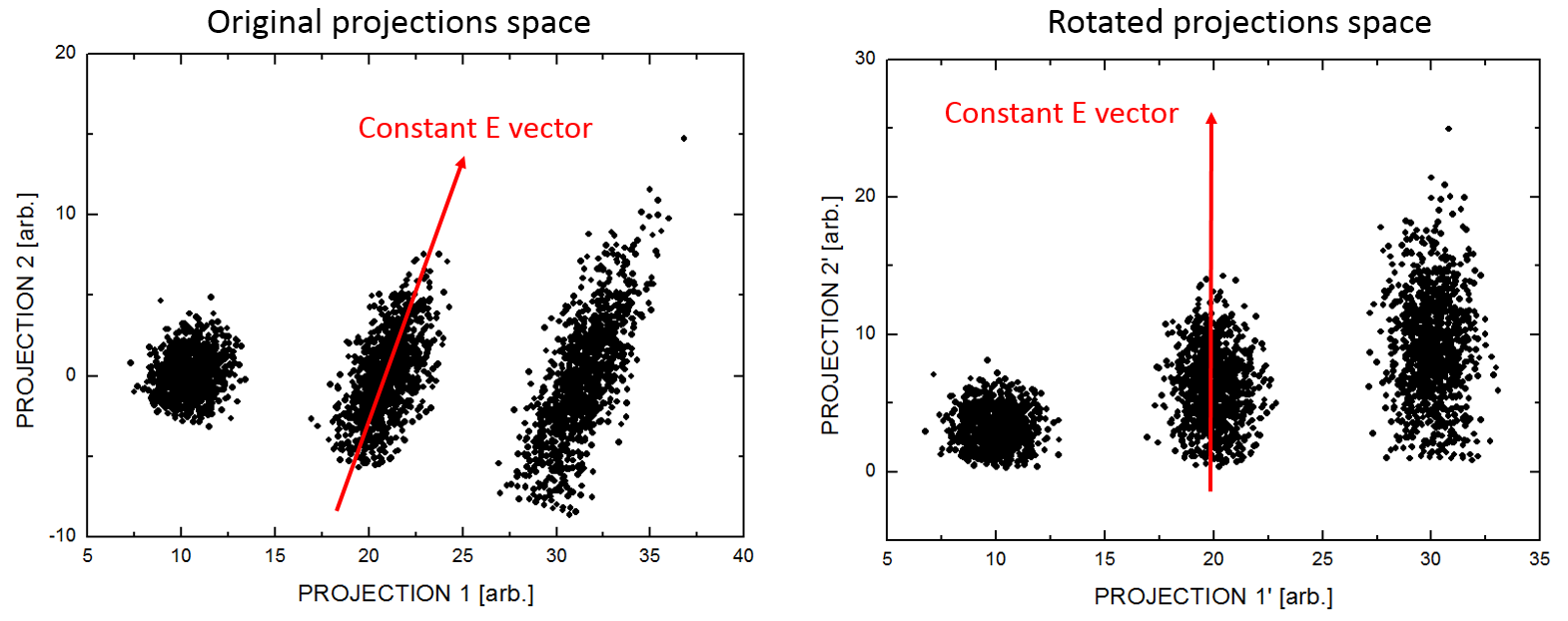}
\caption{\textit{(Left:)} 2D scatter plot of the simulated pulses in the projections space over the first two eigenvectors of the covariance matrix. The red arrow indicates, as labeled, the costant energy direction, obtained by linear fitting the central points cluster. \textit{(Right: )} New space obtained rotating the original one, in order to put the costant E vector vertical.}
\label{PCA_gen2}
\end{figure}

\begin{figure}[H]
\centering
\includegraphics[width=0.9\textwidth]{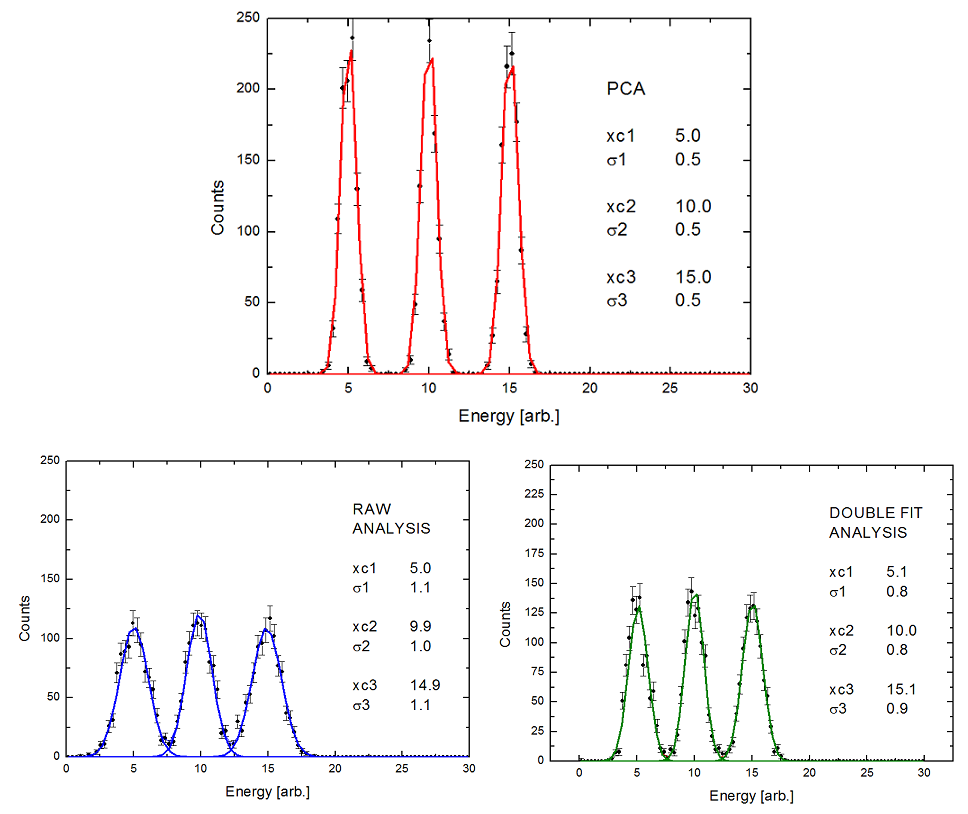}
\caption{\textit{(Top:)} Energy spectrum obtained from the simulated dataset by applying the PCA technique. \textit{(Bottom:)} Energy spectra obtained from the same dataset by means of our standard raw pulse analysis \textit{(left)} and the double pulse fitting procedure \textit{(right}).}
\label{PCA_gen3}
\end{figure}

The dataset has been processed following the same procedures show in the previous section (par. \ref{PCAproclist}). Also in this case we have found that only the first two eigenvectors of the pulses covariance matrix encoded the pulse shape, and thus we have used them to define the projections space from which  to extract the pulses energy spectrum. The pulses projection onto the two selected eigenvectors are shown in Fig. \ref{PCA_gen2} - Left, where three clusters corresponding to the different pulses energies are clearly visible. We have then used the central cluster to identify the direction of constant energy in the scatter plot, and we have finally defined the rotated projections space where this direction correspond to the y-axis (Fig. \ref{PCA_gen2} - Right). We have checked again that in the final projections space the area of the pulses were totally given by the new vector X', thus we have used the projections of the pulses onto this vector to obtain the energy spectrum in Fig. \ref{PCA_gen3} - Top, which has been calibrated in the energy scale fixing the value of the central line at 10 [a.u.]. 

The PCA is able to properly identify the three expected energy lines with an energy resolution of 0.5 a.u., roughly corresponding to the simulated white noise level in the energy scale. This is an evidence of the effectiveness of the PCA method, and a positive validation of our implementation procedures.

Finally, we have processed the same simulated dataset also by our standard \virg pulse raw analysis'' and \virg double pulse fitting procedure'', in order to compare the PCA with these methods of pulse processing. The spectra obtained in these ways are shown in Fig. \ref{PCA_gen3} - Bottom. Both the method are able to properly identify the three expected energy lines, but the achieved energy resolutions are worse than the one obtained by PCA (factor $\sim$ 1.6 of degradation for the double fitting procedure and $\sim$ 2.0 for the raw analysis). 
 
In conclusion, the results reported in this chapter have shown that the PCA can be effectively used to process the CryoAC pulses, and that this analysis is able to provide energy spectra with an improved energy resolution with respect to our others pulse processing methods. This is a good starting point and future activities could be focused to generalize the developed procedure in order to process an higher number of principal components, and to implement advanced PCA algorithms for real-time pulse processing.

\chapter{Improvement of the test setup towards the Demonstration Model}
\chaptermark{Improvement of the test setup}
\label{testsetupch}

The development plan of the CryoAC Demonstration Model foresees to characterize and test the detector in the IAPS HE CryoLab before its delivery to the FPA development team for the integration with the TES array DM. In this context, I report in this chapter the main results of two activities carried out to improve our cryogenic test setup in preparation to the DM integration. 

The first activity has mainly consisted in the development of a cryogenic magnetic shielding system for the ADR cryostat. The performances of TES microcalorimeters are indeed strongly influenced by the external magnetic field, and in order to properly characterize and operate these detector it is important to reach in the sample zone a residual static magnetic field lower than $\mu$T, as it will be done in the X-IFU FPA DM.

The second activity has instead concerned the signal filtering issue. In the context of the X-IFU DM, it was indeed initially foreseen to develop a cold filtering stage at 2K, in order to filter every signal entering the FPA (including those of the CryoAC DM) and thus prevent EMI propagation through wirings towards the TES array. We have therefore first performed a study to understand how this filtering would have affected the CryoAC DM SQUID dynamic, and then developed a mechanical system to implement this filtering stage also in our cryogenic test setup. At system level, it has been finally decided not to implement the cold filtering stage anymore in the X-IFU DM (all the signals will be anyway filtered at warm, at the entrance of the cryostat Faraday cage). However, this activity has been important to better understand SQUID FLL dynamic, and it could come in handy in the future.

\newpage
\section{Magnetic shielding and noise analysis}

In this section I will report the integration in our ADR cryostat of a new magnetic shielding system, assessing its effectiveness by means of FEM simulation and a measurement at warm. Then, I will show the preliminary performance test of this shield carried out with the AC-S8 CryoAC pre-DM prototype, focusing on the effect of the Pulse Tube operations on the detector noise spectra. 

\subsection{The new cryogenic magnetic shielding system}

\begin{figure}[H]
\centering
\includegraphics[width=0.85\linewidth]{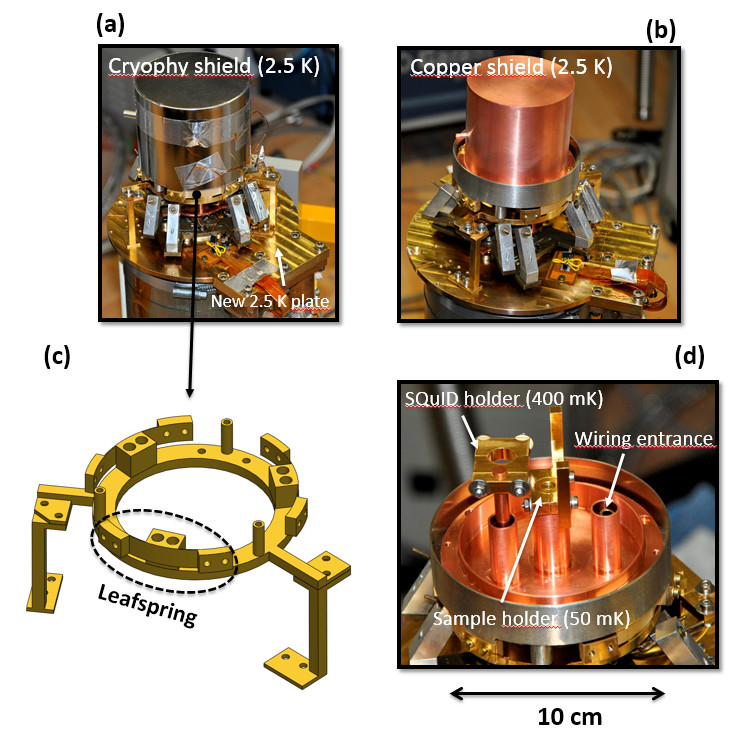}
\caption{New cryostat setup with the cryogenic magnetic shielding system mounted on the 2.5 K plate. (a) The external Cryophy shield. (b) The inner copper shield. (c) CAD model of one mechanical structure supporting the shields. (d) The shielded sample area.}
\label{fig1}
\end{figure}

It is generally known that the performances of TES microcalorimeters are strongly influenced by external magnetic field down to the $\mu$T level \cite{ishisaki}. An appropriate magnetic shielding is therefore required to properly operate these detectors. In this respect, we have upgraded our ADR cryostat (see Sect. \ref{ADRsection}) inserting a cryogenic shielding system at the 2.5 K stage. The ADR superconducting magnet (shielded by the manufacturer) generates a maximum of $10^{-5}$ T in the experimental volume \cite{vericold}, so the dominant field to be shielded is the geomagnetic one ($\sim$ $ 0.5 \cdot 10^{-4}$ T). We have developed a shielding system constituted by three main parts (Fig.~\ref{fig1}).

The first part is a ferromagnetic cylindrical shield made of Cryophy (Fig.~\ref{fig1}a), a nickel-iron soft alloy with high magnetic permeability ($\mu_r \sim 10^5$ at cold), suitable for static and low-frequencies magnetic shielding at cryogenic temperature \cite{cryophy}. The shielding factor S of this shield (i.e. the ratio of the internal magnetic field to the external one) can be roughly estimated as \cite{hoburg}:
\begin{equation}
\label{Sfactoreq}
S \sim \left( \frac{D}{\mu_r \cdot d} \right)^{-1} \sim 500
\end{equation}
where D = 100 mm and d = 1 mm are the diameter and the thickness of the shield, respectively, and $\mu_r = 0.5\cdot 10^5$ @ 4K is the initial permeability measured by the manufacturer on a Cryophy sample ring. We therefore expect a residual static magnetic field inside the shield of about $10^{-7}$ T.

The second element in the system is an OFHC copper shield (Fig.~\ref{fig1}b) that, if necessary, will be lead plated to operate as a superconducting shield. It is placed inside the Cryophy in order to find an appropriate low field environment during the superconductive phase transition, avoiding flux trapping in the shield openings.

Third, we have developed a gold-plated copper structure (Fig.~\ref{fig1}c) to mimimize the mechanical stress on the Cryophy shield during the cooling phase, thus balancing the thermal contractions. This is important for an efficient shielding, as the magnetic permeability of the material rapidly decreases with the stress level \cite{henk}. The core of this structure are three leafspring-like elements with flexible arms (300 $\mu$m thick). 

The shields present three openings to provide the access into the experimental area (Fig.~\ref{fig1}d). Two of these openings are used to host two rods connected to the first ($\sim$ 500 mK) and the second ($\sim$ 50 mK) ADR stages, where the SQUID holder and the sample holder structures have been anchored. The third aperture is instead designed for the wiring entrance (i.e. for TES bias, SQUID read-out and thermometry).

Finally, we have also upgraded the 2.5 K plate of the cryostat (2nd PTR stage), moving from Aluminum to gold-plated Copper, in order to minimize thermal gradients within the plate and do not affect the Pulse Tube performances.

\subsection{Magnetic field modeling and measurements}

To model the magnetic field attenuation inside the new shielding system, we have performed a 3D Finite Element Method (FEM) simulation using the COMSOL Multiphysics \textregistered \cite{comsol} software package. The simulation consists in a stationary study including the \textit{Magnetic Fields, No Currents (mfnc)} physics interface of the \textit{AC/DC module} \cite{acdc}.

The system geometry has been implemented importing in the software the CAD drawings of the shields, and the mesh has been refined until no significant variation in the simulation results was observed. The shields have been placed in a spatially uniform background magnetic field $B_{bkg}$ of strenght $10^{-4}$ T, alternately oriented parallel or perpendicular to their main axis. The cryophy relative permeability has been defined as a constant value: $\mu_r = 51000$, equal to the initial permeability measured at 4 K by the manufacter on a Cryophy sample ring subjected to the same annealing process of the shield. Being the annealing performed at zero magnetic field, the magnetization properties of the material are described by the initial magnetization curve \cite{yashchuk}, and so this value is the most representative for the relative permeabilty of the shield in a low magnetic field. 

\begin{figure}[H]
\centering
\includegraphics[width=0.99\linewidth]{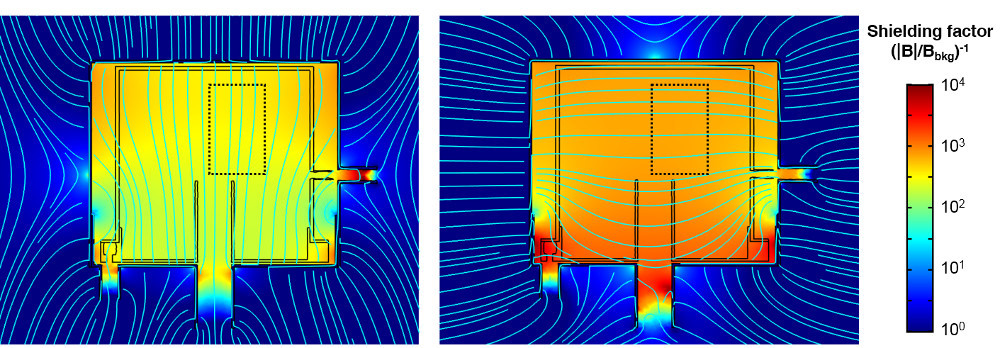}
\caption{Attenuation of the magnetic field inside the new shielding system evaluated by a FEM simulation. The color scale indicates the shielding factor on a cut plane crossing the main shields axis. The magnetic flux density lines are overplotted in light blue, and the detector zone is highlighed by the black dotted line. \textit{Left:} Background magnetic field in the shield axial direction. \textit{Right:} Background magnetic field in the shield transverse direction.}
\label{fig1b}
\end{figure}

The main results of the simulation are shown in Fig.~\ref{fig1b}, where the color scale indicates the shielding factor ($ S = (|B|/B_{bkg})^{-1} $) evaluated on a cut plane crossing the main shields axis. The magnetic flux density lines are overplotted in light blue, showing the field distortion induced by the Cryophy layer (note that for visual purpose the line density is not proportional to the field strenght).

In Tab.~\ref{tab:shieldsf} are reported the minimum and the maximum value of $S$ evaluated inside the shields and in the detector zone (black dotted region in Fig.~\ref{fig1b}), for both the axial and transverse directions of the background field. Note that the results of the simulation agree within a factor $\sim$ 3 with the rough estimation in eq.~(\ref{Sfactoreq}) (S $\sim$ 500).

\begin{table}[H]
\centering
\caption{Shielding factors evaluated by FEM simulation}
\label{tab:shieldsf}
\begin{tabular}{cccccc}
\hline\noalign{\smallskip}
Background field  & \multicolumn{2}{c}{Inside the shields} & \multicolumn{2}{c}{In the detector zone} \\
direction & $S_{MIN}$ & $S_{MAX}$ & $S_{MIN}$ & $S_{MAX}$ \\
\noalign{\smallskip}\hline\noalign{\smallskip}
Axial & 174 & 501 & 251 & 398 \\
Transverse & 501 & 1573 & 631 & 794 \\ 
\noalign{\smallskip}\hline
\end{tabular}
\end{table}

Finally, to preliminary check the simulation results we have performed a measurement of the axial shielding factor in the detector zone by a flux gate magnetometer: a HP Clip on DC Milliammeter 428B equipped with a magnetometer probe. The measurement has been performed at room temperature, fixing the position of the probe and measuring the axial ambient magnetic field with and without the shields around it. The procedure has been then repeated orienting the probe in the opposite direction, in order to correct the systematics due to the instrument zero-setting calibration. We remark that the Cryophy initial permeability does not show any significant degradation from room to liquid helium temperature \cite{masuzawa}, so we expect this measurement to be representative of the shielding effectiveness at the working temperature. We obtained:

\begin{equation}
S_{\textit{\scriptsize axial, measured at detector zone}} = 315 \pm 15
\end{equation}

\smallskip

\noindent The measured value is well consistent with the simulation results reported in Tab.~\ref{tab:shieldsf}.

Given the earth magnetic field ($\sim$ $ 0.5 \cdot 10^{-4}$ T), we therefore expect in the detector zone a residual static magnetic field $< 2 \cdot 10^{-7}$ T.

\subsection{ADR and sample holder thermal performance, and preliminary test with AC-S8}

\begin{figure}[H]
\centering
\includegraphics[width=0.75\linewidth]{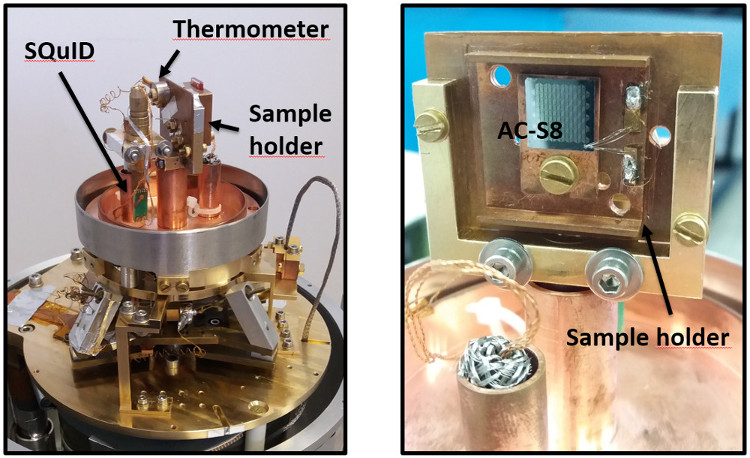}
\caption{\textit{Left}: Setup for the preliminary performance test of the new shielding system. \textit{Right:} Detail of the AC-S8 prototype mounted on the sample holder (before being covered by a copper sheet).}
\label{fig2}
\end{figure}

We have performed a preliminary test of the new cryogenic setup with the CryoAC pre-DM prototype AC-S8 (for details about AC-S8 see Chapt. \ref{preDMchapter}). The detector has been coupled to a commercial SQUID array (Magnicon C6X216FB) and operated with a FLL electronics (Magnicon XXF-1). The test setup is shown in Fig.~\ref{fig2}. 

\begin{figure}[H]
\centering
\includegraphics[width=0.65\linewidth]{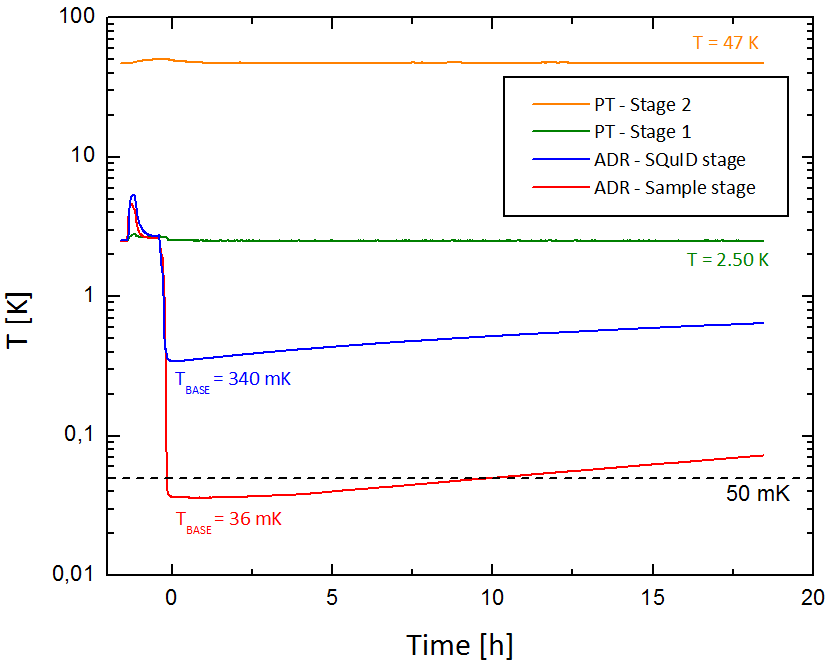}
\caption{Temperature profiles of the cryostat stages measured during an ADR cycle.}
\label{cooling}
\end{figure}

First, we have performed a cooling test to verify the minimum temperature and the hold time achievable with the new setup. The result is shown in Fig.~\ref{cooling}, where are reported the temperature profiles of the cryostat stages acquired during an ADR cycle. 

\noindent Note that the sample stage (red curve in the plot) is able to reach a base temperature of 36mK, with an hold time at T $<$ 50 mK of about 10 hours. The absolute temperature accuracy is due to thermometer calibration uncertainty, which is evaluated to be $\sim$ 2 mK @ T $<$ 50 mK.  The cryostat and sample holder cooling performances are therefore fully suitable for our planned test activities, which foresee to operate the CryoAC DM with the thermal bath at 50 mK.

\begin{figure}[H]
\centering
\includegraphics[width=0.65\linewidth]{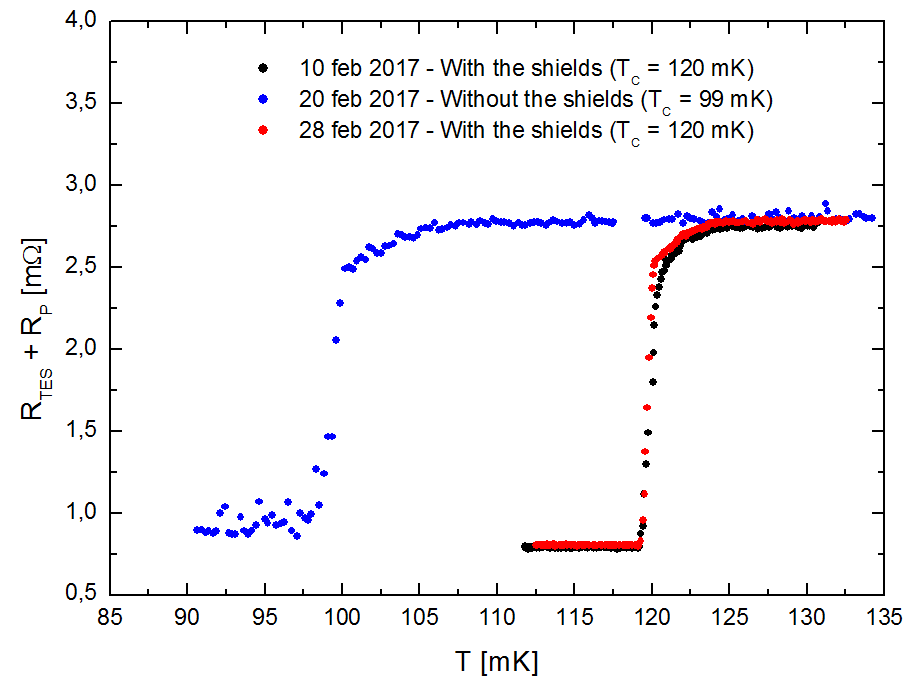}
\caption{Comparison between the AC-S8 transition curves measured with and without the shields. The critical temperature are evaluated at the half of the transition. (Color figure online)}
\label{rts}
\end{figure}

\bigskip
We have then tested how the presence of the new shields could affect the TES superconductive transition. We have performed three consecutive measurements of the R vs T curve: with the shields, without the shields and again with the shields (to replicate the initial conditions). The results are reported in Fig.~\ref{rts}, which shows that with the shields the critical temperature of the sample results about 20 mK higher (from 99 mK to 120 mK). We interpret this as an evidence of the reduction of the static magnetic field perpendicular to the detector surface (which strongly influence the superconductive transition \cite{ishisaki}). We remark that the three measurements have been performed with the same method (SQUID readout, modulation technique with a low amplitude sine wave at 22 Hz) and that the magnetic shielding does not influence the thermal radiation on the AC-S8 sample (which is totally covered by the holder).

\begin{figure}[H]
\centering
\includegraphics[width=0.7\linewidth]{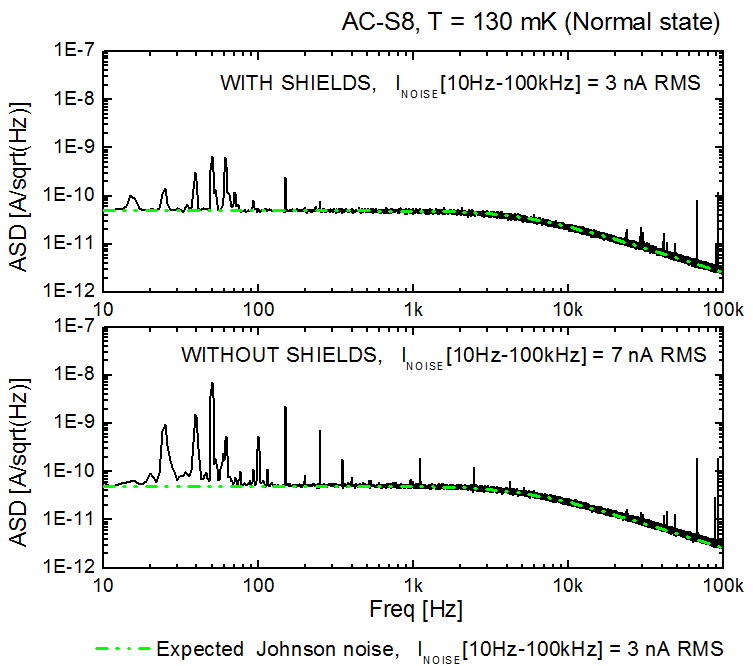}
\caption{Current noise Amplitude Spectral Density measured for AC-S8 in the normal state and the PTR turned OFF, with the shields (Top) and without them (Bottom). The dashed green line represents the expected Johnson noise.}
\label{fig4}
\end{figure}

Lastly, we have performed a noise measurement in order to roughly quantify the effectiveness of the new shielding system also for non-static magnetic fields. In Fig.~\ref{fig4} are reported the noise spectra measured with the detector in the normal state (T = 130 mK), with and without the shields. Both the spectra have been acquired switching off the PTR compressor, in order to minimize the effect of the induced vibrations (in the next section we will focus on this point). The expected Johnson noise, filtered by the L/R of the TES circuit, is overplotted for comparison. 

\noindent The plots show that the introduction of the new shielding system lead to a reduction of the spectral noise lines, especially in the low-frequency region (as expected for Cryophy, which acts up to some hundred Hz). To quantify this improvement of the setup, we have integrated the noise spectra to obtain the equivalent current noise I$_{noise}$ in the 10 Hz - 100 KHz band. The results are reported in the plot, showing that the shields are able to reduce I$_{noise}$ by a factor $\sim$ 2 (from 7 nA to 3 nA RMS), reaching the theoretically expected Johnson value.

\subsection{The effect of the Pulse Tube operation}

\begin{figure}[H]
\centering
\includegraphics[width=0.65\linewidth]{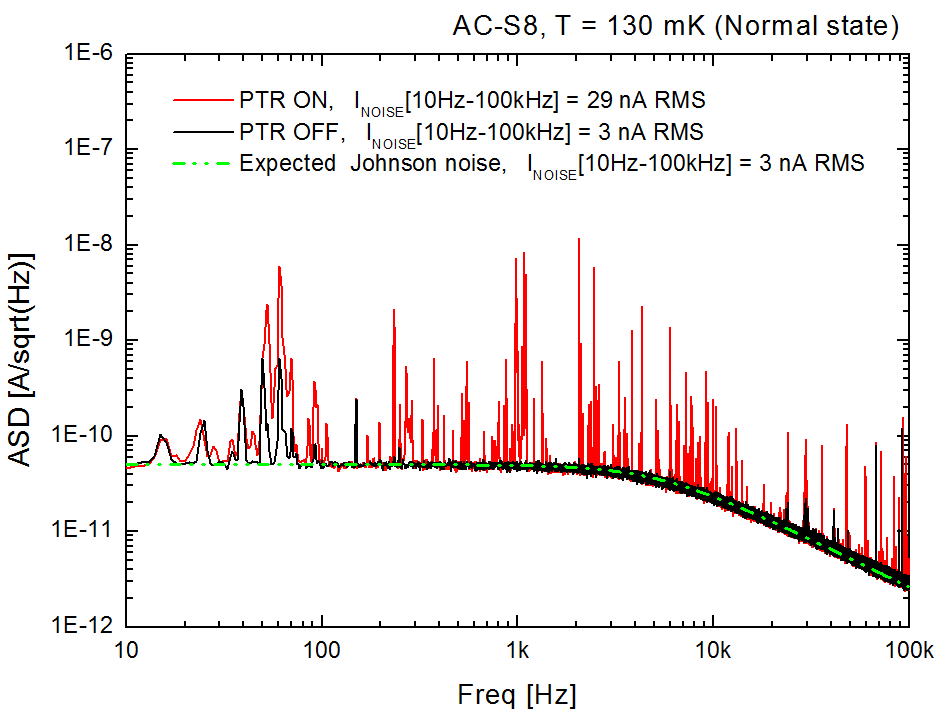}
\caption{Current noise Amplitude Spectral Density measured for AC-S8 in the normal state and both the shields mounted, with the PTR compressor switched ON (red curve) and turned OFF (black curve).}
\label{fig5}
\end{figure}

Despite the good results obtained with the introduction of the new magnetic shielding system, we have found that the PTR operation has a dominant impact on the detector noise spectrum. In Fig.~\ref{fig5} are reported the spectra acquired with AC-S8 in the normal state (T = 130 mK) and the PTR compressor switched on (red line) and then turned off (black line). In both cases the shields were mounted. The noise spectrum degradation is evident, with an increment of the equivalent current noise of one order of magnitude (from 3 nA to 29 nA RMS). This is probably due to an electromagnetic component and to the mechanical vibrations induced by the PTR on the cold stages of the cryostat, causing microphonic disturbance in the TES system. This kind of effects are reported in literature, with an influence on the noise spectrum up to tenths of kHz and characteristics lines around 1 KHz \cite{hollerith}, as in our case. 

We remark that we are, however,  able to switch off the PTR and still operate the detector for a while, typically about 15 min with the bath temperature at 50 mK. Therefore, this problem does not prevent us to perform the planned DM characterization activities.

\subsection{Conclusions}

I have here presented the main improvements of the ADR cryogenic setup that we have achieved in preparation to the CryoAC DM integration activity. 

We have developed a cryogenic magnetic shielding system, and assessed by means of FEM simulation and a test at warm that it is able to provide a shielding factor $S > 250$ in the detector region. Given the earth magnetic field ($\sim$ $ 0.5 \cdot 10^{-4}$ T), we therefore expect in the detector zone a residual static magnetic field $< 2 \cdot 10^{-7}$ T. After the system integration, the cryostat and sample holder thermal performances are fully suitable for the planned DM activity (hold time at 50 mK higher than 10 hours). 

We have performed a preliminary performance test of the magnetic shielding by means of the CryoAC AC-S8 prototype. We have found that inside the new shields the sample shows a critical temperature $\sim$ 20 mK higher than the one measured without the shields (T$_C$ from 99 mK to 120 mK). This is an evidence of the static magnetic field reduction over the detector surface. As expected, the new shielding system has also led to a reduction of the spectral noise lines at low frequency (up to some hundred Hz), improving by a factor $\sim$ 2 the measured equivalent current noise. 

Finally, we have shown the dominant effect of the PTR operations on the noise spectra, concluding that we are able to properly operate the detector only switching off the PTR (T$_{BATH} = 50$ mK typically maintained for $\sim$15 min).

\bigskip
In conclusion, I report that after the installation of the Dilution Refrigerator (see Sect. \ref{DRsection}), we have moved the magnetic shielding system into the new cryostat, developing ad-hoc mechanical structures to mount the system on the mixing chamber plate (so moving the shields from 4K to 50 mK). In the context of the CryoAC DM integration, this will allow us to have more available cooling power (450 $\mu$W @ 100 mK) and could also operate for more time with the PTR switched off (several hours with the cold stage at 50 mK).

\begin{figure}[H]
\centering
\includegraphics[width=0.5\linewidth]{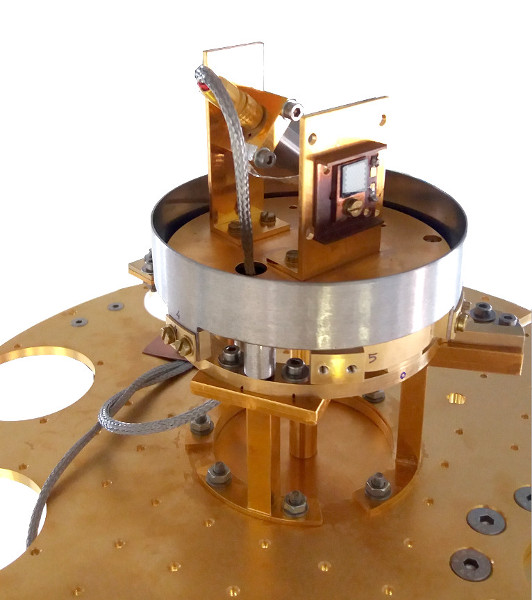}
\caption{The cryophy shield and its mechanical support structure mounted on the Mixing Chamber of the new Dilution Refrigerator. Inside the shield it is possible to see the AC-S8 prototype integrated in a new sample holder.}
\label{shielddilution}
\end{figure}

\newpage
\section{Signal filtering at cold}

Since the X-IFU DM initially foresaw a cold filtering stage at 2K, which regarded also the CryoAC DM signal lines, we have performed a study to analyze how the introduction of electronic filters influence the dynamic of a SQUID operating in Flux Locked Loop mode, in order to evaluate the maximum applicable filtering that do not compromise the stability of our readout system. In this section, I will first report this study, and then I will show a mechanical system that we have developed to integrate a cold filtering stage also in our cryogenic test setup.

\subsection{Filter influence on the FLL SQUID dynamic}

The aim of this study has been to analyse the stability of a filtered SQUID FLL readout configuration. Note that the effect of the filters on the reduction of EMI will not be analyzed here, but it will require a different study. 

The stabilty of a negative feedback system can be evaluated studying its open-loop phase margin (PM), which is the difference between the phase of the output signal at zero dB gain (i.e. at the unit gain frequency) and 180 degrees. A zero or negative PM guarantees instability, while a positive PM represents a \virg safety margin'' to properly operate the circuit. In the following paragraphs I will first derive the filtered system transfer function, finally evaluating the phase margin from the system Bode plots.

\subsubsection{The typical SQUID FLL configuration}

\begin{figure}[H]
\centering
\includegraphics[width=0.6\textwidth]{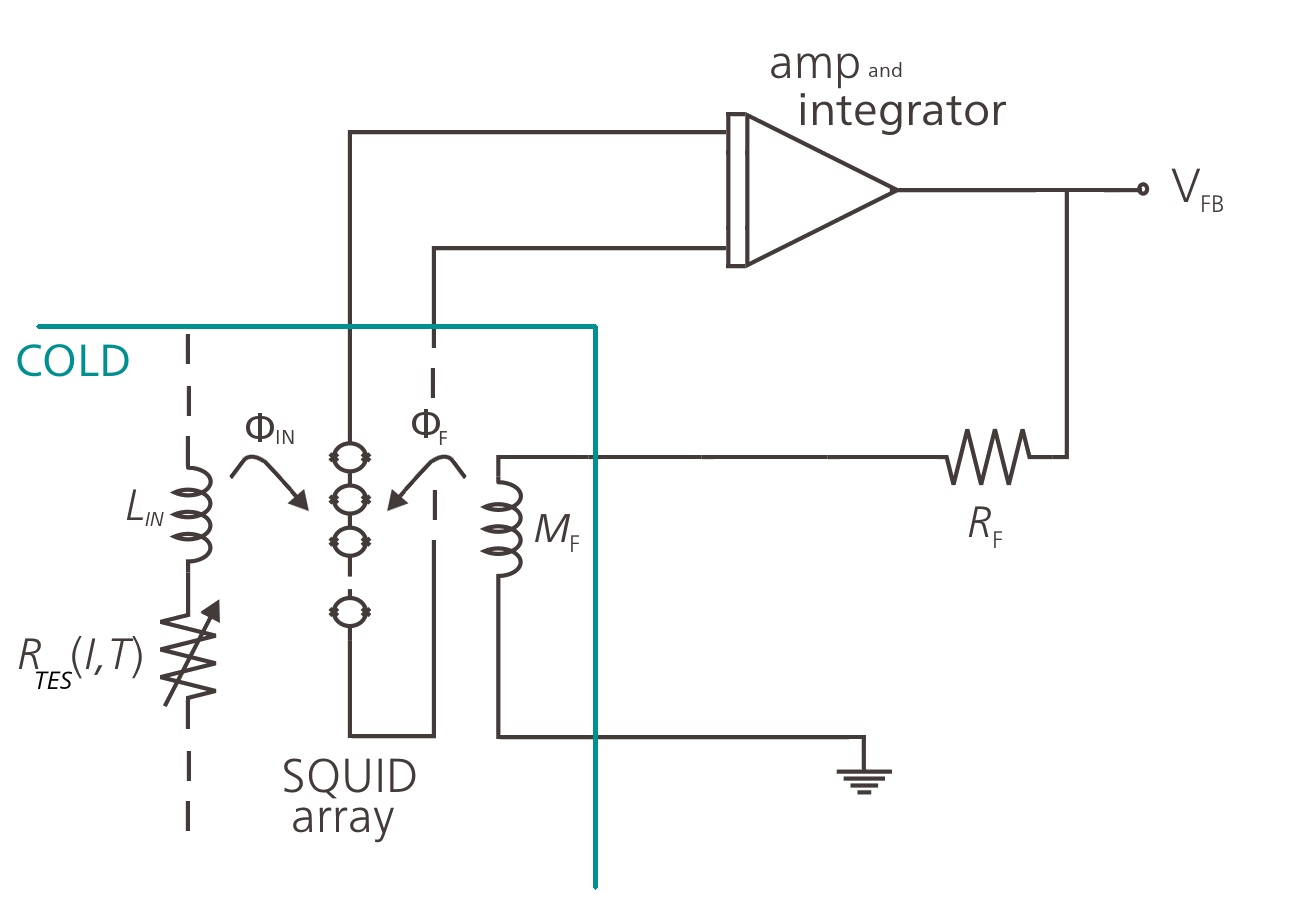}
\caption{Scheme of a typical SQUID FLL configuration for TES readout}
\label{filter1}
\end{figure}

The scheme of a typical SQUID FLL configuration for TES readout is reported in Fig.~ \ref{filter1}.

\begin{figure}[H]
\centering
\includegraphics[width=0.6\textwidth]{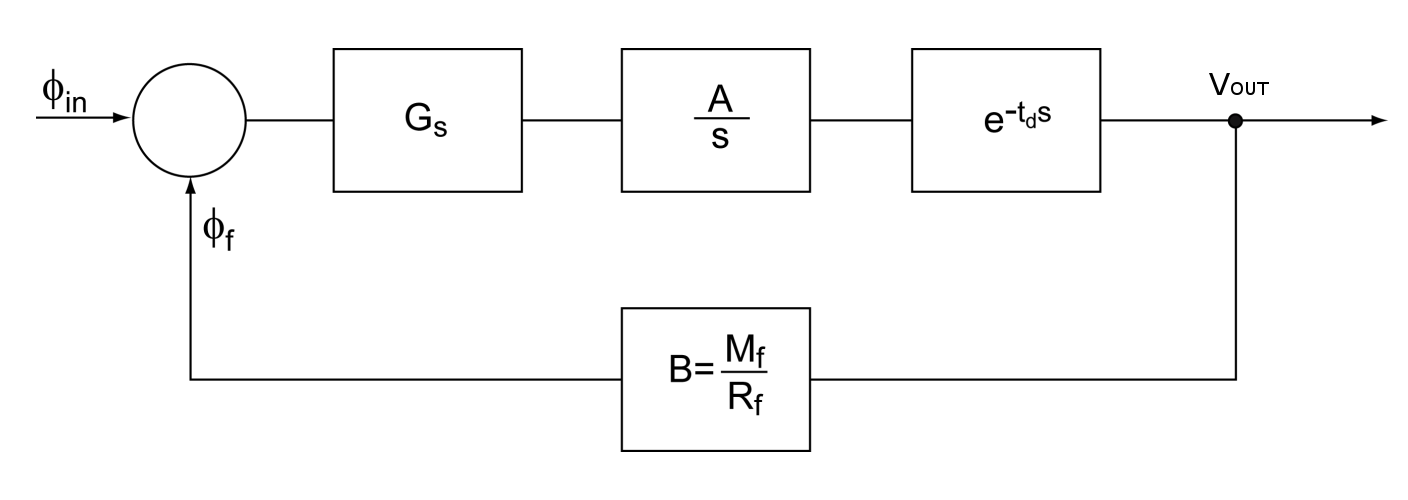}
\caption{Block diagram of a typical SQUID FLL configuration for TES readout}
\label{filter2}
\end{figure}

\noindent L$_{in}$ is the SQUID input coil, R$_f$ the feedback resistance and M$_f$ the mutual inductance between the feedback coil and the SQUID. The corresponding box diagram is shown in Fig. \ref{filter2} 

\noindent G$_S$ represents the SQUID gain (dV/d$\Phi$), A/s is the amplification and integration block, B=M$_f$/R$_f$ is the feedback block and $e^{-t_D\cdot s}$ is a generic delay block (since the stability study is performed at open loop, it is possible to consider only one delay block
without loss of generality).

\bigskip
The reference values of the block for the CryoAC readout has been defined in \cite{torriolitn}, and are here reported in Tab. \ref{tab:blocksvalue}.

\begin{table}[H]
\centering
\caption{Reference values for the CryoAC readout.}
\label{tab:blocksvalue}
\begin{tabular}{ll}
\hline\noalign{\smallskip}
Block & Value \\
\noalign{\smallskip}\hline\noalign{\smallskip}
$G_s$ & 1.2 mV/$\Phi_0$ \\
$A$ & 1.31 $\cdot$ 10$^9$ s$^{-1}$ \\
$t_d$ & 20 ns \\
$M_F^{-1}$ & 100 $\mu$A/$\Phi_0$ \\
$R_F$ & 5 k$\Omega$\\
$B$ & 2$\Phi_0$/V \\
\noalign{\smallskip}\hline
\end{tabular}
\end{table}

\noindent The open loop and the closed loop transfer functions ($H_{OL}$ and $H_{CL}$) of the system are \cite{torriolitn}:

\begin{equation}
H_{OL} = \frac{\Phi_F}{\Phi_{IN}}= G_S \cdot B \cdot  A/s \cdot e^{-t_d s}
\end{equation}

\begin{equation}
H_{CL} = \frac{V_{OUT}}{(\Phi_{IN} + \Phi_{F})} = \frac{1}{B} \cdot \frac{1}{1 + s \tau_1 e^{t_d s}}
\end{equation}

\noindent where $\tau_1$ = $1/(G_S \cdot A \cdot B)$. Note that the related cutoff frequency is $f_1$ = $1/(2 \pi \cdot \tau_1)$,  i.e. $f_1$ = 500 kHz assuming the values in Tab. \ref{tab:blocksvalue}.

\subsubsection{The filtering system}

\begin{figure}[H]
\centering
\includegraphics[width=0.7\textwidth]{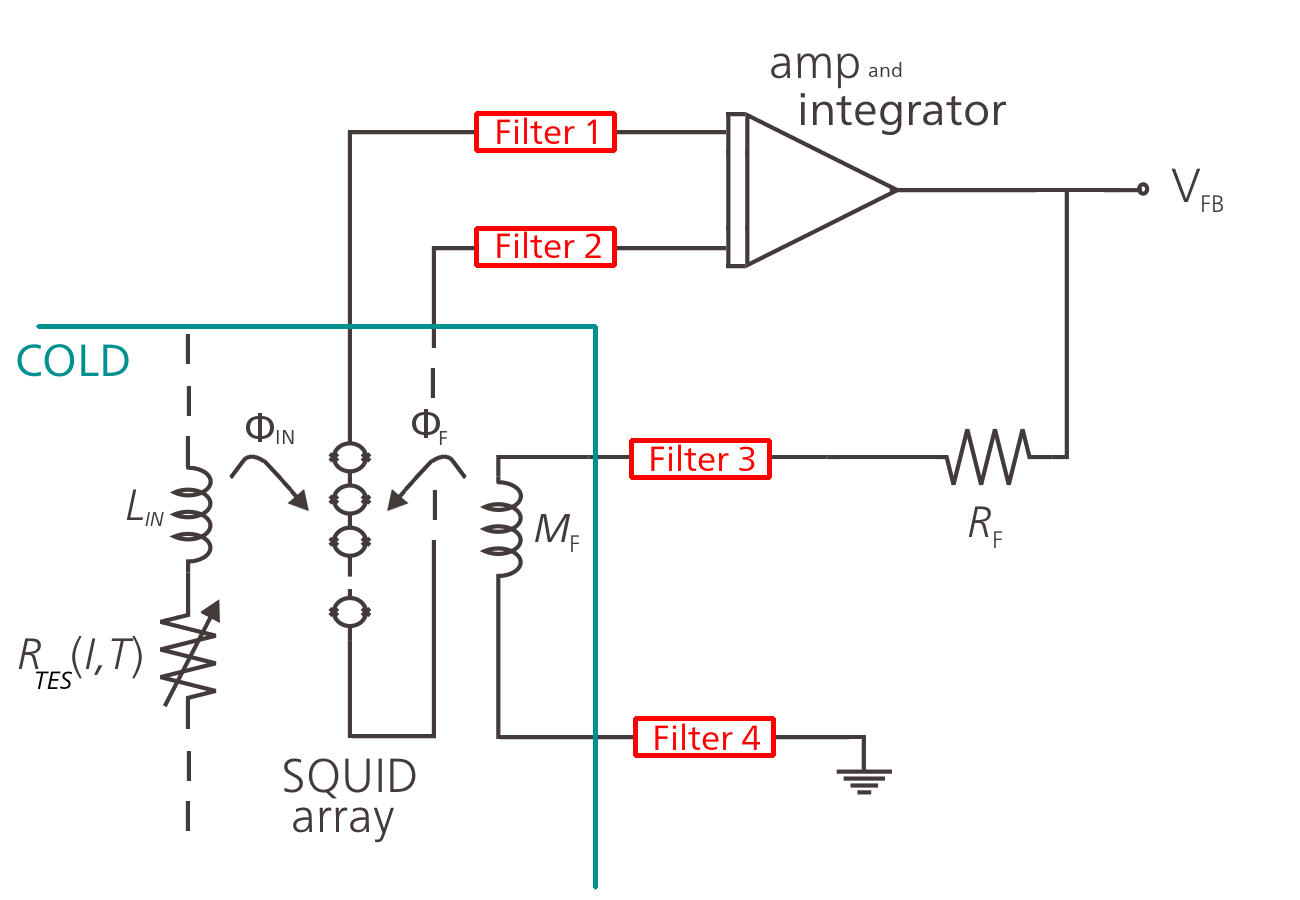}
\caption{Scheme of the SQUID FLL configuration integrated with 4 LP filters.}
\label{filter3}
\end{figure}

Since every signal line connecting the focal plane (cold) and the warm electronics (at room temperature) should be filtered, it is necessary to implement in the SQUID FLL configuration a filtering system composed by four different low-pass filters, as shown in Fig. \ref{filter3}.

\begin{figure}[H]
\centering
\includegraphics[width=0.7\textwidth]{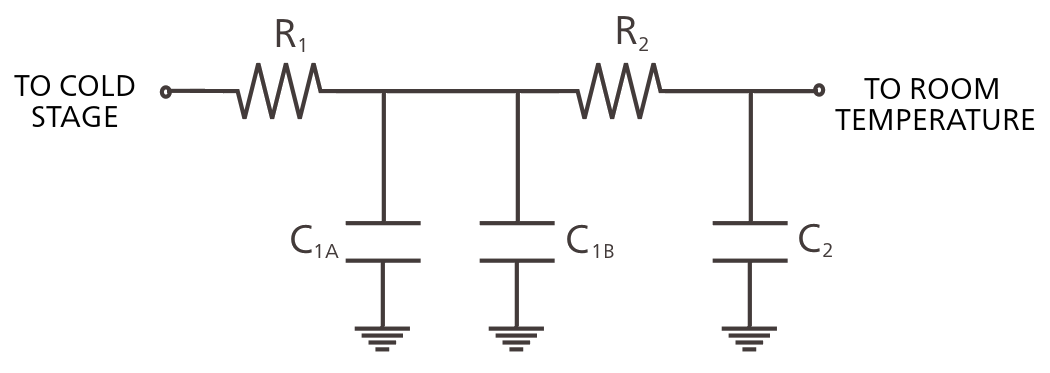}
\caption{Scheme of the LP filters used by the SRON FPA development team.}
\label{filter4}
\end{figure}

In this study I will refer to the LP filter scheme shown in Fig. \ref{filter4}. This scheme has been already used and tested at SRON by the FPA development team, which in the past has implemented analogous filters operating at cold (4K stage \cite{eureca}).

\noindent The reference values for the filter components are reported in Tab. \ref{tab:filvalue}

\begin{table}[H]
\centering
\caption{Reference values for the CryoAC}
\label{tab:filvalue}
\begin{tabular}{ll}
\hline\noalign{\smallskip}
Component & Value \\
\noalign{\smallskip}\hline\noalign{\smallskip}
$R_1$ & 1.5 $\Omega$\\
$R_2$ & 1.5 $\Omega$\\
$C_{1,A}$ & 47 pF\\
$C_{1,B}$ & 47 pF\\
$C_{2}$ & 1 nF\\
\noalign{\smallskip}\hline
\end{tabular}
\end{table}

\subsubsection{The new transfer function}

In order to evaluate the influence of the filters on the SQUID dynamics, it is necessary to study the new open loop transfer function of the system. The study has been performed treating separately the amplification and the feedback branchs of the circuit, which are decoupled each other.

\begin{figure}[H]
\centering
\includegraphics[width=0.7\textwidth]{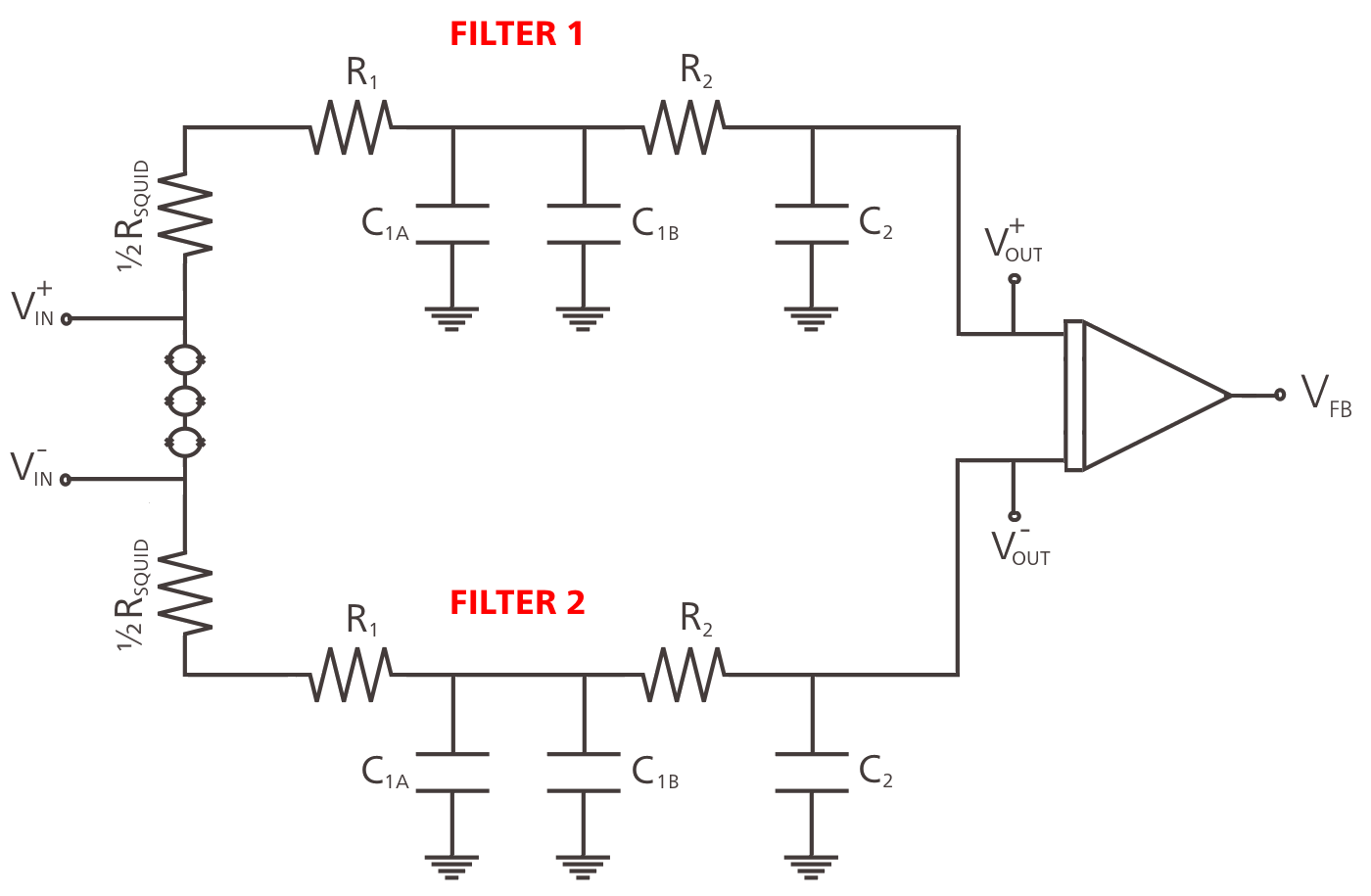}
\caption{Scheme of the filtering system operating on the amplification branch of the circuit.}
\label{filter6}
\end{figure}

\bigskip
The amplification branch of the circuit includes two filters placed between the SQUID array and the Amplifier (Filters 1 and 2 in Fig. \ref{filter3}). Its electrical scheme is shown in Fig. \ref{filter6}, where R$_{SQUID}$ is the output impedance of the SQUID series array ($\sim 20 \Omega$ \cite{torriolitn}) and the input impedance of the pre-amplifier is assumed sufficiently high to be neglected.

\begin{figure}[H]
\centering
\includegraphics[width=0.8\textwidth]{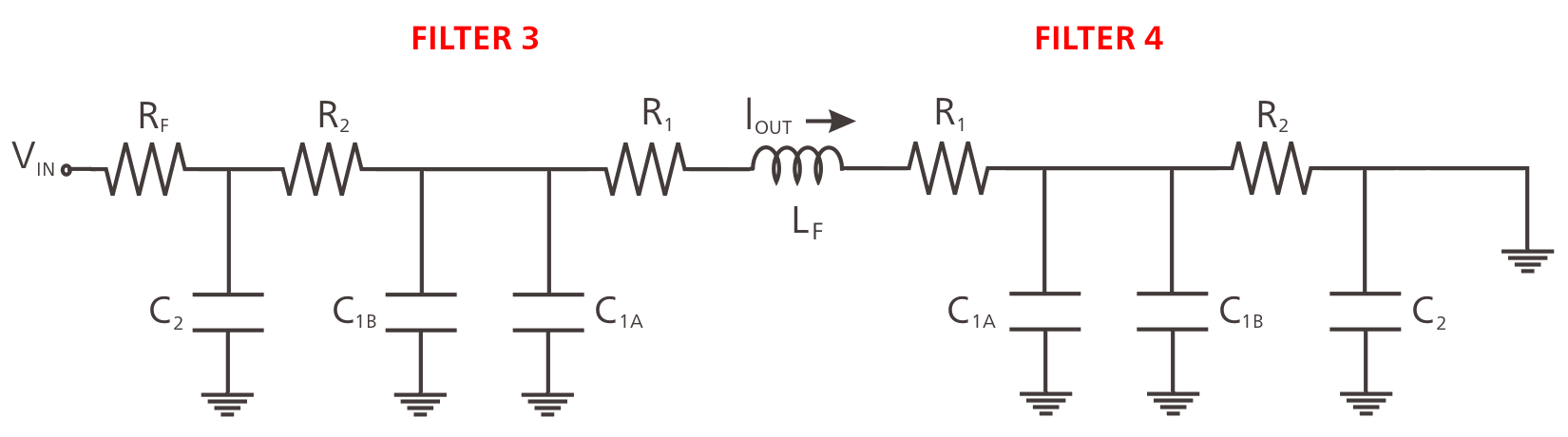}
\caption{Scheme of the filtering system operating on the feedback branch of the circuit.}
\label{filter7}
\end{figure}

\noindent Assuming the filter 1 and 2 identical, the transfer function of this branch is:

\begin{equation}
\label{tfa}
H_{fA} = \frac{V^+_{OUT}-V^-_{OUT}}{V^+_{IN}-V^-_{IN}} =\frac{V^+_{OUT}}{V^+_{IN}} = \frac{V^-_{OUT}}{V^-_{IN}} = \frac{1}{(1+s\tau_{A1})(1+s\tau_{A2})}
\end{equation}

\noindent where the two poles are:

\begin{equation}
\tau_{A1} = \dfrac{2 C_1 C_2 R_1^* R_2}{(C_1 + C_2)R_1^* + C_2R_2 + \sqrt{C_2^2 R_2^2 + (2C_2^2 - 2C_1C_2)R_1^* R_2 + (C_1 + C_2)^2 (R_1^*)^2}}
\end{equation}

\begin{equation}
\tau_{A2} = \dfrac{2 C_1 C_2 R_1^* R_2}{(C_1 + C_2)R_1^* + C_2R_2 - \sqrt{C_2^2 R_2^2 + (2C_2^2 - 2C_1C_2)R_1^* R_2 + (C_1 + C_2)^2 (R_1^*)^2}}
\end{equation}

\noindent whith $C_1 = C_{1A} + C_{1B}$ and $R_1^* = R_1 + R_{SQUID}/2$. The corresponding cutoff frequencies evaluated assuming the filter components values in Tab. \ref{tab:filvalue} and $R_{SQUID}$ = 20 $\Omega$ are:

\begin{equation}
f_{A1} = \frac{1}{2\pi \tau_{A_1}} = 11.4 MHz
\end{equation}
\begin{equation}
f_{A2} = \frac{1}{2\pi \tau_{A_2}} = 1.37 GHz
\end{equation}

$$ \; $$
In the feedback brach there are the other two filters (Filters 3 and 4 in Fig. \ref{filter3}), placed  to the extremities of the feedback coil (Fig. \ref{filter7}).

\noindent The system input voltage corresponds to the amplifier output voltage, whose output impedance is neglegible with respect to R$_F$. The feedback impedance (L$_F$) is of the order of nH, thus neglegible with respect to R$_1$ and R$_2$ for operating frequency around the MHz. In this case the tranfer function results having one zero and three poles:

\begin{equation}
\label{tff}
H_{fF}=\frac{I_{OUT}}{V_{IN}}=\frac{1}{(2R_1 + 2R_2 + R_F)}\cdot \frac{1+s\cdot T_F}{(1+s \cdot \tau_{F1})(1+s \cdot \tau_{F12})(1+s \cdot \tau_{F3})}
\end{equation}

\noindent The zero is due to the fact that increasing the frequency, the current through the feedback coil increases due to the decrease of the impedance of the capacitors in the filter 4 (connected to ground). In this case the cutoff frequencies have been numerically evaluated assuming the components values in Tab. \ref{tab:filvalue}. They are:

\begin{equation}
f_{TF} = \frac{1}{2\pi T_F} = 1.13 GHz
\end{equation}
\begin{equation}
f_{F1} = \frac{1}{2\pi \tau_{F_1}} = 25.1 MHz
\end{equation}
\begin{equation}
f_{F2} = \frac{1}{2\pi \tau_{F_2}} = 1.18 GHz
\end{equation}
\begin{equation}
f_{F3} = \frac{1}{2\pi \tau_{F_3}} = 2.29 GHz
\end{equation}

$$ \; $$
The final whole system open loop tranfer function is:

\begin{equation}
\label{finalres}
H_{TOT} = H_{fA}G_S\frac{A}{s}e^{-t_ds}H_{fF}Mf
\end{equation}

\noindent where $H_{fA}$ and $H_{fF}$ are the transfer functions reported in eq. (\ref{tfa}) and (\ref{tff}).

\subsubsection{Stability analysis}

\begin{figure}[H]
\centering
\includegraphics[width=0.7\textwidth]{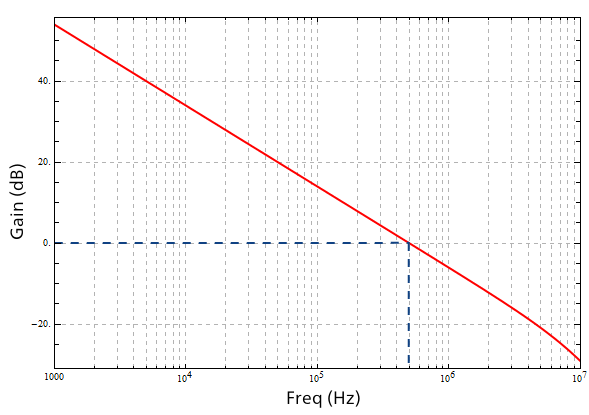}
\caption{Bode diagram of the system operating at open loop - Amplitude.}
\label{filter8}
\end{figure}

\begin{figure}[H]
\centering
\includegraphics[width=0.7\textwidth]{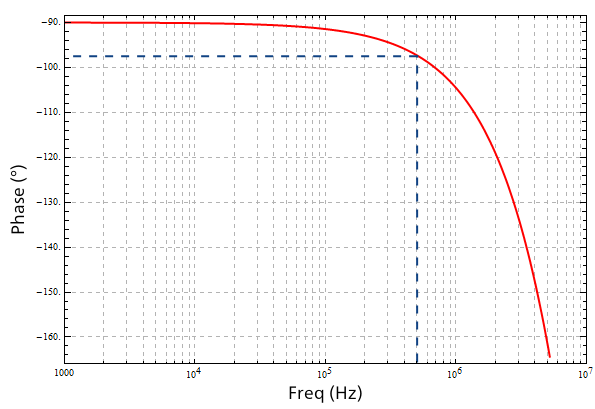}
\caption{Bode diagram of the system operating at open loop - Phase.}
\label{filter9}
\end{figure}

The Bode diagrams of the system, obtained from eq. (\ref{finalres}) are reported in Fig. \ref{filter8} and Fig. \ref{filter9}.

\noindent The unity gain frequency is $f_{0 dB} = 500 kHz$, and the corresponding phase shift is $\phi_{0 dB} = -97.3$ degrees, so there is a phase margin of: 

\begin{equation}
\phi_{M} = 180^{\circ} - \phi_{0 dB} = 82.7^{\circ}
\end{equation}

\noindent Given the filters scheme in Fig. \ref{filter4} and the components value in Tab. \ref{tab:filvalue}, the SQUID FLL configuration is therefore stable assuming the reference block values for the CryoAC readout (Tab. \ref{tab:blocksvalue}).

\bigskip
The stability study has been then repeated with different values of the filters component, in order to evaluate the change in the poles position and in the global phase margin. The results of this analysis are reported in Tab. \ref{tab:endfilter}, showing that it is possible to decease the cutoff frequencies up to about one order of magnitude without compromising the stability of the system.

\begin{table}[htbp]
\centering
\caption{Cutoff frequencies and global phase margin obtained with different values for the filter components.}
\label{tab:endfilter}
\begin{tabular}{ccccccccc}
\hline\noalign{\smallskip}
R$_1$=R$_2$ & C$_1$ & C$_2$ & f$_{A1}$ & f$_{A2}$ & f$_{F1}$ & f$_{F2}$ & f$_{F3}$ & Phase margin\\
$\Omega$ & pF & nF & MHz & GHz & MHz & GHz & GHz & degrees\\
\noalign{\smallskip}\hline\noalign{\smallskip}
\multicolumn{8}{c}{No filters \cite{torriolitn}} & 86 \\
\noalign{\smallskip}\hline\noalign{\smallskip}
1.5 & 2$\cdot$47 & 1 & 11 & 1.4 & 25 & 1.2 & 2.4 &  83\\
\noalign{\smallskip}\hline\noalign{\smallskip}
1.5 & 2$\cdot$68 & 1 & 11 & 0.98 & 24 & 0.83 & 1.7 &  83\\
1.5 & 2$\cdot$68 & 1.5 & 7.6 & 0.95 & 17 & 0.82 & 1.6 &  81\\
1.5 & 2$\cdot$100 & 2.2 & 5.2 & 0.64 & 11 & 0.56 & 1.1 &  78\\
1.5 & 2$\cdot$150 & 3.3 & 3.5 & 0.43 & 7.6 & 0.37 & 0.73 &  75\\
1.8 & 2$\cdot$220 & 4.7 & 2.3 & 0.25 & 4.4 & 0.21 & 0.41 &  68\\
2.2 & 2$\cdot$330 & 6.8 & 1.5 & 0.14 & 2.5 & 0.12 & 0.22 &  59\\
\noalign{\smallskip}\hline
\end{tabular}
\end{table}

\subsection{Filters integration in the IAPS setup}

Following these studies, we have started an activity aimed to integrate a cold filtering stage also in our cryogenic setup, in order to test the CryoAC DM in an environment as representative as possible of the X-IFU FPA DM. In this context, a custom multilayer filtering PCB has been developed by the Bruno Kessler Foundation (FBK), following the design used by SRON. The PCB contains 8 filter to low pass the CryoAC DM SQUID and TES lines (SQUID bias + and -; SQUID signal + and -; SQUID feedback + and -; TES bias + and - )

\bigskip
\begin{figure}[H]
\centering
\includegraphics[width=0.3\textwidth]{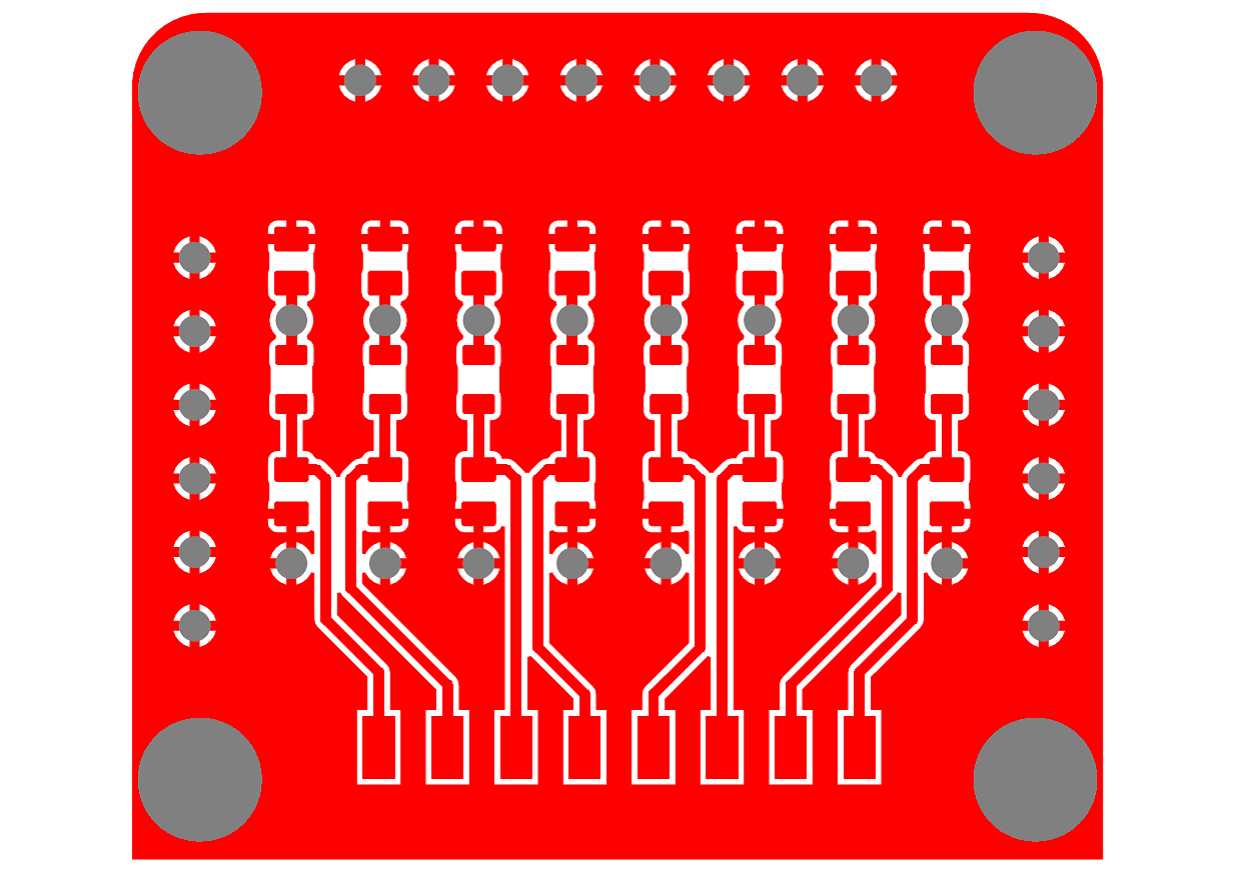}
\includegraphics[width=0.3\textwidth]{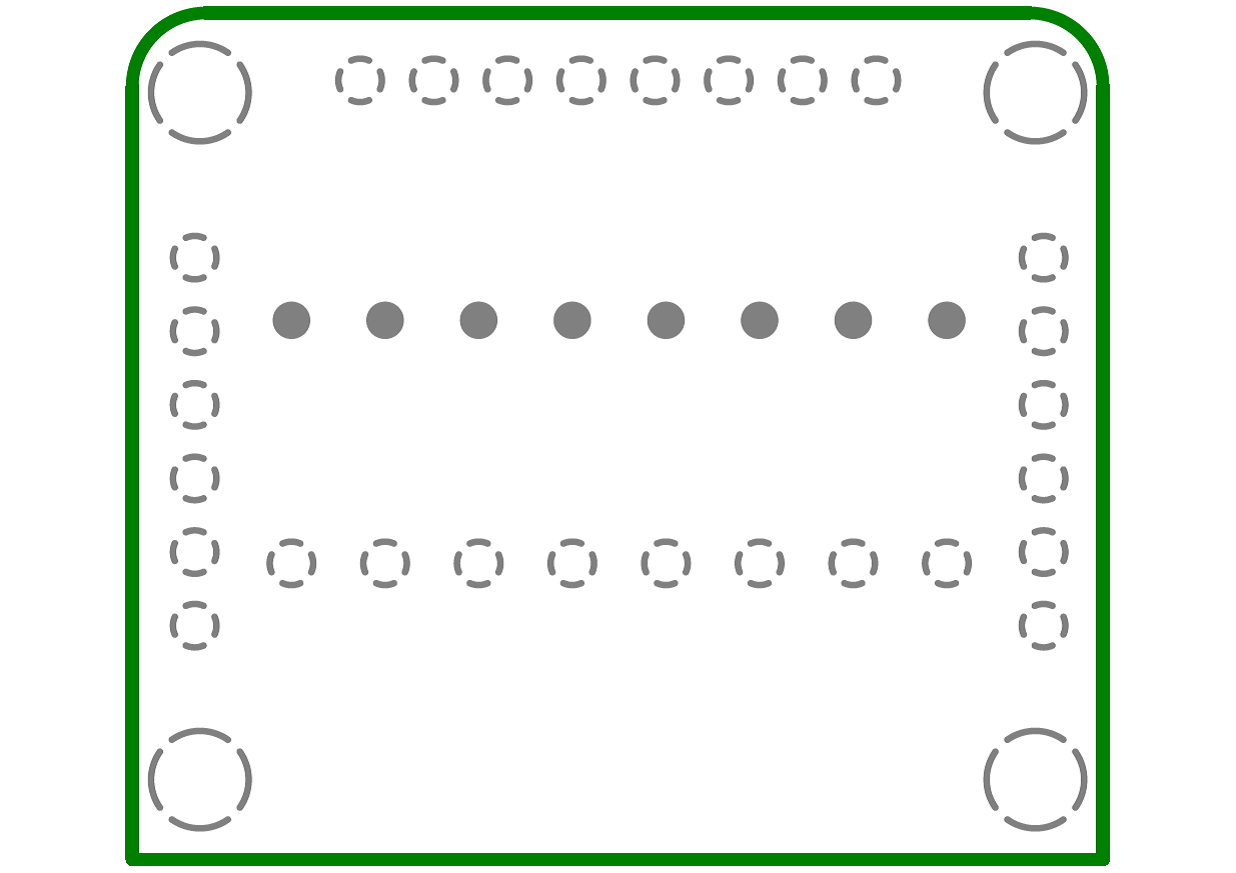}
\includegraphics[width=0.3\textwidth]{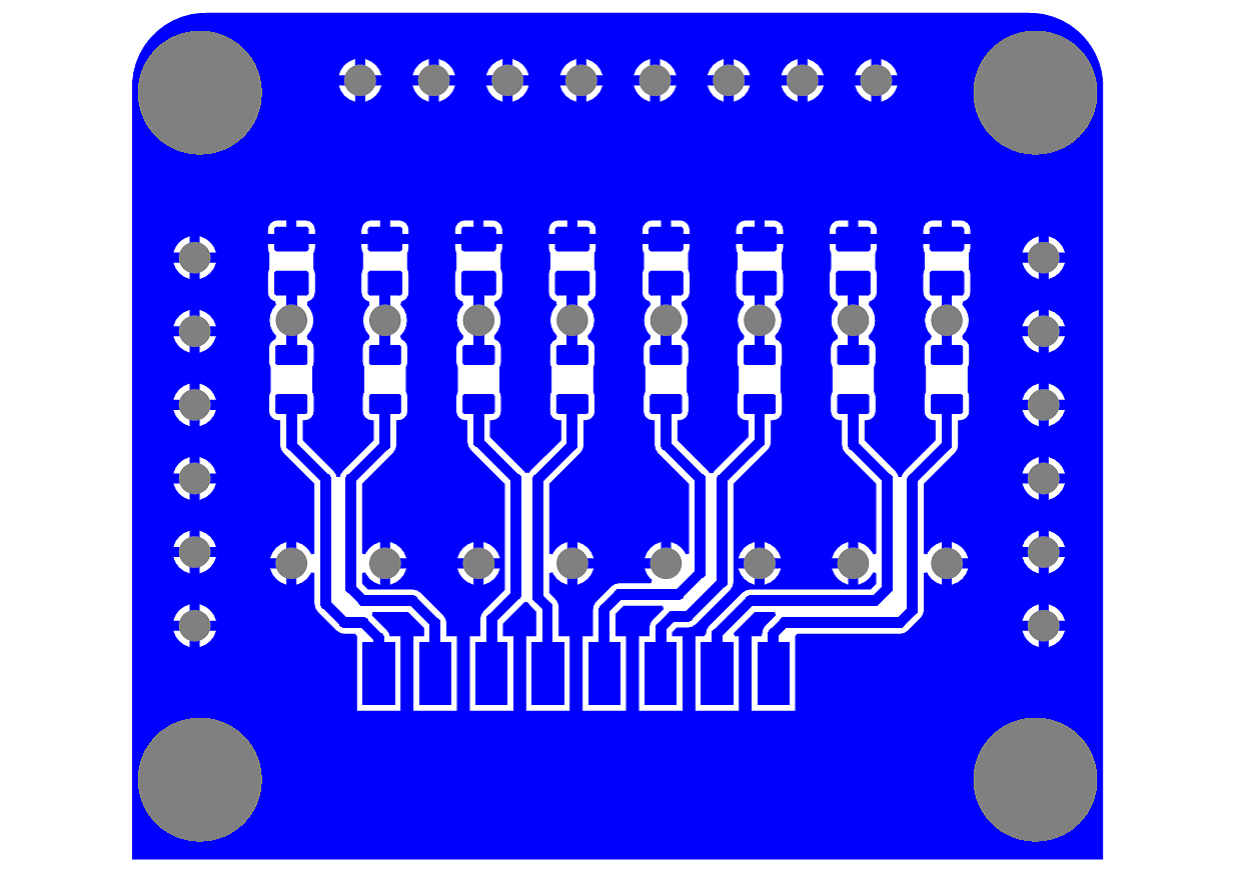}
\caption{Layout of the multilayer filtering PCB developed by FBK for IAPS. \textit{Left:} the top layer, to which are soldered the wires coming from room temperature.\textit{Center:} the inner layer (ground plane). \textit{Right:} the bottom layer, to which are soldered the wires going towards the cold stage).}
\label{fbkpcb}
\end{figure}

\begin{figure}[H]
\centering
\includegraphics[width=0.8\textwidth]{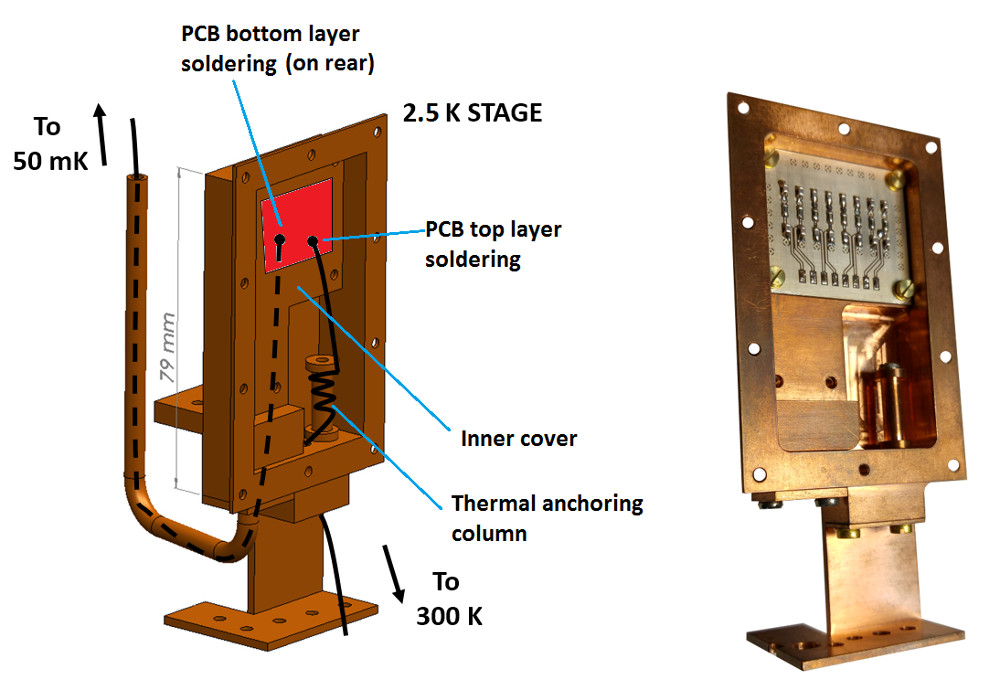}
\caption{Holder box designed to integrate the filtering PCB at the 2.5 K stage of the IAPS ADR. \textit{(Left :)} CAD drawing of the box where the components of the system and the signal wires path are highlighted. \textit{(Right :)} The manufactured holder box with the filtering PCB produced by FBK mounted on the inner cover.}
\label{filtersintegration}
\end{figure}

\begin{figure}[H]
\centering
\includegraphics[width=0.6\textwidth]{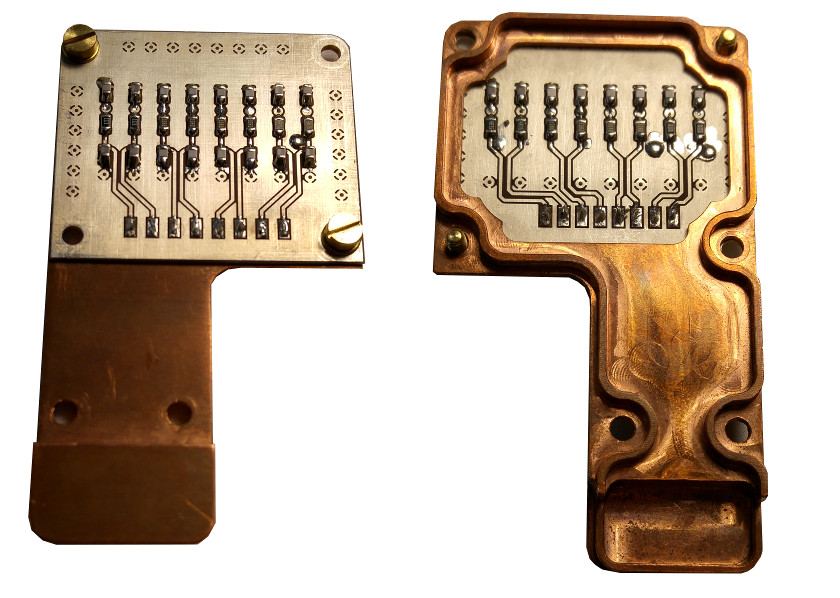}
\caption{Detail of the filtering PCB produced by FBK mounted on the inner cover of the holder box.}
\label{filtersintegration2}
\end{figure}

Simultaneously, we have developed an holder box to integrate the filtering PCB at the 2.5 K stage of the ADR cryostat (Fig. \ref{filtersintegration} and Fig. \ref{filtersintegration2}). The box, made of OFHC copper, contains a column for the wires thermal anchoring and an inner cover aimed to separate the \virg filtered signal zone'' (towards the 50 mK stage)  from the \virg not filtered signal zone'' (towards the 300 K stage). Note that the exit from the box in direction of the 50 mK stage take place through a tube designed to enter the new ADR cold magnetic shields. 

Before ending this section, I remark that, at system level, it has been finally decided not to implement the cold filtering stage in the X-IFU FPA DM (all the signals will be anyway filtered at warm, at the entrance of the cryostat Faraday cage). Following this decision, we have decided not to integrate the cold filters neither in our setup, in order to maintain the maximum representativity of the test environment.

\chapter{The CryoAC Demonstration Model}
\label{DMchapt}

The X-IFU development program foresees to build and characterize an instrument Demonstration Model (DM) before the mission adoption. In this respect, we are now developing the CryoAC DM, which will be delivered to the Focal Plane Assembly (FPA) development team at SRON for the integration with the TES array DM. This will be the first compatibility test for the two detectors, representing a milestone on the path towards the X-IFU development. 

\bigskip
\noindent At subsystem level, the CryoAC DM is part of a wider program aimed to reach TRL 5. This will be performed developing two representative detector models:
\begin{itemize}
\item The Demonstration Model (DM): a suspended square single pixel able to probe the detector critical technologies;
\item The Structural Thermal Model (STM): a mechanical model of the CryoAC 4-pixels array to be vibrated and cooled in a representative environment.
\end{itemize}

In this context, the main requirements to be satisfied by the CryoAC DM to enable the detector critical technology are:
\begin{itemize}
\item Pixel size (absorber area) = 10$\times$10 mm$^2$, to be representative of the Flight Model (FM) pixel dimensions (4 pixels, $\sim$1.2 cm$^2$ area each one);
\item Low energy threshold $\leq$ 20 keV, necessary to meet the requirement on the CryoAC geometric rejection efficiency ($>$ 98\% for primaries);
\item Operation at T$_{B}$ = 50-55 mK, corresponding to the X-IFU thermal bath;
\item Reproducible thermal conductance between the absorber and the thermal bath, in order to control the thermal decay time of the detector, which defines its dead time. At present, this implies for the design a suspended absorber.
\end{itemize}

Here I will first introduce the CryoAC DM, describing the detector design and its fabrication process. I will present also the DM test setup and the characterization measurements performed on the SQUID used in the DM Cold Front End Electronics (CFEE). Then I will report the main results obtained with AC-S9, a DM prototype that we have integrated and tested before performing the final etching process to suspend the silicon absorber. Finally, I will show the preliminary results obtained with the first proper CryoAC DM prototype, namely AC-S10.

\newpage
\section{Design and Fabrication}

In the following paragraphs I will describe the CryoAC DM design and its fabrication processes. I remark that the final DM design has been fixed thanks to the experience gained with the AC-S7 and AC-S8 pre-DM prototypes, whose test activity has been described in the Chapt. \ref{preDMchapter} of this thesis.

\subsection{Design}
\begin{figure}[H]
\centering
\includegraphics[width=0.6\textwidth]{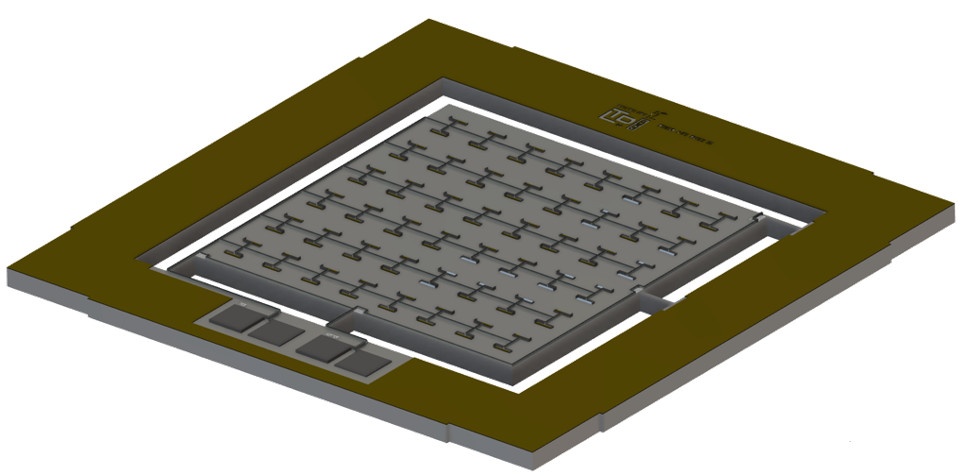}
\caption{Sketch of the CryoAC DM design.}
\label{DM_sketch}
\end{figure}

The CryoAC DM is based on a wide area silicon absorber (1 cm$^2$, 525 $\mu$m thick) sensed by a network of 96 iridium:gold TES in parallel configuration, and readout by a SQUID operating in the standard Flux Locked Loop (FLL) configuration. To obtain a well-defined and reproducible thermal conductance towards the thermal bath, the absorber is connected to a silicon rim (gold plated and in strong thermal contact with the bath) through four narrow silicon bridges (100 $\times$ 1000 $\mu$m$^2$), achieving a suspended free-standing structure (Fig. \ref{DM_sketch}). 

The TES network is designed to ensure a uniform absorber surface coverage for an efficient athermal phonons collection, while constraining down the heat capacity contribution of the metal film thanks to the quite small TES size (50x500 $\mu$m$^2$). All these design solutions are necessary to speed up the detector thus enabling the high rejection efficiency of particle background imposed by mission requirements. The bias line on board of the absorber are anti-inductive overlapping Nb wirings, to limit the magnetic coupling with the TES array.

Platinum heaters are also embedded on the absorber to increase its temperature if necessary. This has been foreseen to deal with the high current needed to bias the TES network with T$_B$=50 mK, that has been measured to be of the order of mA in the previous CryoAC prototype (see Chapt. \ref{preDMchapter}). By means of the heater it will be possible to operate the detector by lower bias current, locally increasing the absorber temperature, thus limiting magnetic coupling effects to the TES array. Furthermore, increasing the absorber temperature could be useful to increase the heat capacity of the detector, extending its maximum detectable energy ($E_{MAX} \sim C\cdot\Delta T_C$). Alternatively, the heater can also be switched on to drive the TES into the normal state and bias it, thus avoiding to overcome the high critical current by injecting on the detector several mA, and then switched off during the detector operations once the working point is got.

\bigskip

\bigskip

\noindent The specifications of the CryoAC DM are summarized in Tab. \ref{tab:DM_spec}. 

\begin{table}[H]
\centering
\caption{CryoAC DM specifications.}
\label{tab:DM_spec}
\begin{tabular}{ll}
\hline\noalign{\smallskip}
Parameter & Value \\
\noalign{\smallskip}\hline\noalign{\smallskip}
Silicon chip thickness & 525 $\mu$m \\
Silicon chip resistivity &  $>$10 k$\Omega\cdot$cm \\
Chip overall area & (16.6 $\times$ 16.6) mm$^2$\\ 
Absorber area & (10.0 $\times$ 10.0) mm$^2$ \\
Beam dimensions & (1000 $\times$ 100) $\mu$m$^2$\\
Rim width & 2.3 mm \\
Ir:Au TES size ($\times$ 96) &  (50 $\times$ 500) $\mu$m$^2$ ($\sim$ 150 nm thick)\\
Ir:Au T$_C$ & 100 mK (TBC)\\
\noalign{\smallskip}\hline
\end{tabular}
\end{table}

\subsection{Fabrication}

\begin{figure}[H]
\centering
\includegraphics[width=0.6\linewidth]{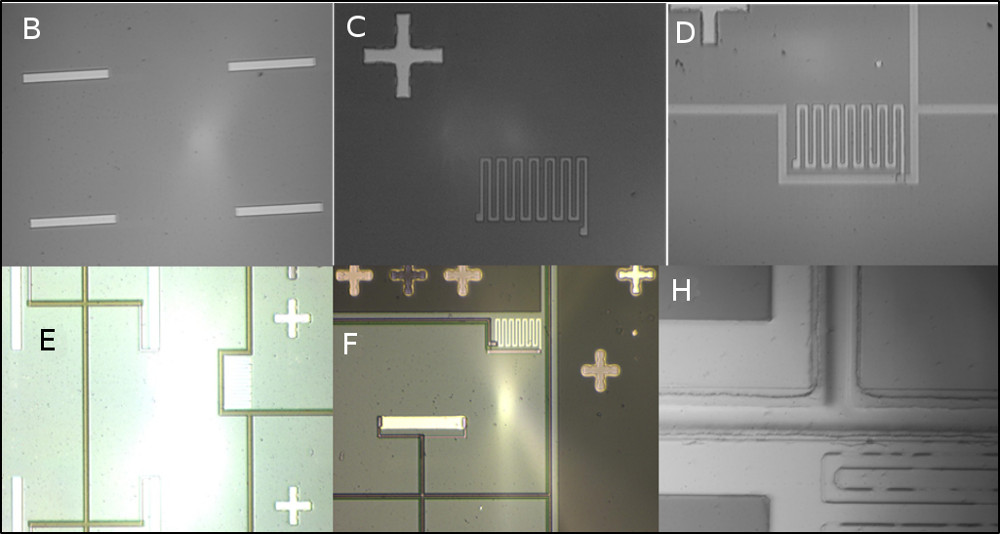}
\includegraphics[width=0.33\linewidth]{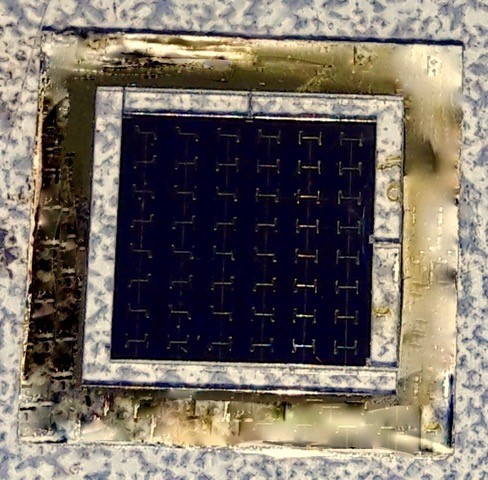}
\caption{\textit{(Left)} CryoAC DM fabrication steps. See text for details. \textit{(Right)} Sample at the end of the fabrication process (not working prototype due to an issue on the bias pads).}
\label{DMfabr}
\end{figure}

The detector is produced from a commercially available silicon wafer at the Genoa University (Phys. department). The fabrication is divided in several step (Fig. \ref{DMfabr}). 

\bigskip

\bigskip
\noindent The fabrication processes are:

\begin{itemize}
\item[A.]	Ir:Au film deposition on silicon chip by Pulsed Laser Deposition;
\item[B.]	Ir:Au film patterning using positive photolithography and dry etching process (Fig. \ref{DMfabr} B);
\item[C.]	Pt heaters deposition by e-beam evaporation, patterning performed using negative photolithography and lift-off process (Fig. \ref{DMfabr} C);
\item[D.]	Nb wiring (lower strip) deposition by RF-sputtering, pattering performed using negative photolithography and lift-off process (Fig. \ref{DMfabr} D);
\item[E.]	SiO2 insulation layer deposition by e-beam evaporation, patterning performed using negative photolithography and lift-off process (Fig. \ref{DMfabr} E);
\item[F.]	Nb Wiring (upper strip) deposition by RF-sputtering, patterning performed using negative photolithography and lift-off process ((Fig. \ref{DMfabr} F);
\item[G.]	Au deposition on the rim by thermal evaporation, patterning performed using negative photolithography and lift-off process;
\item[H.]	Si Deep RIE-ICP etching using bosch process with aluminum hard-mask deposition by thermal evaporation, patterning by negative photolithography and final lift-off process (Fig. \ref{DMfabr} H);
\end{itemize}

All the processes have been defined and tested. During the processes setting, some defective DM sample has been produced (Fig. \ref{DMfabr} - Right) and used to define and test the handling and integration procedures. About the last step, after the production of the first test samples the Si RIE-ICP final etching (Step H) turned out to be poisoned by the RIE-ICP of Ir:Au (step B), with decreases of etching rate and quality with respect to our first tests. We have then setted the two etching processes (step B and step H) in two different reactors to avoid cross-contamination, and we have etched dummy samples to recover the original RIE-ICP process quality. These dummy samples have been thermal cycled in the dilution unit to test their robustness and the handling/integration procedure.

In the meantime of this recovery action, we have decided to characterize and test a DM-like sample (namely AC-S9) before the final etching, thus seeing if the process will affect the DM. This has allowed us also to verify all the previous fabrication steps and to preliminary test the detector integration and our cryogenic setup. The AC-S9 test activity will be presented later in this chapter.

\newpage
\section{DM test setup at IAPS}

The CryoAC DM test plan foresees to characterize and test the detector in our CryoLab at IAPS before its delivery to SRON for the integration in the Focal Plane Assembly (FPA) DM. We have therefore designed and developed a series of mechanical components to integrate and to test the CryoAC DM in our cryogenic systems.

\bigskip

\begin{figure}[H]
\centering
\includegraphics[width=0.69\linewidth]{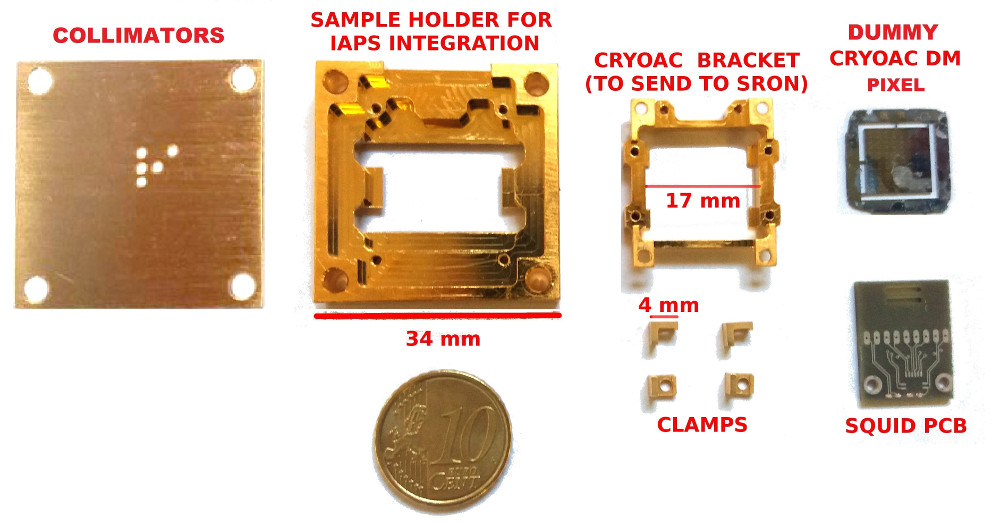}
\includegraphics[width=0.3\linewidth]{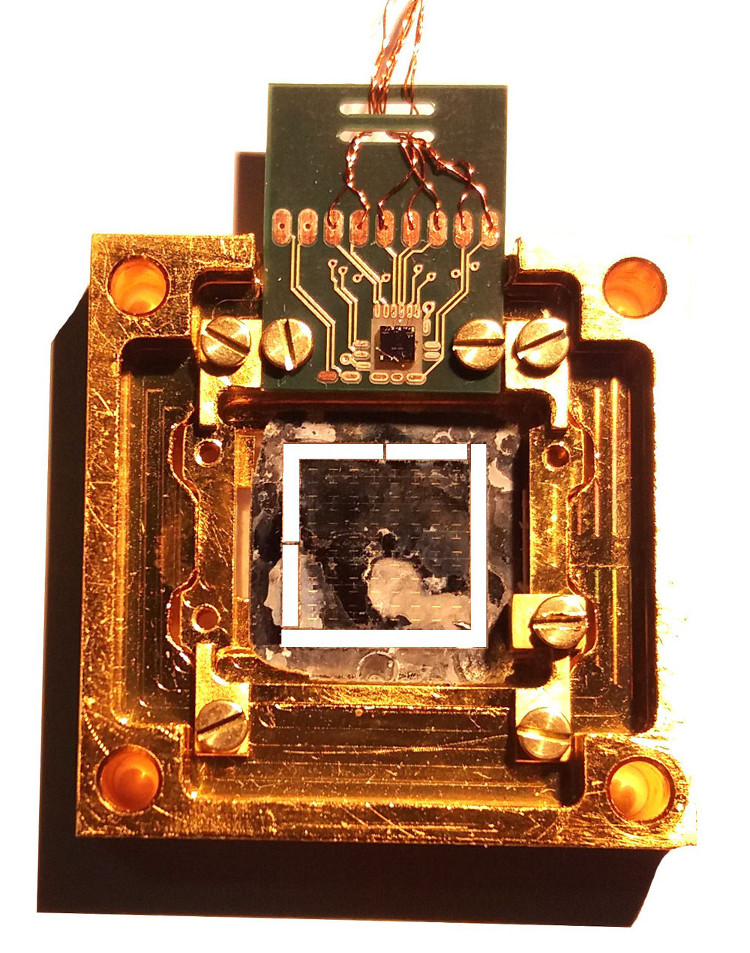}
\caption{\textit{Left:} Mechanical parts developed to integrate and test the CryoAC DM at IAPS.\textit{Right:} Assembly of all the components during a test performed with a CryoAC DM dummy pixel.}
\label{DMparts}
\end{figure}

\bigskip
\noindent The main developed components are:

\begin{itemize}
\item The \virg CryoAC bracket'', a gold-plated OFHC copper structure in which the CryoAC DM pixel and the PCB with its cold readout electronics are integrated. This assembly will be sent to SRON for the joined test with the TES array in the X-IFU FPA DM;

\item The \virg clamps'', which are needed to fasten the CryoAC DM pixel inside the bracket, compensating the different thermal contractions of copper and silicon;

\item The \virg sample holder'', which is the support designed to integrate the bracket in the IAPS cryogenic setup, before its delivery to SRON;

\item The \virg collimators'', that are used to illuminate the pixel in different regions, in order to test the detector response homogeneity. This is an important measurement to preliminary investigate the detector rejecting efficiency. 

\end{itemize}

\begin{figure}[H]
\centering
\includegraphics[width=0.47\linewidth]{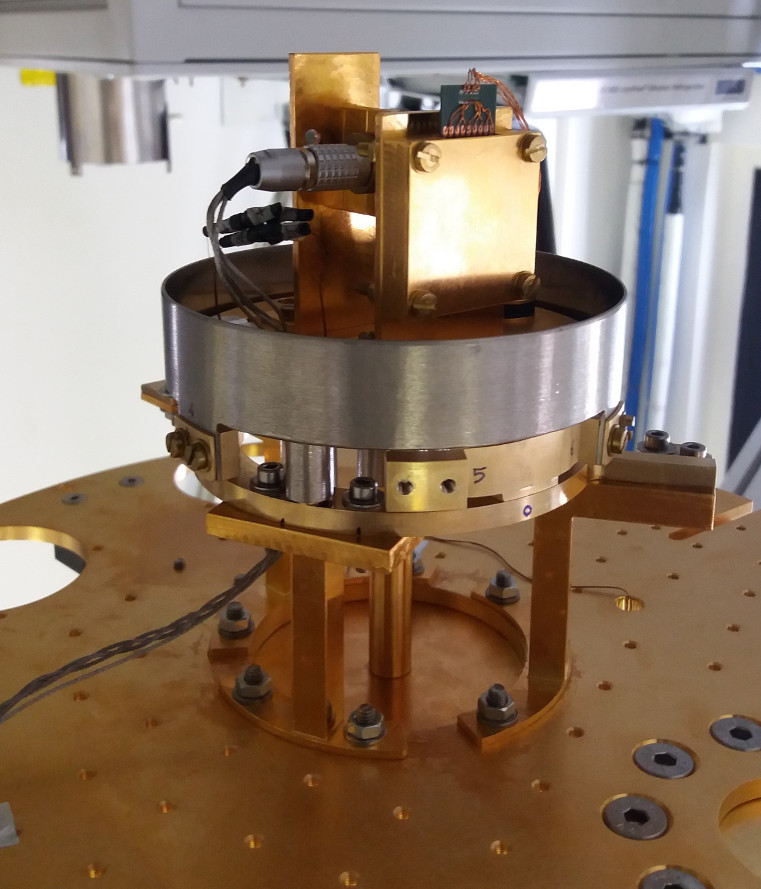}
$\;\;\;$
\includegraphics[width=0.415\linewidth]{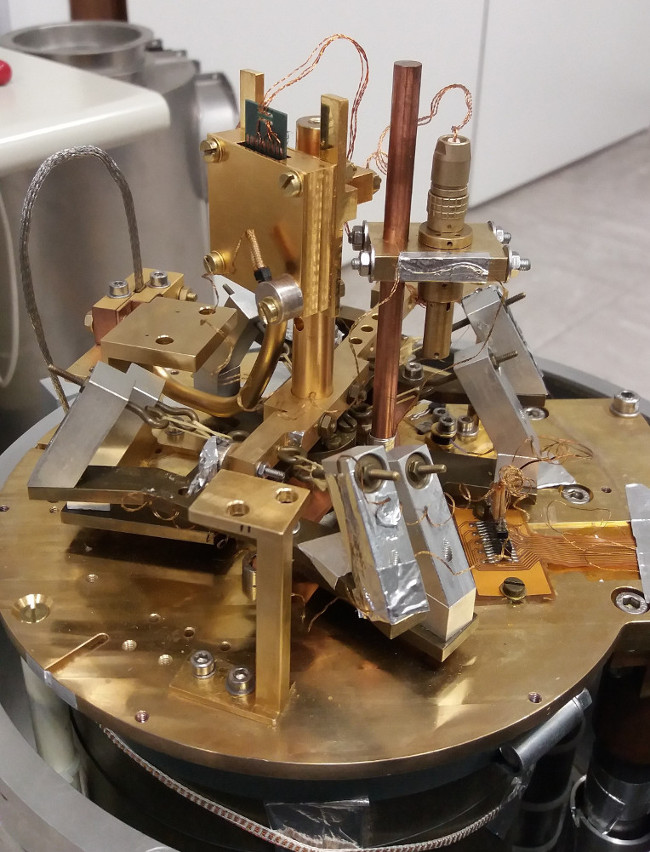}
\caption{Integration of the DM test setup in both the IAPS cryostat. \textit{Left:} The dilution refrigerator. \textit{Right:} The ADR.}
\label{DM_DR_ADR}
\end{figure}

\noindent In Fig. \ref{DM_DR_ADR} the integration of the system in the two IAPS cryogenic systems is shown. We have choosen to prepare both the cryostat for the DM integration in order to have a backup solution in the event of problems with the dilution refrigerator.

\bigskip

\begin{figure}[H]
\centering
\includegraphics[width=0.9\linewidth]{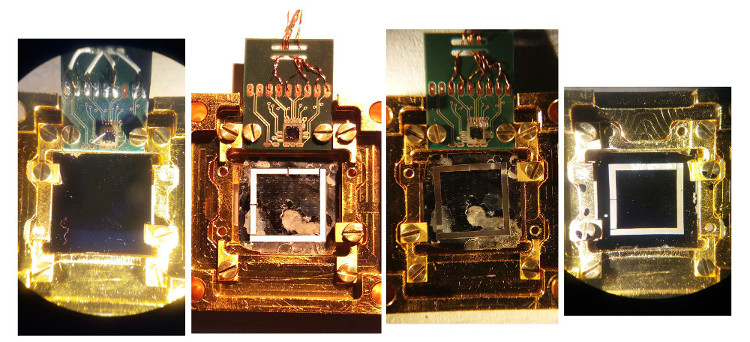}
\caption{CryoAC DM dummy pixels integrated in the holder structures.}
\label{DM_dummies}
\end{figure}

We have performed several cooling test (from room temperature down to 50 mK) with dummy CryoAC DM pixels in order to verify the assembly mechanical strength and its response to thermal cycles (Fig. \ref{DM_dummies}). No critical issues have been found during this activity (craks, damages, deformations...), validating the mechanical setup and the handling and integration procedures.

\bigskip

\newpage
\section{SQUIDs characterization}

The CryoAC DM cold read-out electronics is based on a SQUID that will be operated in a standard FLL configuration (see par. \ref{squidpar}), and that have on-board also a R$_{S}$ = 0.5 m$\Omega$ shunt resistor. The SQUID is a custom model specifically produced for the CryoAC DM by the VTT Technical Research Centre of Finland (namely VTT K4 SQUID). In this section I report the characterization measurements performed on two K4 SQUID chips (namely \textit{ret G} and \textit{ret M}) that we have receive from VTT.

\subsubsection{SQUID VTT K4 ret G}
\label{vttk4retg}

\begin{figure}[H]
\centering
\includegraphics[width=0.45\linewidth]{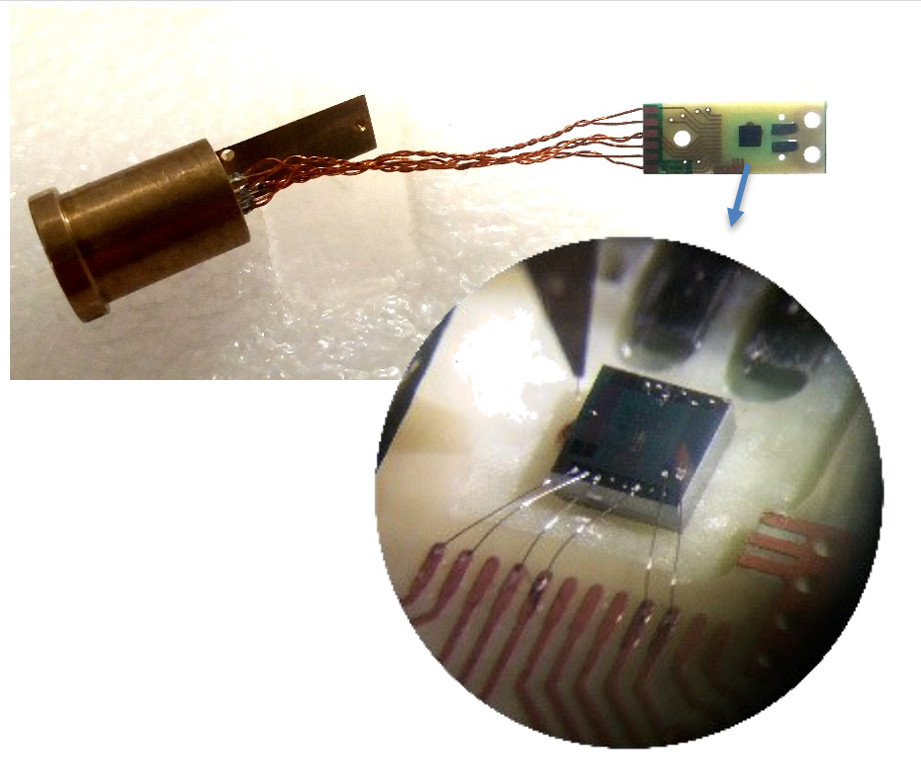}
\includegraphics[width=0.45\linewidth]{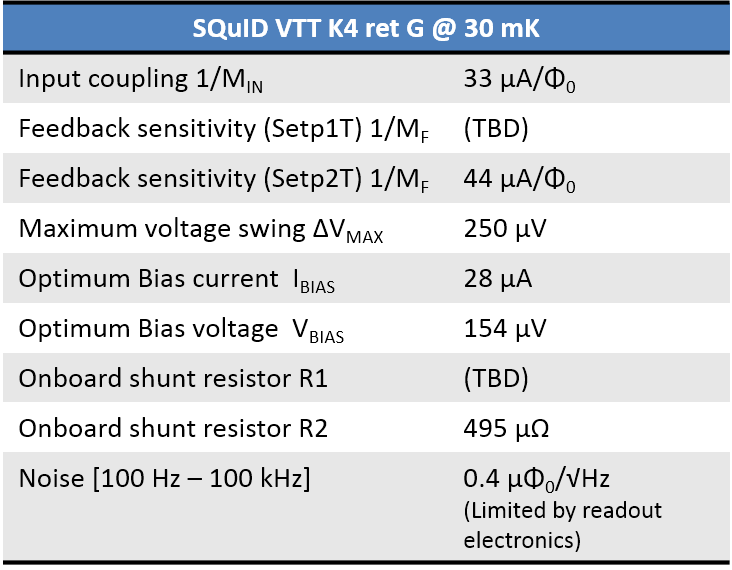}
\includegraphics[width=0.45\linewidth]{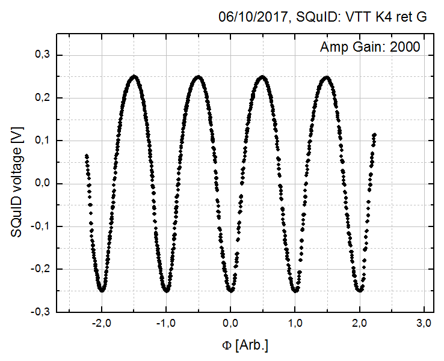}
\includegraphics[width=0.45\linewidth]{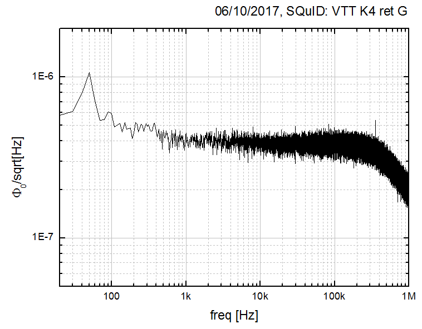}
\caption{SQUID VTT K4 ret G characterization. \textit{Top Left:} The SQUID mounted on a commercial Magnicon package.  \textit{Top Right:} Table containing the main measured SQUID parameters. \textit{Bottom Left:} The SQUID characteristic V-$Phi$ curve. \textit{Bottom Right:} SQUID noise.}
\label{retG}
\end{figure}

We have mounted this chip on a commercial Magnicon package and performed the test using a Magnicon XXF-1 FLL Electronics. The chip has been cooled down by the Dilution Refrigerator, and tested at $\sim$ 30 mK. The sample space was magnetically shielded with a 1 mm thick Cryophy shield. The main results of this activity are reported in Fig. \ref{retG}, where it is also incapsulated a table containing the main measured SQUID parameters.

\bigskip

\noindent The characteristic SQUID V-$\Phi$ curve is very well-shaped. It has been acquired sending a 200 $uA$ peak-to-peak triangular wave to the \virg Sept 2T'' feedback coil, and so it shows about 4.5 periods (= 200 $uA$/44 $uA$/$\Phi_0$). Finally, note that the measured white noise of 0.4 $\mu \Phi_0 / \sqrt(Hz)$  corresponds to a voltage noise of $\sim$ 0.3 $nV / \sqrt(Hz)$, compatible with the expected white noise due to the pre-amplifier of the Magnicon XXF-1 FLL Electronics (0.32 $nV / sqrt(Hz)$ \cite{magniconxxf1}).

\subsubsection{SQUID VTT K4 ret M}

\begin{figure}[H]
\centering
\includegraphics[width=0.45\linewidth]{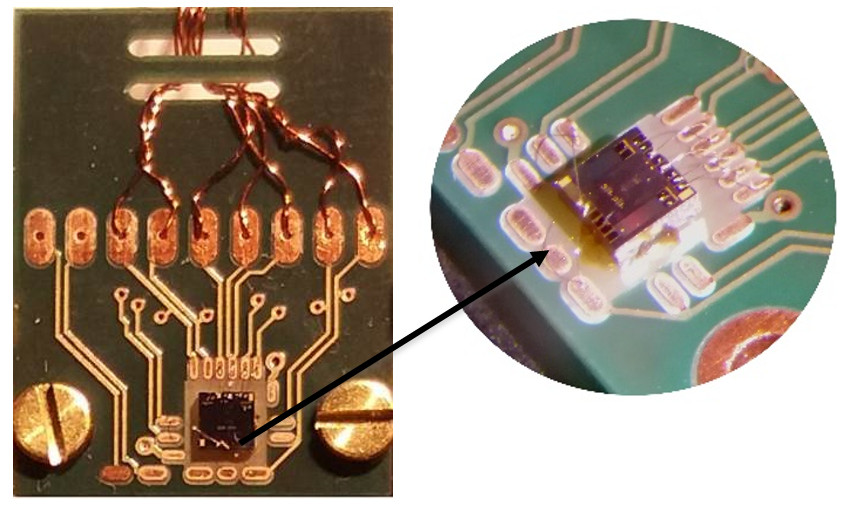}
\includegraphics[width=0.45\linewidth]{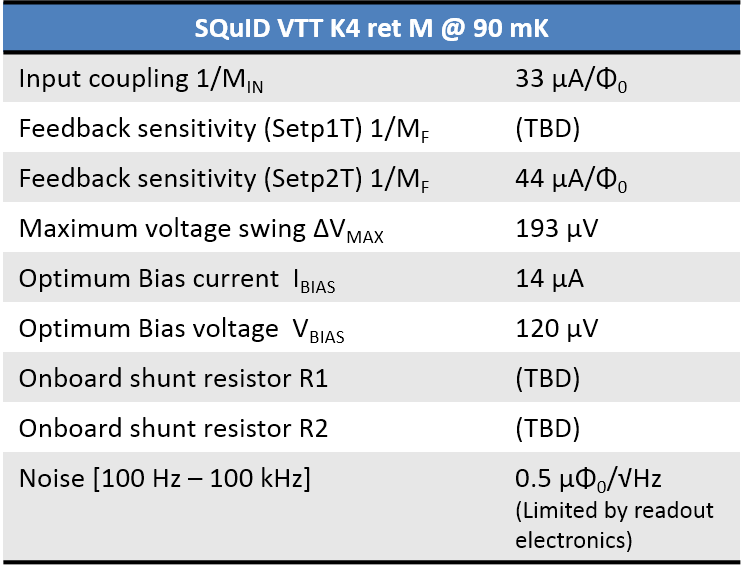}
\includegraphics[width=0.45\linewidth]{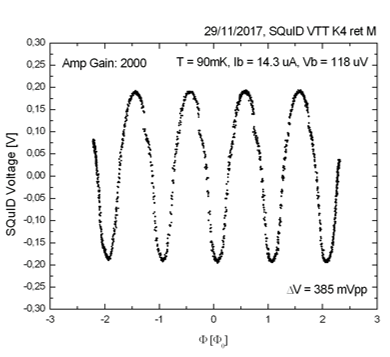}
\includegraphics[width=0.45\linewidth]{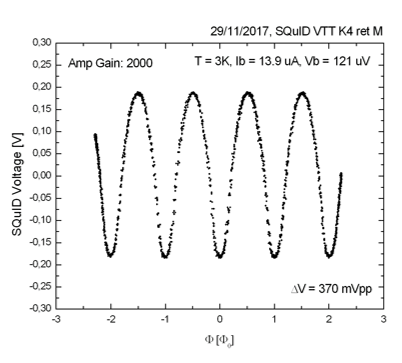}
\caption{SQUID VTT K4 ret M characterization. \textit{Top Left:} The SQUID mounted on the CryoAC DM custom PCB holder.  \textit{Top Right:} Table containing the main measured SQUID parameters. \textit{Bottom Left:} The SQUID characteristic V-$Phi$ curve measured at 90 mK. \textit{Bottom Right:} The SQUID characteristic V-$Phi$ curve measured at 3 K.}
\label{retM}
\end{figure}

We have mounted this chip on our custom PCB holder developed for the CryoAC DM and performed the test using the same Magnicon XXF-1 FLL Electronics. The chip has been cooled down by the Adiabatic Demagnetization Refrigerator, and tested at $\sim$ 90 mK. In this case the sample space was not magnetically shielded, but we have been able to de-flux the SQUID  (removing the magnetic field trapped in it) by applying current to its Josephson Junctions (69 mA for 0.100 s). The main results of the characterization activity are reported in Fig. \ref{retM}, where it is also incapsulated a table containing the main measured SQUID parameters.

\bigskip

\noindent Note that in this case the SQUID V-$\Phi$ curve shows some \virg kinks'' (the small instabilities around the 0.12 V level), which however do not affect the proper SQUID operation. These \virg kinks'' are not present in the V-$\Phi$ curve acquired at more high temperature (3K). We are now investigating this issue. Finally, note that also in this case the SQUID noise is dominated by the readout electronics.

\newpage
\section{Test of a pre-etching DM sample (AC-S9)}

\begin{figure}[htbp]
\centering
\includegraphics[width=0.8\linewidth]{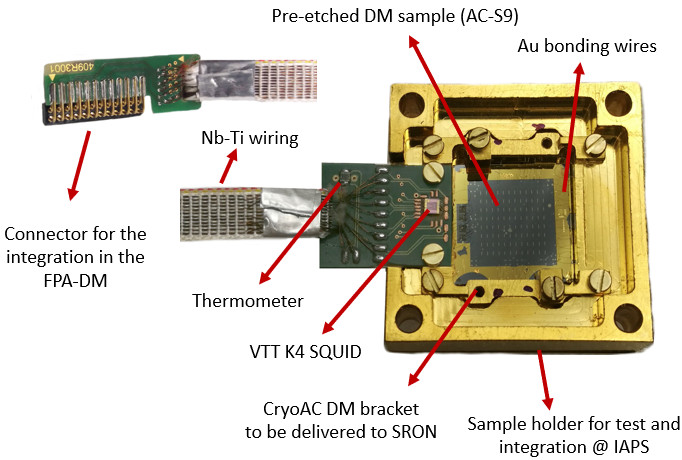}
\caption{The pre-etching DM sample (AC-S9) integrated and tested in a DM-like setup. The DM CryoAC will appear as this sample but etched.}
\label{ACS9setup}
\end{figure}

The pre-etching AC-S9 DM sample has been integrated and tested in a DM-like setup (Fig. \ref{ACS9setup}).

The proper CryoAC DM assembly consists of the detector sample and the PCB with its cold read out electronics (including the VTT K4 ret M SQUID and also a SMD Ruthenium Oxide thermometer). Both the detector and the PCB are integrated in the dedicated OFHC copper gold-plated bracket which will be delivered to SRON for the integration in the FPA DM. In the SRON setup the CryoAC DM assembly will be placed on the backside of the TES array mount plate, leaving a clearance of 0.5 mm between the TES array and the CryoAC DM detector. In our test setup the bracket is instead inserted in the dedicated gold plated OFHC copper sample holder (Fig. \ref{ACS9setup}). A Nb-Ti loom runs from the cold electronics to a PCB connector designed by SRON to fit the FPA DM setup. For thermalization purpose, gold bonding wires are applied from the gold-plated detector rim to the supporting bracket. 
We remark that for AC-S9, since it has been not etched, these Au wire bondings represent the main contribution to the thermal conductance between the silicon absorber and the thermal bath (i.e. the bracket and the sample holder assembly).

\subsection{AC-S9 preliminary characterization}

The AC-S9 DM-like setup has been integrated and tested in the IAPS Dilution Refrigerator, where the system has been magnetically shielded at cold by the Cryophy shield anchored to the Mixing Chamber plate. The detector has been operated with a commercial 2-channels Magnicon XXF-1 electronics, which has been used to operate both the TES/SQUID system (channel 1) and the on-board heater (channel 2 current generator).

\begin{figure}[htbp]
\centering
\includegraphics[width=0.49\linewidth]{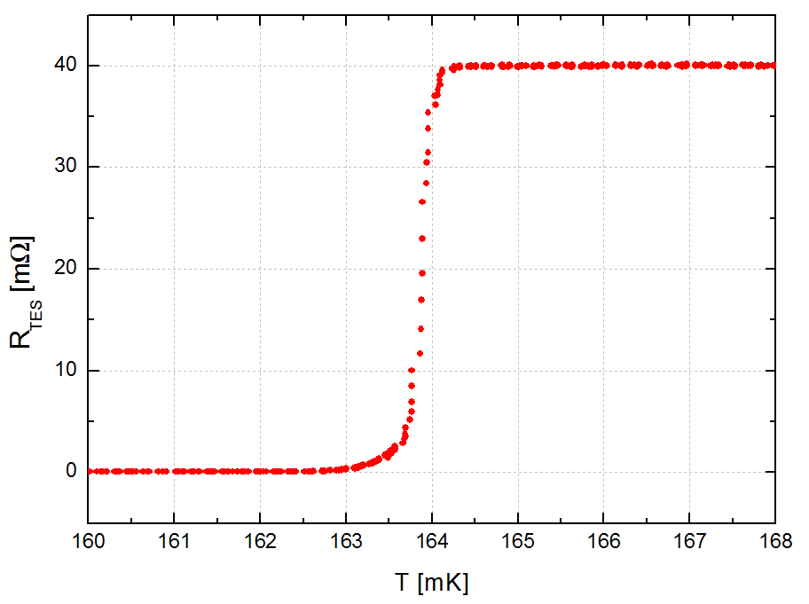}
\includegraphics[width=0.49\linewidth]{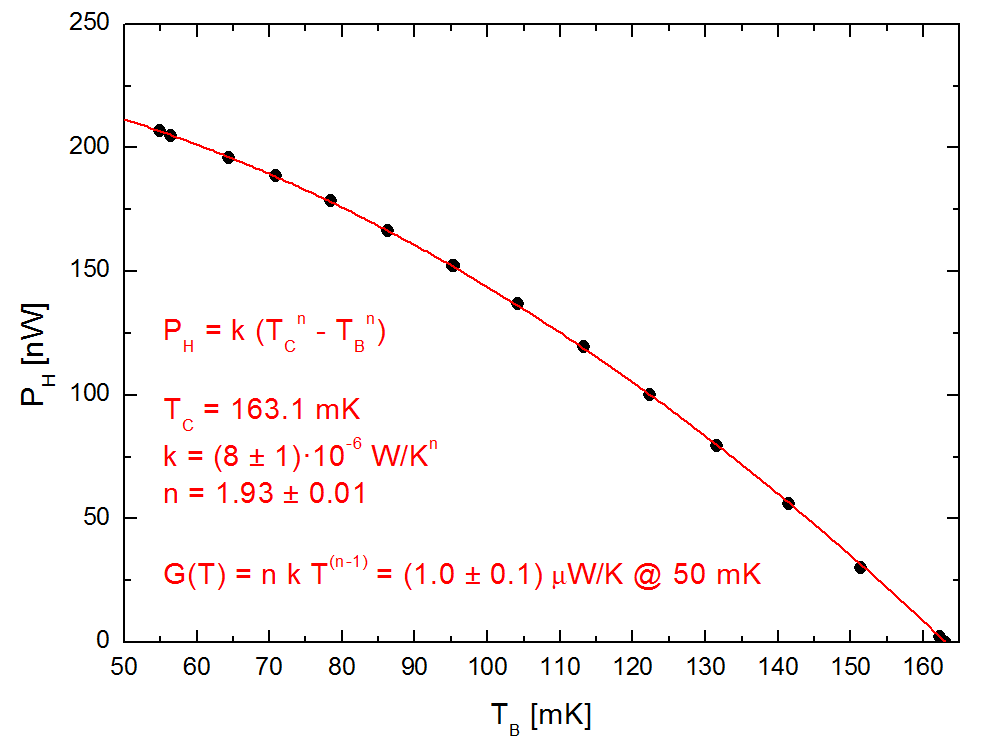}
\caption{Preliminary thermoelectric characterization of the sample. \textit{(Left)} Transition curve. \textit{(Right)} Power injected on the absorber by the on-board heater to bring the TES into the transition (R$_{TES}$/R$_N$ = 5\%) as a function of the thermal bath temperature.}
\label{ACS9char}
\end{figure}

The transition curve of the sample is shown in Fig. \ref{ACS9char}-Left. The narrow transition ($\Delta T_C < $1 mK), obtained with 96 TES connected in parallel, demonstrates the uniformity of the TES network. The critical temperature (T$_C$ = 163 mK) is instead significantly higher than the expected value (T$_{C, target}$ $\sim$ 100 mK), revealing a problem occurred during the TES deposition procedure. A subsequent investigation revealed that this was due to an incorrect calibration of PLD process to deposit the Ir:Au film, the TES are indeed thinner than expected (global thickness $\sim$ 50 nm instead of the $\sim$150 nm goal). As reported in literature \cite{bogorin}, the critical temperature of thin Ir film significantly increases with respect to the bulk value (T$_{C, bulk Ir}$ $\sim$ 140 mK) for thickness lower than $\sim$100 nm. This has increased our TES T$_C$, even though the proximization of the Ir film. A dedicated activity has then started to re-calibrate the PLD process by producing and testing some Ir:Au test structures.

We have then operated the on-board heater to heat the absorber and drive the TES into the transition (at the fixed level R$_{TES}$/R$_N$ = 5\%), at different thermal bath temperatures (Fig. \ref{ACS9char}-Right). The experimental points have been fitted with the general equation describing the power flow to the heat bath \cite{irwin}:

\begin{equation}
P_H = k \cdot \left( T_c^n - T_B^n \right)
\end{equation}

\noindent obtaining the best-fit values n = (1.93 $\pm$ 0.01) and k = (8 $\pm$ 1)$\cdot$ 10$^{-6}$ W/K, and a respective thermal conductance:

\begin{equation}
G_{measured} = n \cdot k \cdot T_B^{(n-1)} = (1.0 \pm 0.1) \mu W/K @ T_B = 50 mK
\end{equation}

\noindent As expected, the thermal link between the absorber and the thermal bath is dominated by a metallic component (n $\sim$ 2, see par. \ref{Gestim}), corresponding to the gold wire-bondings between the gold plated detector rim and the supporting bracket. It is possible to estimate the thermal conductance of the bonding wires by the formula \cite{pyle}:

\begin{equation}
G_{Au Bondings} = n_B \cdot (v_f \cdot d_e)/3 \cdot \Gamma_e \cdot S/L \cdot T = 1.5 \mu W/K @ T = 50 mK
\end{equation}
where we used the parameters listed in Table \ref{tab:Au_bonding}. The measured and the expected values are roughly consistent, validating our description of the thermal conductance between the detector rim and the thermal bath.

To complete the description of the thermal properties of the detector, in Table \ref{tab:ACS9_est} it is reported the estimation of its heat capacity at 50 mK and 160 mK, evaluated by taking into account the different contributions from TES, absorber and rim gold plating (the parameters are given in Table \ref{tab:ACS9_est_2} ).

\begin{table}[htbp]
\centering
\caption{Gold bonding wires properties.}
\label{tab:Au_bonding}
\begin{tabular}{lll}
\hline\noalign{\smallskip}
Parameter & Description & Value \\
\noalign{\smallskip}\hline\noalign{\smallskip}
n$_B$ & Bondings number & 20\\
v$_{f,Au}$ & Fermi velocity & 1.4$\cdot$ 10$^6$ m/s \cite{pyle}\\
d$_{e,Au}$ & Electron diffusion length & 1 $\mu$m \cite{pyle}\\
$\Gamma_{e,Au}$ & Electronic heat capacity coefficient & 66 J/m$^3$ $\cdot$ K$^2$ \cite{pyle}\\
S & Bondings section & n$_{B}$ $\cdot$ $\pi$ $\cdot$ (12.5 $\mu$m)$^2$\\
L & Bondings lenght & $<$ 1 cm\\
\noalign{\smallskip}\hline
\end{tabular}
\end{table}

\begin{table}[htbp]
\centering
\caption{AC-S9 heat capacity evaluation.}
\label{tab:ACS9_est}
\begin{tabular}{lllll}
\hline\noalign{\smallskip}
Parameter & Source & Formula (see par. \ref{hcest}) & @ 50 mK & @ 160 mK \\
\noalign{\smallskip}\hline\noalign{\smallskip}
C$_{TES, el}$ & TES e$^-$ & 2.43$\cdot$ $\rho$/A$\cdot$V$\cdot$ $\gamma_e$ $\cdot$T & 6 pJ/K & 19 pJ/K\\
C$_{ABS, el}$ & Absorber e$^-$ & $\rho$/A$\cdot$V$\cdot$ $\gamma_e$ $\cdot$T &  32 pJ/K & 96 pJ/K\\
C$_{ABS, ph}$ & Abs Phonons & 12/5$\cdot$ $\pi^4$ $\cdot$R$\cdot$ $\rho$/A$\cdot$V$\cdot$T$^3$/$\Theta_D^3$ &  11 pJ/K & 352 pJ/K\\
C$_{GOLD}$ & Rim gold e$^-$ & $\Gamma_e$ $\cdot$V$\cdot$ T & 126 pJ/K & 403 pJ/K\\
\textbf{C$_{AC-S9}$} & \textbf{Total} & & \textbf{175 pJ/K }& \textbf{870 pJ/K}\\
\noalign{\smallskip}\hline
\end{tabular}
\end{table}

\begin{table}[htbp]
\centering
\caption{Parameters for the AC-S9 heat capacity evaluation (parameters marked with * are estimated from measurements on a Silicon sample).}
\label{tab:ACS9_est_2}
\begin{tabular}{ll}
\hline\noalign{\smallskip}
Parameter & Value \\
\noalign{\smallskip}\hline\noalign{\smallskip}
$\rho_{TES}$ & 22.56 g/cm$^3$ (Ir) \\
A$_{TES}$ & 192.217 g/mol (Ir) \\
$\gamma_{e, TES}$ & 3.20 $\cdot$ 10$^{-3}$ J/(K$^2$ mol) \cite{furukawa} \\
$V_{TES}$ & 96 $\cdot$ 500 $\mu$m $\cdot$ 50 $\mu$m $\cdot$ 50 nm \\
& \\
$\rho_{ABS}$ & 2.329 g/cm$^3$ (Si) \\
A$_{ABS}$ & 28.085 g/mol (Si) \\
$\gamma_{e, Si}$ & 5.06 $\cdot$ 10$^{-8}$ J/(K$^2$ mol) (*) \\
$V_{ABS}$ & 525 $\mu$m $\cdot$ 16.6 mm $\cdot$ 16.6 mm \\
& \\
R & 8.314472 J/(mol K)\\
$\Omega_{D, ABS}$ & 645 K [K] (Si)\\
& \\
$\Gamma_{Gold}$ & 66 J m$^-3$ K$^-2$ \cite{pyle} \\
V$_{Gold}$ & 300 nm $\cdot$ 127.52 mm$^2$\\
\noalign{\smallskip}\hline
\end{tabular}
\end{table}

\subsection{Thermal and Athermal responses}

We have stimulated the detector and studied its response in two different ways:
\begin{itemize}
\item Injecting energy into the absorber by sending fast square current pulses (width = 1 $\mu$s) to the on-board heater (Injected energies: 115 keV, 170 keV, 380 keV, 460 keV, 680 keV, 1000 keV, 1830 keV)
\item Illuminating the absorber with X-ray photons by means of collimated radioactive sources (6 keV photons from a $^{55}$Fe source illuminating the absorber surface with the TES network, and 60 keV photons from a shielded 241Am source) 
\end{itemize}

\begin{figure}[htbp]
\centering
\includegraphics[width=0.49\linewidth]{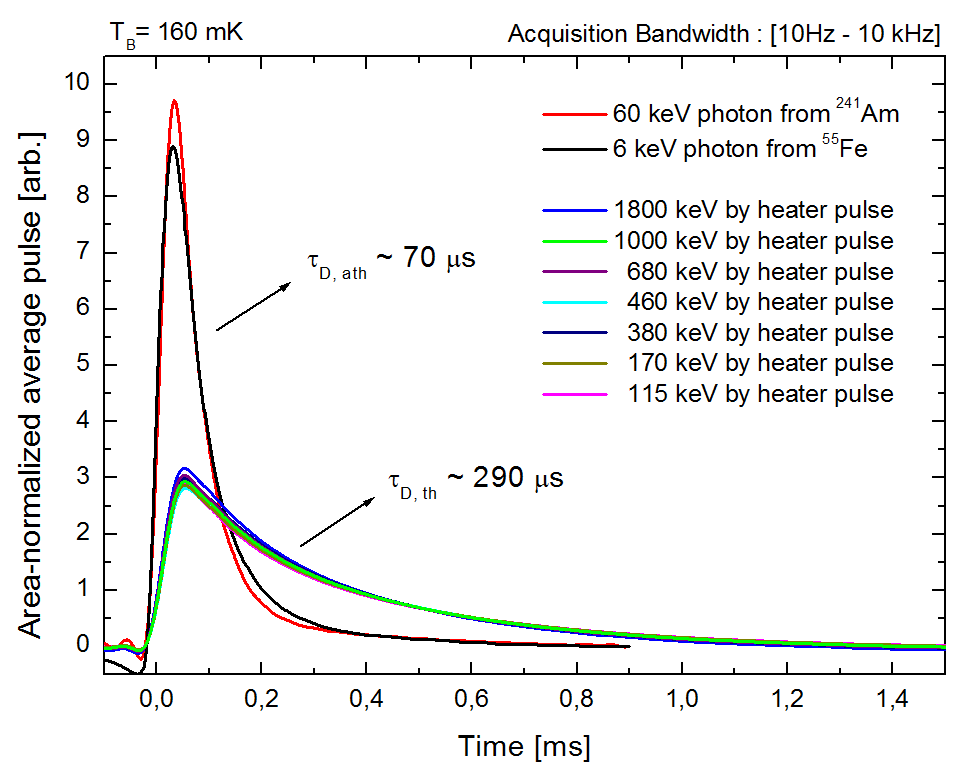}
\includegraphics[width=0.49\linewidth]{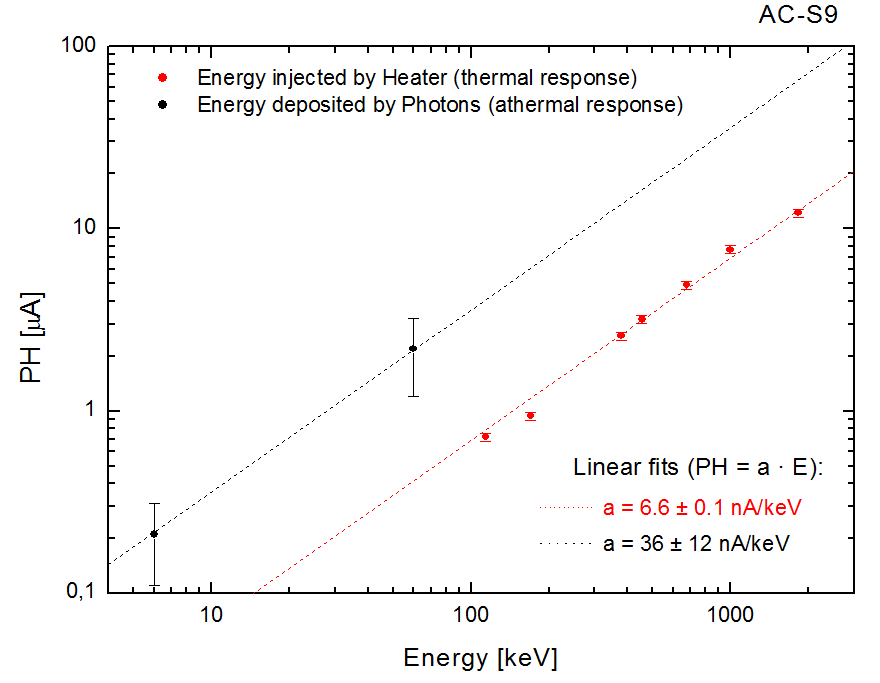}
\caption{Study of the detector thermal and athermal responses. \textit{(Left)} Shape of the average pulses acquired stimulating the detector with X-ray photons and fast heater pulses. \textit{(Right)} Average Pulse Height as a function of the energy injected by the heater (red points) or deposited by the X-ray photons (black points).}
\label{ACS9thath}
\end{figure}

All the acquisitions have been made fixing T$_B$ = 160 mK and biasing the detector at R$_{TES}$/R$_N$ $\sim$ 5 \% (I$_{TES}$ = 325 $\mu$A), with negligible loop gain factor. The SQUID output has been filtered by a low noise band-pass filter 1Hz-10kHz, and acquired with an ADC sampling rate of 500 ks/s. 
In Fig. \ref{ACS9thath}-Left it is reported the shape of the acquired average pulses normalized by their areas. Note that the pulse shape strongly depends on the method to inject energy into the absorber (photon absorption / heater injection), whereas - fixed the method - it is quite independent from the energy amount. 

The detector response to the heater energy injections is purely thermal, showing a decay time $\tau_{D, TH} \sim$ 290 $\mu$s roughly consistent with the expected C/G ratio at 160 mK (C/G $\sim$ 870 [pJ/K] / 3.2 [$\mu$W/K] $\sim$ 270 $\mu$s), thus showing the negligible ETF loop gain.
 
As expected, the response to the photon absorption is instead faster than the thermal one, demonstrating that the detector is able to work in an athermal regime. The TES network is indeed able to collect the first population of phonons generated when a particle or photon deposits energy into the absorber, before their thermalization into the silicon. Note that in this case the pulses do not show any significant thermal component, differently from what we observed in the  pre-DM prototypes (Chapter \ref{preDMchapter}). We are now investigating this point.

In Fig. \ref{ACS9thath}-Right the Average Pulse Height is plotted as a function of the energy, with the error bars showing the width of the PH distributions (1$\sigma$). In both the thermal (red points) and athermal (black points) regimes the detector shows a linear response in the explored energy range, as highlighted by the respective linear fits (constrained to pass through the origin). Note that working in the athermal regime allows to increase the detector response, thus lowering the low energy threshold on the detector. On the other hand, the athermal PH dispersion is higher with respect to the thermal one, reducing the spectroscopic capabilities of the detector. 

\subsection{Energy spectra}

\begin{figure}[htbp]
\centering
\includegraphics[width=0.6\linewidth]{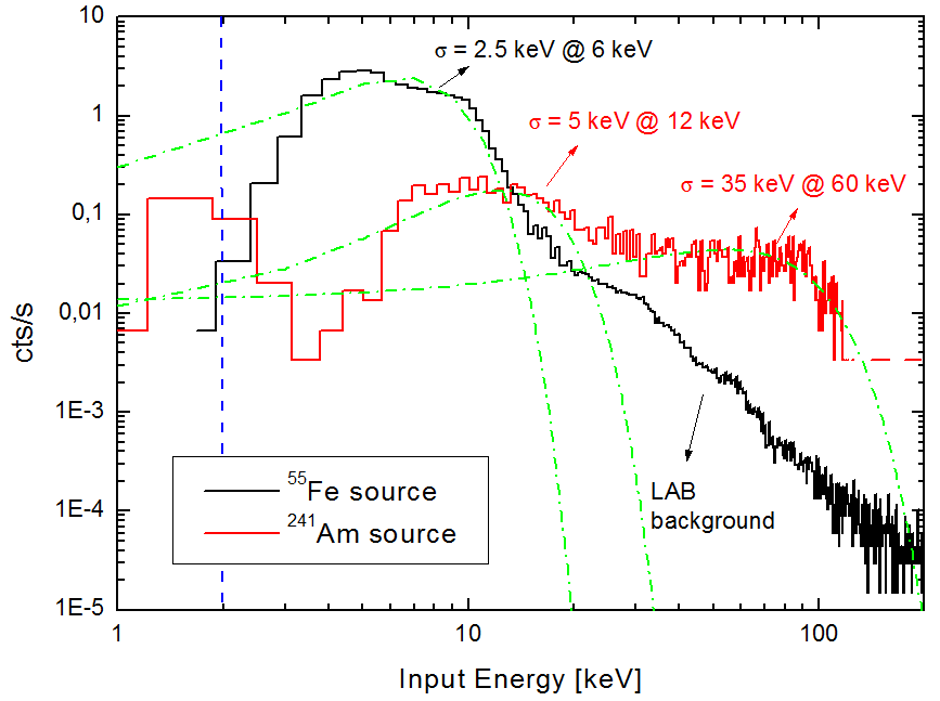}
\caption{Raw energy spectra acquired illuminating the detector with the $^{55}$Fe (black line) and the $^{241}$Am (red line) sources. The dashed blue line represents the trigger threshold (5$\sigma$ baseline). The green lines are Gaussian fits of the main spectral bumps (6 keV, 12 keV, 60 keV)}
\label{ACS9spectra}
\end{figure}

The energy spectra acquired illuminating the detector with the $^{55}$Fe and the $^{241}$Am sources are shown in Fig. \ref{ACS9spectra}. They have been obtained from the raw PH spectra, where the energy axis has been calibrated assuming an athermal detector response of 36 nA/keV (from Fig. \ref{ACS9thath}-Right). The dashed blue line in the plot represents the trigger threshod (5$\sigma$ baseline level), whereas the green lines are Gaussian fits of the main spectral bumps.

The $^{241}$Am spectrum (red curve) refers to a 5 minutes acquisition, with a count rate of $\sim$ 9 cts/s. It shows two main bumps: the first around 12 keV is due to Compton-scattered 60 keV photons (E$_{Compton Edge, 60 keV}$ = 11.3 keV) and probably copper and gold fluorescences (E$_{Cu,k\alpha}$ = 8 keV, E$_{Au,k\alpha}$ = 10 keV). The second one corresponds to the 60 keV photopeak, which is largely spread by the instrumental response.  

The ``black'' spectrum is reconstructed from two different acquisitions. The first one, 5 minutes duration at $\sim$ 36 cts/s and low trigger threshold ($\sim$ 2 keV), has been used to detect the main 6 keV line of $^{55}$Fe. The second one has been instead performed with an higher trigger threshold ($\sim$ 20 keV) and a much longer acquisition time ($\sim$ 19 hours with a count rate $\sim$0.07 cts/s), in order to show the laboratory background level. This was also a stability test of the detector bias, which did not show any significant drift during the long acquisition time.
We remark that being able to detect the $^{55}$Fe 6 keV line, the detector demonstrates a low energy threshold fully compatible with our requirement (20 keV), thus representing a promising step towards the CryoAC DM. 

From the spectroscopic point of view, AC-S9 shows instead a very poor energy resolution ($\Delta$E/E $\sim$ 1). This has probably due to the several limitations due to the absence of the final etching (high thermal capacity, high thermal conductance, partial TES surface coverage). From the DM we expect an heat capacity about 5 times lower than the AC-S9 one, an halved thermal conductance and an optimal absorber TES coverage, thus a better detector response. Anyway, we remark that the CryoAC is aimed to operate as anticoincidence particle detector, and it is not conceived to perform X-ray spectroscopy (i.e. there are no requirements on the CryoAC spectral resolution).

\subsection{Operation at Tb = 50 mK}

Thanks to the onboard heater, we have been able to operate the detector also with the thermal bath at T$_B$ = 50 mK, as required for the DM. We have first fixed the bath temperature, and then used the heater to inject power into the absorber (P$_H$ = 195 nW), increasing its temperature. Finally, we have injected current into the TES circuit until we reached a good working point ($I_B$ = 1000 $\mu$A). We can estimate the absorber temperature as:

\begin{equation}
T_{ABS} = \left[ \left( P_H + P_{TES} \right)/k + T_B^n \right]^{1/n} \sim 155 mK
\end{equation}
where P$_{TES}$ is negligible and the k and n parameters are those evaluated in the previous sections.

\begin{figure}[htbp]
\centering
\includegraphics[width=0.9\linewidth]{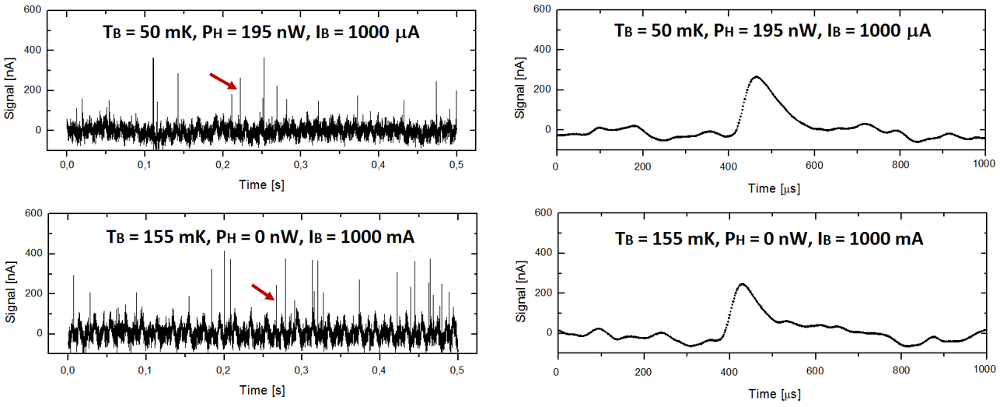}
\caption{\textit{(Left)} Signals acquired with the detector working in different conditions, with (top) and without (down) operate the onboard heater. In both the cases the absorber has been stimulated with the $^{55}$Fe source. See text for details. \textit{(Right)} Zoom on the 6 keV pulses indicated on the left by the red arrows.}
\label{ACS950mK}
\end{figure}

For comparison we have also performed an acquisition setting directly T$_B$ = 155 mK and biasing the TES with the same bias current (I$_B$ = 1000 $\mu$A), without operating the heater. Two signal strips acquired in these different conditions are shown in Fig. \ref{ACS950mK}-Left. In both the cases, the absorber has been stimulated with the $^{55}$Fe source illuminating the surface with the TES network. The 6 keV pulses indicated by the red arrows are zoomed on the Right of the Figure. 

The AC-S9 response results very similar in the two different operation conditions, demonstrating that the detector can be operated at T$_{B}$ = 50 mK without noticeable issues. This is another promising step towards the CryoAC DM.

\subsection{Fe 55 front/rear side illumination}

\begin{figure}[htbp]
\centering
\includegraphics[width=0.6\linewidth]{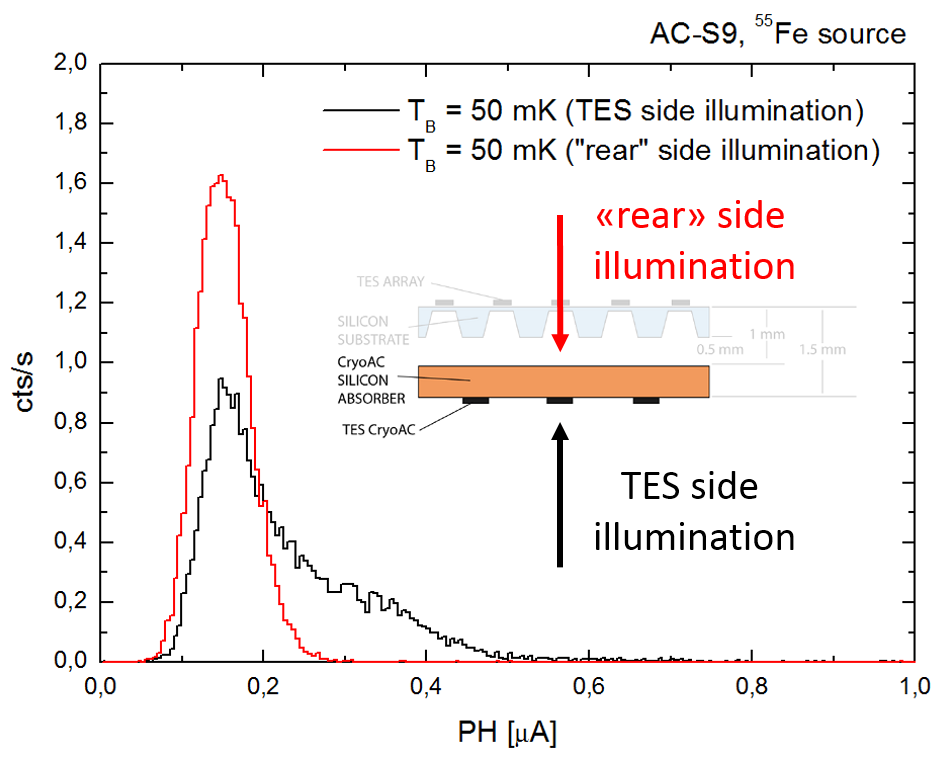}
\caption{Pulse Height spectra acquired stimulating the detector by means of a $^{55}$Fe source (6 keV photons), illuminating the absorber from the TES (red line) and the ``rear'' (black line) side.}
\label{ACS9rearfront}
\end{figure}

Lastly, we have stimulated the detector by means of the $^{55}$Fe source alternatively illuminating the TES and the ``rear'' side of the absorber (see the sketch inside Fig. \ref{ACS9rearfront}). We remark that 6 keV photons are completely absorbed in $\sim$ 100 $\mu$m of silicon (transmission $<$ 5\%), so we can expect different paths for the phonons propagation towards the TES in the two cases, thus a different detector response. The Pulse Height spectra acquired with the detector operated in the same working point (T$_B$ = 50 mK, P$_H$ = 195 nW, I$_B$ = 1000 $\mu$A) are shown in Fig. 9.

Note that the spectrum is better shaped with the ``rear'' side illumination, whereas illuminating the absorber from the TES side the Pulse Height distribution shows a long tail after the main peak. We are now carrying on an activity aimed to simulate the phonons propagation in the absorber and investigate the generation mechanism of the thermal and athermal pulses. This is a due activity in the context of the usual Phase A mission program (trade-off studies).

\subsection{Conclusions}

In conclusion, the activity performed on AC-S9, the pre-etching CryoAC DM sample integrated and tested in a DM-like setup, have shown promising steps towards the DM, in particular:

\begin{itemize}

\item We have successfully operated the heater on-board the detector absorber;
\item We have measured the thermal conductance between the DM rim and the thermal bath (due to the Au wire  bondings), validating our description of this thermal coupling;
\item We have observed and studied both the thermal and athermal detector responses, producing the respective pulse templates. 
\item Despite the limitations due to the absence of the final etching (high thermal capacity, high thermal conductance, partial TES surface coverage) we have detected 6 keV photons, thus having a low energy threshold fully compatible with our requirement (20 keV);
\item We have been able to operate the detector at T$_B$ = 50 mK (DM requirement);
\item Illuminating by means of a $^{55}$Fe source the absorber from the rear side (the opposite side with respect to the TES network), a good spectrum shape has been obtained.

\end{itemize}

\newpage
\section{CryoAC DM: preliminary test}

\begin{figure}[H]
\centering
\includegraphics[width=0.4\linewidth]{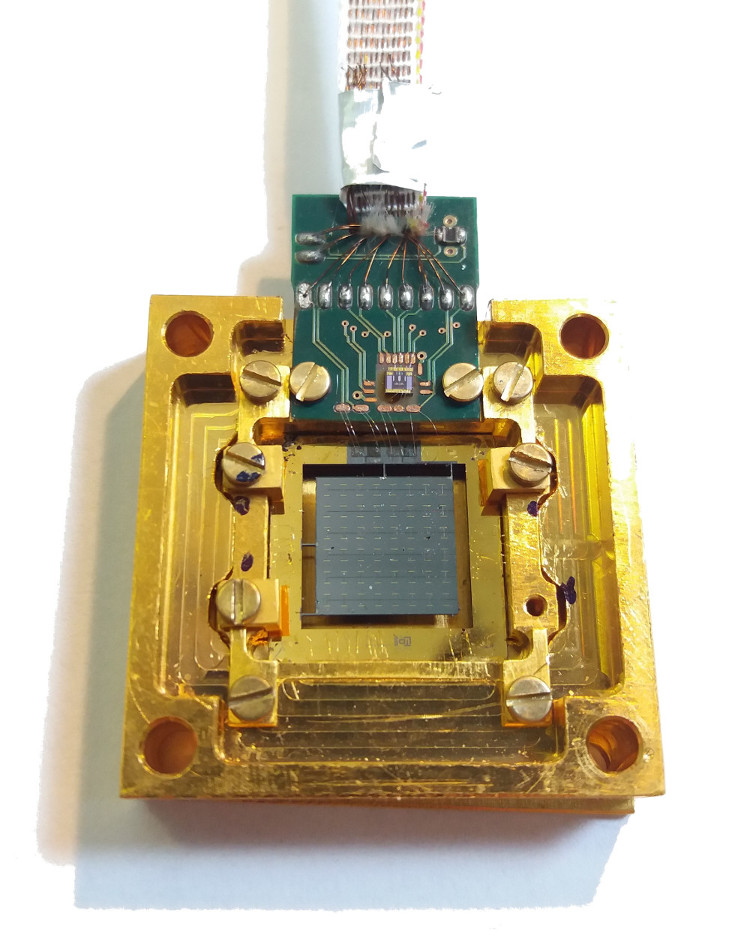}
\caption{The AC-S10 DM sample integrated in the holder structure developed for the test in our cryogenic setup.}
\label{ACS10_integration}
\end{figure}

The first proper CryoAC DM sample, namely AC-S10, has been then finally produced and successfully integrated in the cryogenic setup (Fig. \ref{ACS10_integration}). In the following I will report the results of the preliminary test activity performed on this sample.

\begin{figure}[H]
\centering
\includegraphics[width=0.47\linewidth]{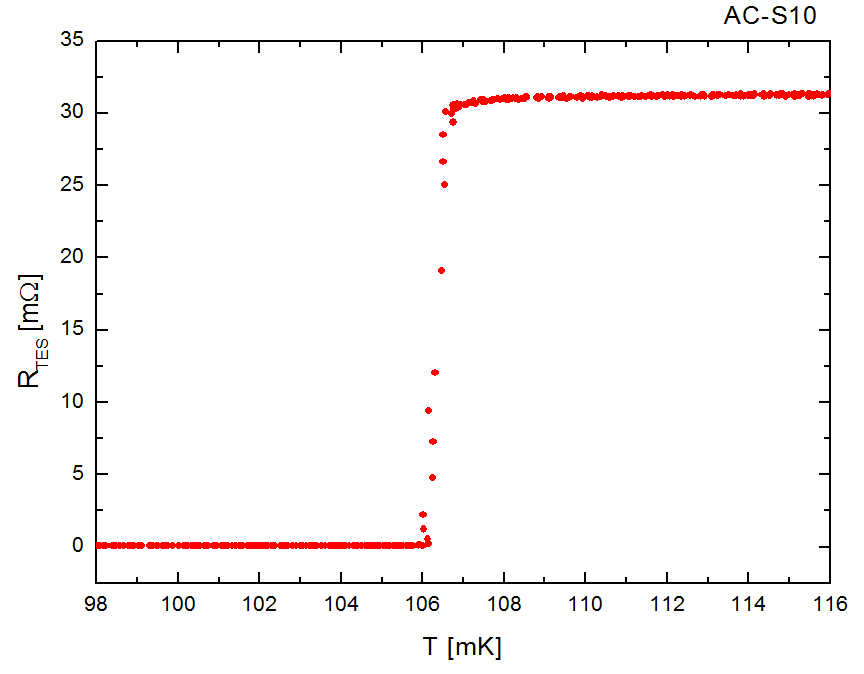}
\includegraphics[width=0.5\linewidth]{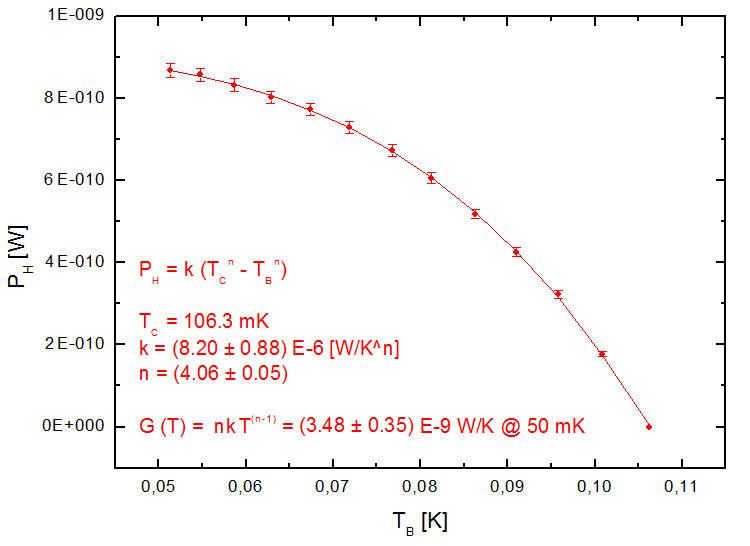}
\caption{Preliminary thermoelectric characterization of the sample. \textit{(Left)} Transition curve. \textit{(Right)} Power injected on the absorber by the on-board heater to bring the TES into the transition (R$_{TES}$/R$_N$ = 5\%) as a function of the thermal bath temperature.}
\label{ACS10_RTandPower}
\end{figure}

The transition curve of the sample is shown in Fig. \ref{ACS10_RTandPower} - Left. The narrow transition ($\Delta$T$_C$ $\sim$ 0.5 mK) confirms again the uniformity achieved by large TES network. Note that, differently from AC-S9, in this case the transition temperature T$_C$ = 106 mK is close to our goal value (T$_C$ $\sim$ 100 mK). This is a demonstration that the re-calibration of the PLD process (that caused the AC-S9 high transition temperature) has been properly carried out.

\newpage
We have then repeated on AC-S10 the thermal conductance measurement performed on AC-S9, operating the on-board heater to heat the absorber and drive the TES into the transition for different thermal bath temperatures (Fig. \ref{ACS10_RTandPower} - Right). Note that in this case the thermal conductance is realized by the four narrow silicon beams that connect the absorber to the gold plated rim, which now represents the thermal bath of the detector. By fitting the experimental points we found:

\begin{equation}
G_{AC-S10} = n \cdot k \cdot T_B^{(n-1)} = (3.5 \pm 0.3) nW/K \;\;\; @ \;\;\; T_B = 50 mK
\end{equation}

\noindent where k = (8.20 $\pm$ 0.88) $\mu$W/$K^n$ and n = (4.06 $\pm$ 0.05). The resulting power-law index n is perfectly compatible with the value expected for a phonon coupling in a crystal (n$_{\text{phonons}}$ = 4, see par. \ref{hcest}), confirming that the measured conductance is really due to the silicon beams.

\begin{figure}[H]
\centering
\includegraphics[width=0.7\linewidth]{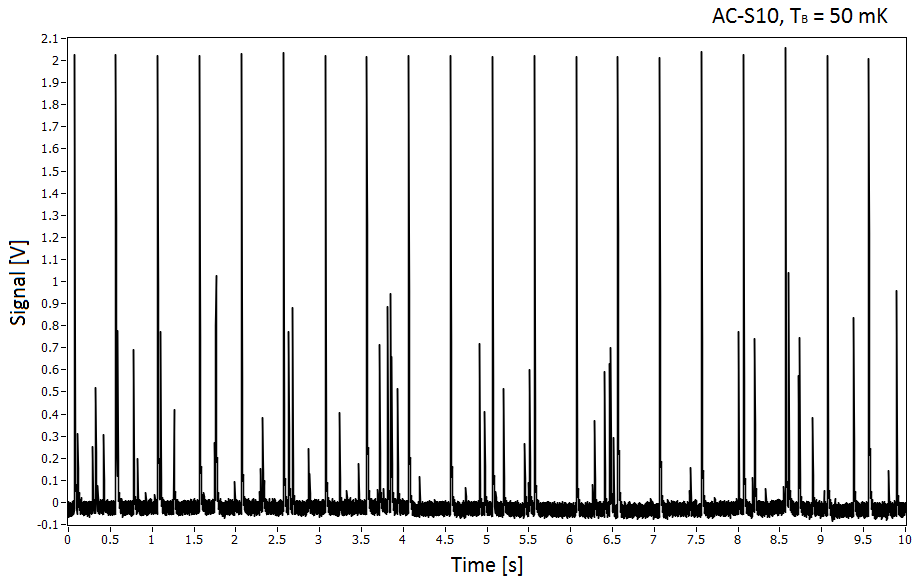}
\caption{Functional test of AC-S10 operated at a bath temperature of 50 mK (CryoAC DM requirement). The sample has been stimulated injecting 200 keV of energy by the on-board heater with a 2 Hz rate (2 V pulses) and simultaneously illuminating the detector with 60 keV photons by an $^{241}$Am source ($\sim$ 0.6 V pulses). }
\label{ACS10_strip}
\end{figure}

\bigskip
Finally, we have been able to operate the detector with the thermal bath at 50 mK (DM requirement), by injecting via heater a power of 0.86 nW (so driving the absorber close to the TES critical temperature), and by biasing the TES circuit with I$_{bias}$ = 10 $\mu$A. In these conditions, the sample has been stimulated injecting 200 keV of energy by the on-board heater with a 2 Hz rate, and simultaneously illuminating the detector with 60 keV photons by an $^{241}$Am source. A strip of the acquired signal is shown in Fig. \ref{ACS10_strip}, where both the 200 keV (PH $\sim$ 2 V) and the 60 keV pulses (PH $\sim$ 0.6 V) are clearly visible. 

\noindent In conclusion, with the preliminary test activity performed on AC-S10 we have: 

\begin{itemize}
\item measured a good transition (T$_{C}$ = 106 mK, R$_N$ = 31 m$\Omega$, $\Delta$T$_C$ $\sim$ 0.5 mK);
\item measured for the first time the thermal conductance due to the narrow silicon beams connecting the suspended absorber and the thermal bath ( G $\sim$ 3 nW/K @ 50 mK). This is an important result, giving us a reference value also for the CryoAC Flight Model design;
\item finally operated the CryoAC DM with the thermal bath at 50 mK (requirement).
\end{itemize}

\bookmarksetup{startatroot}
\chapter*{Conclusions}
\addcontentsline{toc}{chapter}{Conclusions}
\markboth{CONCLUSIONS}{CONCLUSIONS}

In this thesis I have presented the research activity carried out during my PhD at the \textit{Cryogenic Laboratory for X-ray Astrophysics} of the INAF/IAPS Roma. My work concerns the ATHENA space mission, and it has been in particular focused on the development and test of the Cryogenic Anticoincidence Detector (CryoAC) of the X-IFU instrument. Here I will summarize the main results obtained.

\bigskip
The CryoAC plays a fundamental role in the reduction of the X-IFU particle background, allowing the instrument to reach the scientific requirement of a not-focused Non X-ray Background (NXB) $<$ 5 $\times$ 10$^{-3}$ cts cm$^2$ s$^{-1}$ keV$^{-1}$ (between 2 and 10 keV). This is mandatory to enable a significant part of the X-IFU science, mainly concerning the study of faint cluster and outskirts. A first rough evaluation based on the current Mock Observing Plan (MOP) shows that the background reduction should be needed to proper exploit about the 50$\%$ of the X-IFU observing time. In addition, it has also an impact on the discovery science not included in the MOP. I have presented as case study the serendipitously search of distant Compton Thick AGN with the X-IFU, showing that the background reduction will enable to detect $\sim$ 170 CT AGN at z $>$ 2 (limit flux: 4$\times$10$^{-16}$ erg cm$^2$ s). 

\bigskip
The X-IFU particle backgrund is estimated by means of detailed Monte Carlo simulations, which are performed using the Geant4 toolkit. One key element of the simulations is the Mass Model of the instrument, which has been recently updated starting from the new CAD model of the cryostat (developed by CNES) and the Focal Plane Assembly (developed by SRON). I have directly contributed to this activity, being in charge of the simplification of the CAD models before their insertion in Geant4. Note that an accurate mass model of the instrument, especially in the detector proximity, it is crucial to proper estimate the particle background level, since it strongly depends on the materials, their placement, their shapes, and on the total mass shielding the detector from radiation. The new background simulations, including also a new input GCR proton spectrum and the new Space Physiscs List, have been then performed, evaluating an unrejected particle background level for the current X-IFU baseline configuration of 4.64 $\times$ 10$^{-3}$ cts cm$^2$ s$^{-1}$ keV$^{-1}$, i.e. meeting the scientific requirement. (Publications [R4] and [CP3] - Lotti et al.).

\bigskip
In the context of the AHEAD project, I have explored with my research group the possible observational capabilities of the CryoAC in the Hard X-ray band. We have found that in the baseline configuration the CryoAC could operate as hard X-ray detector in the narrow band 10-20 keV, with a limit flux (5$\sigma$, 100 ks, 1 pixel) of 1.6$\cdot$10$^{-12}$ erg/cm$^2$/s ($\sim$ 0.2 mCrab). The energy band is limited by the drop of the optics effective area at high energies, and not by the detector features. Furthermore, we have found that an optimization of the CryoAC energy resolution up to $\Delta$E = 2 keV (FWHM) could have a scientific return in the observation of bright sources with a spectral cut-off in this band, as High Mass X-ray Binaries (HMXB). (Publication [R3] - D'Andrea et al.) 

Still in the AHEAD context, we have also shown that the insertion of six additional CryoAC vertical pixels in the FPA could provide a reduction in the residual background level of about a factor 2 with respect to the X-IFU baseline design. This significant improvement of performance has been positively considered by the X-IFU system team, though the impact of implementation will require a dedicated trade-off study.

\bigskip
With regard to the CryoAC detector physics, I have developed an electrothermal model for athermal phonon mediated microcalorimeters, showing that in the low-signal and low-inductance limits, assuming a strong decoupling between the TES electrons and the absorber phonons, the response to a particle event is given by the sum of two different pulse components: a fast \virg athermal'' and a slow \virg thermal'' one. The model foresees that only the decay time of the athermal component is influenced by the electrothermal feedback, while the decay of the thermal component is independent from the loop gain. It also foresees that the thermal rise time should be equivalent to the athermal decay time. Both these arguments have found a first confirmation with the measurements performed on the AC-S7 and the AC-S8 CryoAC prototypes.

\bigskip
The AC-S7 and AC-S8 pre-DM prototype have been deeply tested, and the results of these activity have allowed us to better understand the detector athermal/thermal dynamic and to collect important feedbacks towards the Demonstration Model development. In particular we have (Publications [R1] and [CP2] - D'Andrea et al., [R5] and [CP6] - Macculi et al., [CP4] - Biasotti et al.): 
\begin{itemize}
\item measured for both the detectors narrow superconductive transitions ($\Delta$T$_{C}$ $\sim$ 2 mK), demonstrating for the first time the uniformity of TES large networks (64 TES connected in parallel) in a planar geometry;
\item shown that the athermal contribution to the pulses energy increased of a factor $\sim$ 4 from AC-S7 (without Al collectors) to AC-S8 (with Al collectors), so obtaining a first evidence of the improvement in the athermal collection efficiency due to the use of the Al collectors.
\item demonstrated that the AC-S8 collectors design is able to prevent the quasi-particles recombination in the Al, assuring a fast pulse rising front;
\item shown with AC-S7 that this kind of detector can achieve a low energy threshold lower than 20 keV, showing an energy resolution of $\sim$11 keV @ 60 keV (FWHM);
\item demonstrated by AC-S8 that this kind of detector can be operated with the thermal bath at 50 mK (CryoAC DM requirement), showing a pulse rise time and athermal decay time within the CryoAC requirements ($\tau_R$ = 30 $\mu$s, $\tau_{D,ATH}$ = 190 $\mu$s);
\end{itemize}
but also:
\begin{itemize}
\item measured for both the detectors very low normal resistances ($R_N \sim$ 2 m$\Omega$), understanding the need to change the TES aspect ratio in the DM, in order to have a higher R$_N$, thus better operating the detector in its voltage bias;
\item observed that this kind of detector requires high bias currents to be operated with a low bath temperature (some mA to the TES bias circuit), so introducing an heater in the DM design to eventually increase the absorber temperature in order to limit magnetic coupling effects to the TES array;
\item measured for both the detectors slow thermal decay time ($>$ 10 ms), understanding the need to better control the heat capacity of the absorber and the thermal conductance towards the thermal bath, and thus starting an activity aimed to better evaluate these quantities for the DM.
\end{itemize}

\bigskip
Regarding data analyis, we have for the first time analyzed CryoAC data with a new pulse processing method based upon Principal Component Analysis. Applying this procedure to the pulses collected by AC-S7, we have been able to improve by a factor 1.5 - 2 the spectral resolution of the detector with respect to our old analysis. Although the CryoAC is not aimed to perform spectroscopy, improving the energy resolution is important to better determine the low threshold energy of the detector.  (Publication [CP2] - D'Andrea et al.).

\bigskip
In preparation for the CryoAC Demonstration Model integration, we have updated our test facility and optimized the experimental setup. We have in particular installed a new dilution refrigerator, integrating in the cryostat the test setup needed to perform the foreseen test activities and developing a series of ad-hoc mechanical components to integrate the CryoAC DM and to test it in this cryogenic system.

We have also developed a cryogenic magnetic shielding system, and assessed by means of FEM simulation and a test at warm that it is able to provide a shielding factor $S > 250$ in the detector region. Given the earth magnetic field ($\sim$ $ 0.5 \cdot 10^{-4}$ T), we therefore expect in the detector zone a residual static magnetic field $< 2 \cdot 10^{-7}$ T. (Publication [R2] - D'Andrea et al.). We have finally prepared our cryogenic setup to eventually host a signal filtering stage a cold, verifying that the foreseen filters configuration doeas not compromise the stability of the SQUID readout system.

\bigskip
Finally, we have focused our activities on the CryoAC DM. Initially we have performed several cooling test with dummy CryoAC DM pixels, in order to verify the DM assembly mechanical strength and its response to thermal cycles, validating the mechanical setup and the handling and integration procedures. Then we have characterized the VTT SQUIDs used in the DM Cold Front End Electronics (CFEE). At this point we have tested AC-S9, a pre-etching CryoAC DM sample integrated in a DM-like setup, that has shown promising steps towards the DM, in particular we have (Publication [CP1] - D'Andrea et al.):
\begin{itemize}
\item successfully operated the heater on-board the detector absorber;
\item measured the thermal conductance between the DM rim and the thermal bath (due to the Au wire  bondings), validating our analytical description of this thermal coupling;
\item observed and studied both the thermal and athermal detector responses, producing the respective pulse templates;
\item operate the detector at T$_B$ = 50 mK (DM requirement);
\item detected 6 keV photons despite the limitations due to the absence of the final etching (high thermal capacity, high thermal conductance, partial TES surface coverage), thus having a low energy threshold fully compatible with our requirement (20 keV);
\end{itemize}
The first proper CryoAC DM sample, namely AC-S10 , has been then finally produced and successfully integrated in the cryogenic setup. With the preliminary test activity performed on this sample we have:
\begin{itemize}
\item measured a good transition (T$_{C}$ = 106 mK, R$_N$ = 31 m$\Omega$, $\Delta$T$_C$ $\sim$ 0.5 mK);
\item measured for the first time the thermal conductance due to the narrow silicon beams connecting the suspended absorber and the thermal bath ( G $\sim$ 3 nW/K @ 50 mK). This is an important results, giving us a reference value also for the CryoAC Flight Model design;
\item finally operated the CryoAC DM with the thermal bath at 50 mK (requirement).
\end{itemize}

\bigskip

\noindent The CryoAC AC-S10 DM sample will be now deeply tested in our cryogenic setup at INAF/IAPS, and will be then delivered to the X-IFU Focal Plane Assembly (FPA) development team at SRON, in order to be integrated in the FPA DM and tested for the first time with the TES array. This step, enabled by the work shown in this thesis, will represent a milestone in the X-IFU development towards the ATHENA mission adoption.

\renewcommand{\bibname}{Publications list}

\renewcommand{\bibname}{Bibliography}

\end{document}